\documentclass[12pt,fullpage]{book}
\usepackage{graphicx}
\usepackage{supertabular}
\begin{document}
\title{The Spectrum of the Th-Ar Hollow-Cathode Lamp Used 
with the {\it 2dcoud\'e} Spectrograph}
\date{October 25, 2001}
\author{Carlos Allende Prieto \\ 
\\
\\
McDonald Observatory and Department of Astronomy \\ 
The University of Texas at Austin}

\maketitle

\tableofcontents

\part{Description}

\section*{Introduction}

Thorium-Argon lamps are the most popular choice for wavelength calibration on
high-resolution 
grating spectrographs. The wealth of emission lines, the lack of hyperfine
splitting for the only isotope of thorium naturally occuring ($^{232}$Th),
and its heavy weight, that results in a very small Doppler broadening, 
makes of this actinide a good choice. Thorium (Z=90)
is responsible for
most of the emission lines observed in the optical spectrum emitted by a 
hollow-cathode Th-Ar lamp.
In this type of lamp, an electric
discharge ionizes Ar atoms, which are accelerated into the cup-shaped 
thorium cathode.
The Ar II ions sputter Th atoms that get excited by collisions and then 
radiatively decay to lower levels.

Using a state-of-the-art  CCD, a short exposure can register with high quality 
the spectrum of a commercial hollow-cathode lamp, even at very high dispersion.
The appearance of the spectrum changes depending on the characteristics of the
lamp (e.g. the vendor), the spectral resolution, the electric current applied, 
and also as a result of aging.
For this reason, atlases of the spectrum of the calibration sources  
are commonly prepared for specific combinations of lamps/spectrographs. In fact,
almost all recently developed spectrographs are provided with maps of the 
spectra of their wavelength-calibration sources. Often, these maps are
accessible through the World Wide Web (WWW), taking advantage of the added flexibility
 compared to printed copies (e.g. UVES on the Very Large 
Telescope, FEROS on the ESO 1.52m, UCLES on the Anglo-Australian Telescope, 
MUSICOS on the Isaac Newton Telescope, the Coud\'e spectrograph on the KPNO 2.1m, 
or UES on the William Herschel Telescope).

To produce one such atlas for the {\it 2dcoud\'e} (Tull, MacQueen, Sneden \& 
Lambert 1995)
at a resolving power $R$
$\sim$ 60,000 is an easy task, as the 
cross-dispersion allows for a large spectral coverage -- the complete optical
range can be acquired in just two exposures. As part of a survey of nearby 
stars, we recently observed a number of 
standard stars that are likely of wide interest for other observers. With little
more effort the Th-Ar exposures obtained for the survey were combined to 
produce an atlas. 

The Th-Ar linelist commonly used for calibration in IRAF,
(Willmarth 1992) provides standard (15$^{\rm o}$ C, 760 torr) dry air 
wavelengths for 3056 features\footnote{A  recent
Th-Ar atlas produced by Willmarth et al. with the KPNO 2.1m coud\'e 
spectrograph is available online from 
{\tt http://www.noao.edu/kpno/specatlas/}.}. The line list seems to be 
a combination of the identifications of Palmer and Engleman (1983) for Th and 
Norl\'en (1973) for Ar. About 90 \% of the lines are attributed to neutral or
ionized Th. Palmer and Engleman used the Fourier Transform Spectrometer (FTS) 
at Kitt Peak National Observatory. They 
quote an average accuracy for their wavenumbers
of 0.002 cm$^{-1}$, which is equivalent to 0.0002 \AA\ at 3000 \AA\, 
and 0.001 \AA\ at 7000 \AA.
It became apparent that the McDonald spectra would provide similar, and in some
cases improved, precision. For this reason, and because errors in the FTS  
measurements have a different dependence with wavelength or line strength than
they do in our grating spectrograph, the wavelengths labeling lines in this
atlas  are those determined in 
the McDonald spectra.

Some users may be happy with this printed version of the {\it 2dcoud\'e} Th-Ar
atlas. However, further advantages can be obtained from the digital version 
that is available through the WWW. With prior knowledge of the 
central wavelength or order, the presentation of the atlas can be 
 changed interactively to match the actual spectral setup.

\section*{Acquisition and Reduction of Data}

The Th-Ar hollow-cathode employed is a Westinghouse lamp, Type WL-23418, 
powered with a 14 mA current. 
This lamp has been
used at McDonald since December 1992. The detector was TK3, 
a thin $2048\times2048$ Tektronix CCD with 24 $\mu$m pixels, which was 
installed at the $R \sim 60,000$ focus F3. We used 
 grating E2, a 53.67 gr mm$^{-1}$ R2 echelle from Milton Roy Co. 
 We used slit \#4, which has a central width of 511 $\mu$m (or 
 approx. 1.2 arcsec on the sky).  We refer the reader to Tull et al. (1995)
 for more details about the instrument.
 
 The exposures used in this atlas were
obtained in three different epochs (December 2000, May 2001, and 
September 2001), as part of a survey of 
nearby stars (Lambert, Allende Prieto \& Cunha 2001). We scaled and 
combined 42 individual exposures obtained centering 5112 \AA\ in order 67 
in the middle of the CCD ({\it blue} setup), and 29 exposures obtained with 
5178 \AA\ (in the same order) at the center of TK3 ({\it red} setup). Figure 1
shows the area covered by the detector on the focal plane for each setup. 
Most of the exposures were 30s long, but some were up to a factor of three longer.

\begin{figure}[ht!]
\centering
\includegraphics[width=9cm,angle=90]{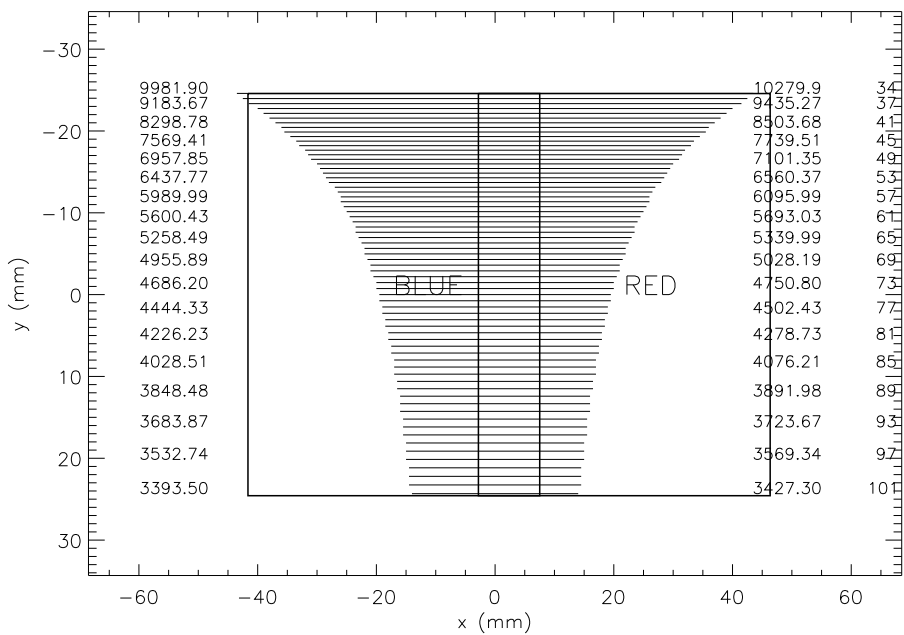}
\protect\caption[ ]{Spectral setups employed in the making of the atlas. 
The boxes labeled as RED and BLUE show the size of TK3 on the focal plane. 
The horizontal 
rules indicate the free spectral range for each order. Order numbers 
and the beginning and ending wavelengths  for some  orders  
(in \AA) are printed.
}
\label{f1}
\end{figure}

Using IRAF tasks, we removed the bias, estimated from the overscan area, and
we extracted the spectra. The  apertures for extraction were defined with the
help of a standard star (tracing of the orders) and the spectrum of a quartz 
lamp that is normally used for flatfield correction (width of the orders). 
Each of the individual spectra was calibrated in wavelength using the Th-Ar 
linelist available within IRAF ({\tt thar.dat}), identifying about 1000 lines per setup,
and fitting a 4$^{\rm th}$ order polynomial in the dispersion direction, and
a 5$^{\rm th}$ order polynomial in the cross-dispersion direction. The
rms scatter was always in the range $0.0016 - 0.0026$ \AA. No attempt was
made to correct the instrumental response or the pixel-to-pixel sensitivity
variations. Mixing observations from three different observing periods, 
with multiple changes between the two setups in each run, smoothes out 
most systematic sensitivity variations between nearby pixels, as the positioning
mechanism for the grating is imperfect. 

\begin{figure}[ht!]
\centering
\includegraphics[width=8cm,angle=90]{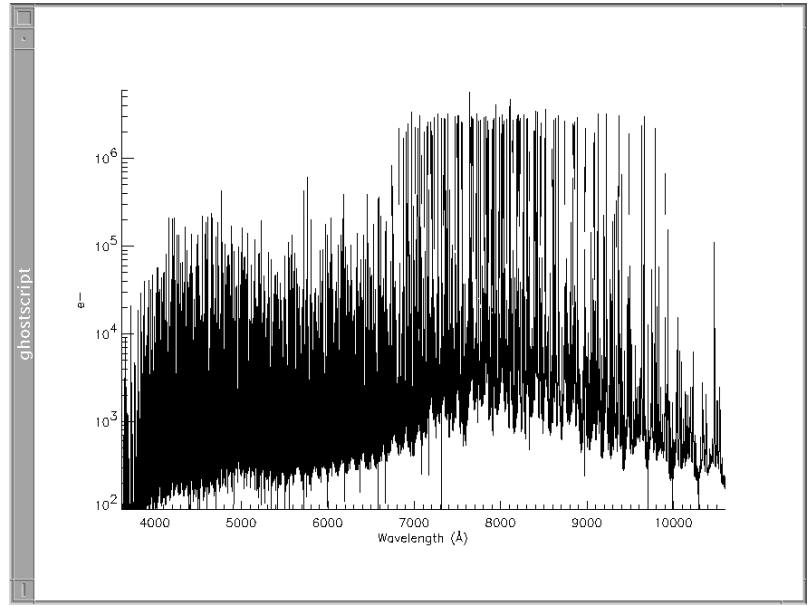}
\protect\caption[ ]{Full spectrum of the Th-Ar hollow-cathode 
used with the {\it 2dcoud\'e} spectrograph.
}
\label{f1}
\end{figure}

It is indeed desirable to keep 
corrections to a minimum, as the main goal of this atlas is helping  
{\it 2dcoud\'e} observers
in the identification of the Th-Ar spectrum.
As an example, there is a number of spurious features in the orders  next
to some very strong features in the red. These artifacts cannot be generally 
matched with wavelengths of thorium or argon features, and they tend to have a 
distinctive shape. Therefore, we could easily {\it clean} the spectra of these 
features, but in doing so we would be altering significantly the aspect of the
lamp spectrum as extracted by an observer for calibration. Most observers
do not removed scattered light from the Th-Ar calibration spectra. At a
resolving power of 60,000
the stray light is significant only in the red (see Fig. 2), and  
it has little impact on the accuracy of the calibration. We have
not removed it from our spectra. Finally, many strong lines present in the red
part of the spectrum are saturated. Observers intentionally overexpose those
features in order to bring out weaker lines in the blue and near infrared.
Very short exposures and a filter can provide more reliable intensities and
wavelengths for those lines, but would alter their aspect compared to typical
observations. Thus, we have not taken any measure to correct this problem.

The center of the
lines identified in the McDonald spectra were determined with the 
{\tt center1d} algorithm  in the IRAF task {\tt ecidentify}. For each of the
lines whose wavelength was measured at least in twice in one of the setups, 
the average and standard error of the mean were determined. Multiple measures
of a line occur because of the availability of multiple exposures for each setup,
the overlap between the two setups, and the wavelength overlap between adjacent 
orders. When average values were derived for a line
measured in both the {\it blue} and {\it red} setups, which happened for half
of the features, they were weighted with the inverse of the square of their 
standard errors, and averaged.

\begin{figure}[ht!]
\centering
\includegraphics[width=8cm,angle=90]{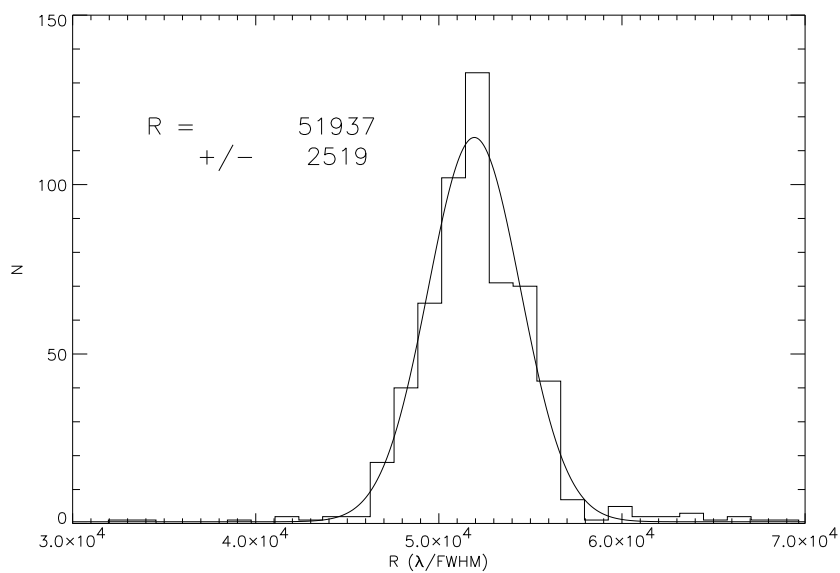}
\protect\caption[ ]{Number of emission lines measured in the Th-Ar hollow-cathode
spectra as a function of the resolving power.
}
\label{f3}
\end{figure}

The final digital atlas available on the 
WWW\footnote{{\tt http://hebe.as.utexas.edu/2dcoude/}}
covers the region between 3611.9 and 10596.4 \AA. 
However,
as the line identification was based on Willmarth's list, 
the pictorial atlas and the list of wavelengths was restricted to the
interval 3642.2 \AA\ -- 9664.7 \AA.
To produce the pictorial atlas, we scaled the individual exposures 
by their level of counts, averaging them out after rejecting deviant pixels. We
plotted the 
resulting spectrum  in segments of 30 \AA, and labeled the lines  
keeping only the relevant number of digits. 
Table 1 (Part III) lists the determined wavelengths and their 
uncertainty (standard error).

To determine the resolution of the Th-Ar atlas we selected 584 lines with
a standard error in their wavelengths $\le$ 0.0002 \AA, and fitted a Gaussian
profile with a constant background to the spectrum in an interval 
0.4 $\lambda({\rm \AA})/3700$ around the line center. Figure 3 displays 
the histogram of the lines as a function of 
the resolving power R$=\lambda/{\rm FWHM}$, which peaks at about 52,000.
The intrinsic width of the lines only contributes  about 10 \% of 
the observed width.


\section*{Acknowledgements}

Sincere thanks to Ed Barker, David Doss, David Lambert, Bob Tull,
and Russel White for help 
and  guidance with the {\it 2dcoud\'e} and in the preparation of this atlas.

\section*{References}

\begin{itemize}

\item Lambert, D. L., Allende Prieto, C., \& Cunha, K. 2001,  in the proceedings
of the 12th Cambridge Workshop on Stars, Stellar Systems, and the Sun, ASP Conf.
Series, in press 

(available at {\tt http://hebe.as.utexas.edu/poster.html})

\item Norl\'en, G. 1973, Physica Scripta, 8, 249

\item Palmer, B. A., \& Engleman, R., Jr. 1983, Atlas of the Thorium Spectrum,
LA-9615, Los Alamos National Laboratory, Los Alamos, New Mexico

\item Tull, R. G., MacQueen, P. J., Sneden, C., \& Lambert, D. L. 1995, PASP, 107, 251

\item Willmarth, D. 1992, A CCD Atlas of Comparison Spectra: Thorium-Argon Hollow 
Cathode 3180 \AA\ -- 9665 \AA.

\end{itemize}

\part{Pictorial Atlas}

\begin{figure}
\centering
\includegraphics[width=10cm,angle=90]{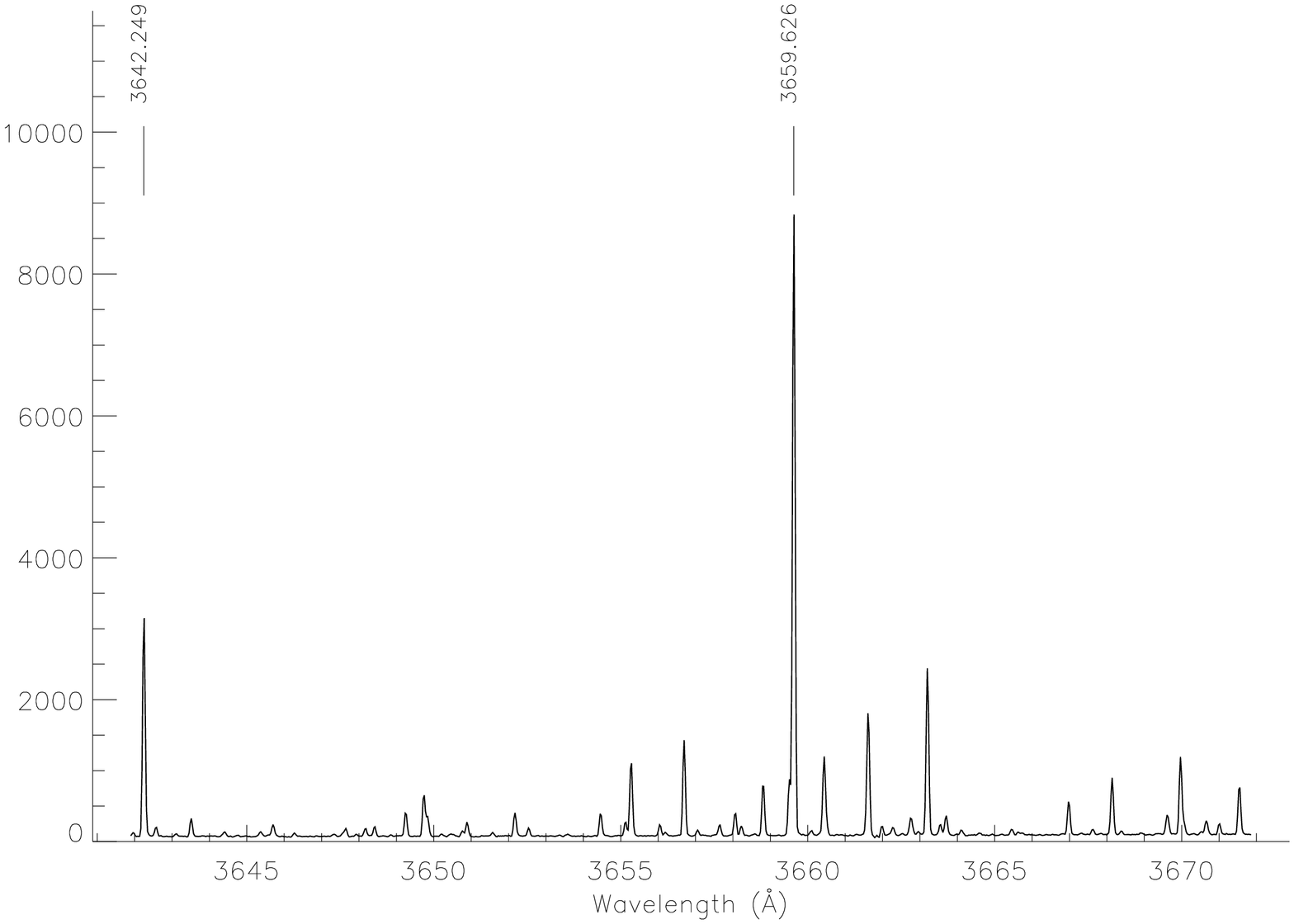}
\includegraphics[width=10cm,angle=90]{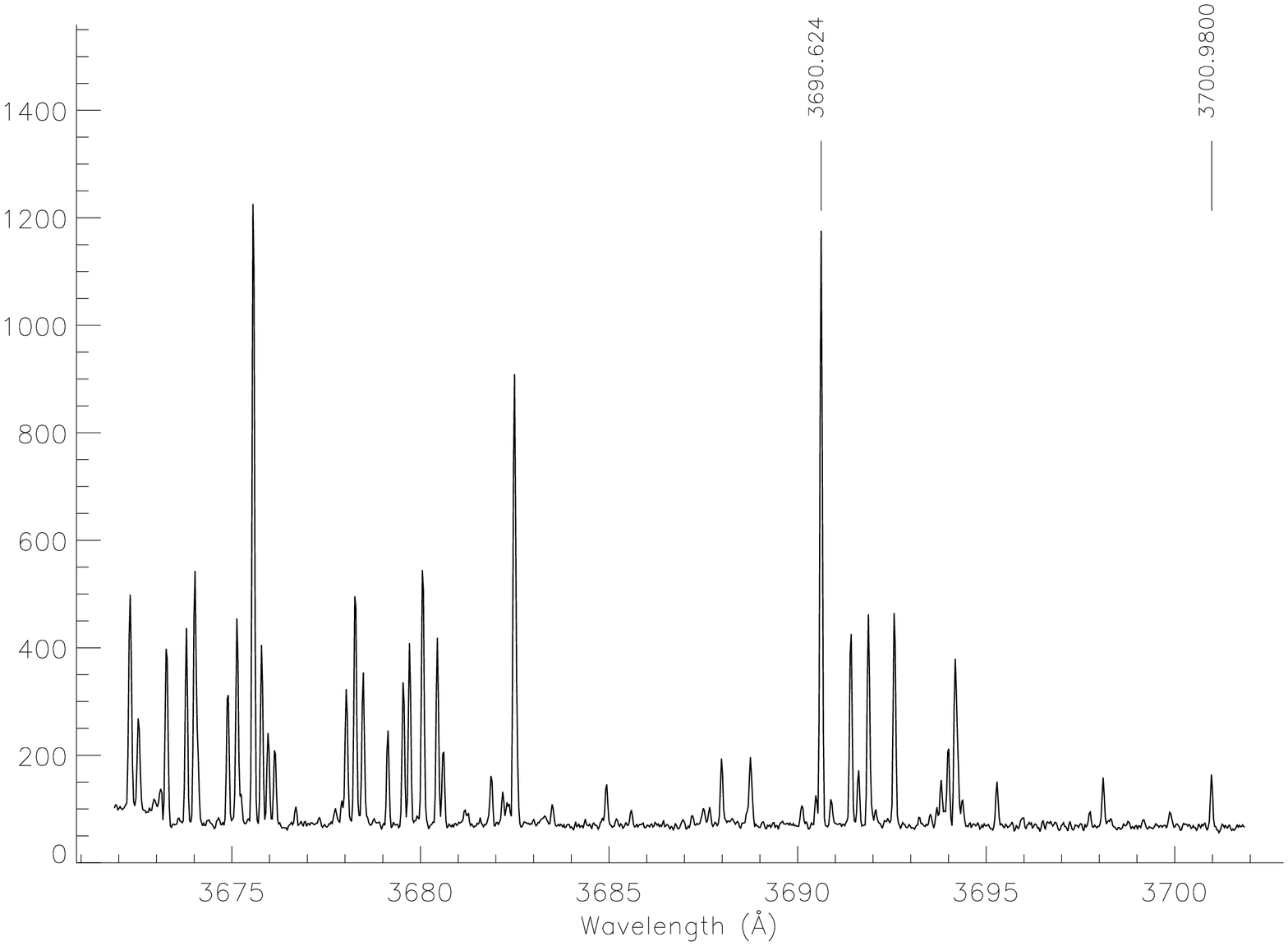}
\end{figure}
\clearpage
   
\begin{figure}
\centering
\includegraphics[width=10cm,angle=90]{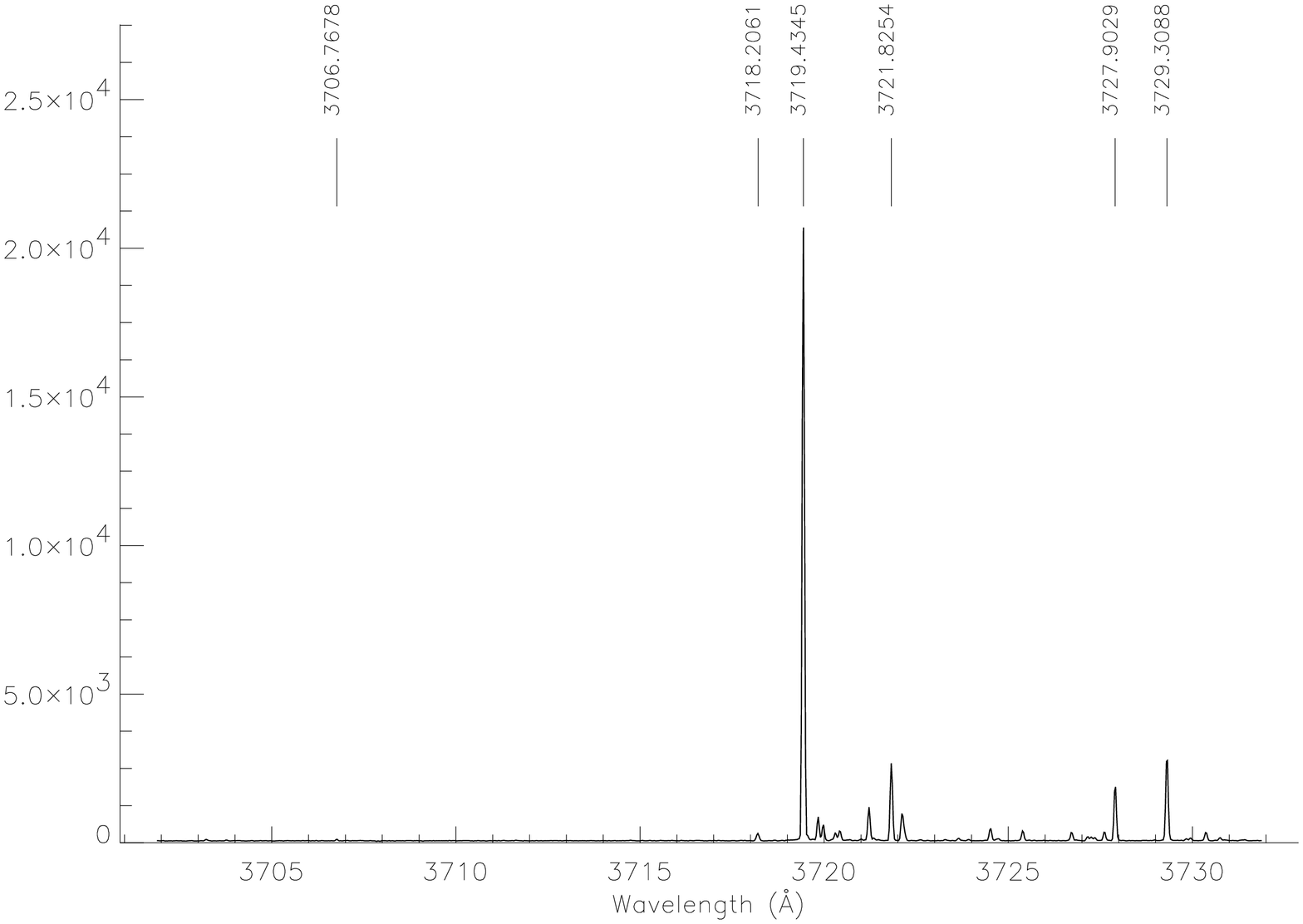}
\includegraphics[width=10cm,angle=90]{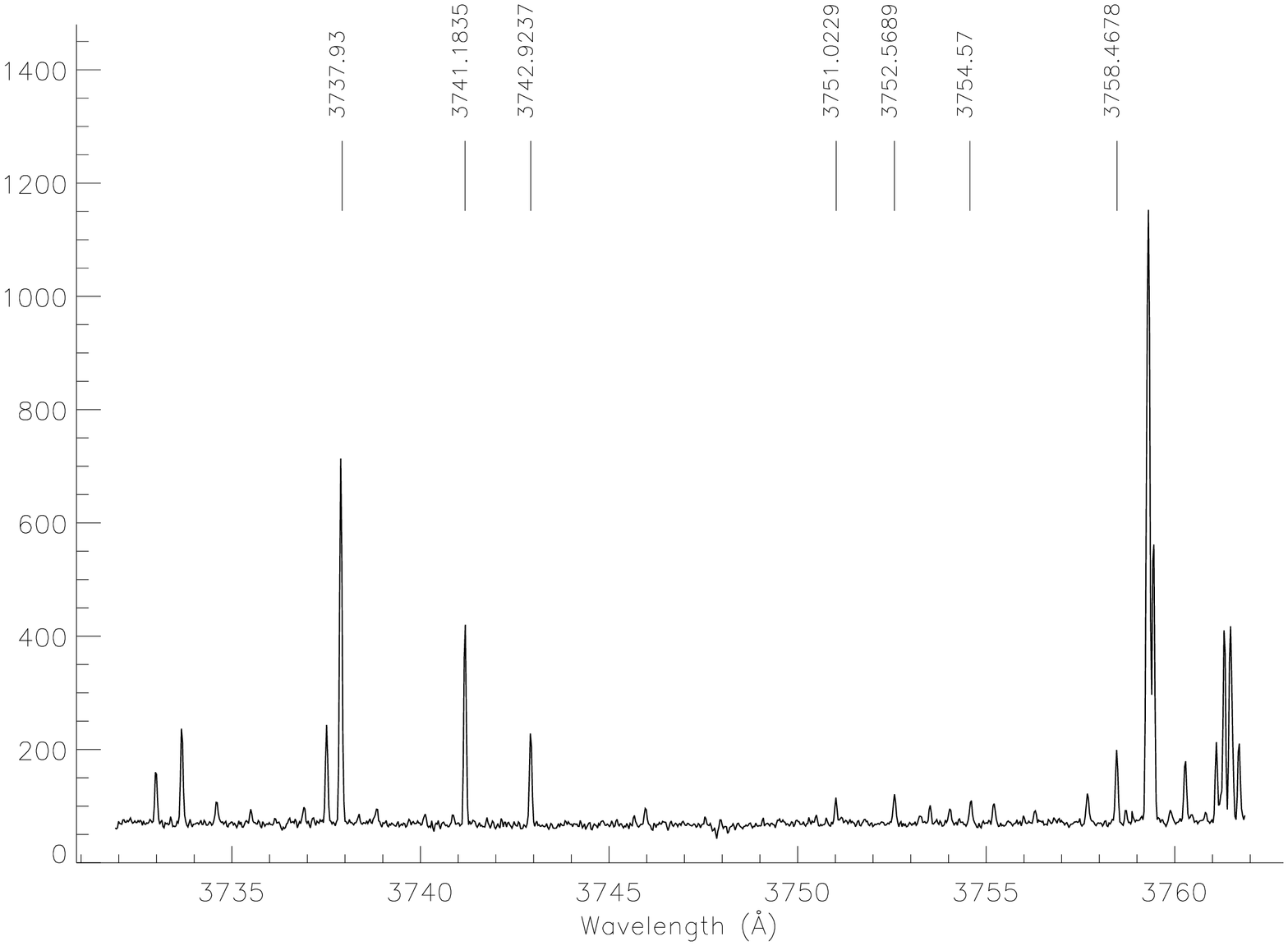}
\end{figure}
\clearpage
   
\begin{figure}
\centering
\includegraphics[width=10cm,angle=90]{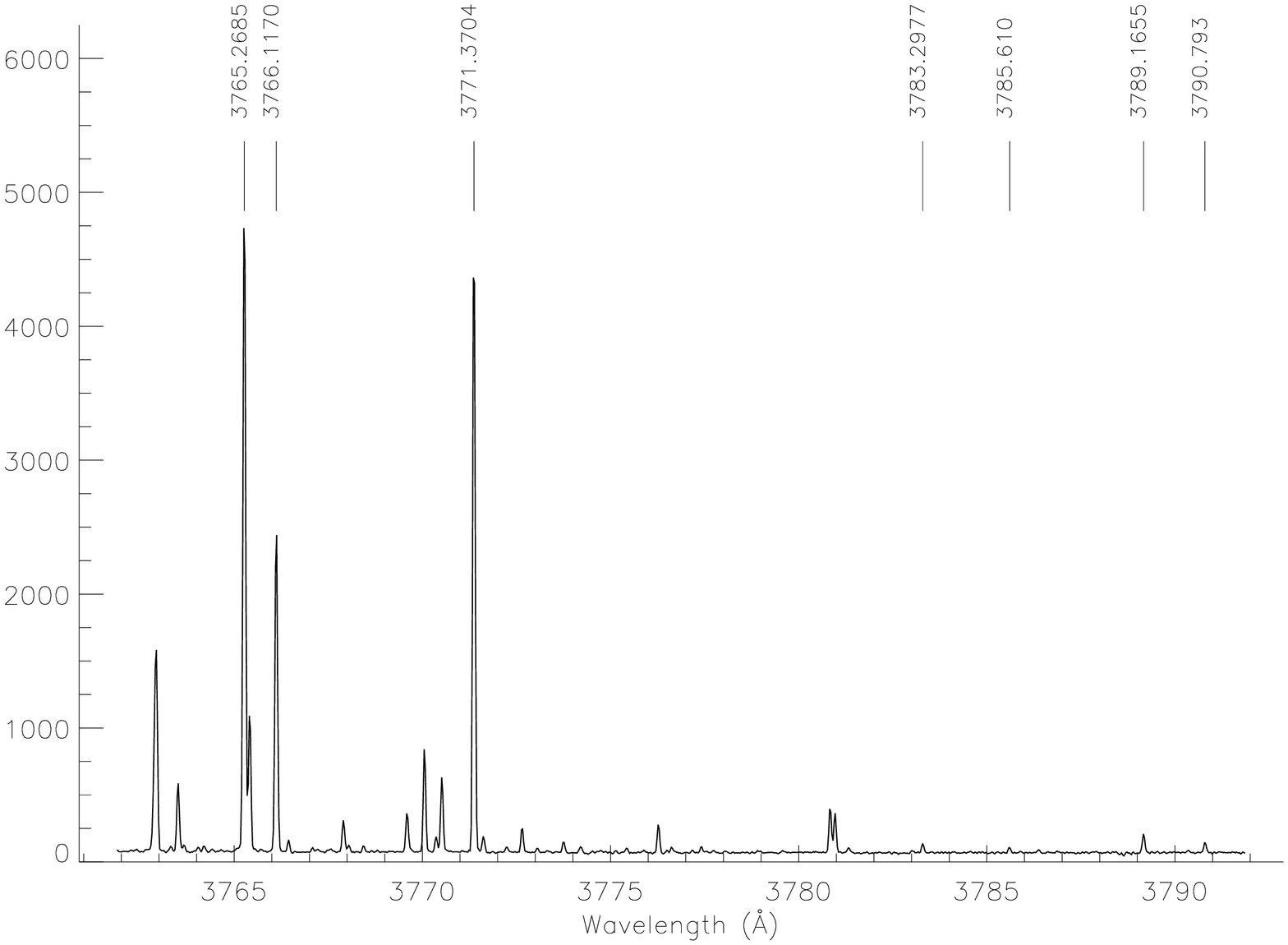}
\includegraphics[width=10cm,angle=90]{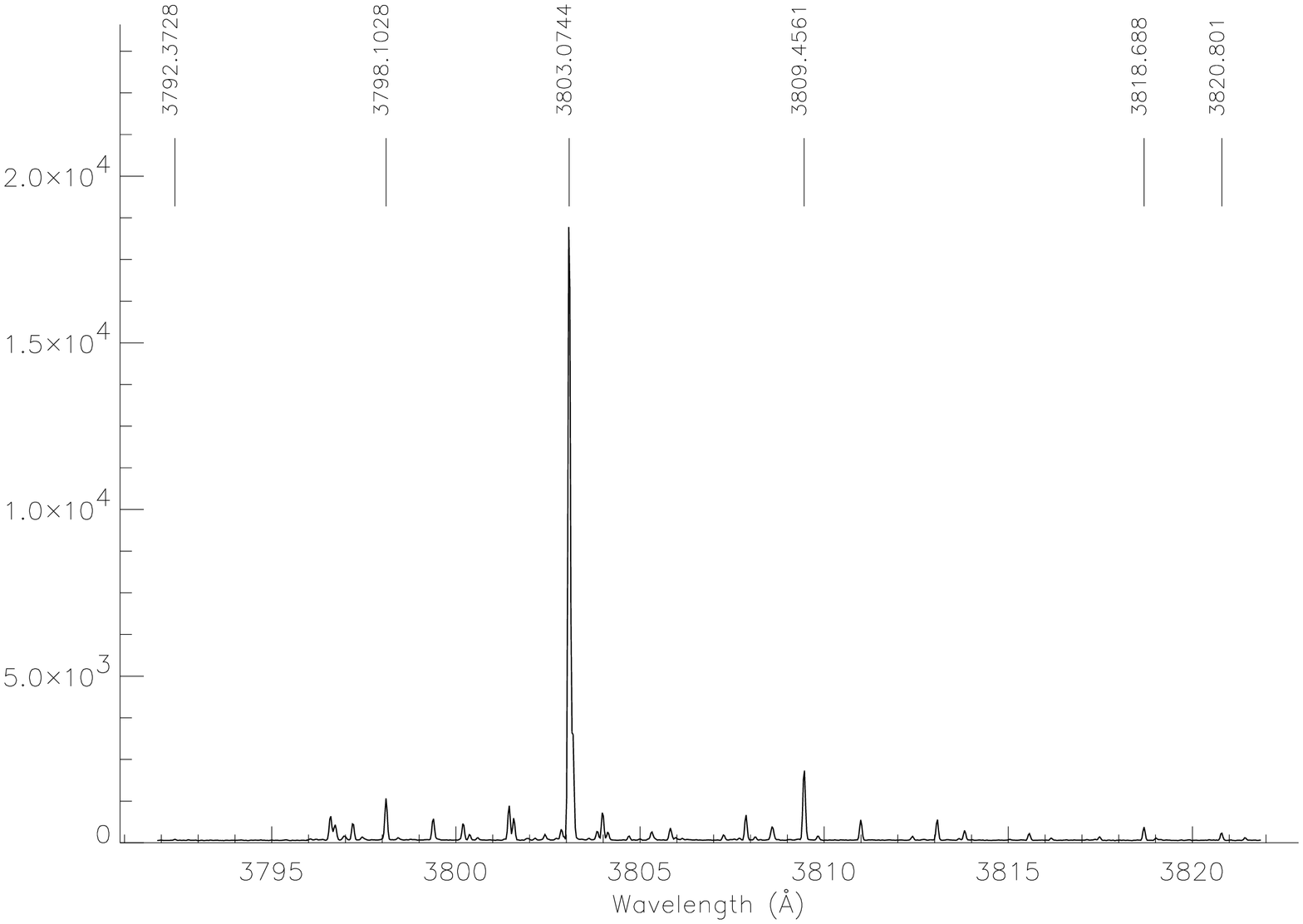}
\end{figure}
\clearpage
   
\begin{figure}
\centering
\includegraphics[width=10cm,angle=90]{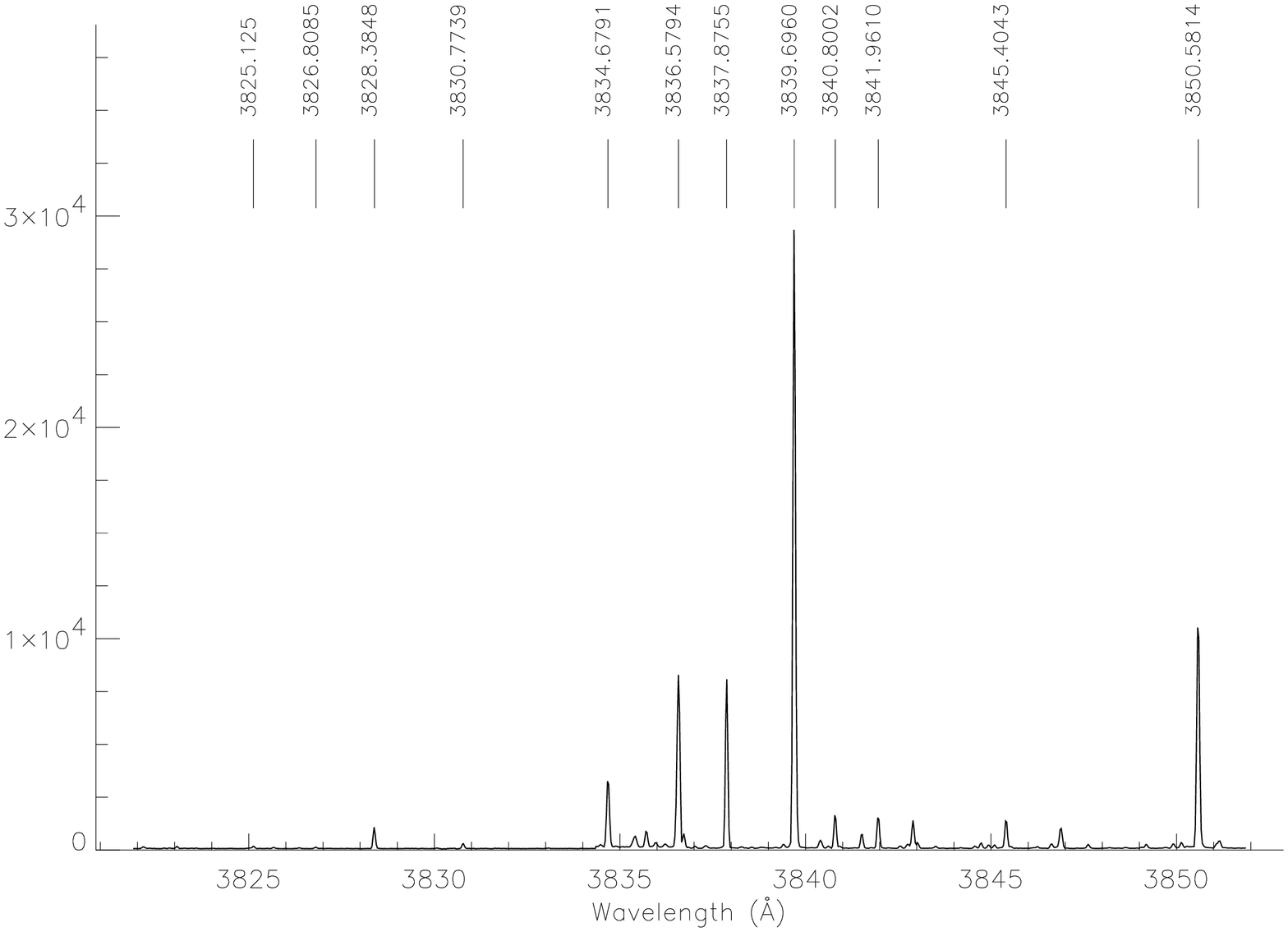}
\includegraphics[width=10cm,angle=90]{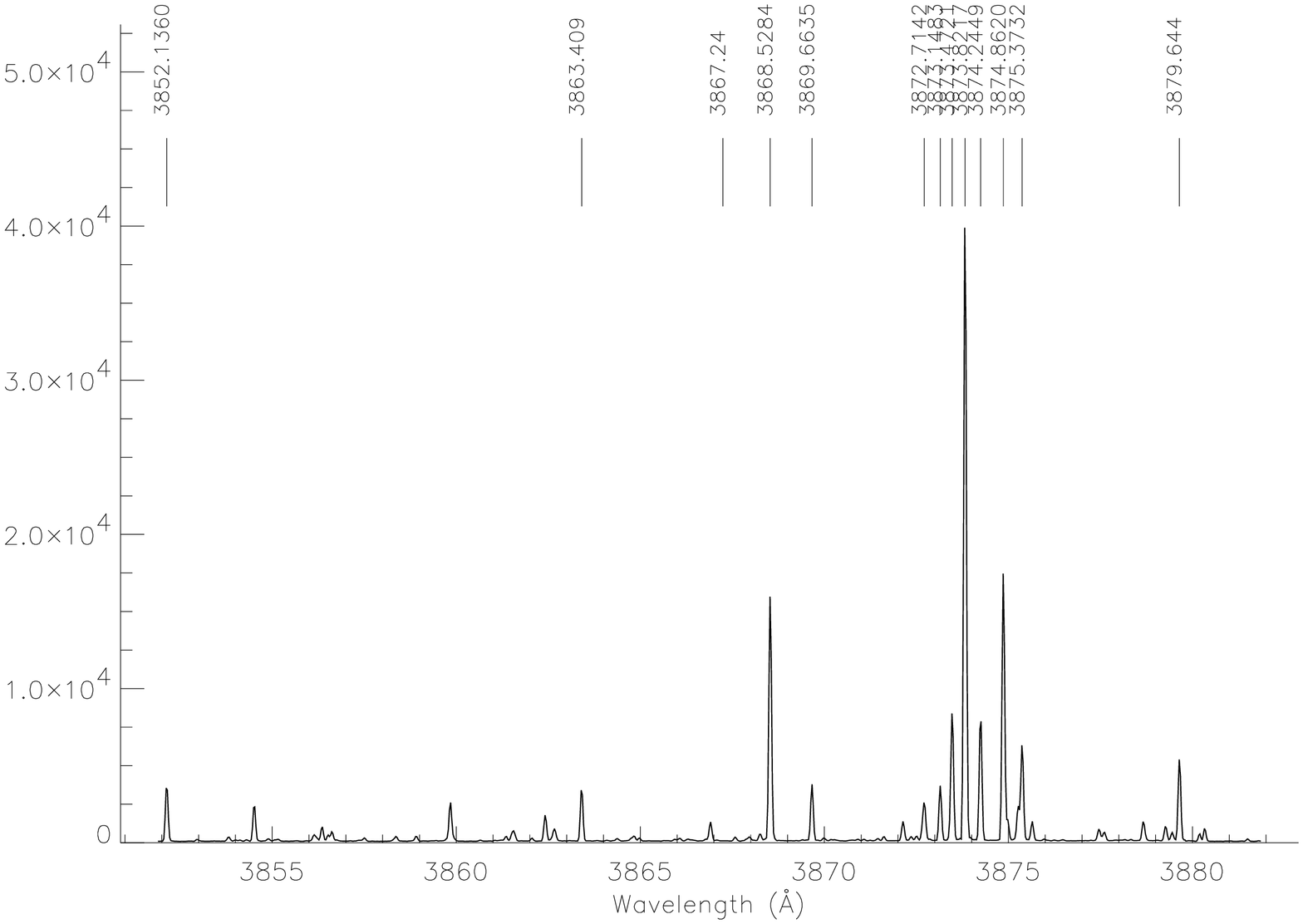}
\end{figure}
\clearpage
   
\begin{figure}
\centering
\includegraphics[width=10cm,angle=90]{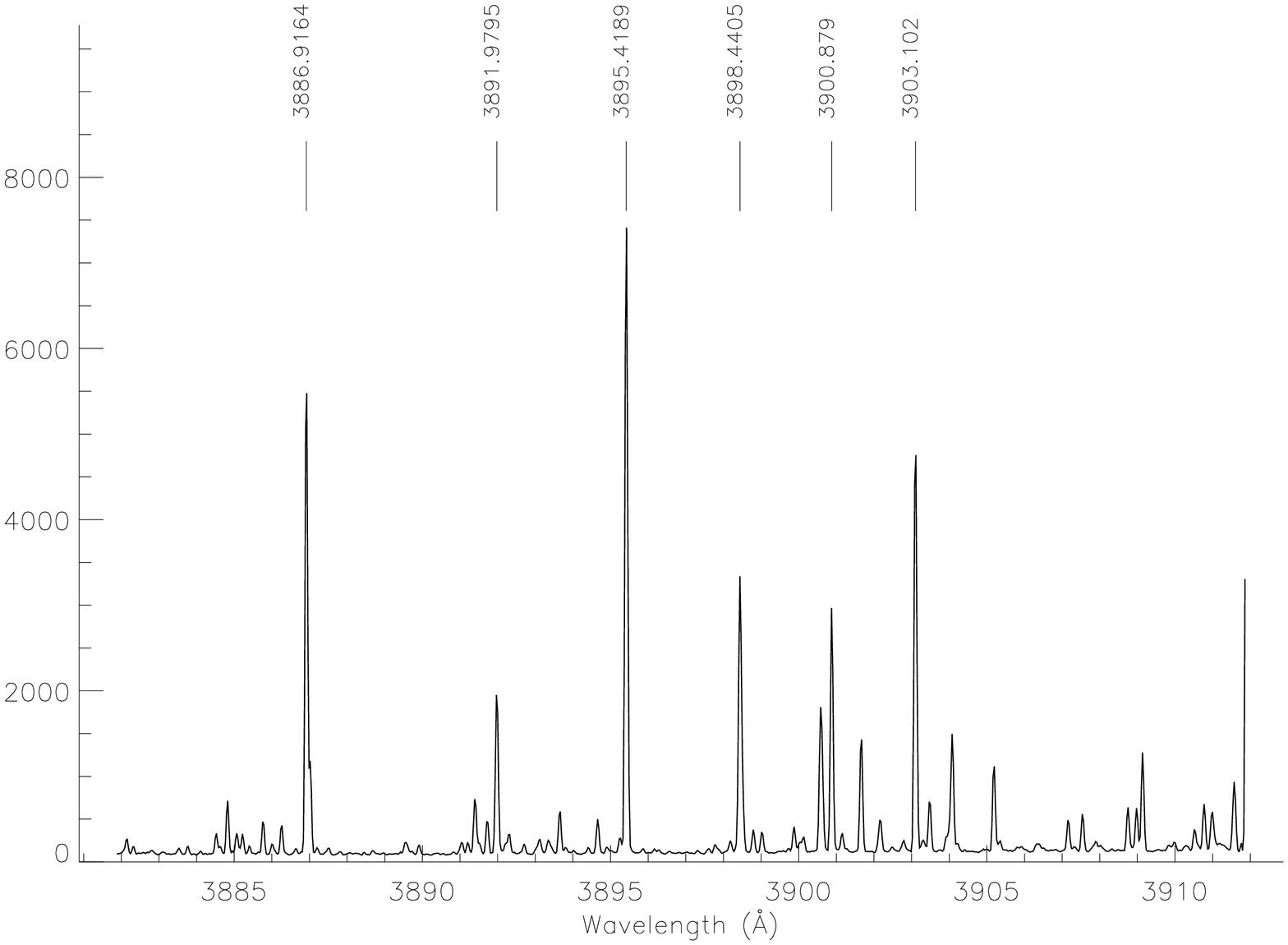}
\includegraphics[width=10cm,angle=90]{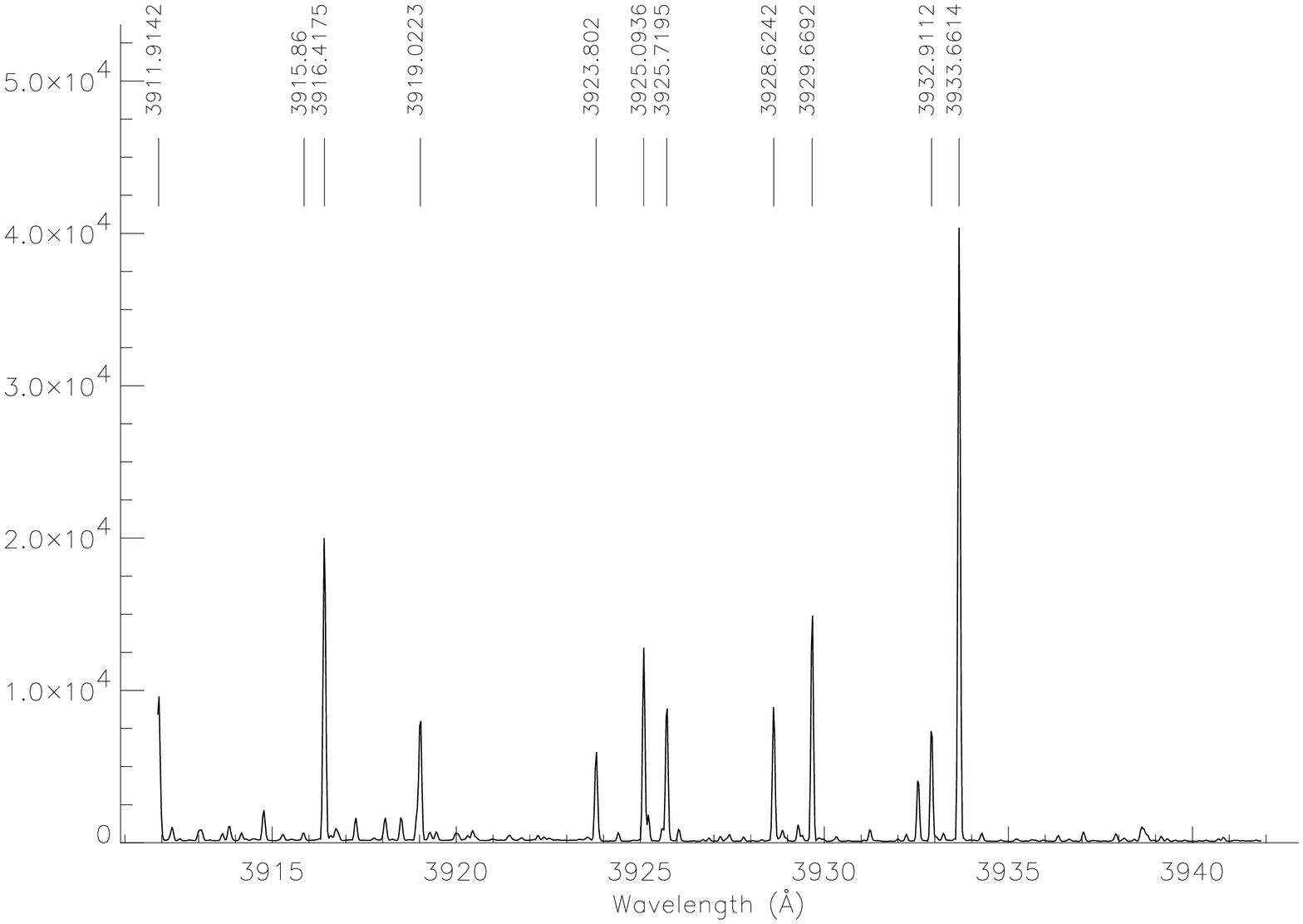}
\end{figure}
\clearpage
   
\begin{figure}
\centering
\includegraphics[width=10cm,angle=90]{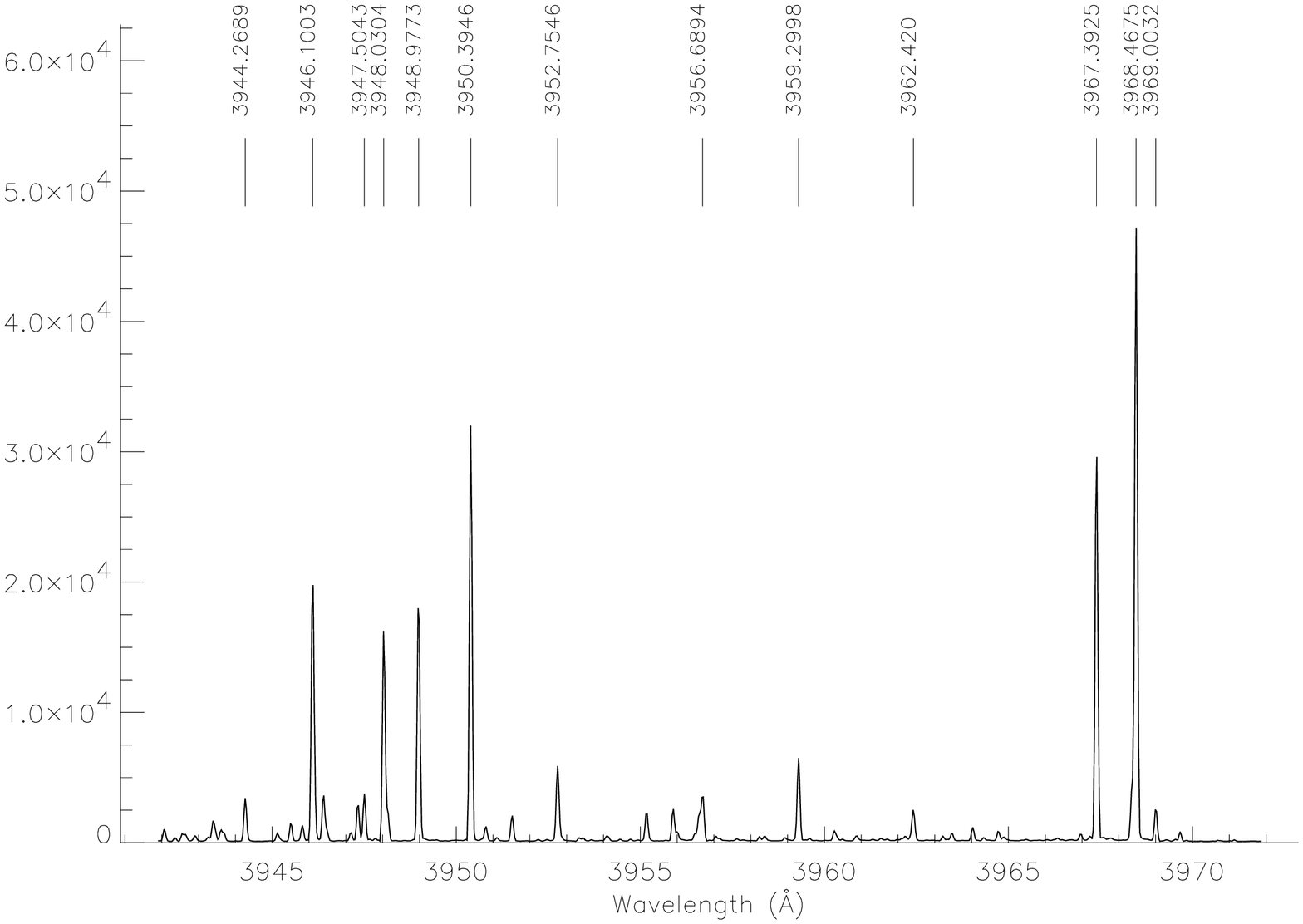}
\includegraphics[width=10cm,angle=90]{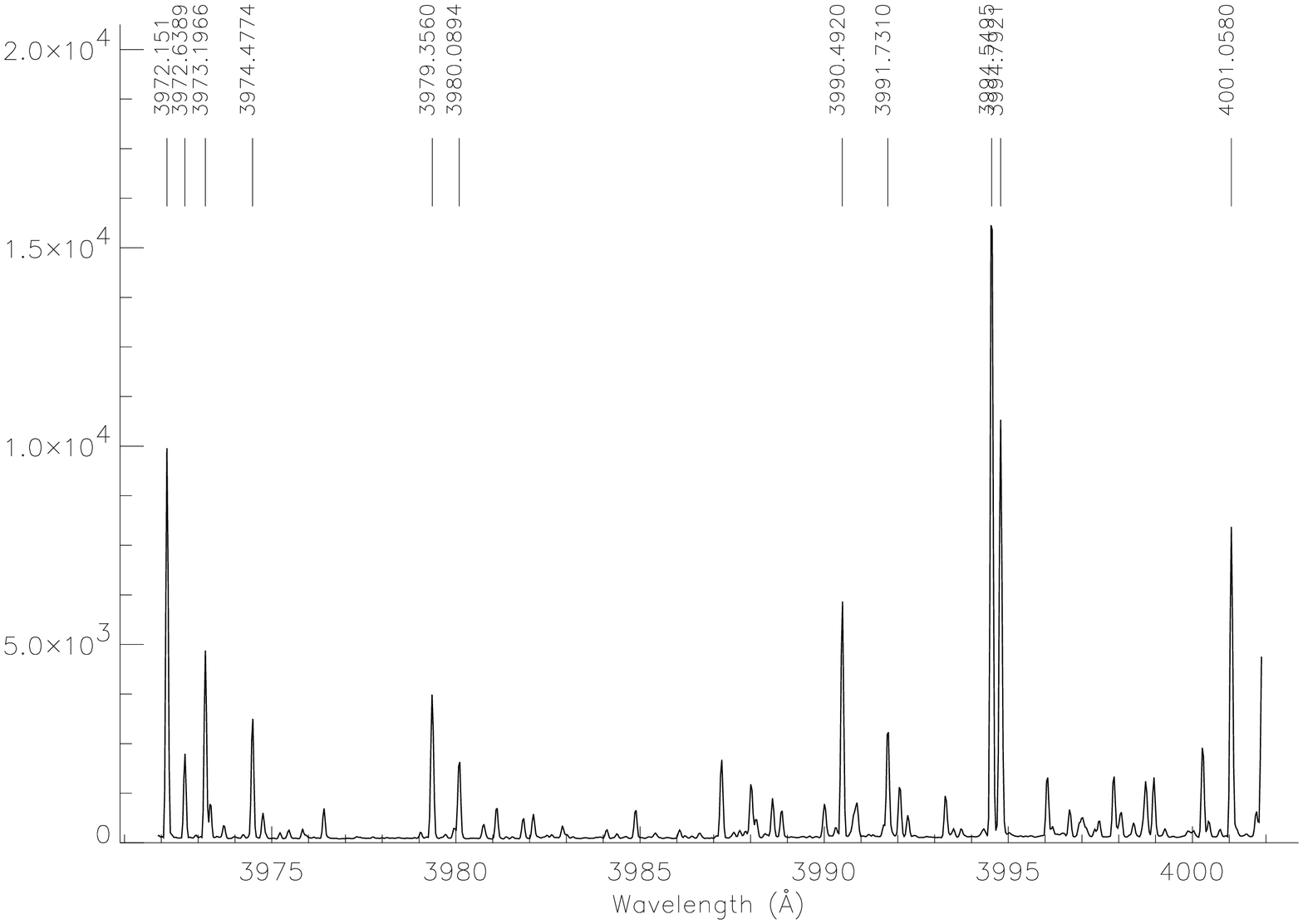}
\end{figure}
\clearpage
   
\begin{figure}
\centering
\includegraphics[width=10cm,angle=90]{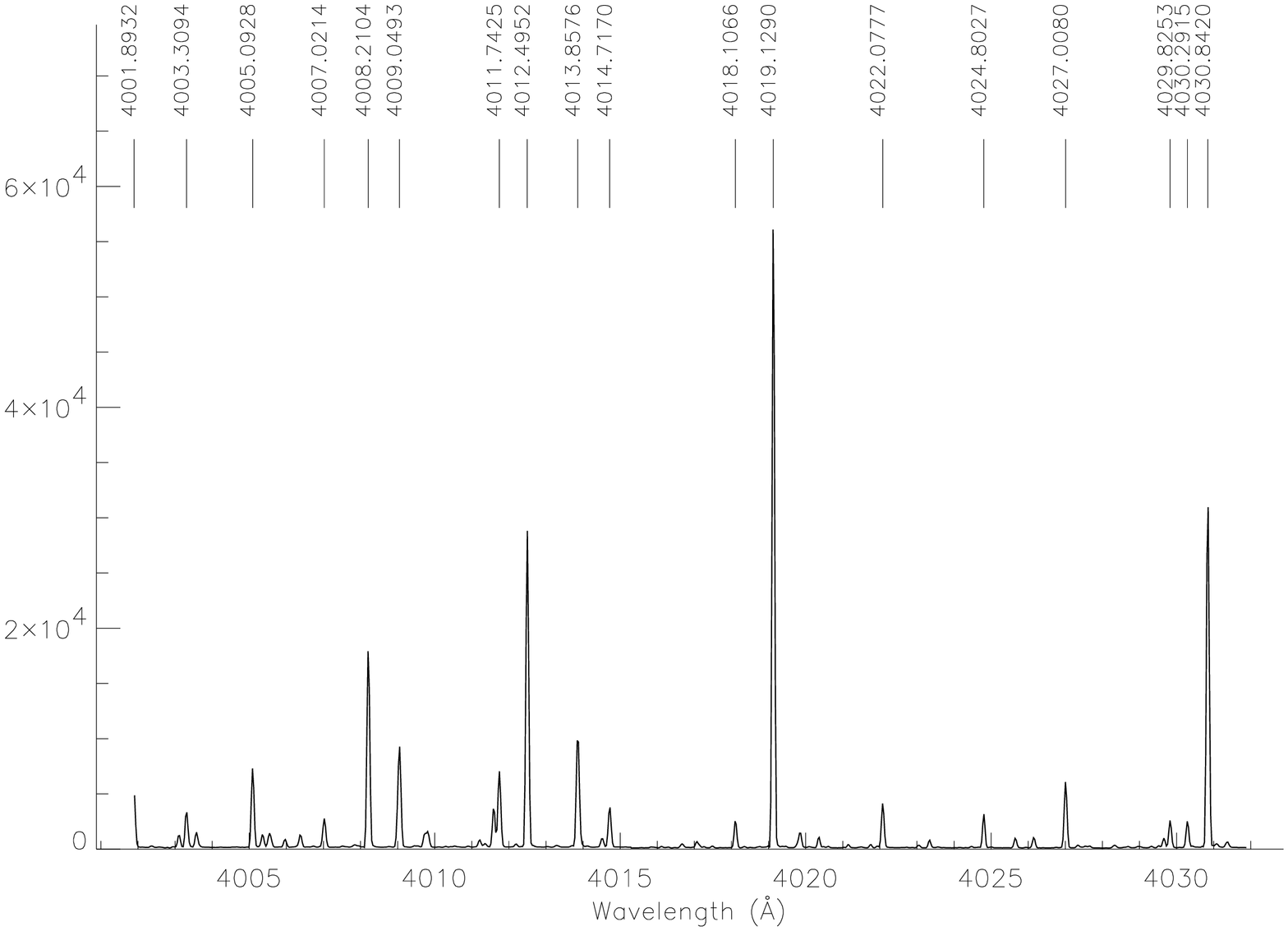}
\includegraphics[width=10cm,angle=90]{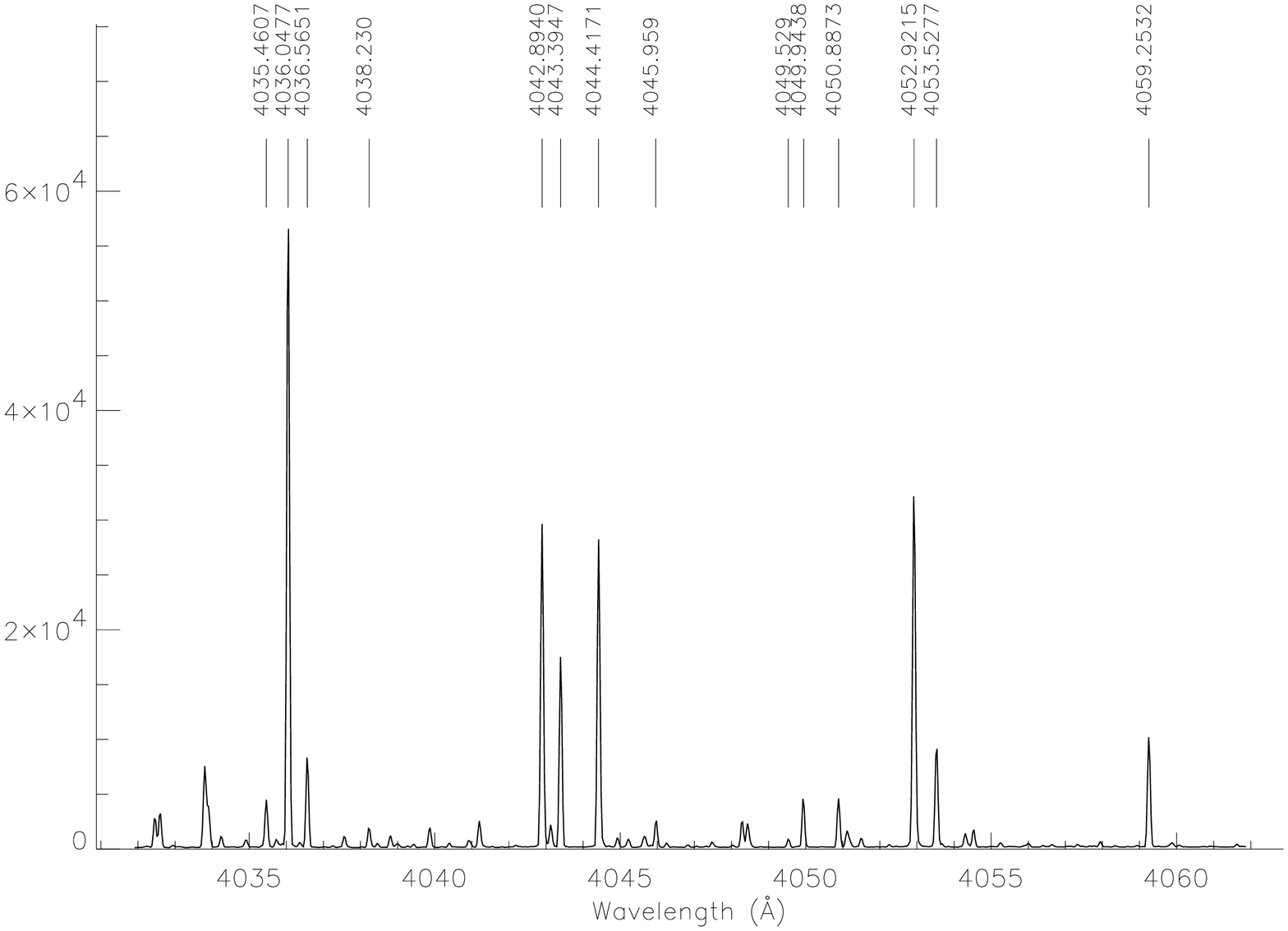}
\end{figure}
\clearpage
   
\begin{figure}
\centering
\includegraphics[width=10cm,angle=90]{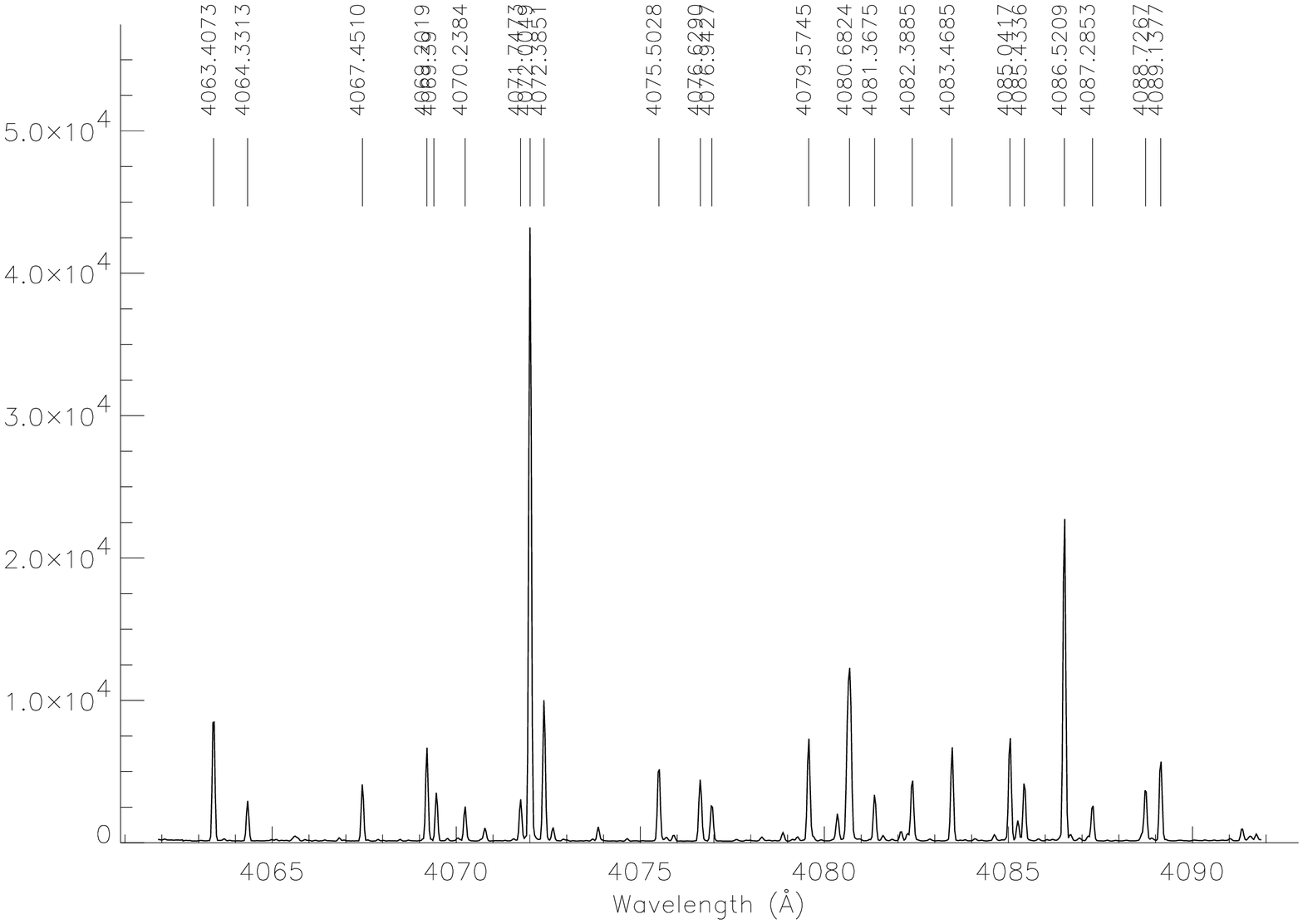}
\includegraphics[width=10cm,angle=90]{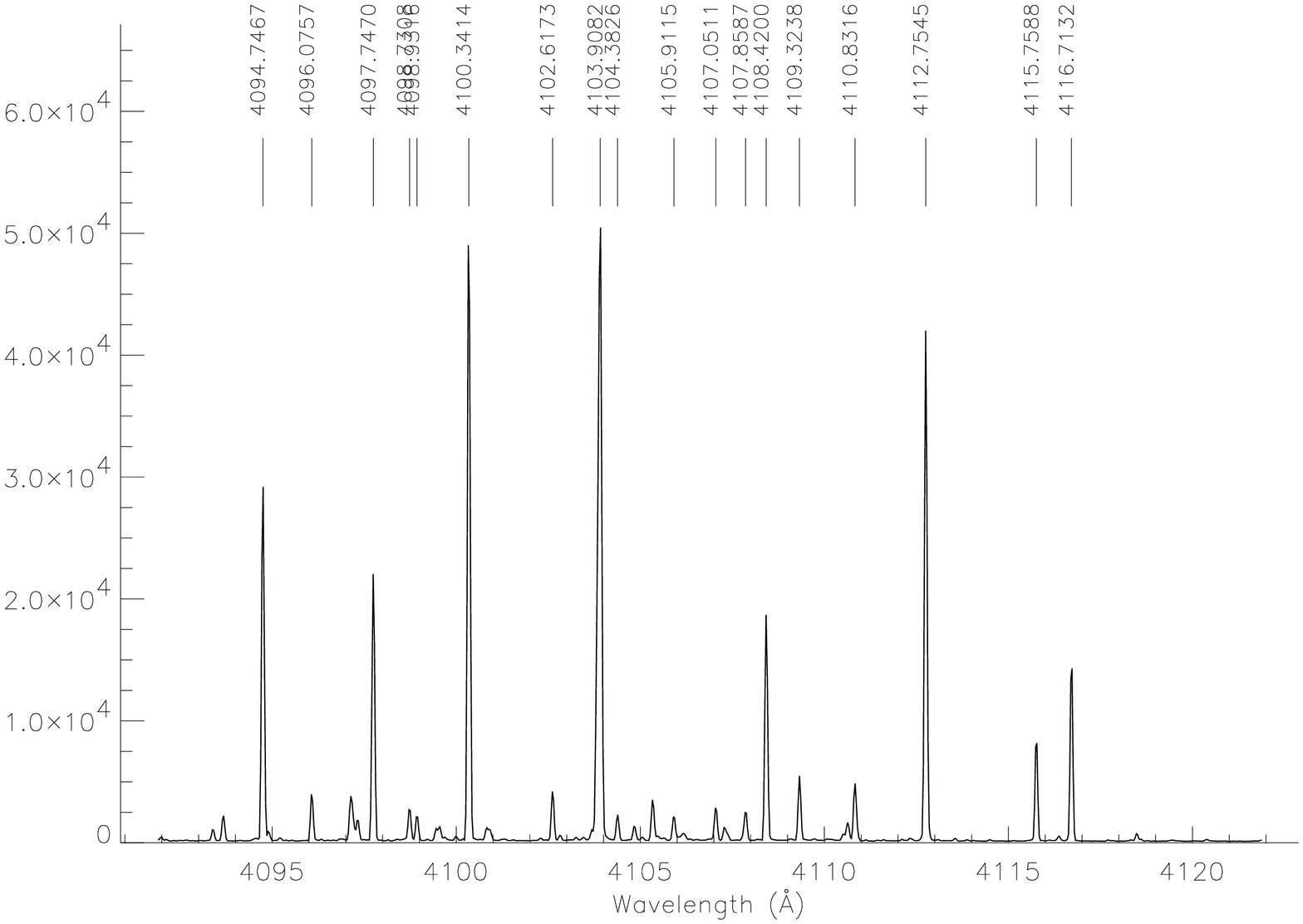}
\end{figure}
\clearpage
   
\begin{figure}
\centering
\includegraphics[width=10cm,angle=90]{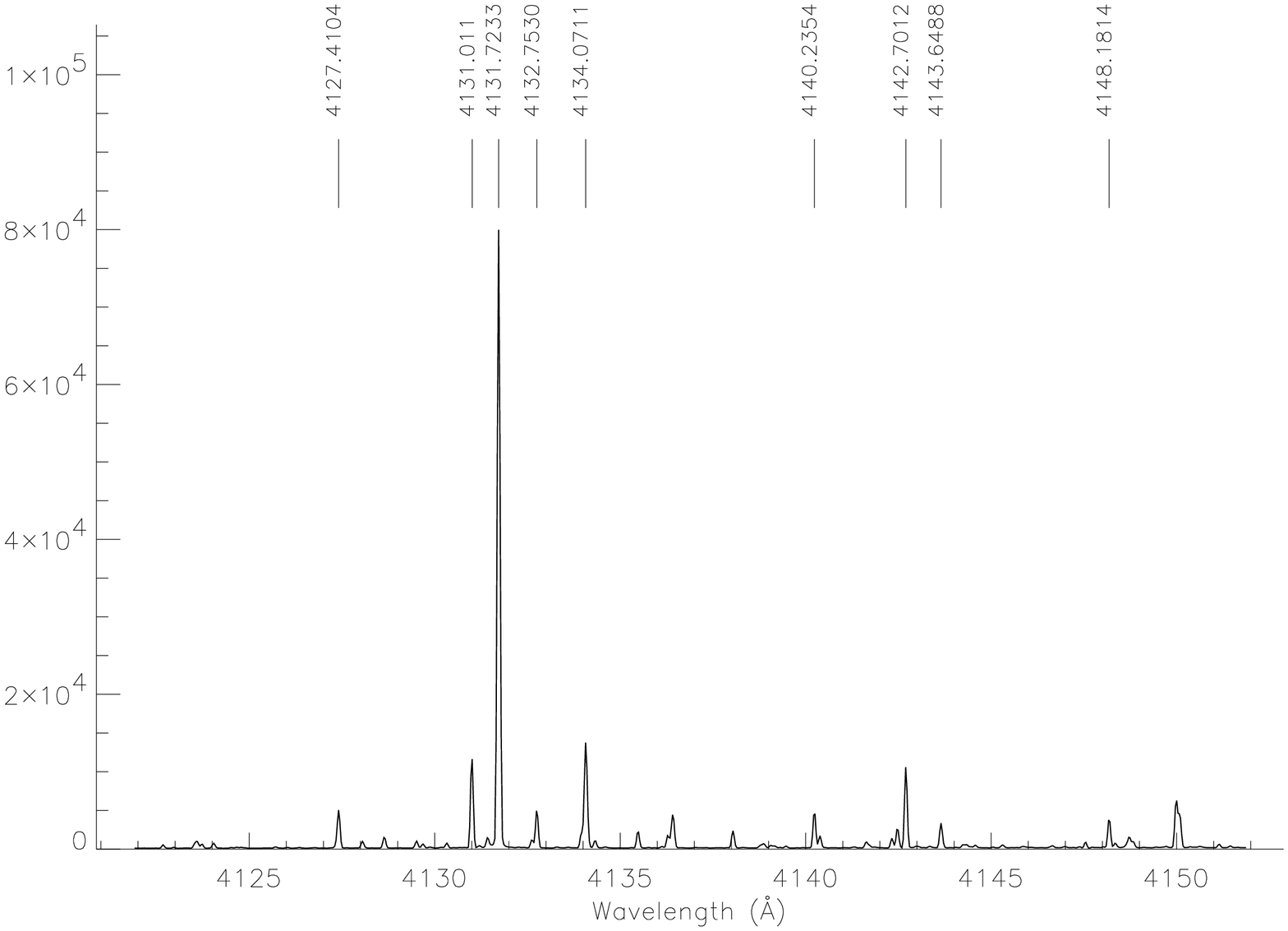}
\includegraphics[width=10cm,angle=90]{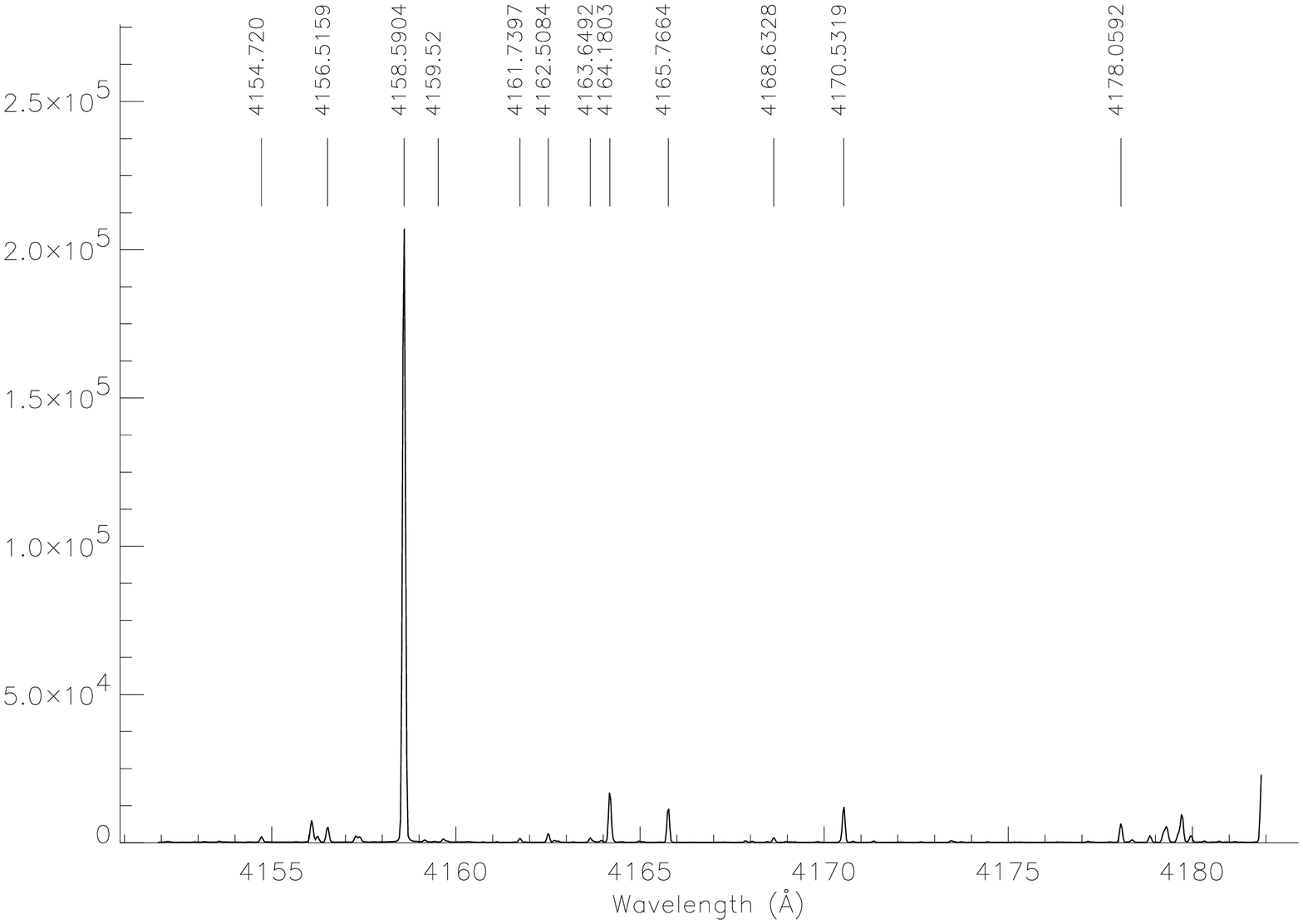}
\end{figure}
\clearpage
   
\begin{figure}
\centering
\includegraphics[width=10cm,angle=90]{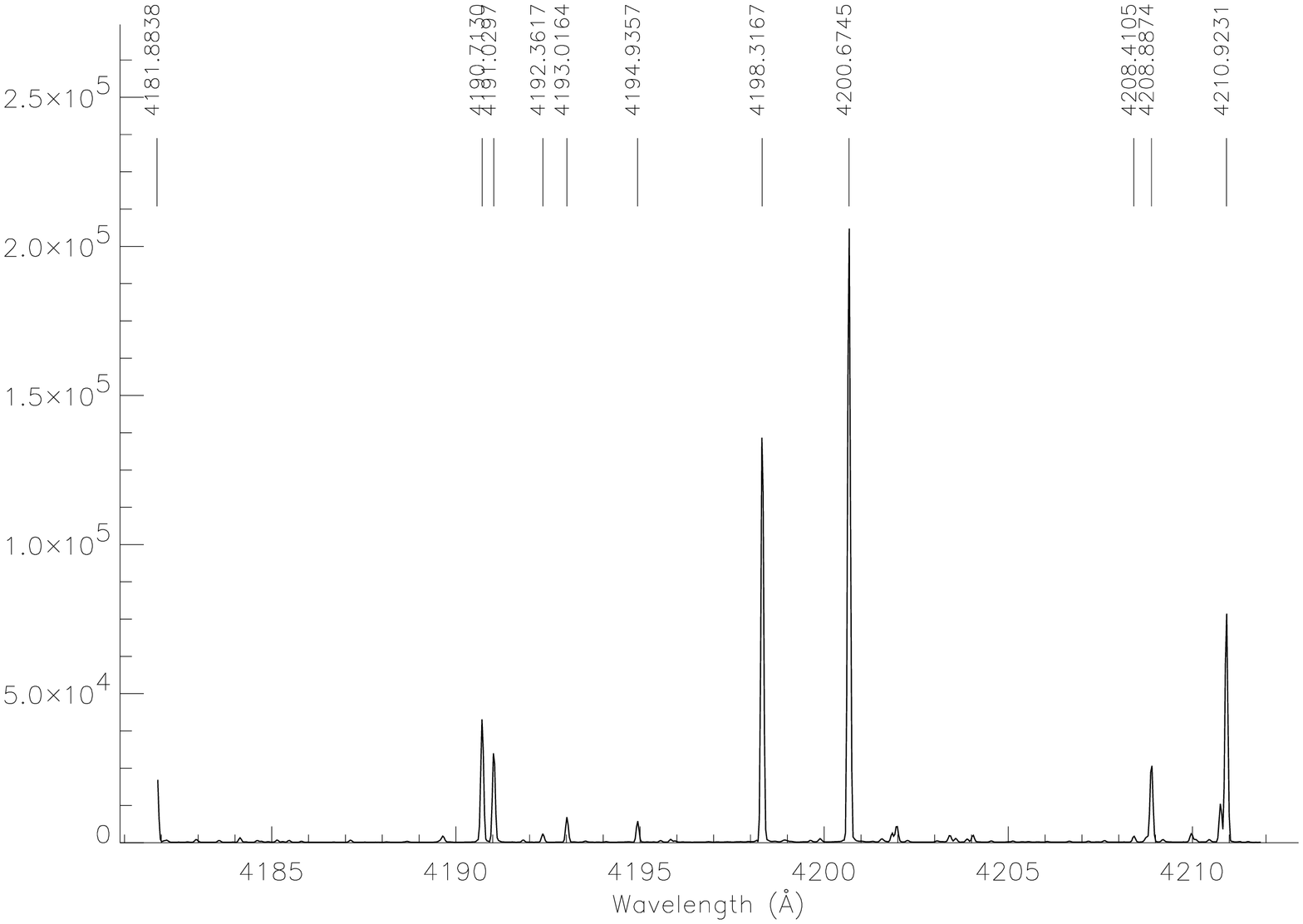}
\includegraphics[width=10cm,angle=90]{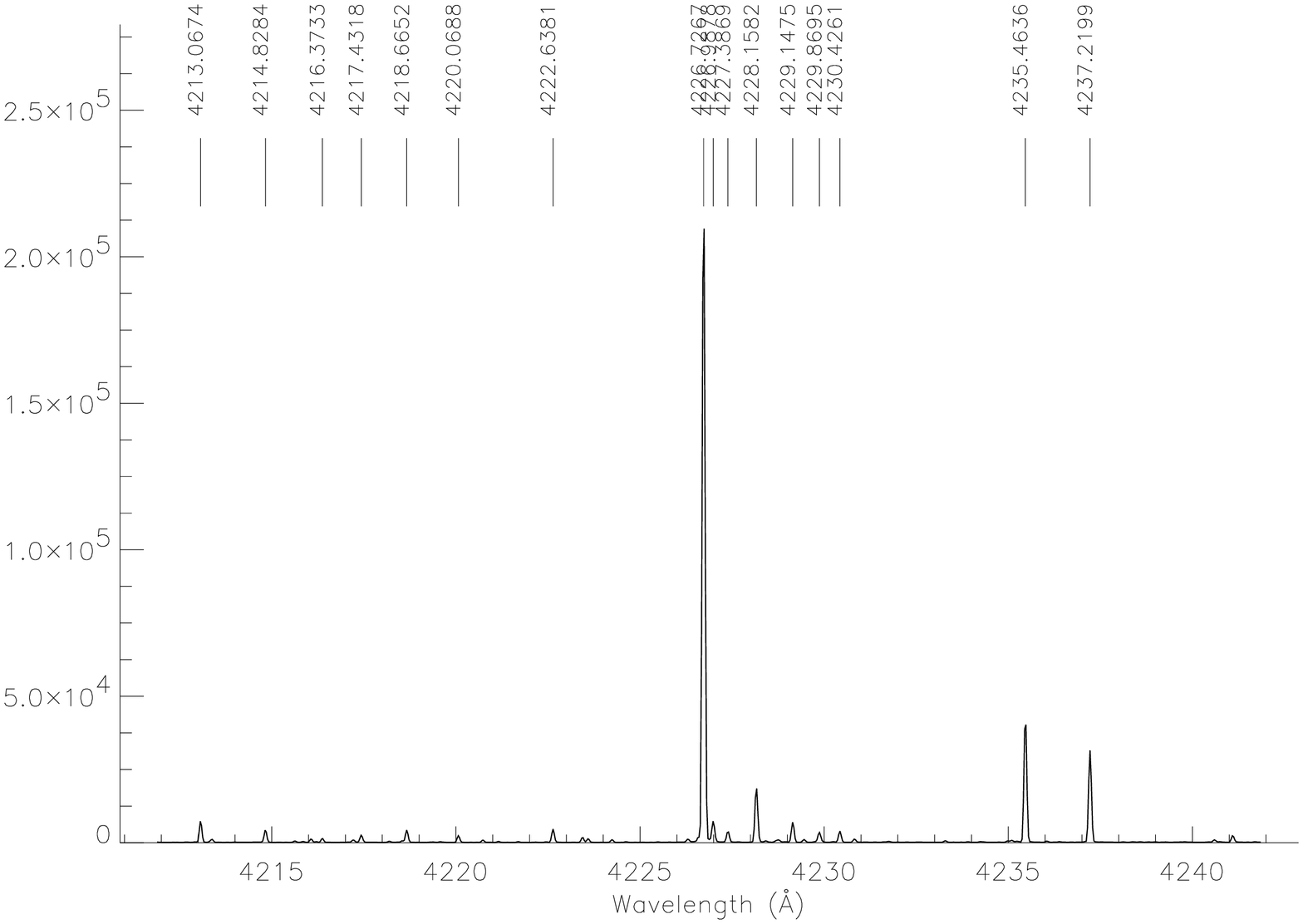}
\end{figure}
\clearpage
   
\begin{figure}
\centering
\includegraphics[width=10cm,angle=90]{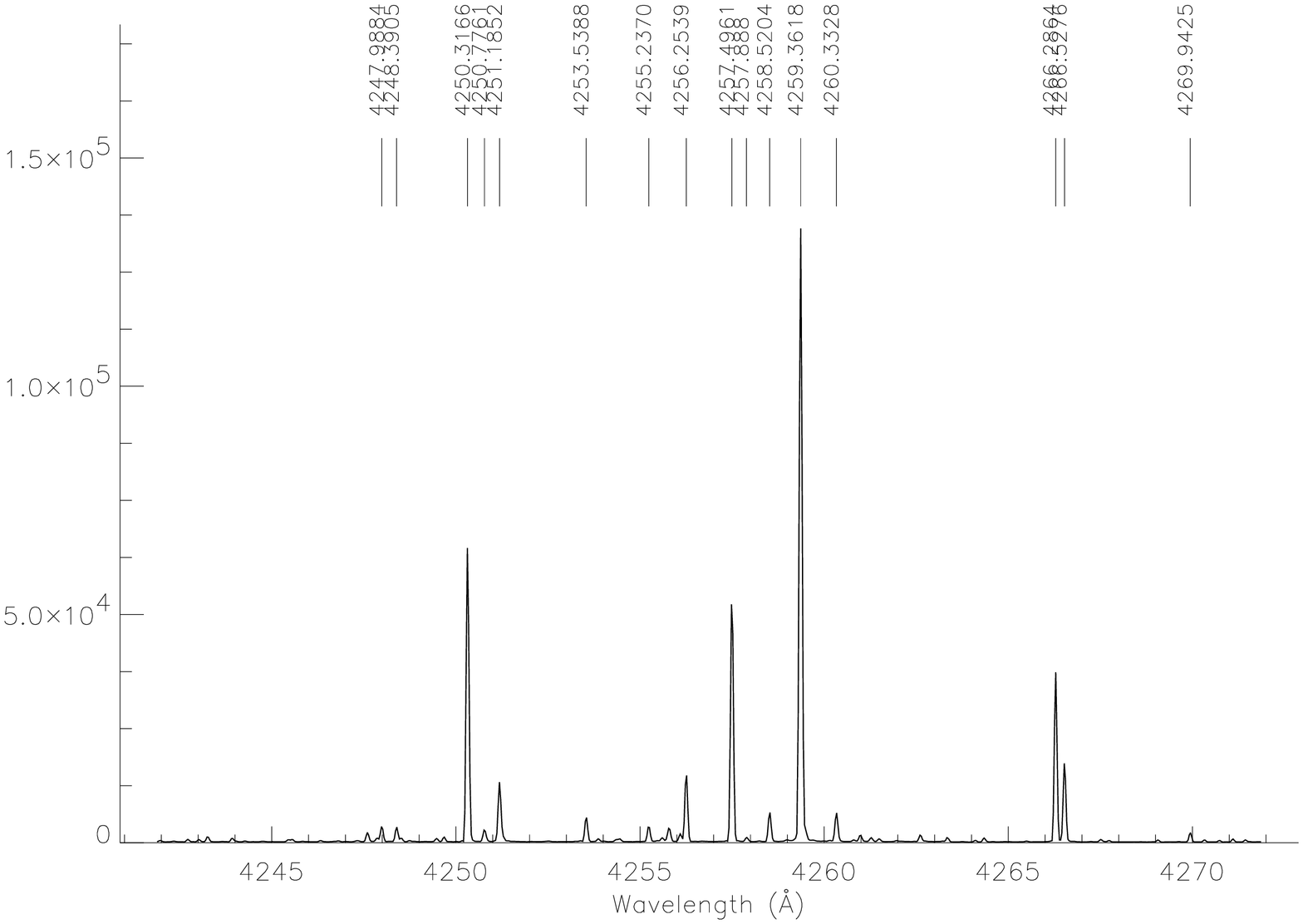}
\includegraphics[width=10cm,angle=90]{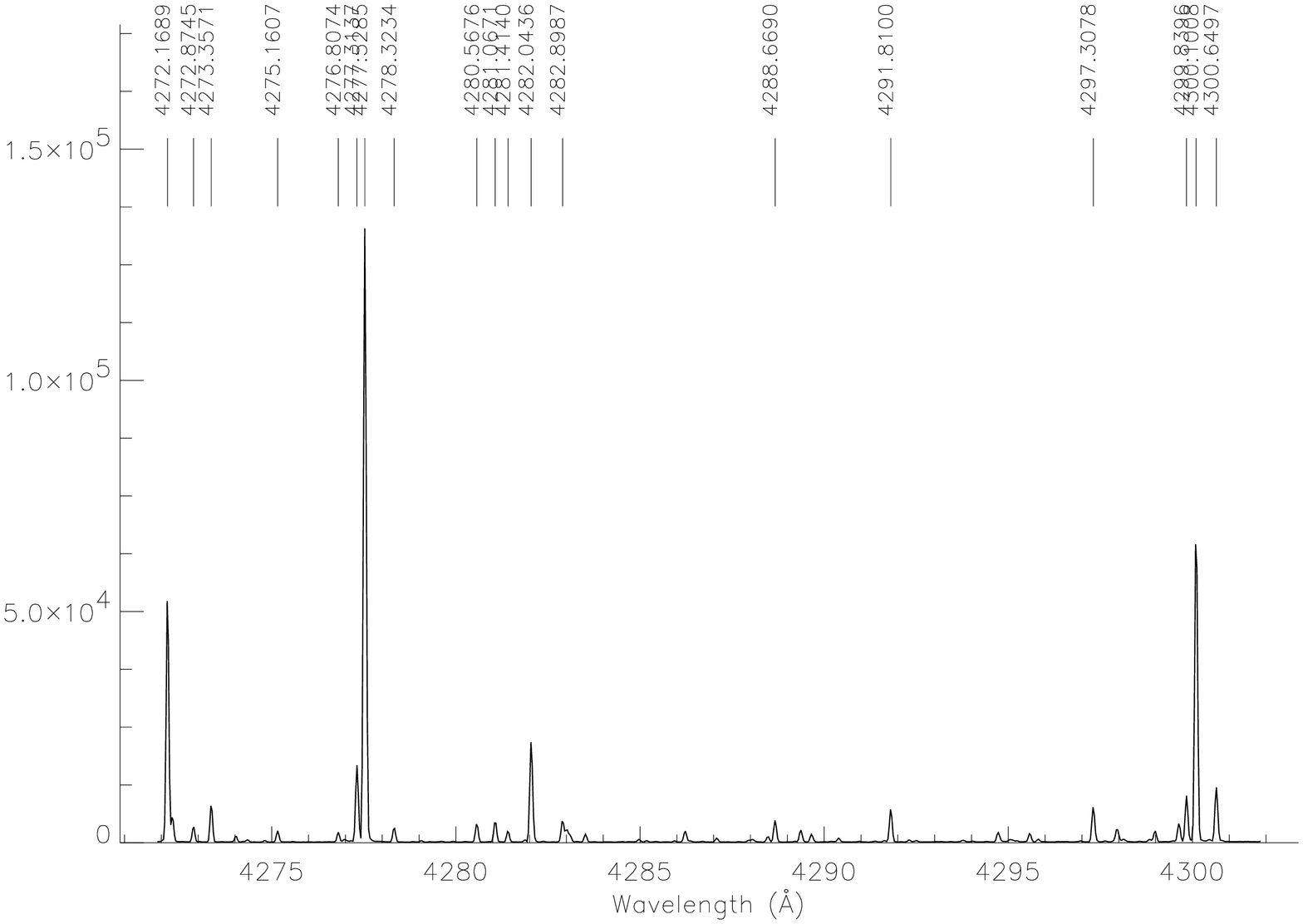}
\end{figure}
\clearpage
   
\begin{figure}
\centering
\includegraphics[width=10cm,angle=90]{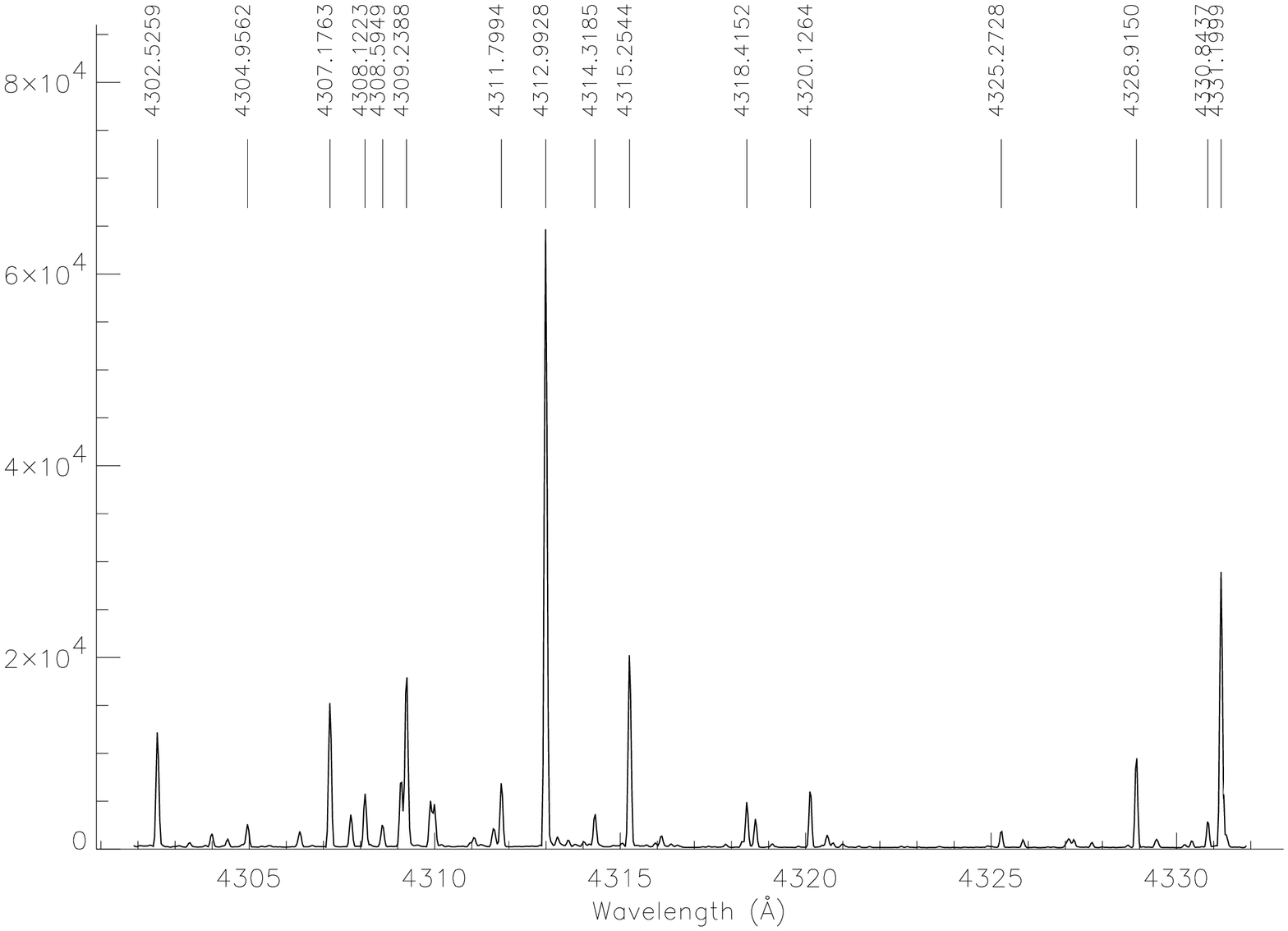}
\includegraphics[width=10cm,angle=90]{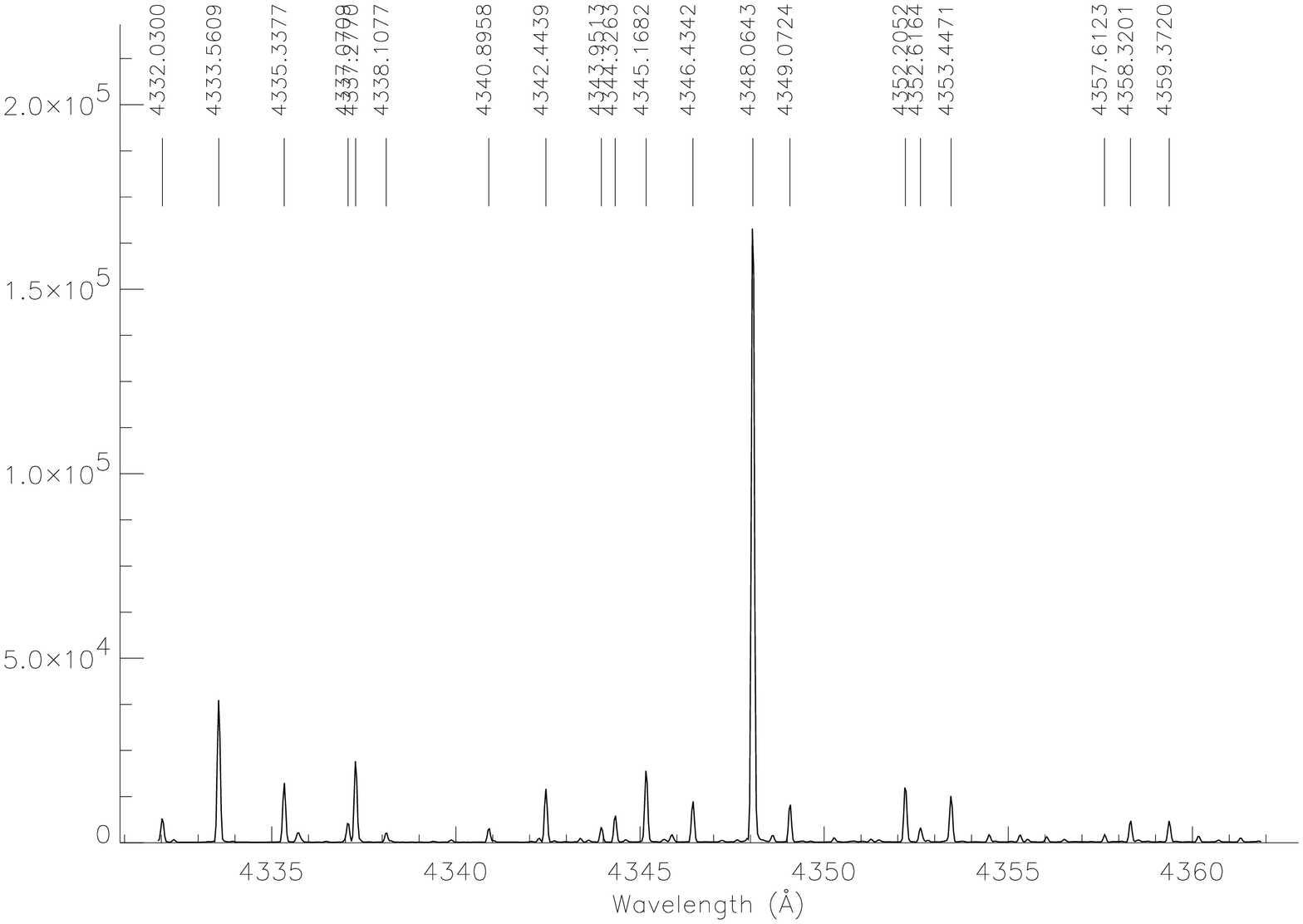}
\end{figure}
\clearpage
   
\begin{figure}
\centering
\includegraphics[width=10cm,angle=90]{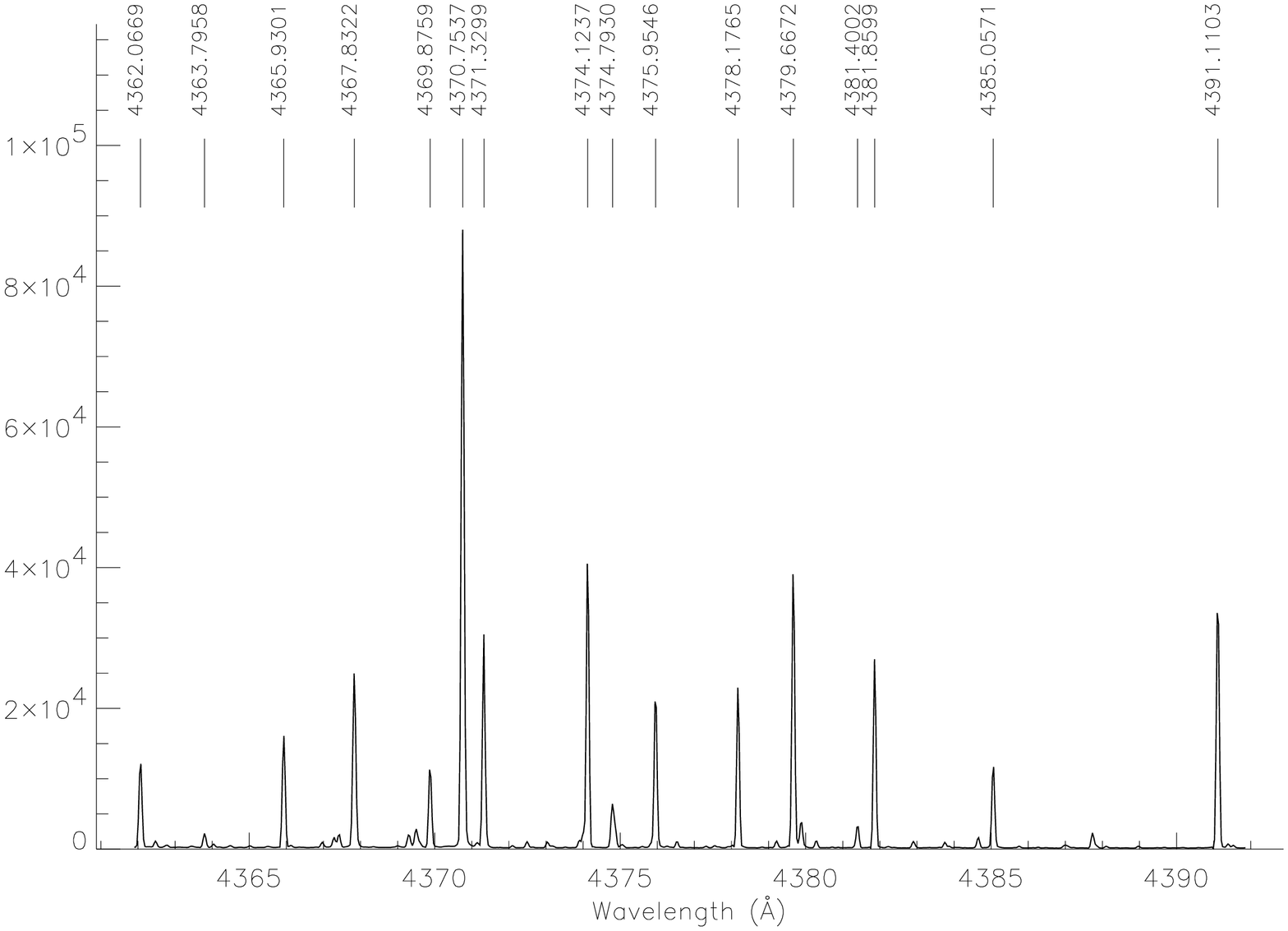}
\includegraphics[width=10cm,angle=90]{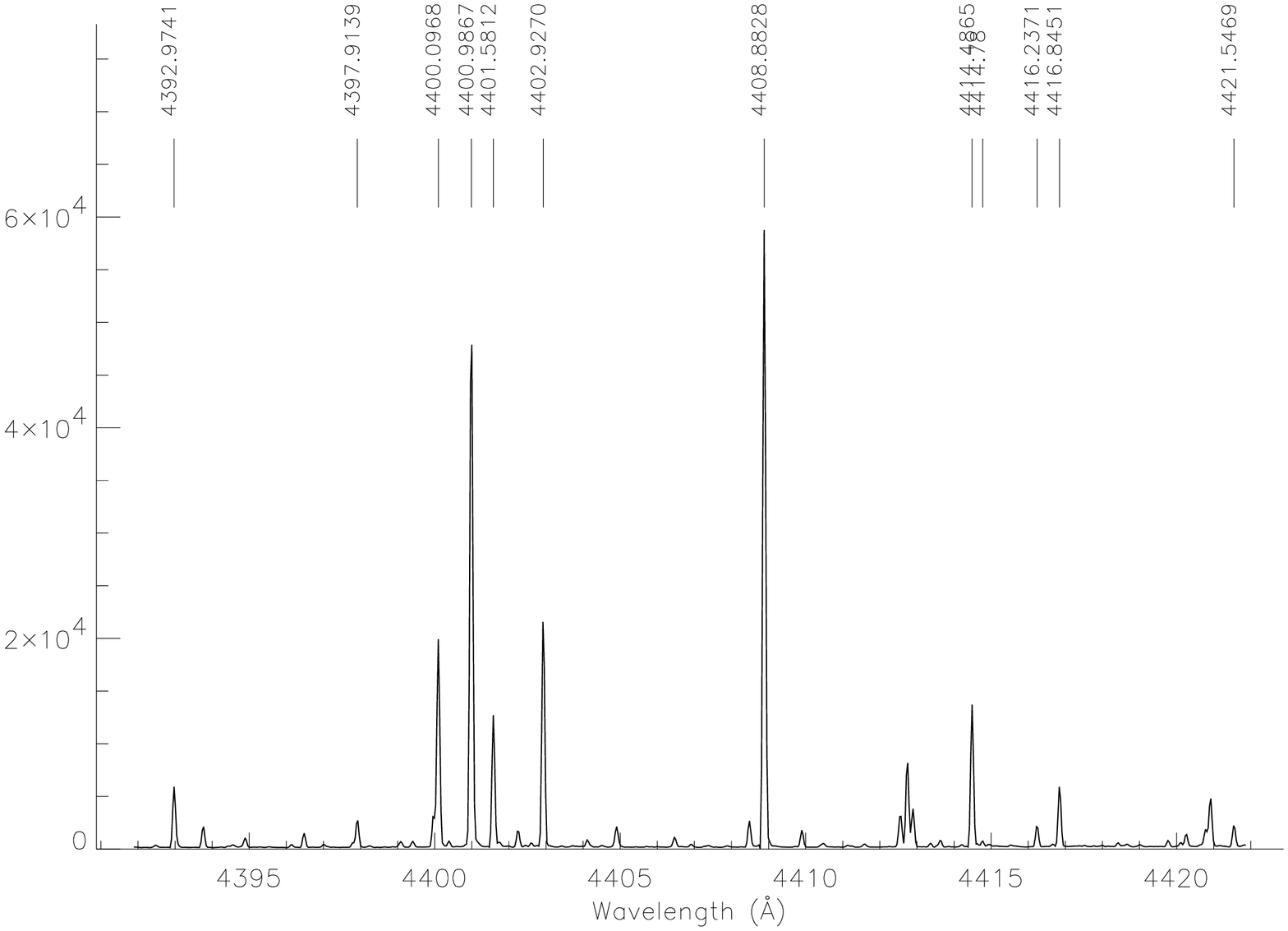}
\end{figure}
\clearpage
   
\begin{figure}
\centering
\includegraphics[width=10cm,angle=90]{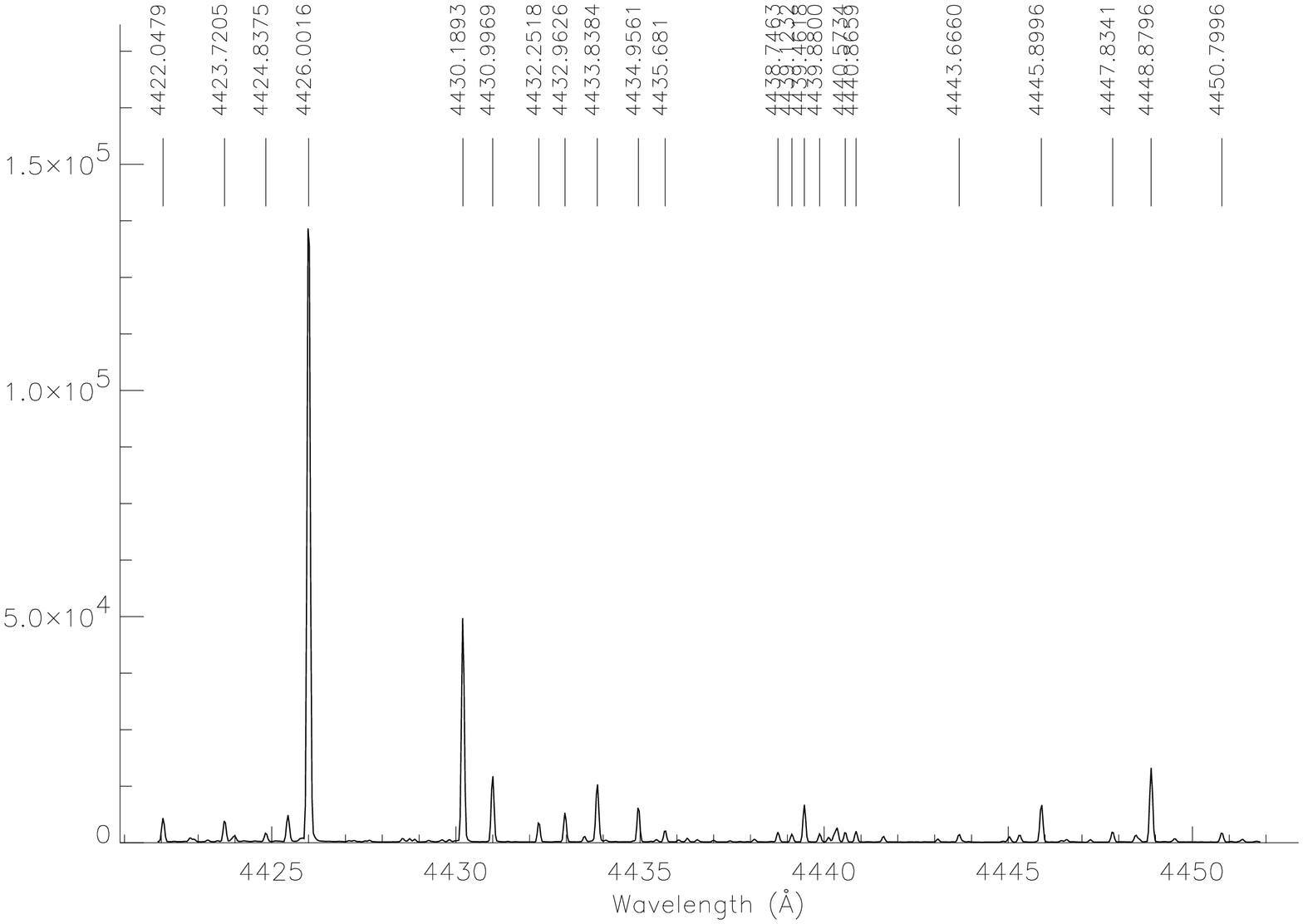}
\includegraphics[width=10cm,angle=90]{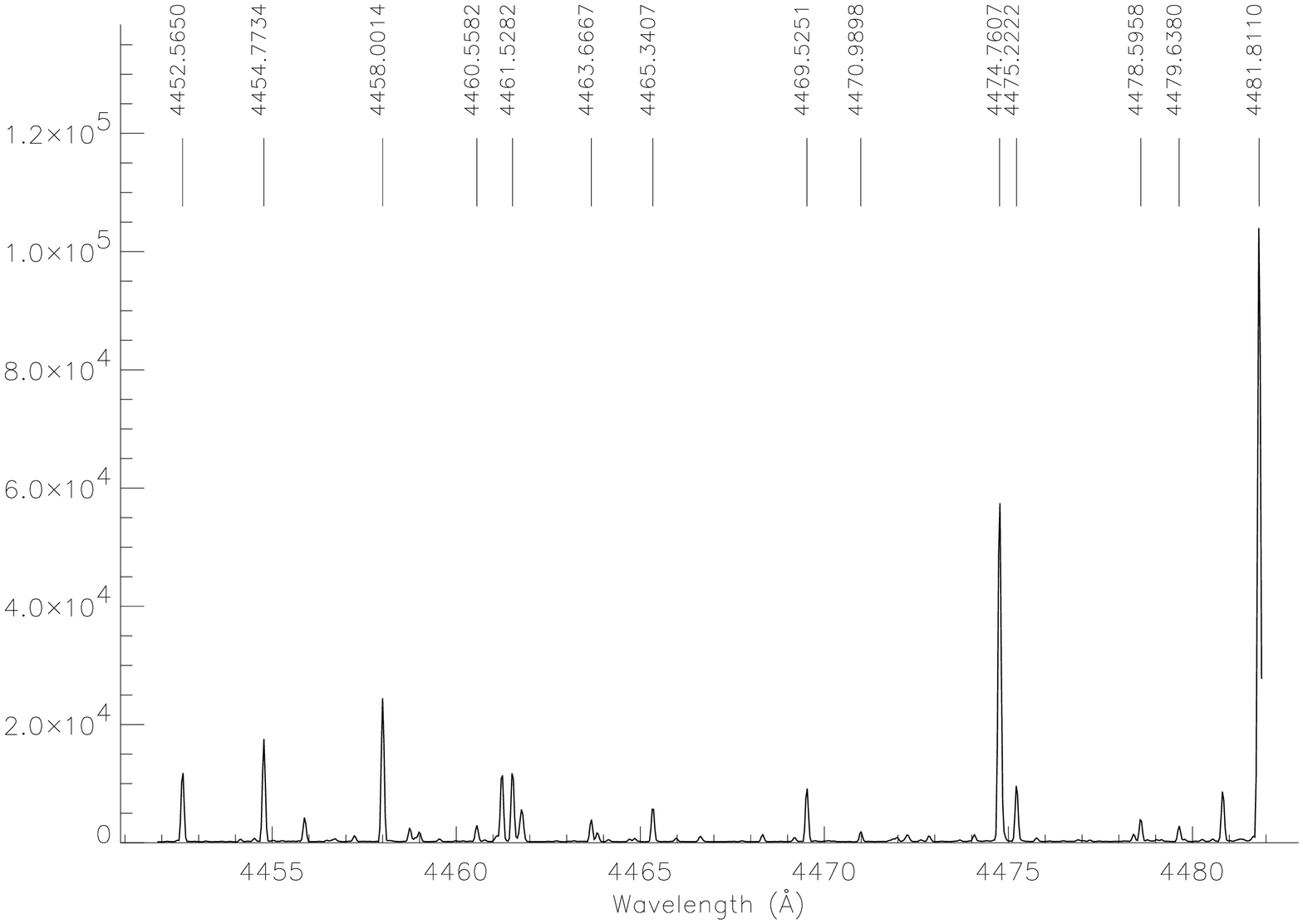}
\end{figure}
\clearpage
   
\begin{figure}
\centering
\includegraphics[width=10cm,angle=90]{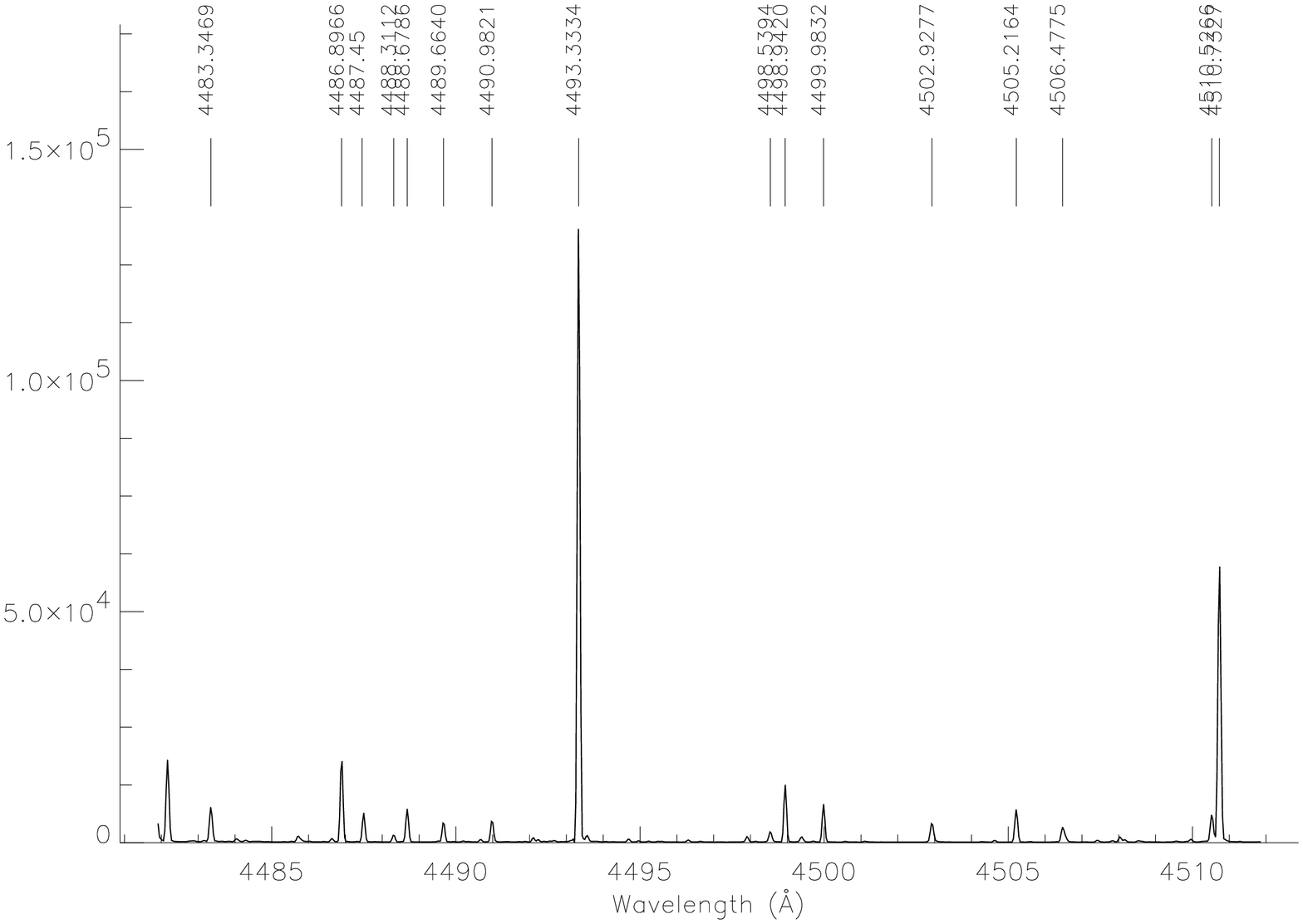}
\includegraphics[width=10cm,angle=90]{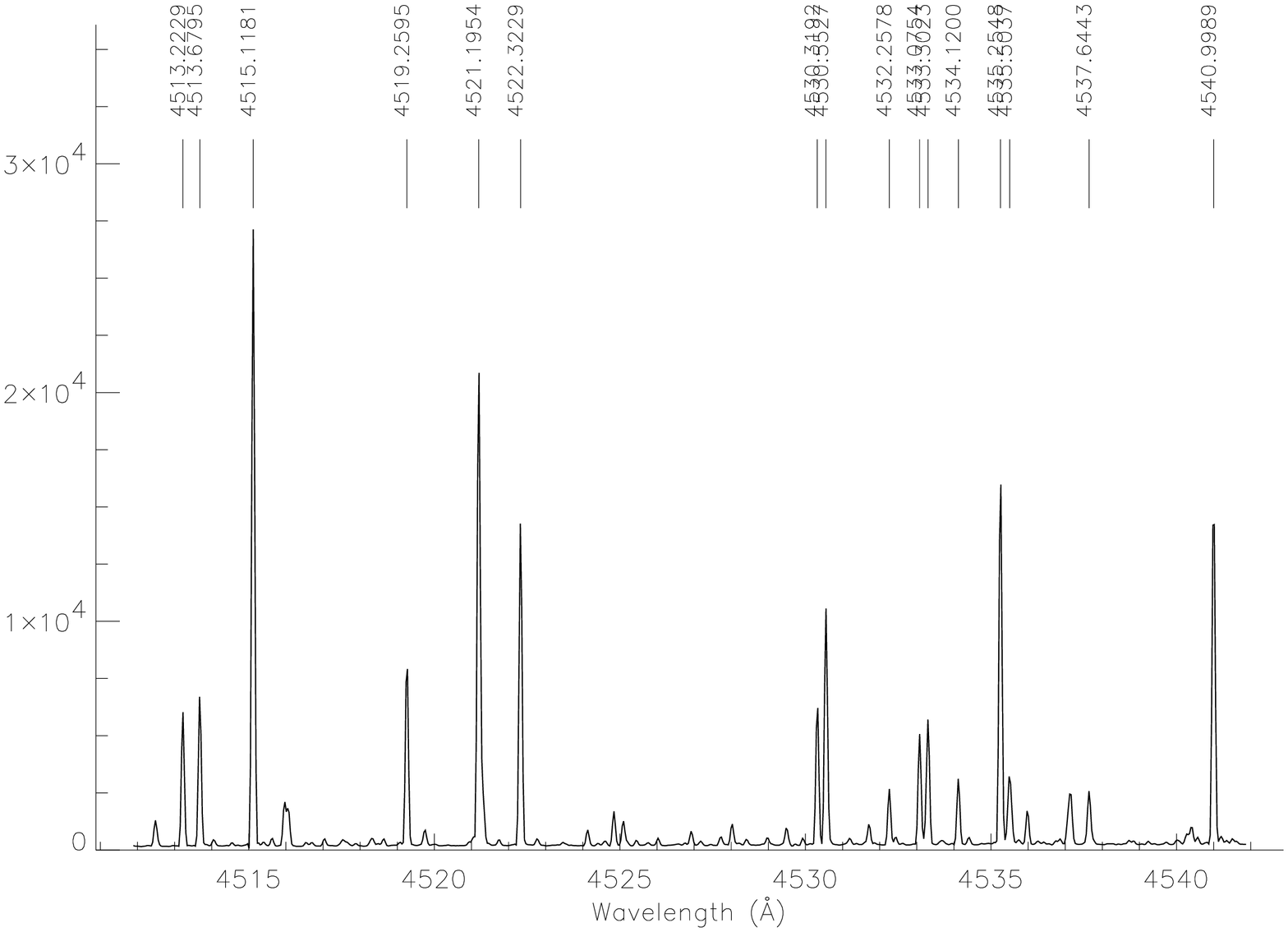}
\end{figure}
\clearpage
   
\begin{figure}
\centering
\includegraphics[width=10cm,angle=90]{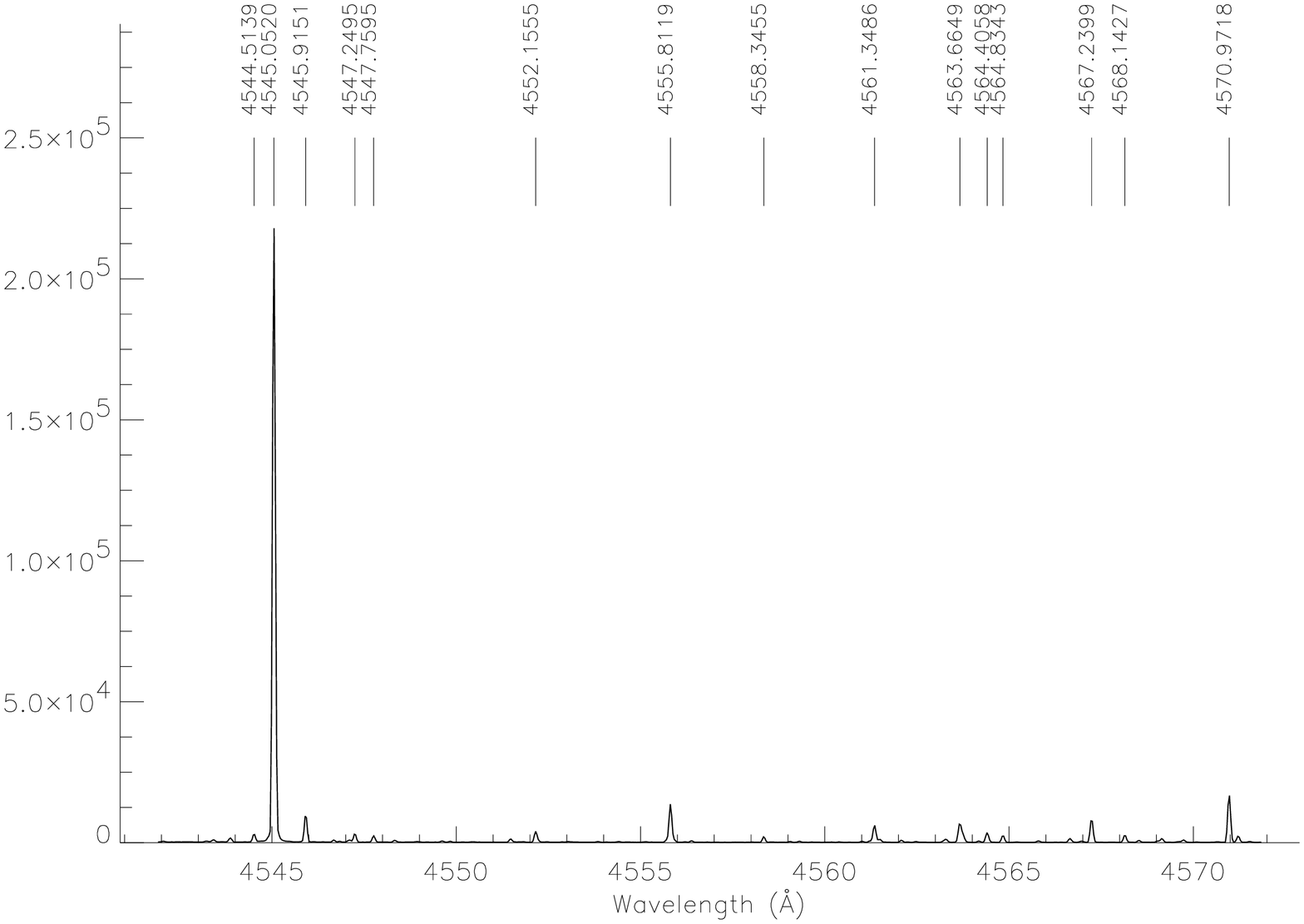}
\includegraphics[width=10cm,angle=90]{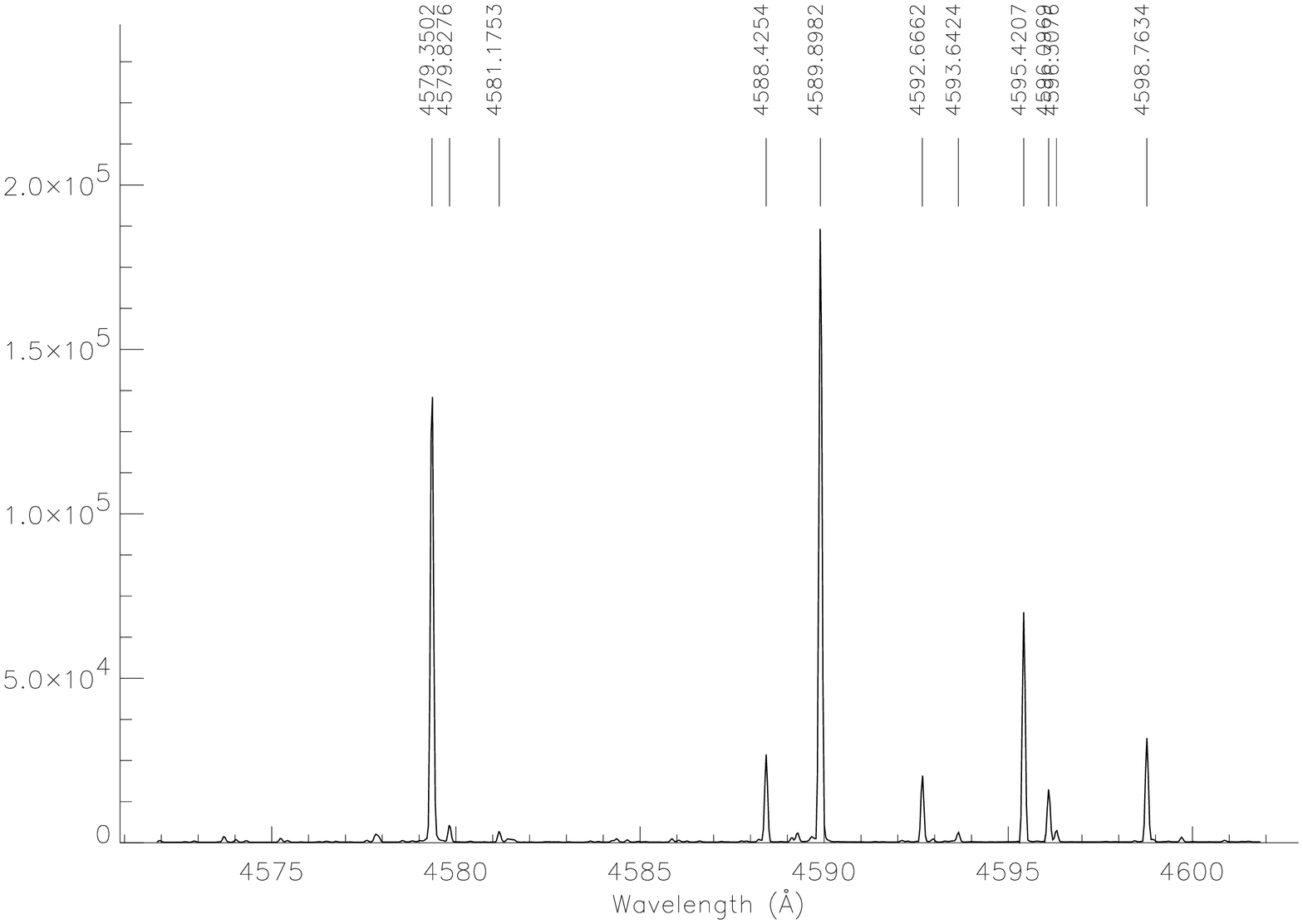}
\end{figure}
\clearpage
   
\begin{figure}
\centering
\includegraphics[width=10cm,angle=90]{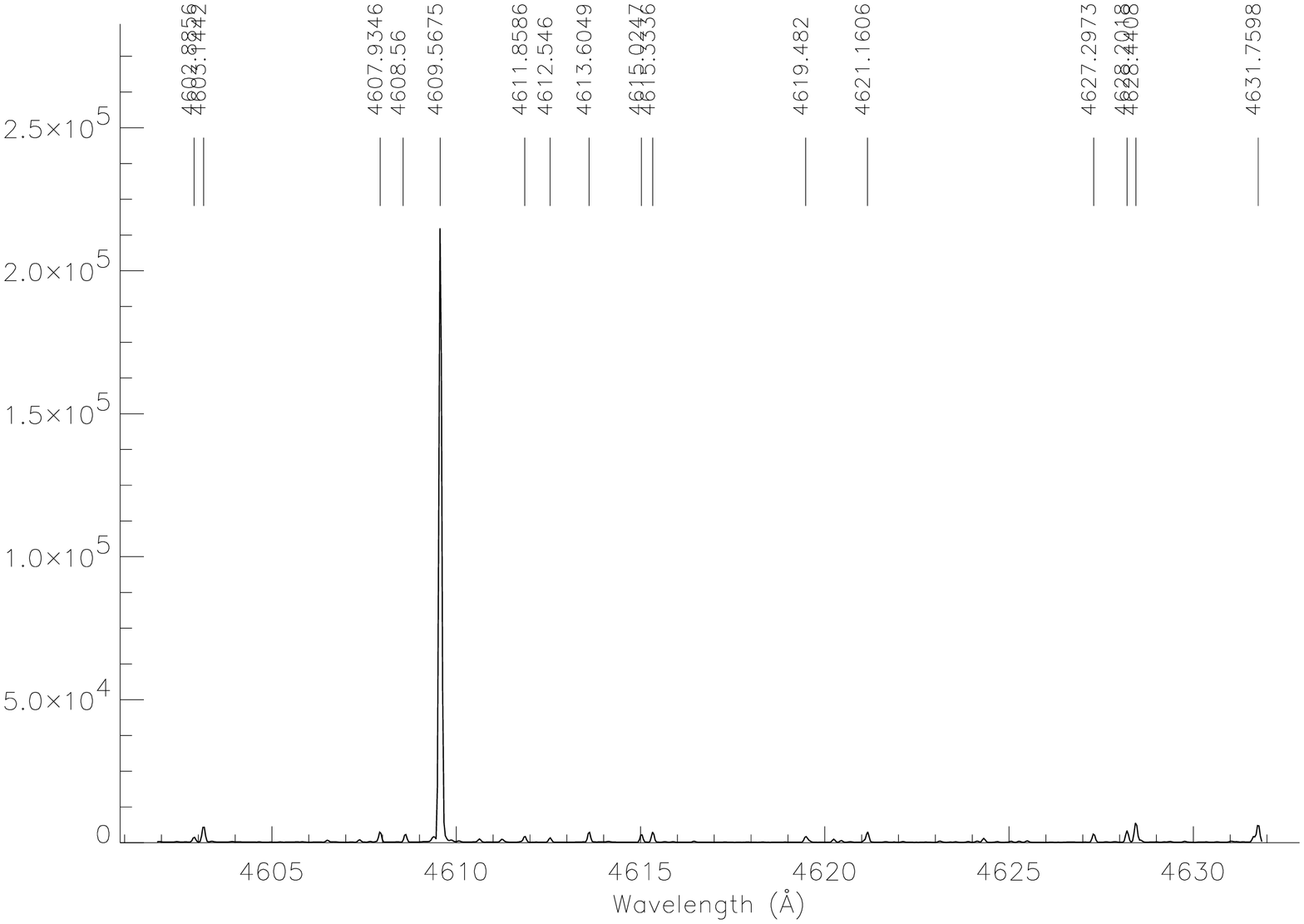}
\includegraphics[width=10cm,angle=90]{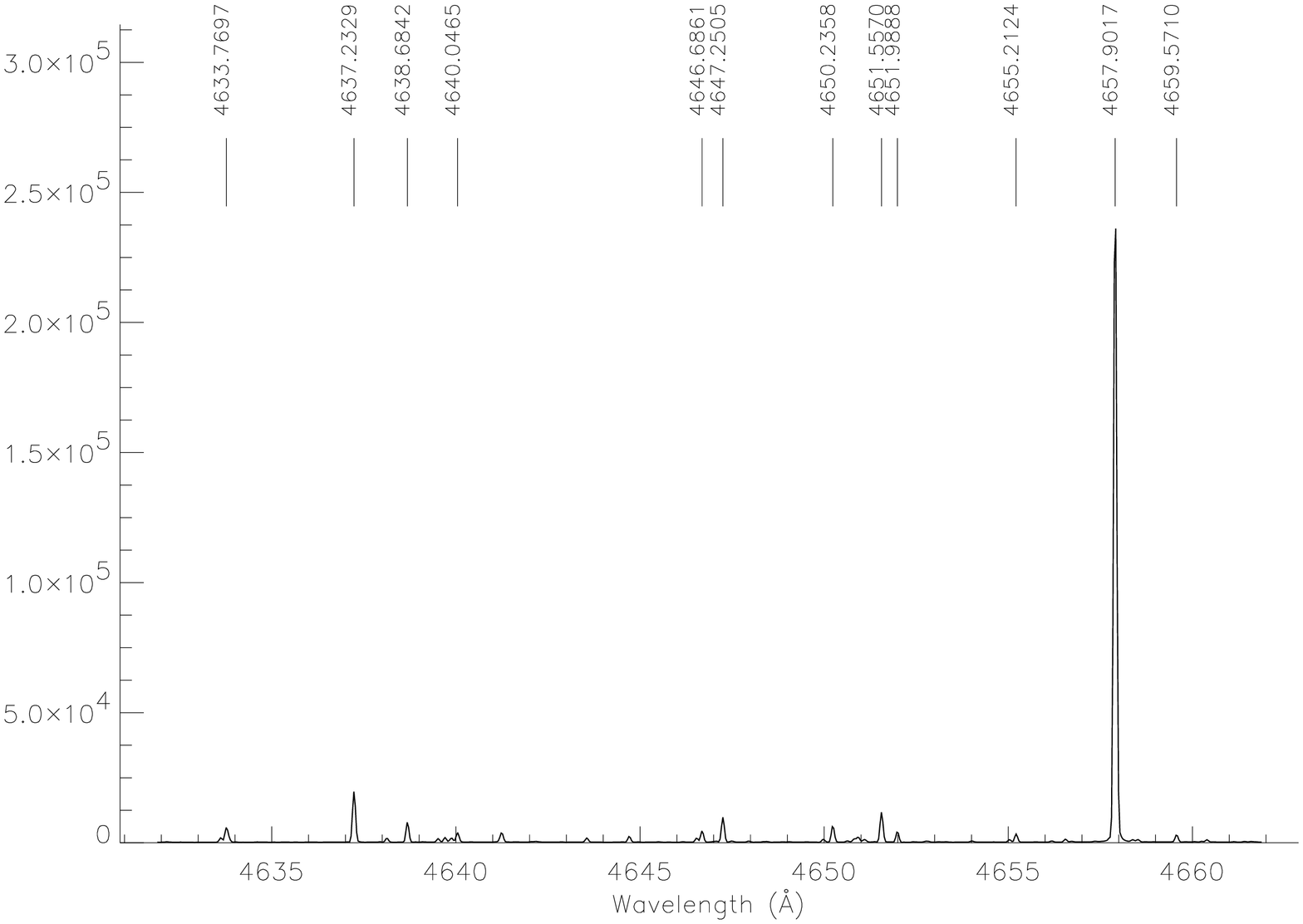}
\end{figure}
\clearpage
   
\begin{figure}
\centering
\includegraphics[width=10cm,angle=90]{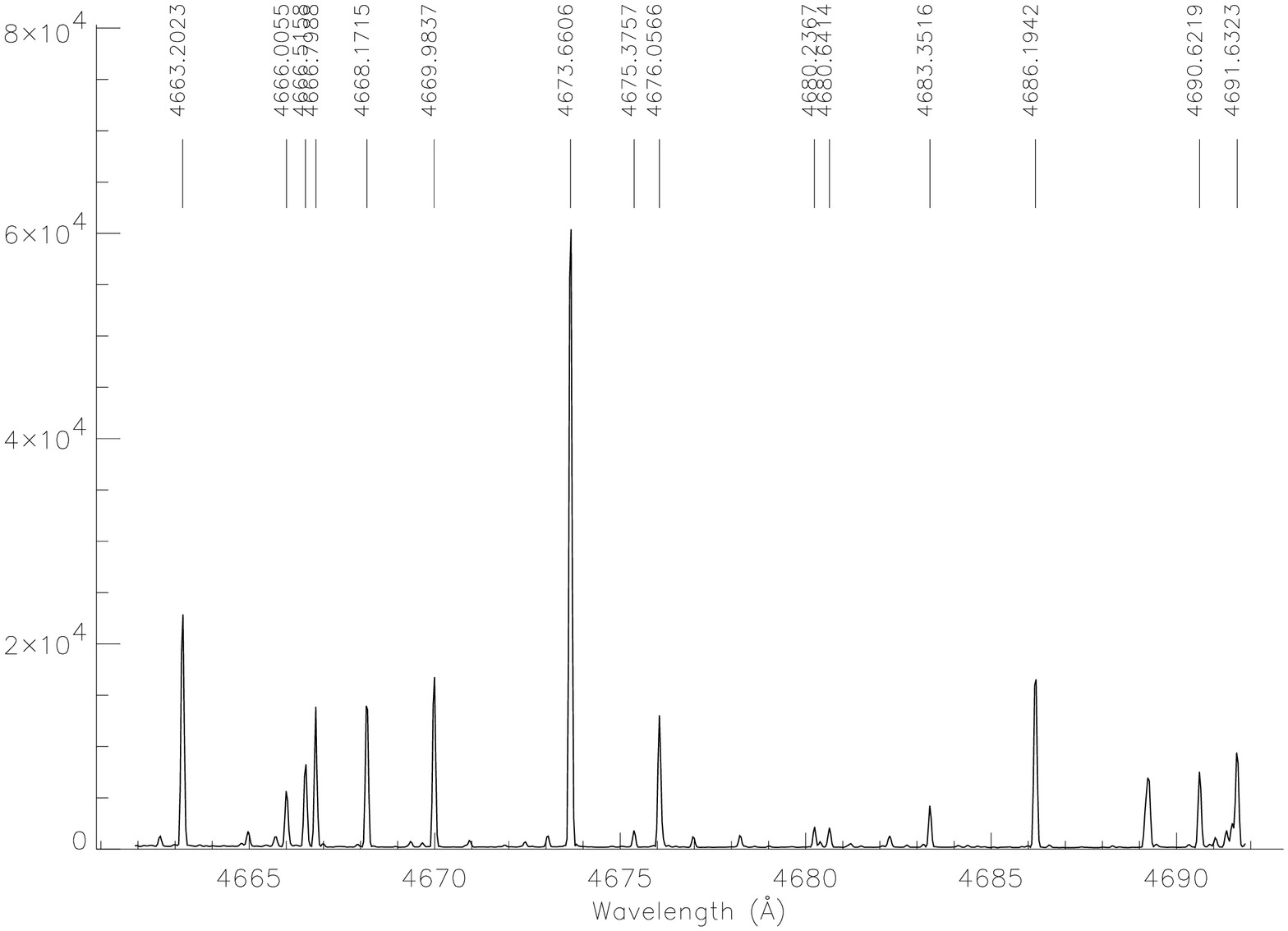}
\includegraphics[width=10cm,angle=90]{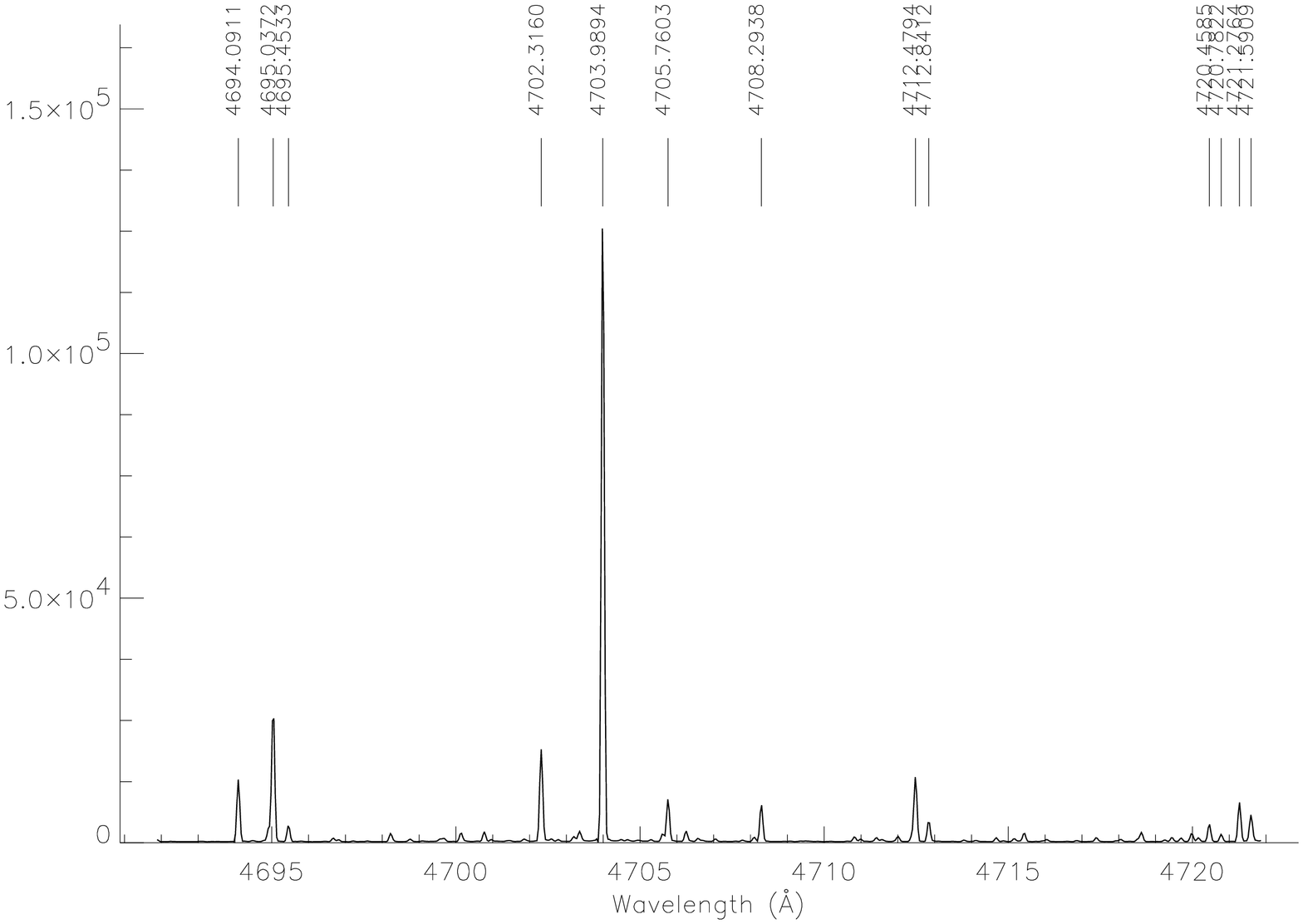}
\end{figure}
\clearpage
   
\begin{figure}
\centering
\includegraphics[width=10cm,angle=90]{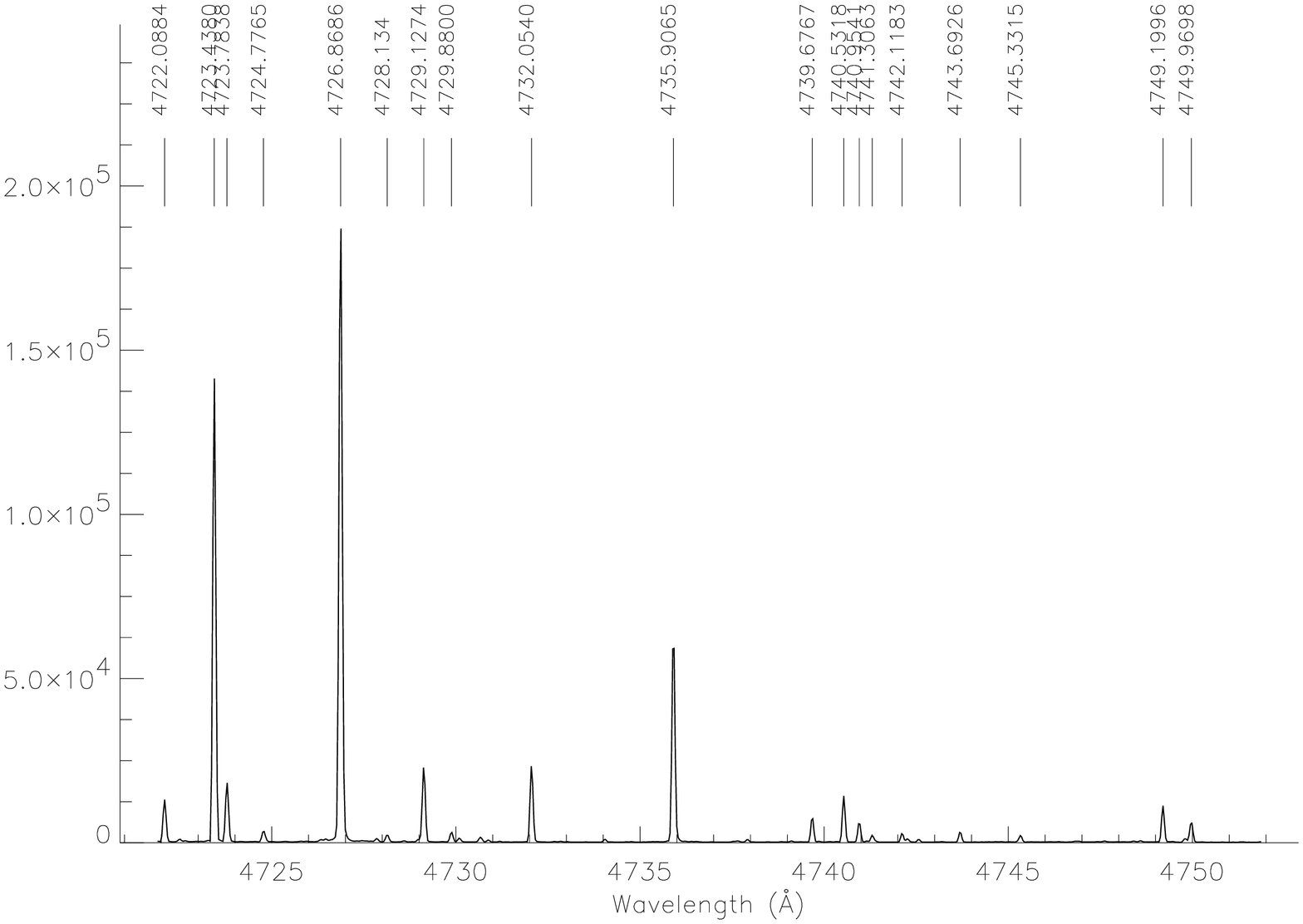}
\includegraphics[width=10cm,angle=90]{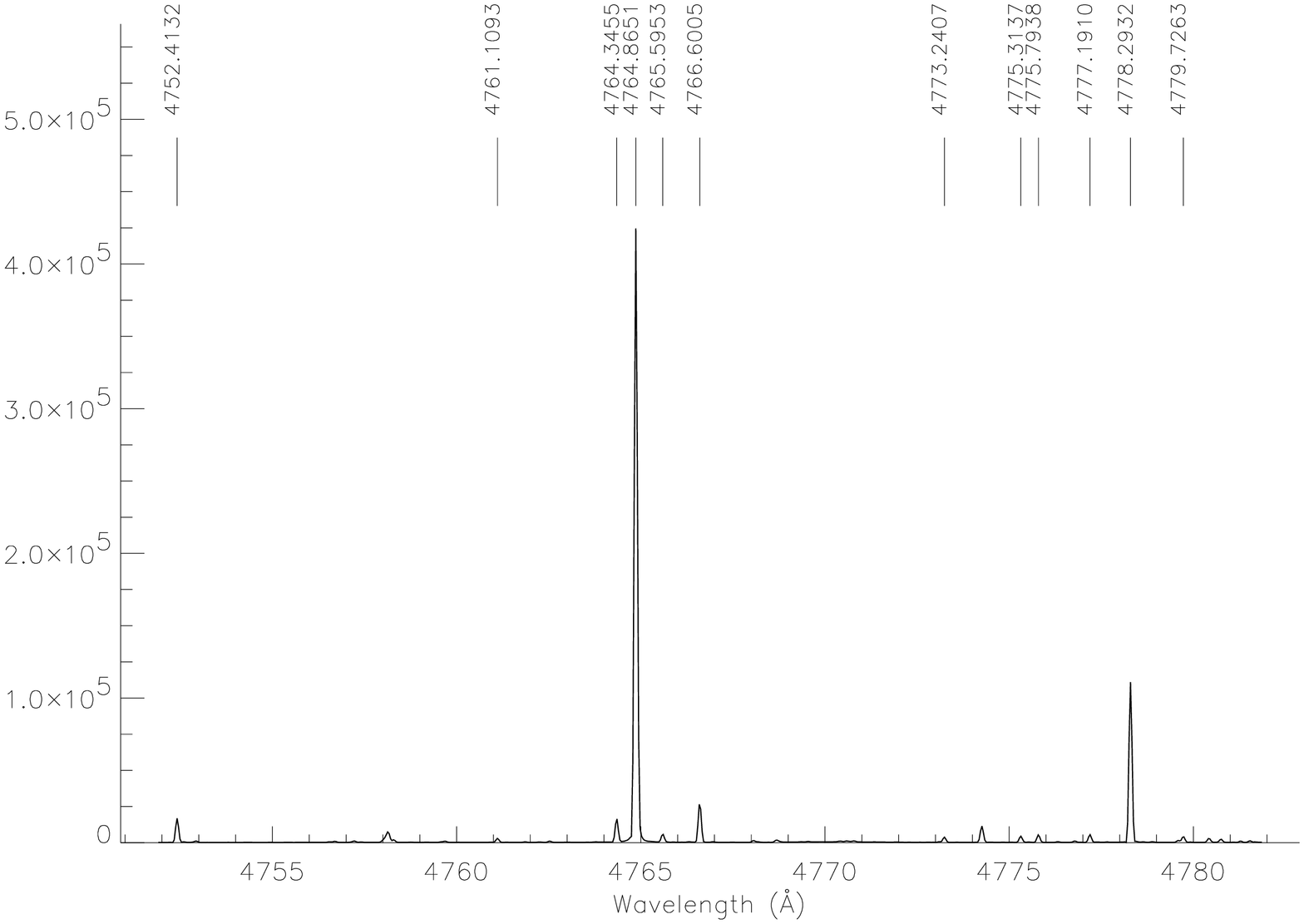}
\end{figure}
\clearpage
   
\begin{figure}
\centering
\includegraphics[width=10cm,angle=90]{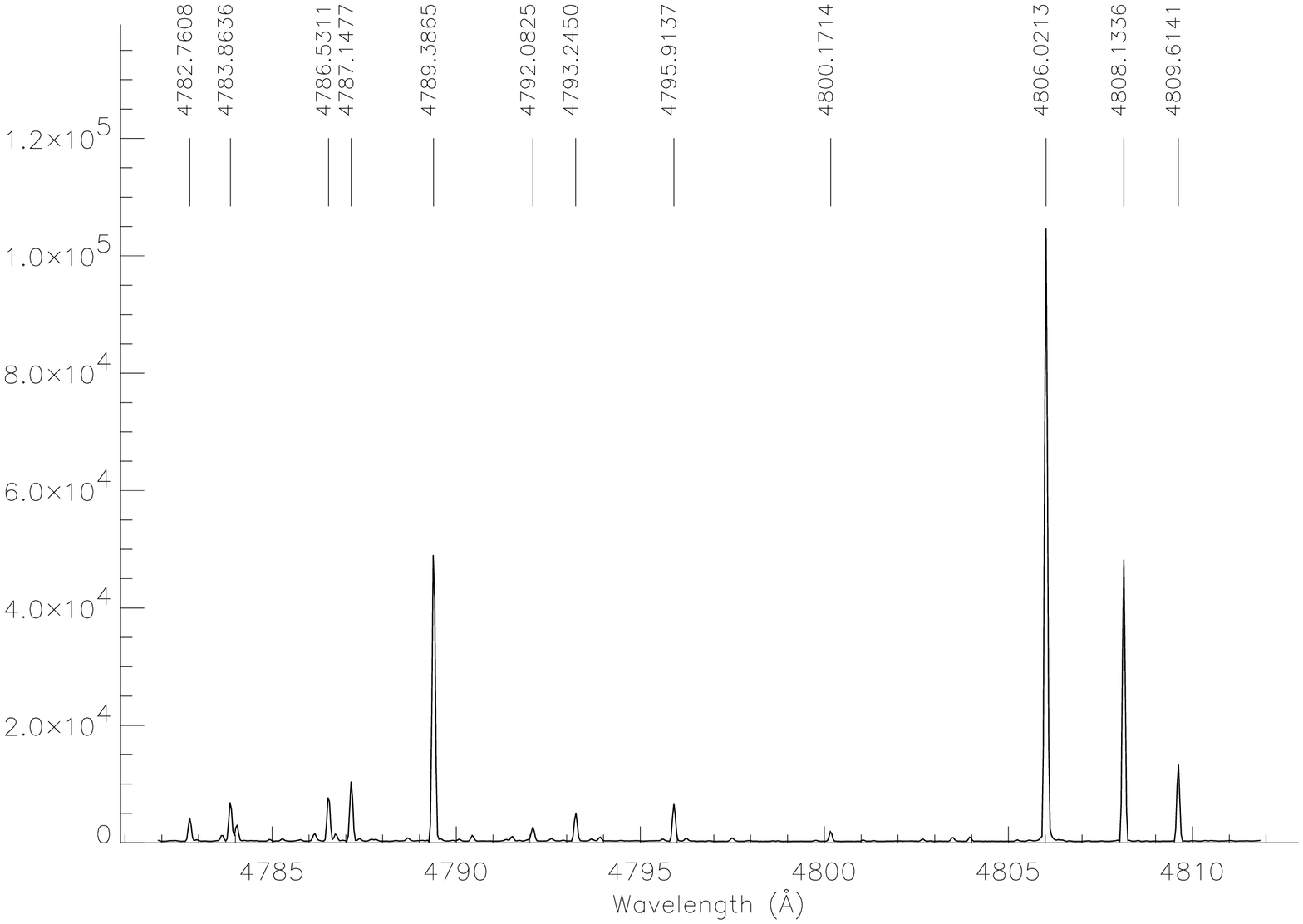}
\includegraphics[width=10cm,angle=90]{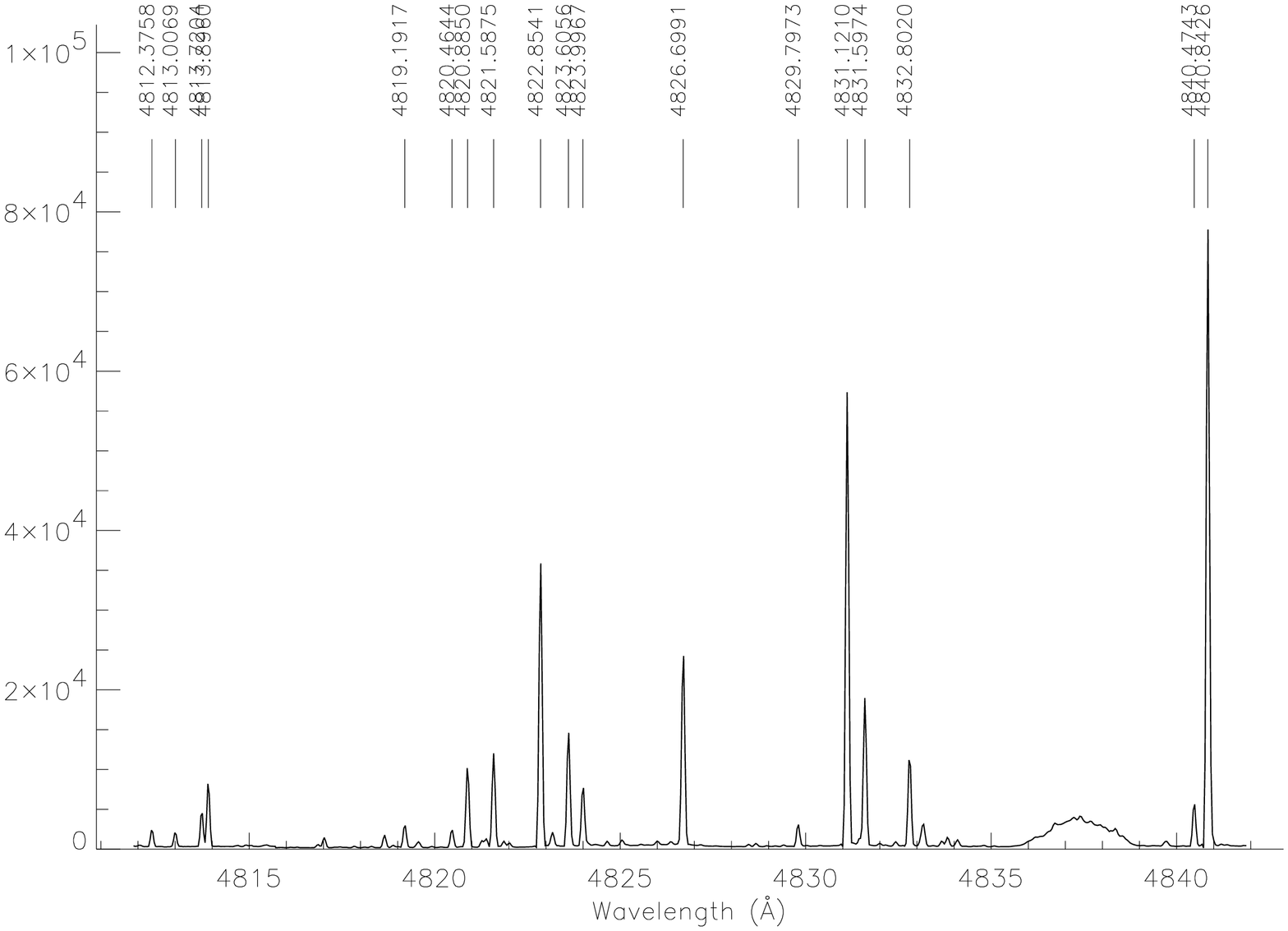}
\end{figure}
\clearpage
   
\begin{figure}
\centering
\includegraphics[width=10cm,angle=90]{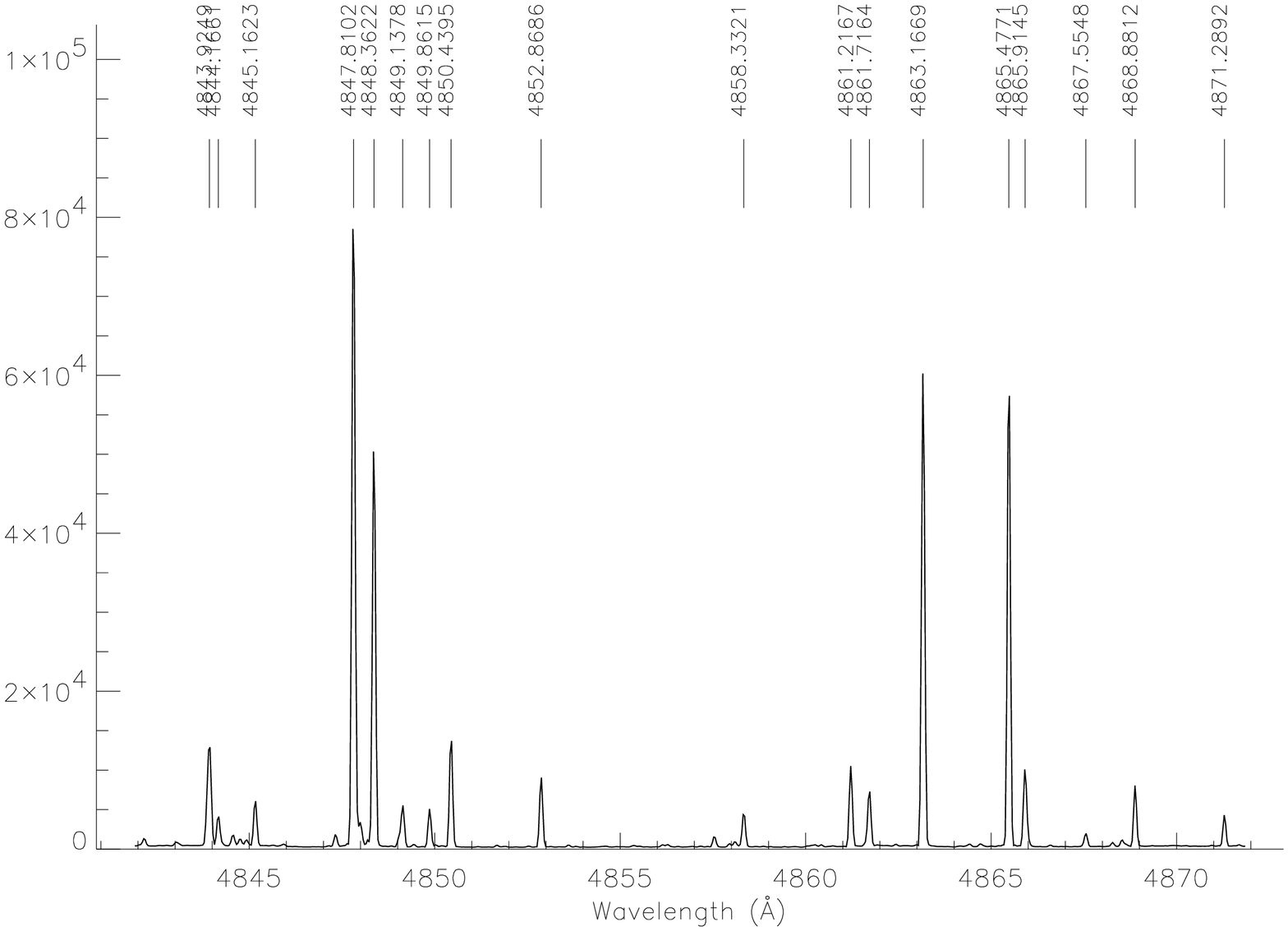}
\includegraphics[width=10cm,angle=90]{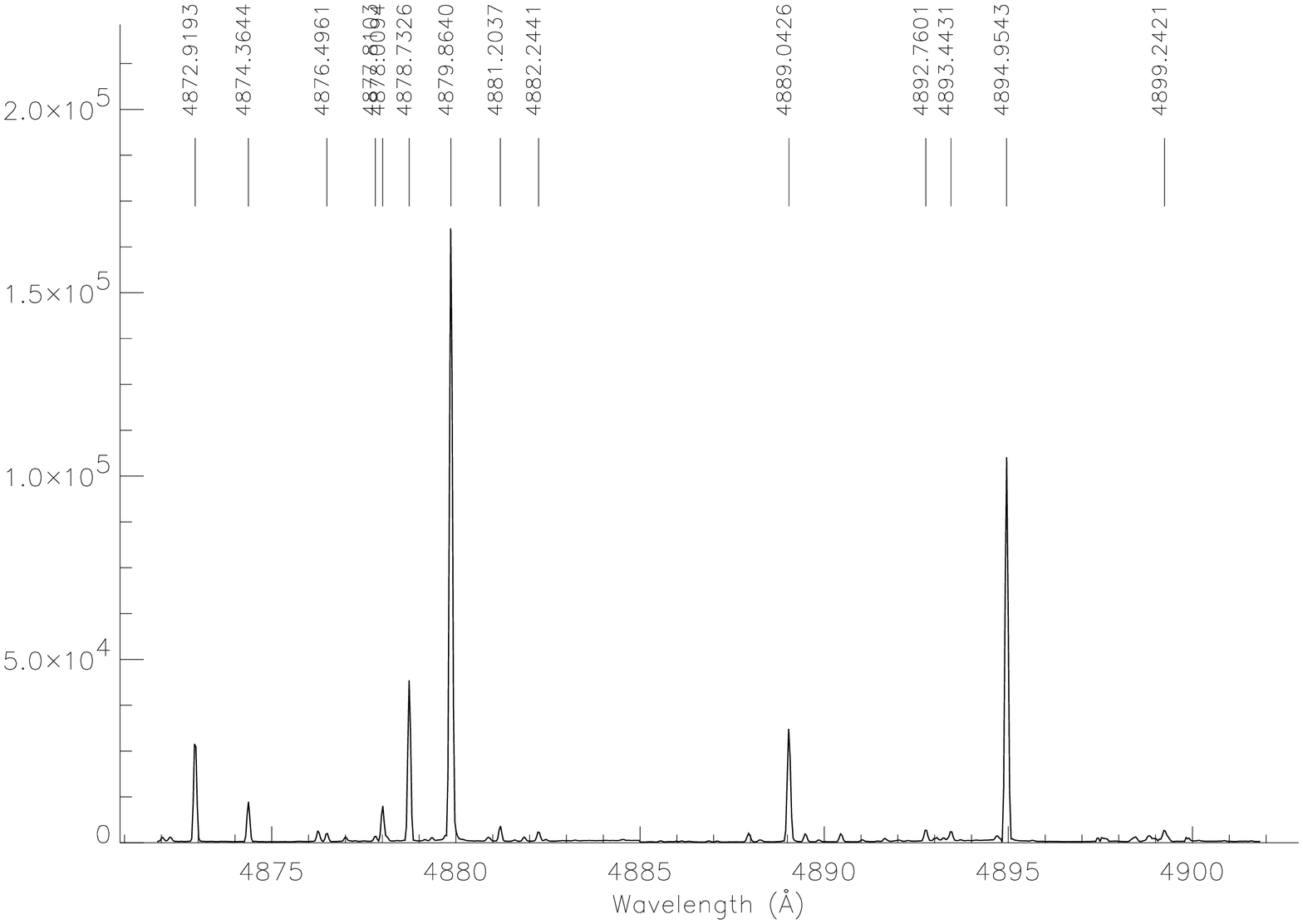}
\end{figure}
\clearpage
   
\begin{figure}
\centering
\includegraphics[width=10cm,angle=90]{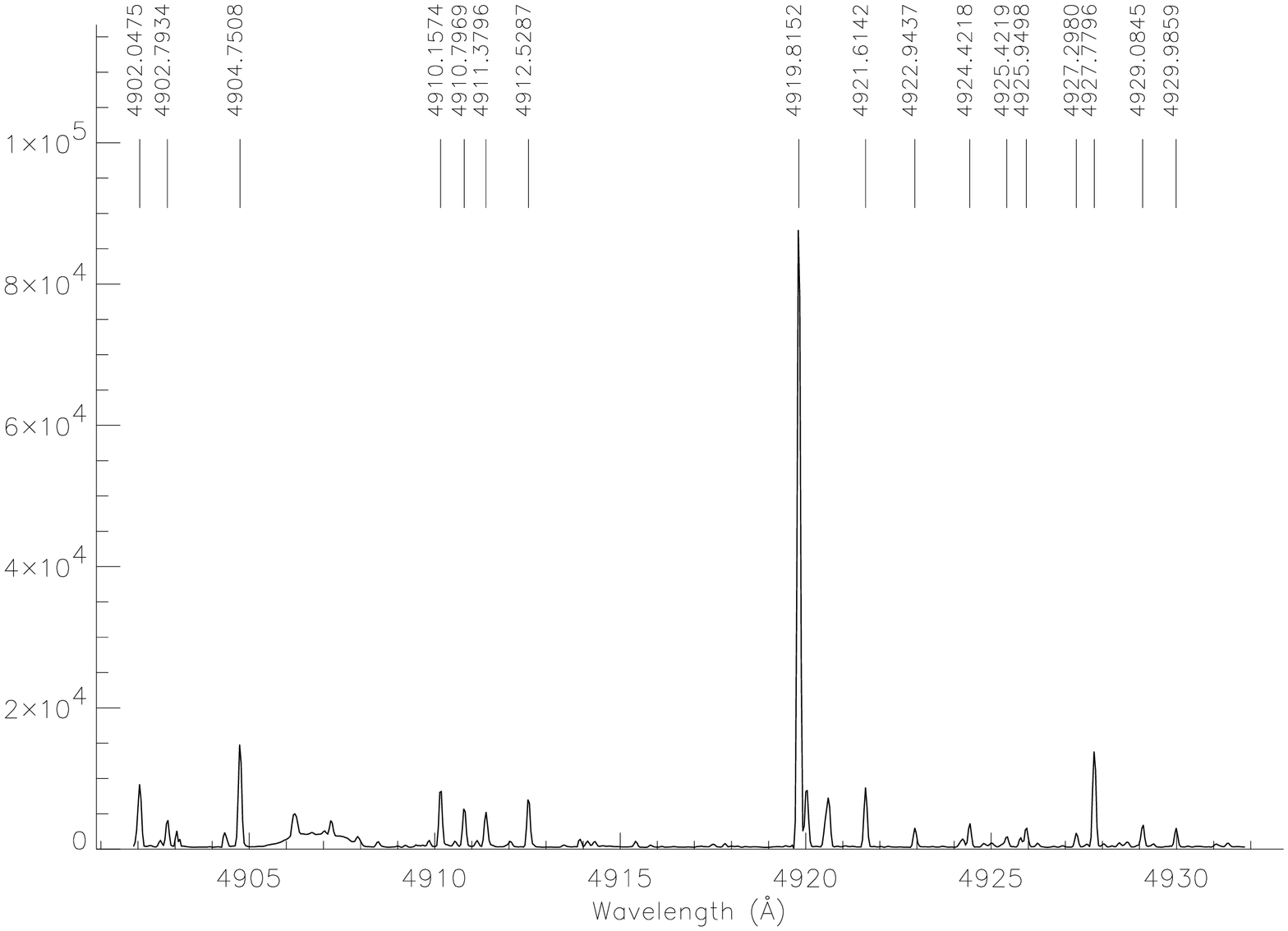}
\includegraphics[width=10cm,angle=90]{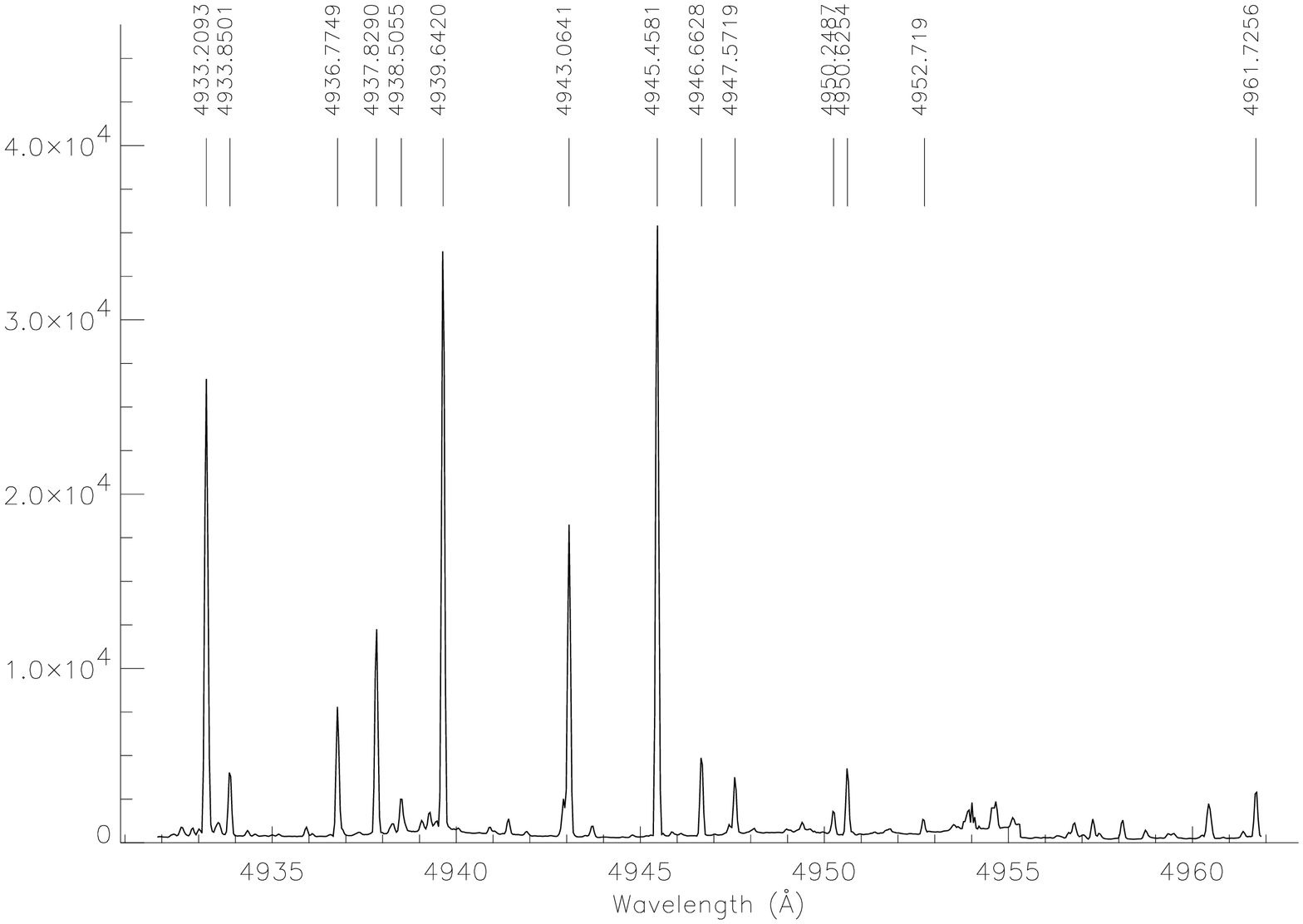}
\end{figure}
\clearpage
   
\begin{figure}
\centering
\includegraphics[width=10cm,angle=90]{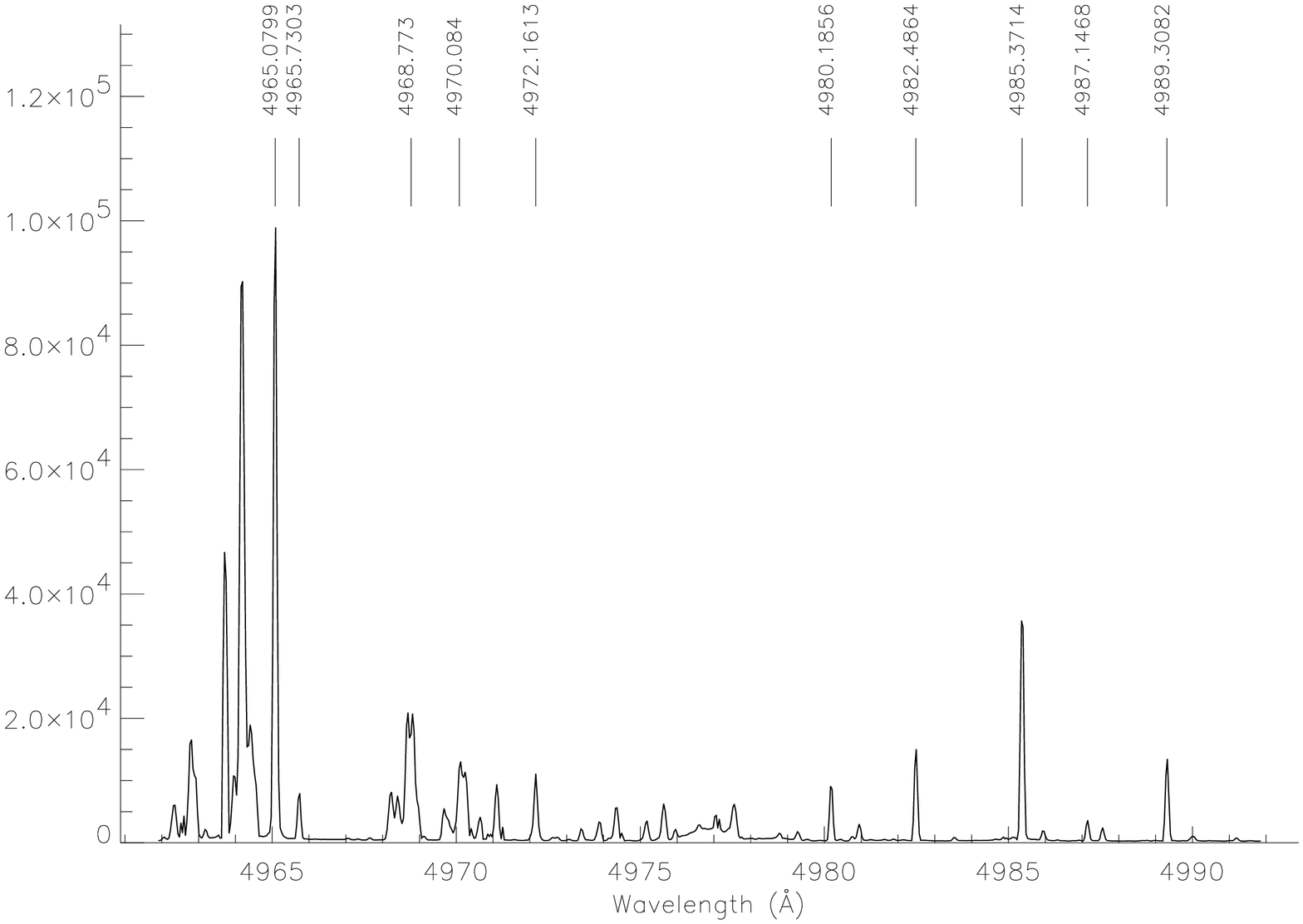}
\includegraphics[width=10cm,angle=90]{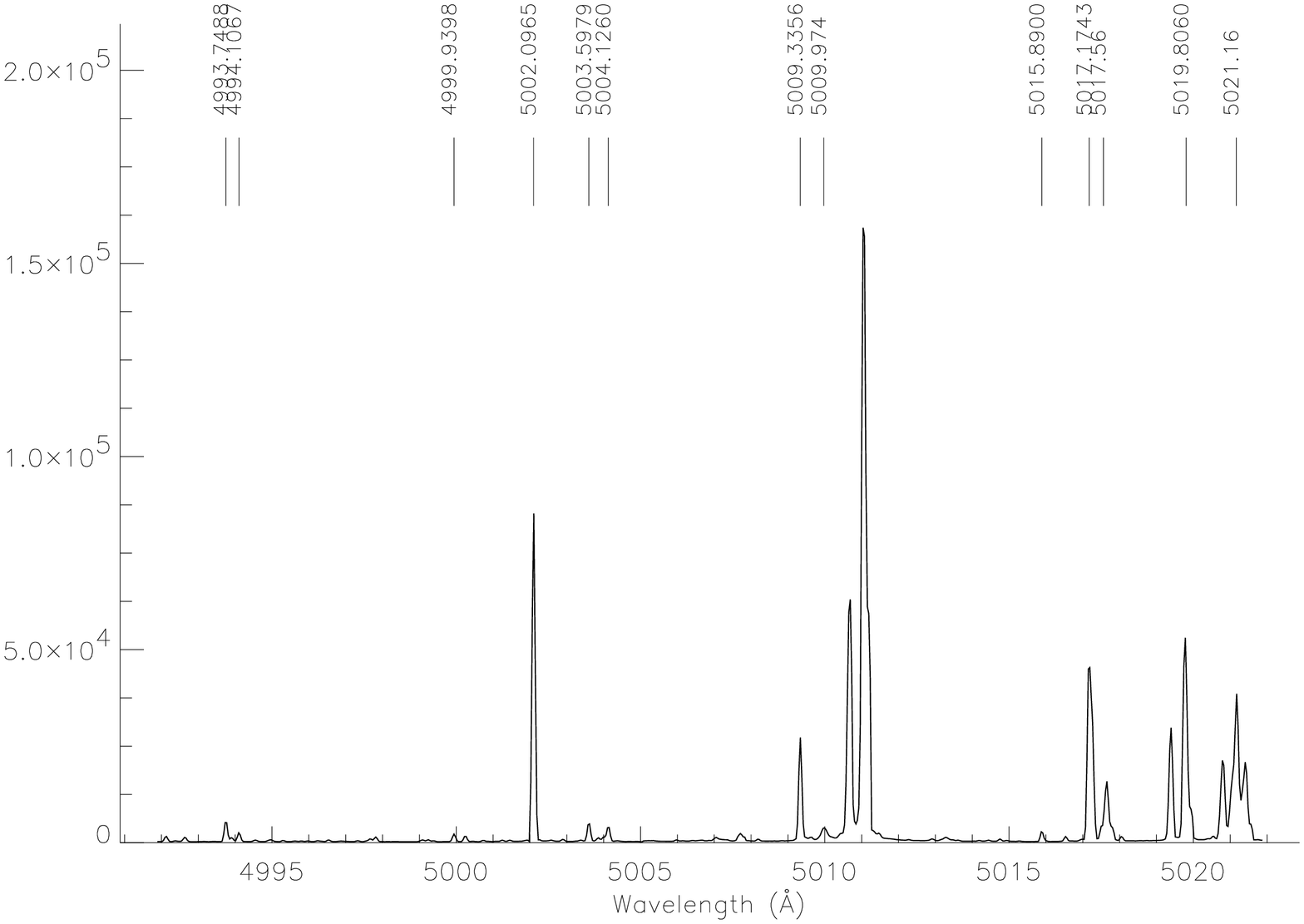}
\end{figure}
\clearpage
   
\begin{figure}
\centering
\includegraphics[width=10cm,angle=90]{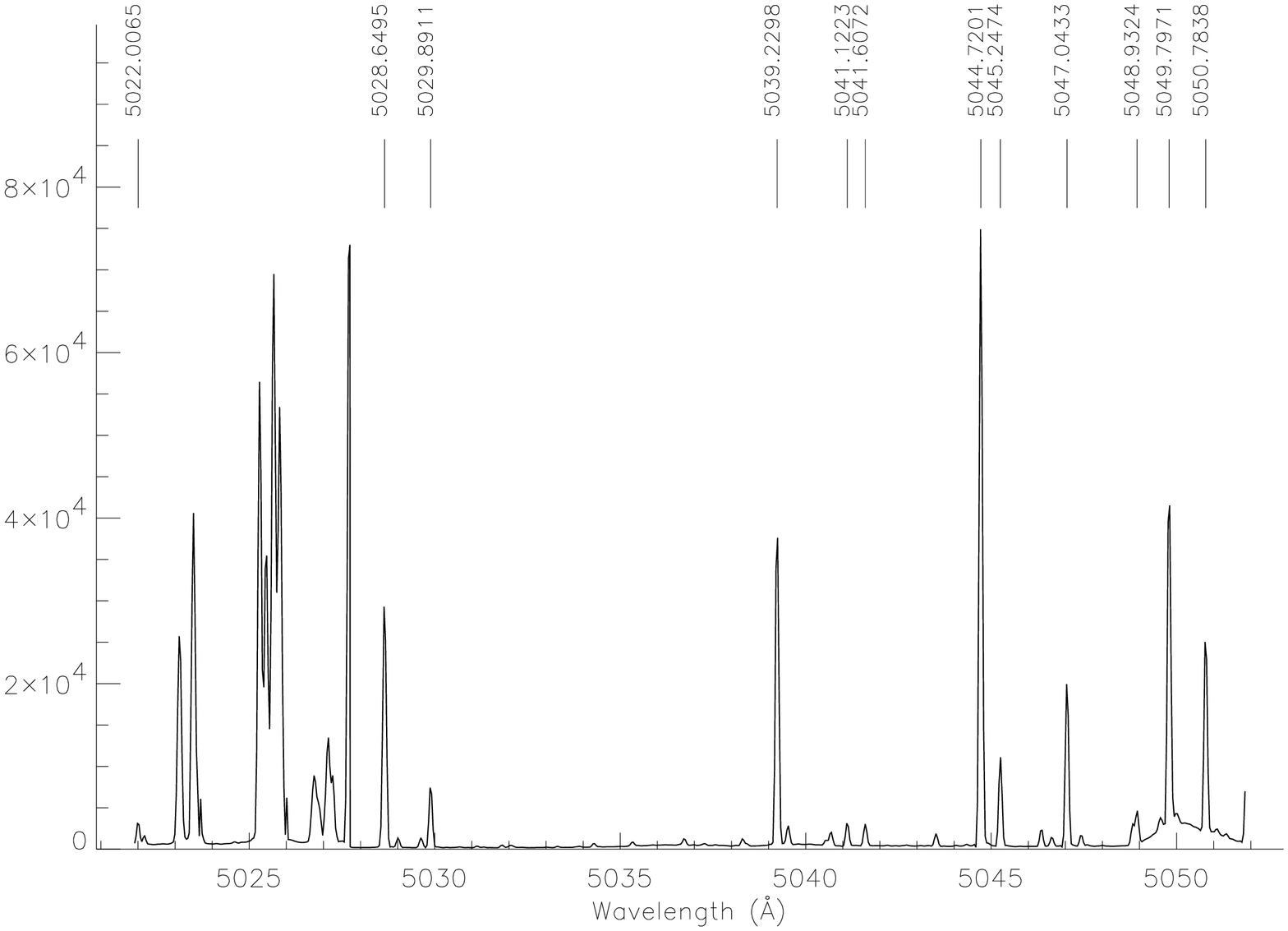}
\includegraphics[width=10cm,angle=90]{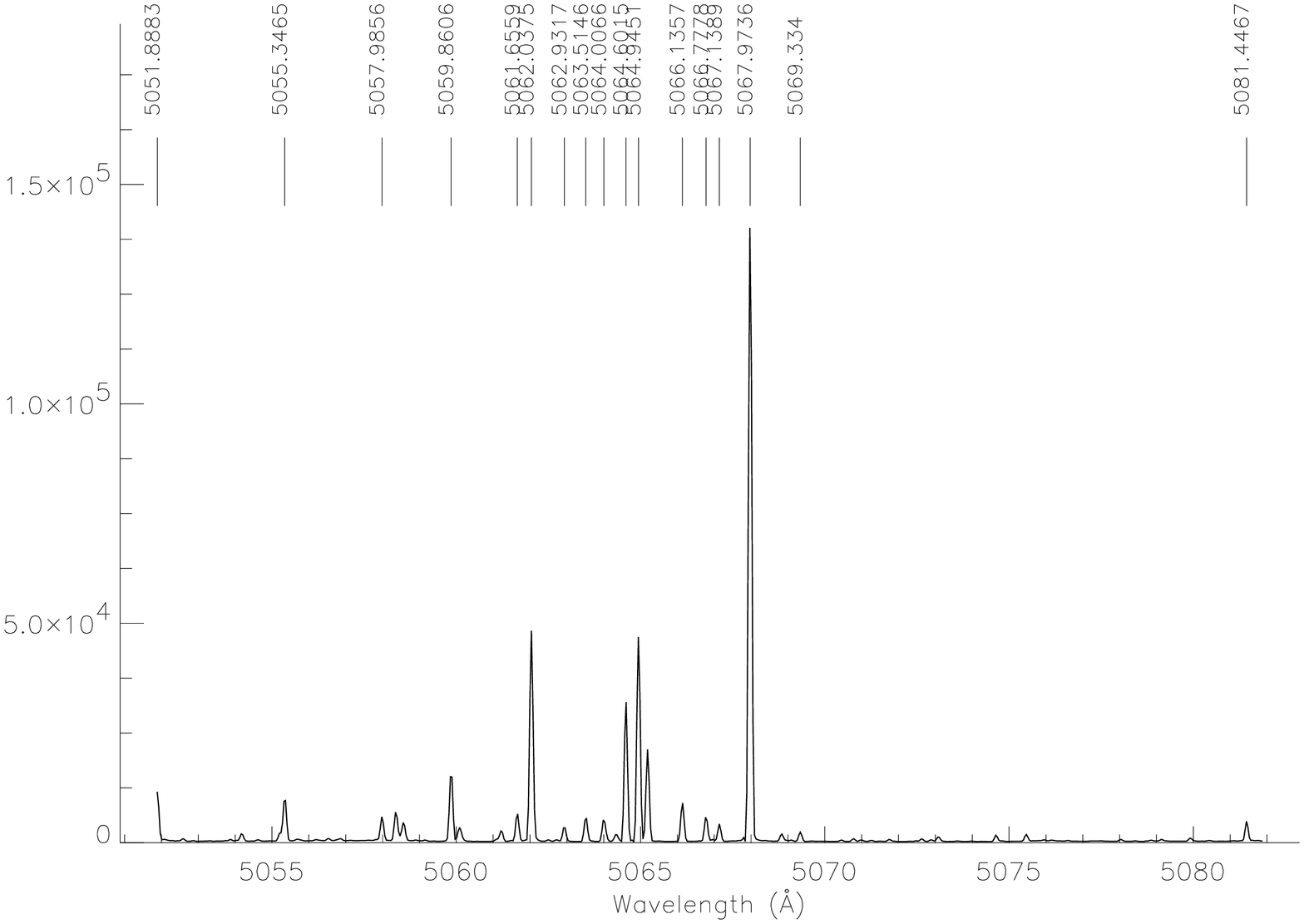}
\end{figure}
\clearpage
   
\begin{figure}
\centering
\includegraphics[width=10cm,angle=90]{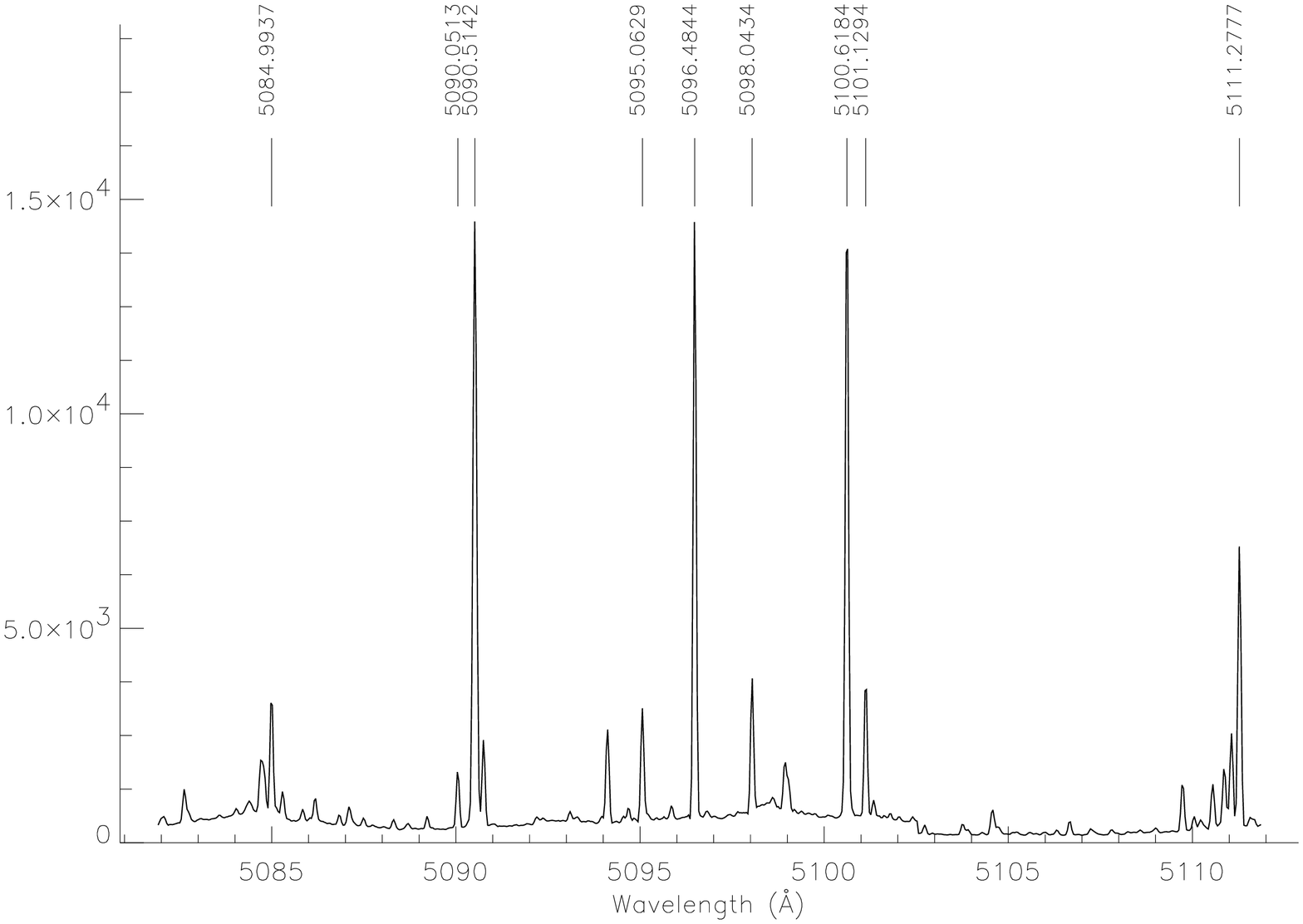}
\includegraphics[width=10cm,angle=90]{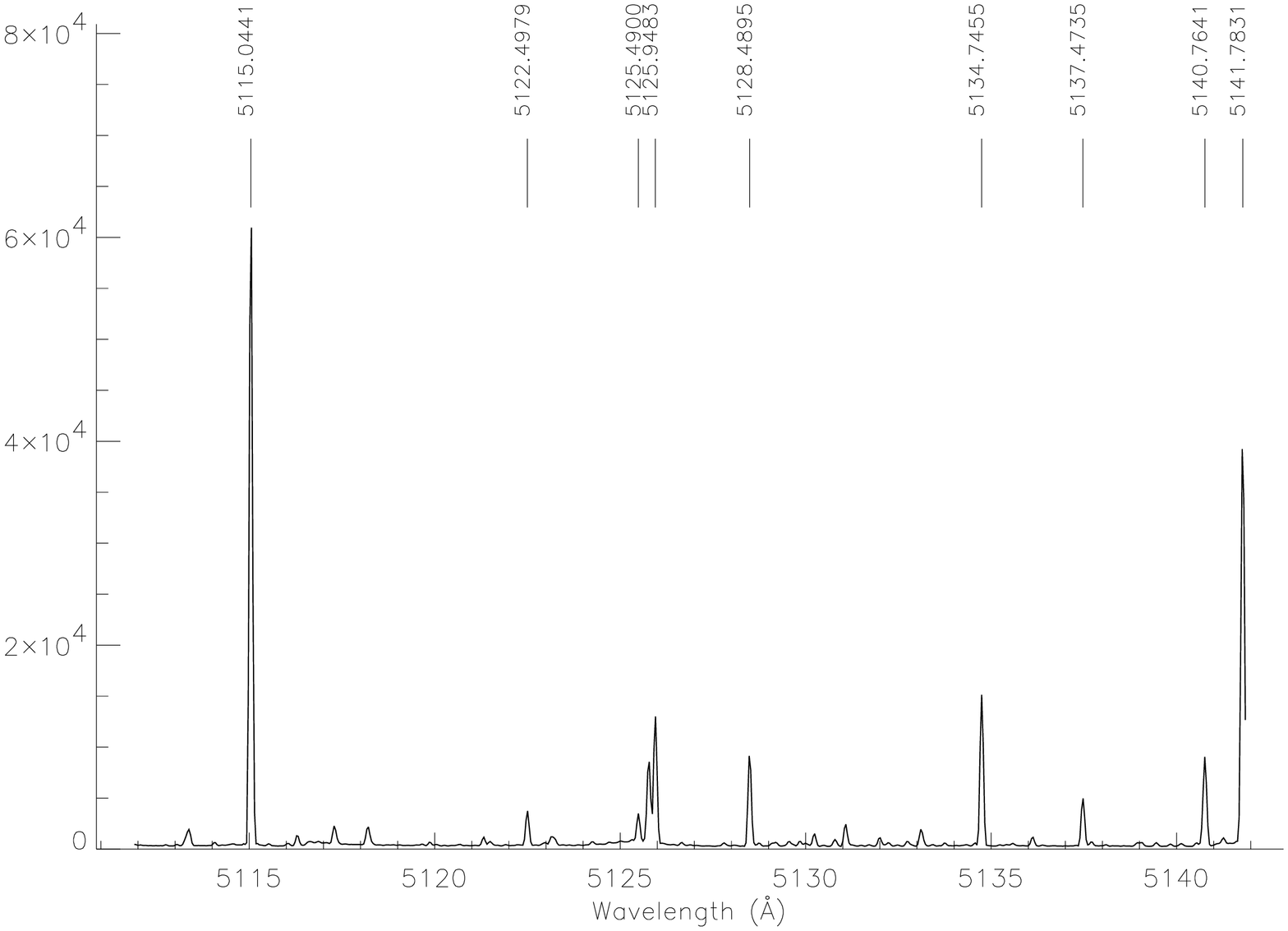}
\end{figure}
\clearpage
   
\begin{figure}
\centering
\includegraphics[width=10cm,angle=90]{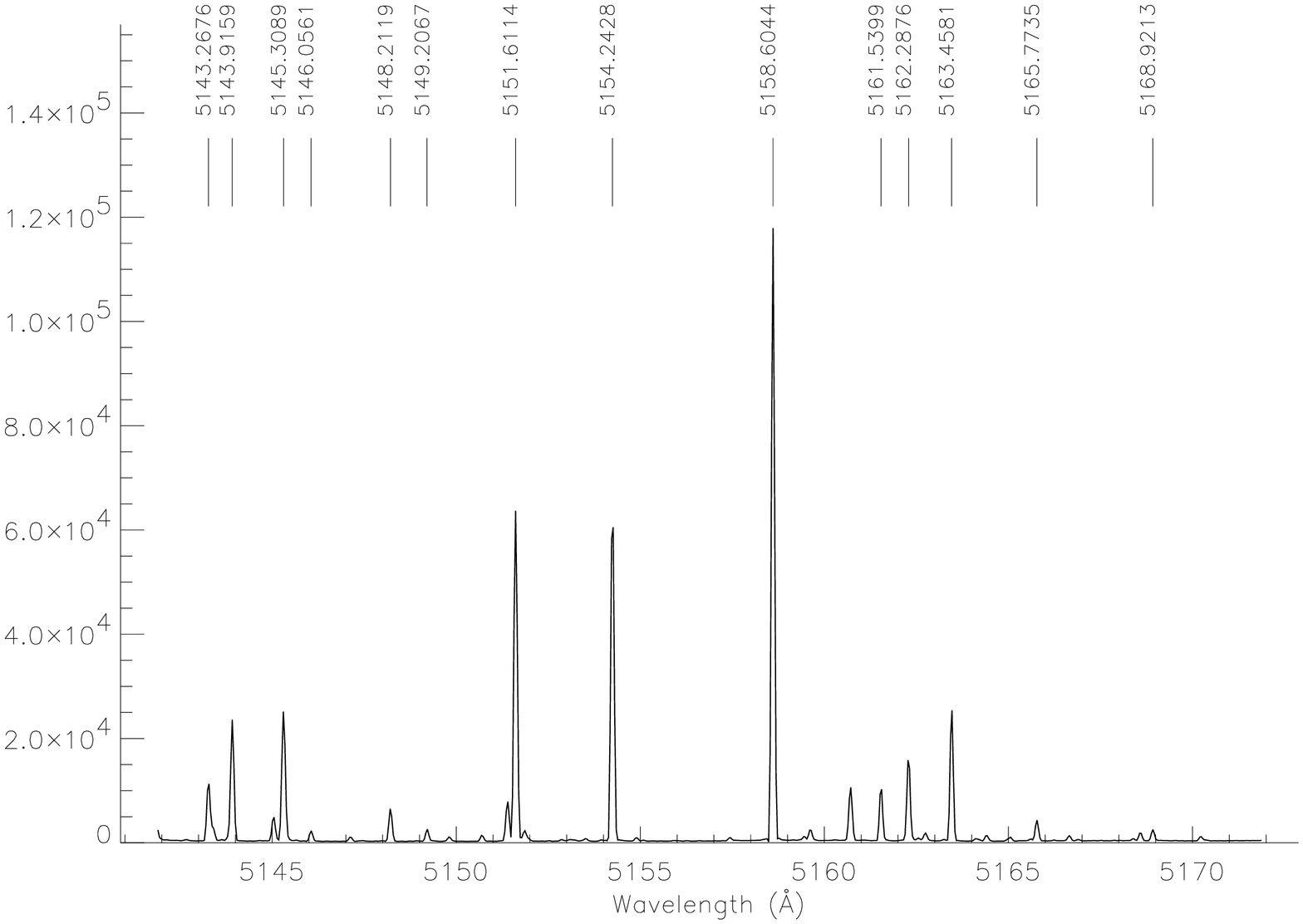}
\includegraphics[width=10cm,angle=90]{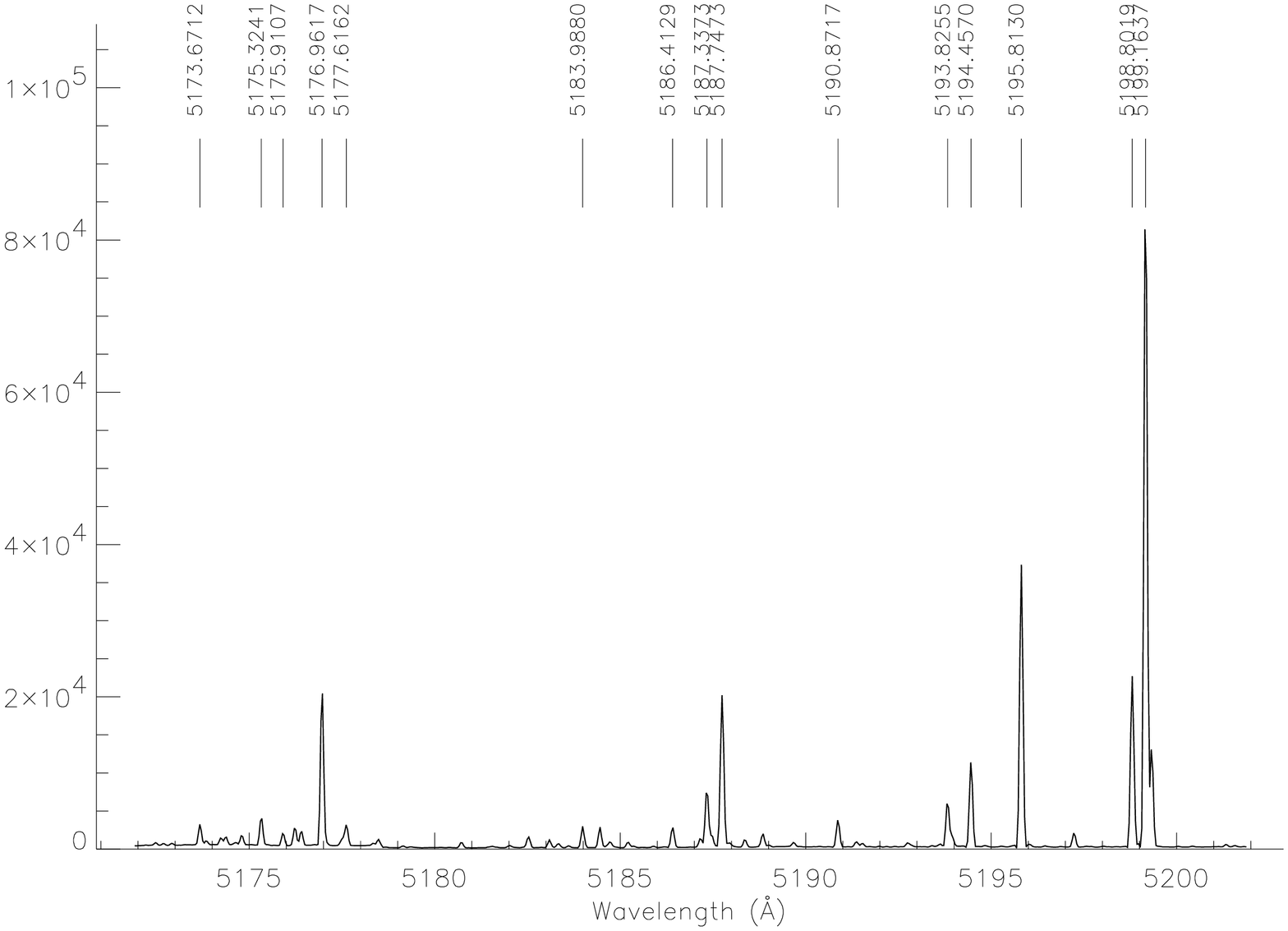}
\end{figure}
\clearpage
   
\begin{figure}
\centering
\includegraphics[width=10cm,angle=90]{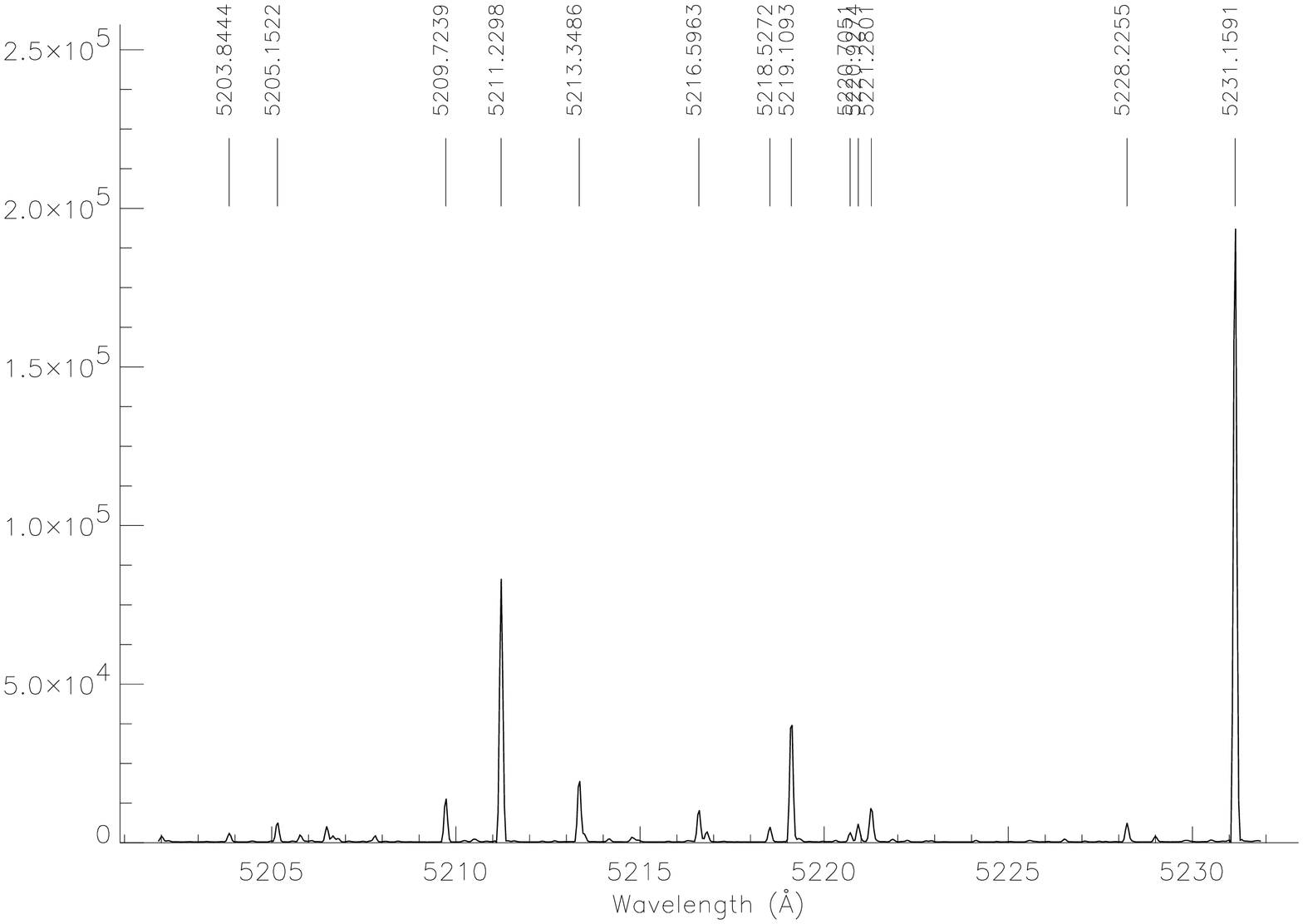}
\includegraphics[width=10cm,angle=90]{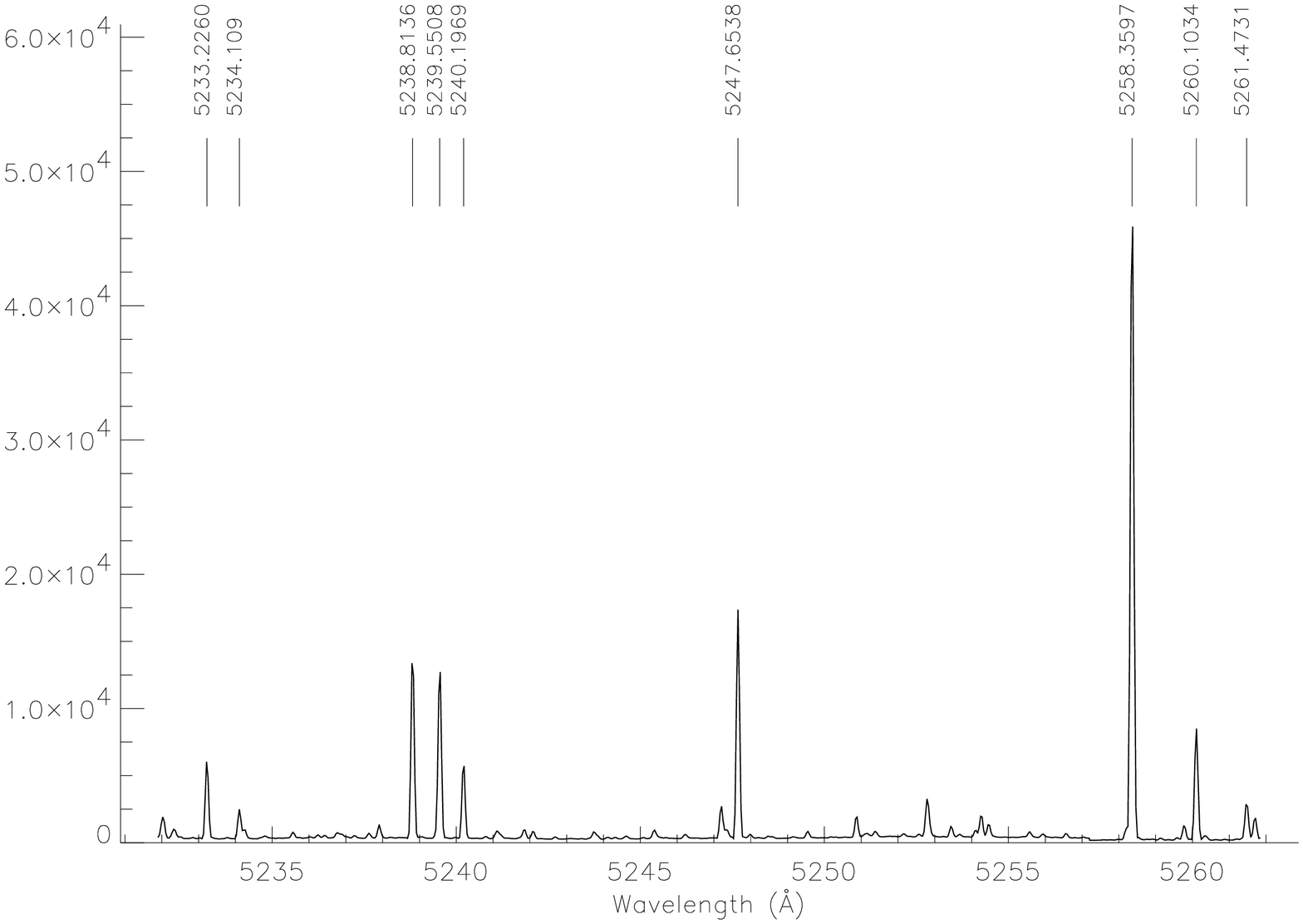}
\end{figure}
\clearpage
   
\begin{figure}
\centering
\includegraphics[width=10cm,angle=90]{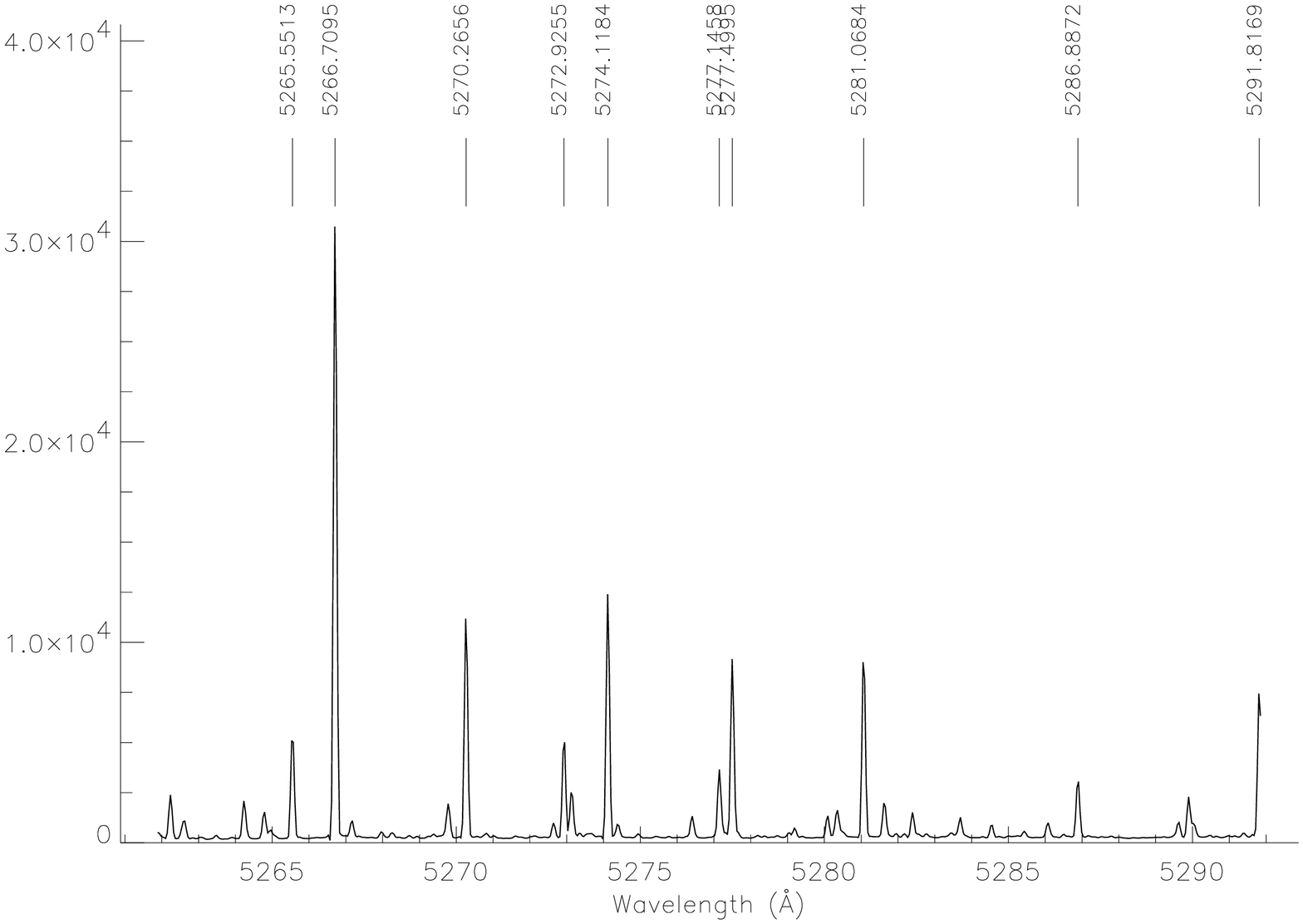}
\includegraphics[width=10cm,angle=90]{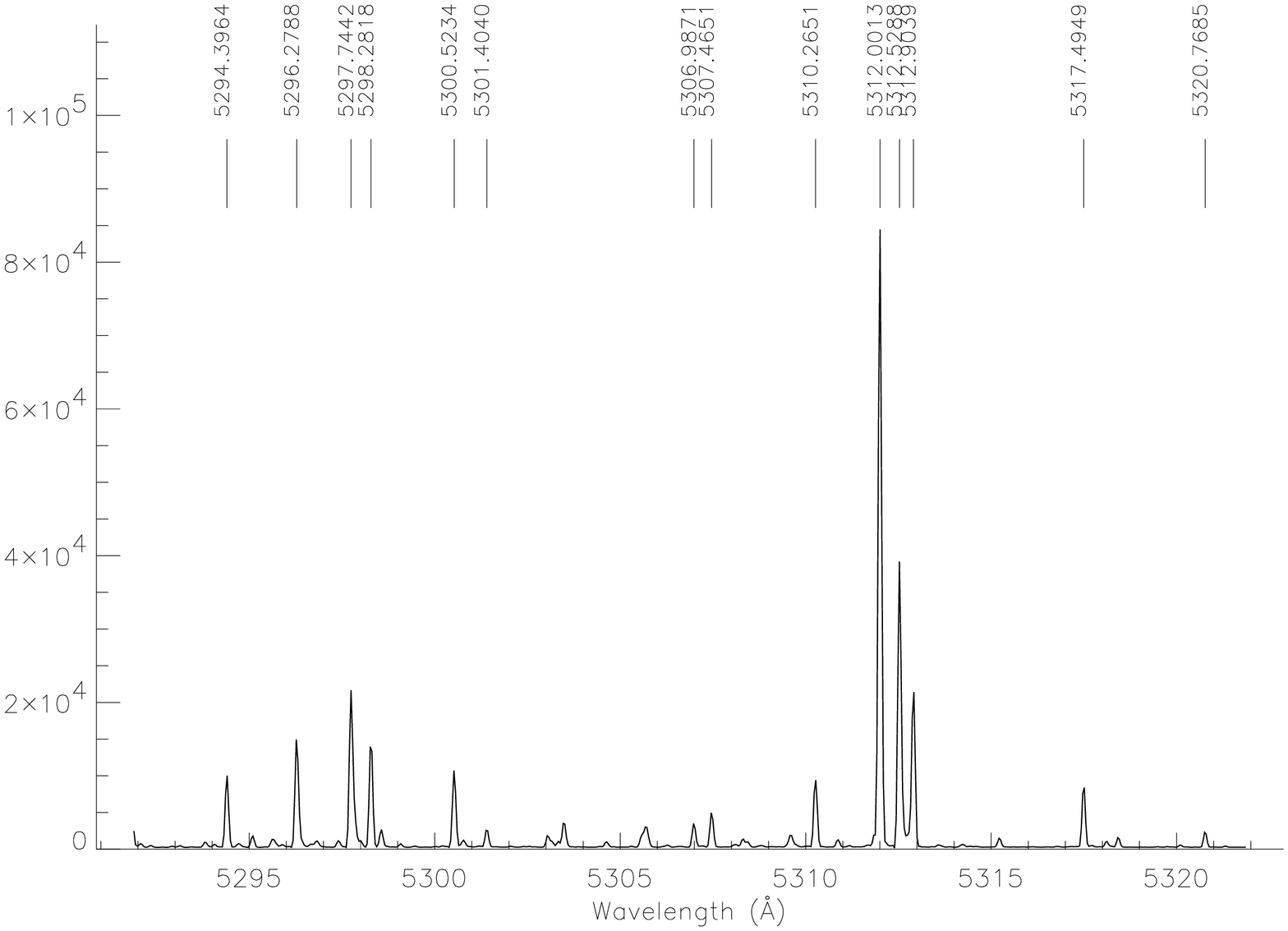}
\end{figure}
\clearpage
   
\begin{figure}
\centering
\includegraphics[width=10cm,angle=90]{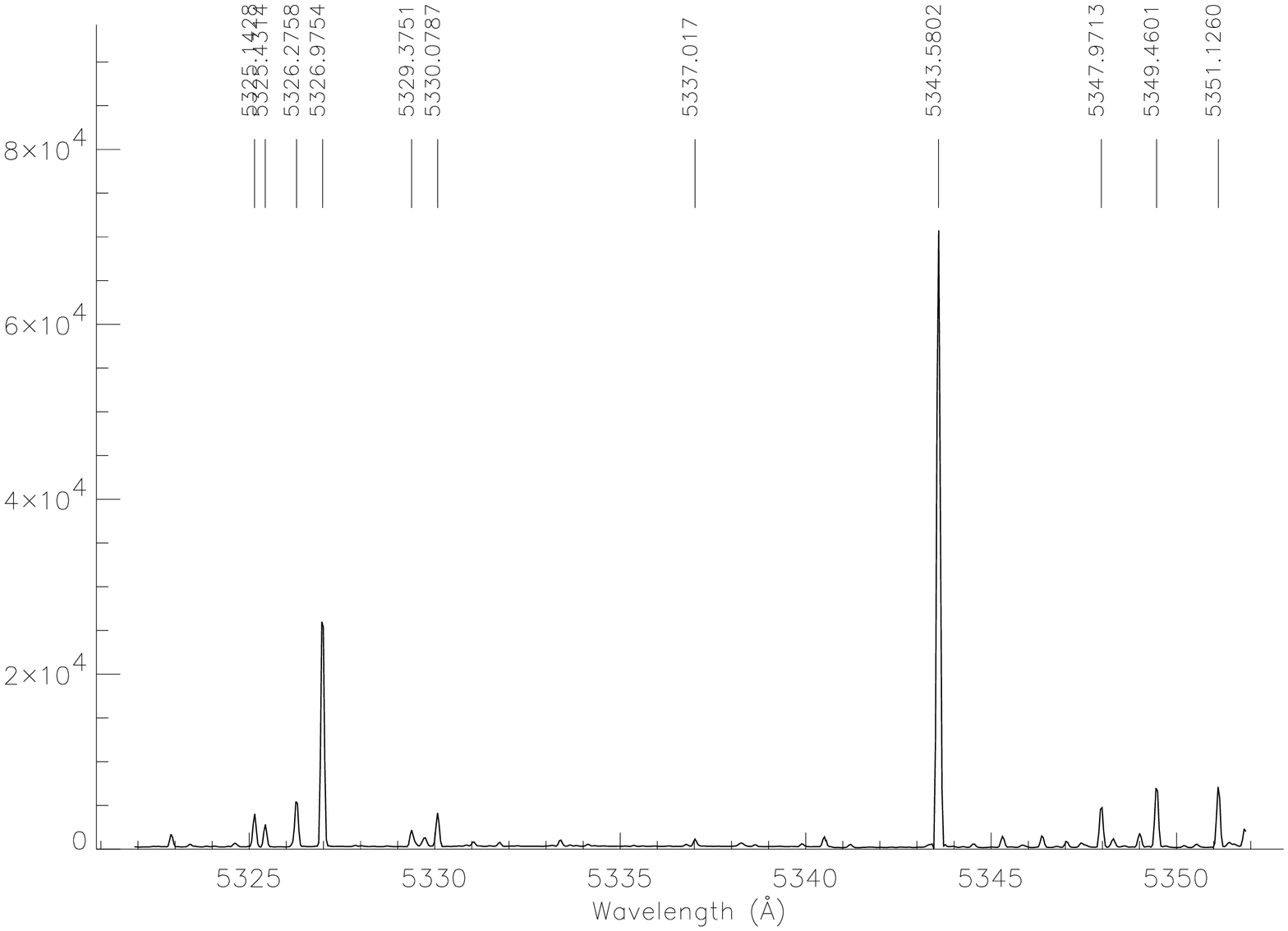}
\includegraphics[width=10cm,angle=90]{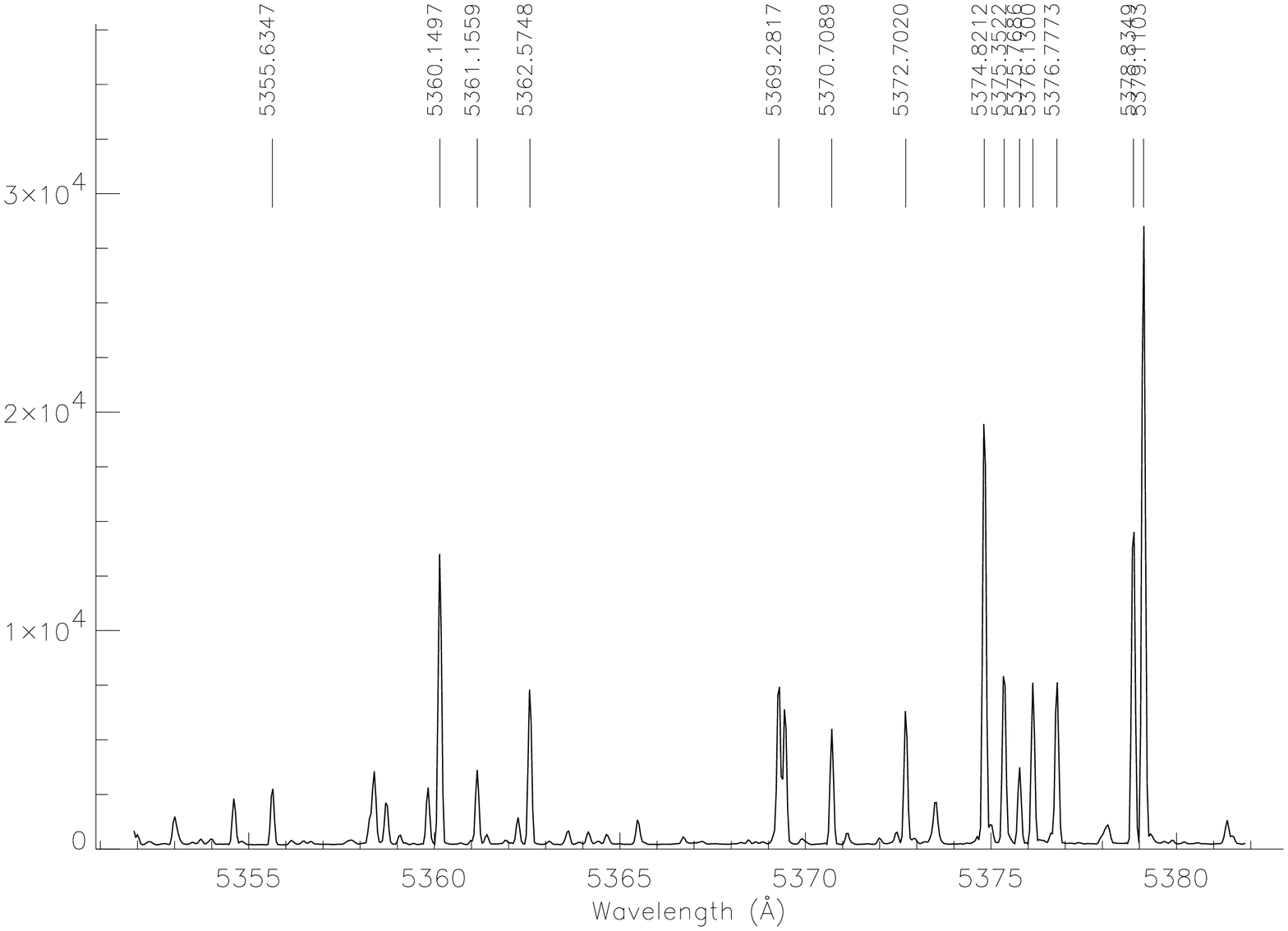}
\end{figure}
\clearpage
   
\begin{figure}
\centering
\includegraphics[width=10cm,angle=90]{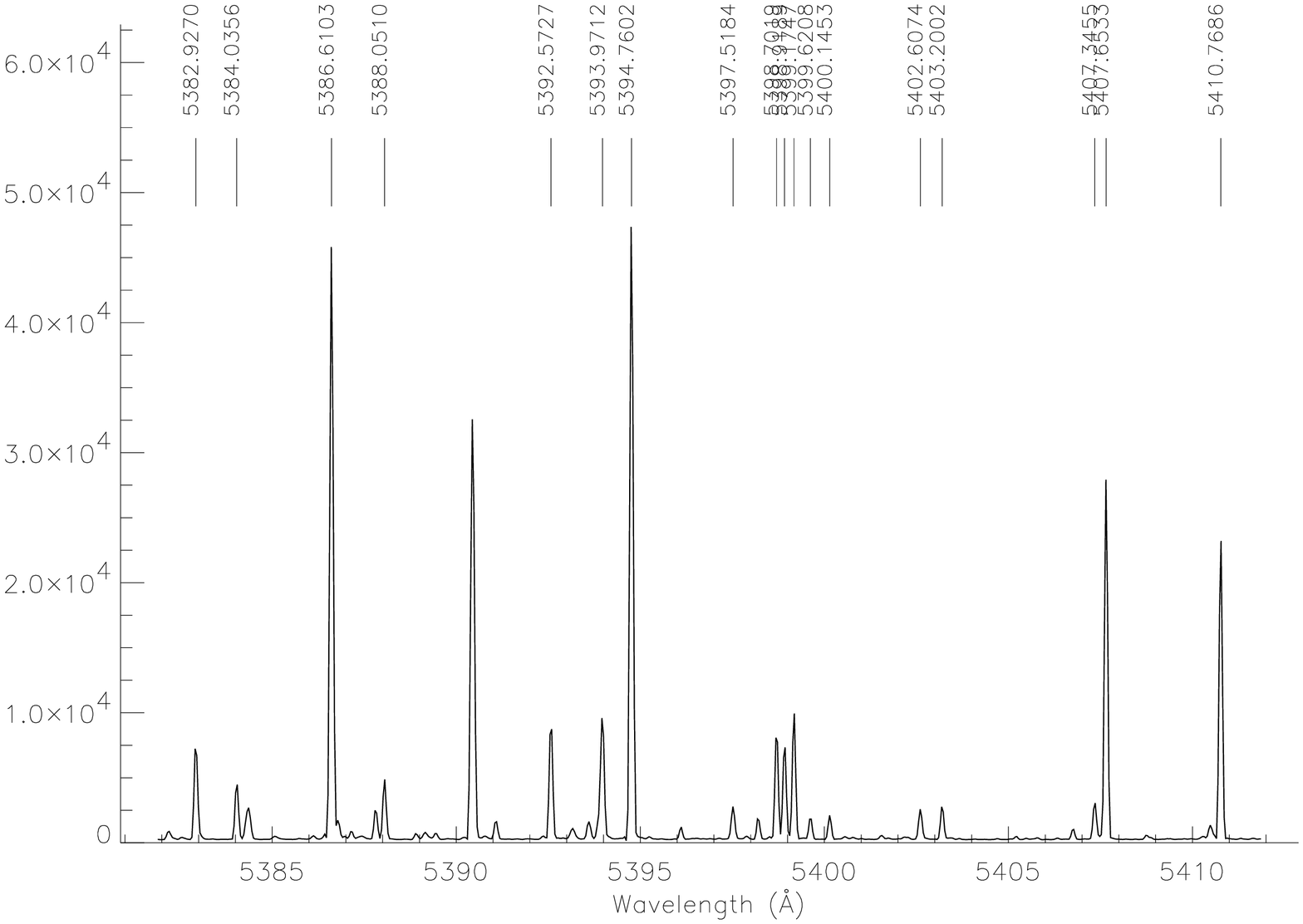}
\includegraphics[width=10cm,angle=90]{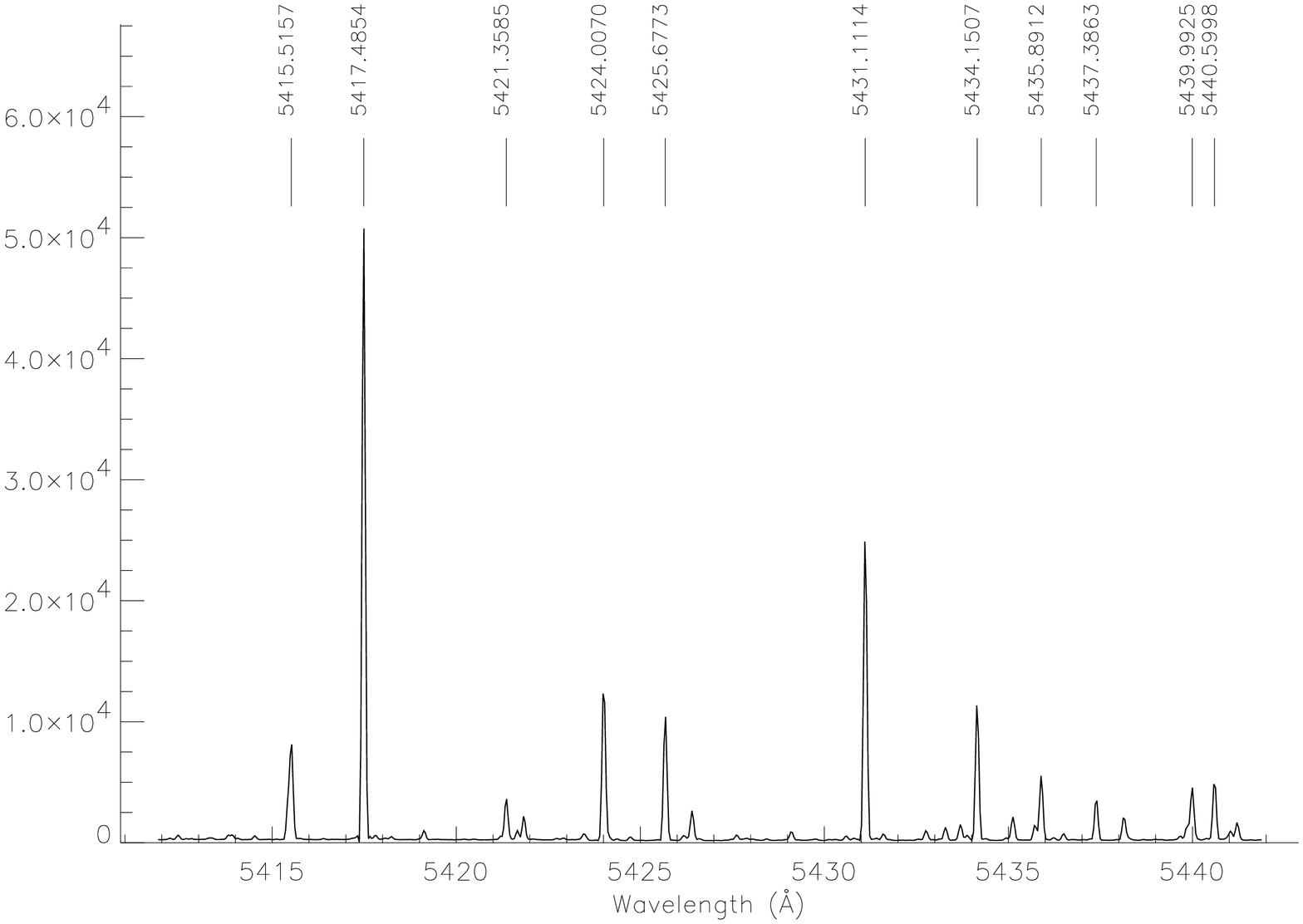}
\end{figure}
\clearpage
   
\begin{figure}
\centering
\includegraphics[width=10cm,angle=90]{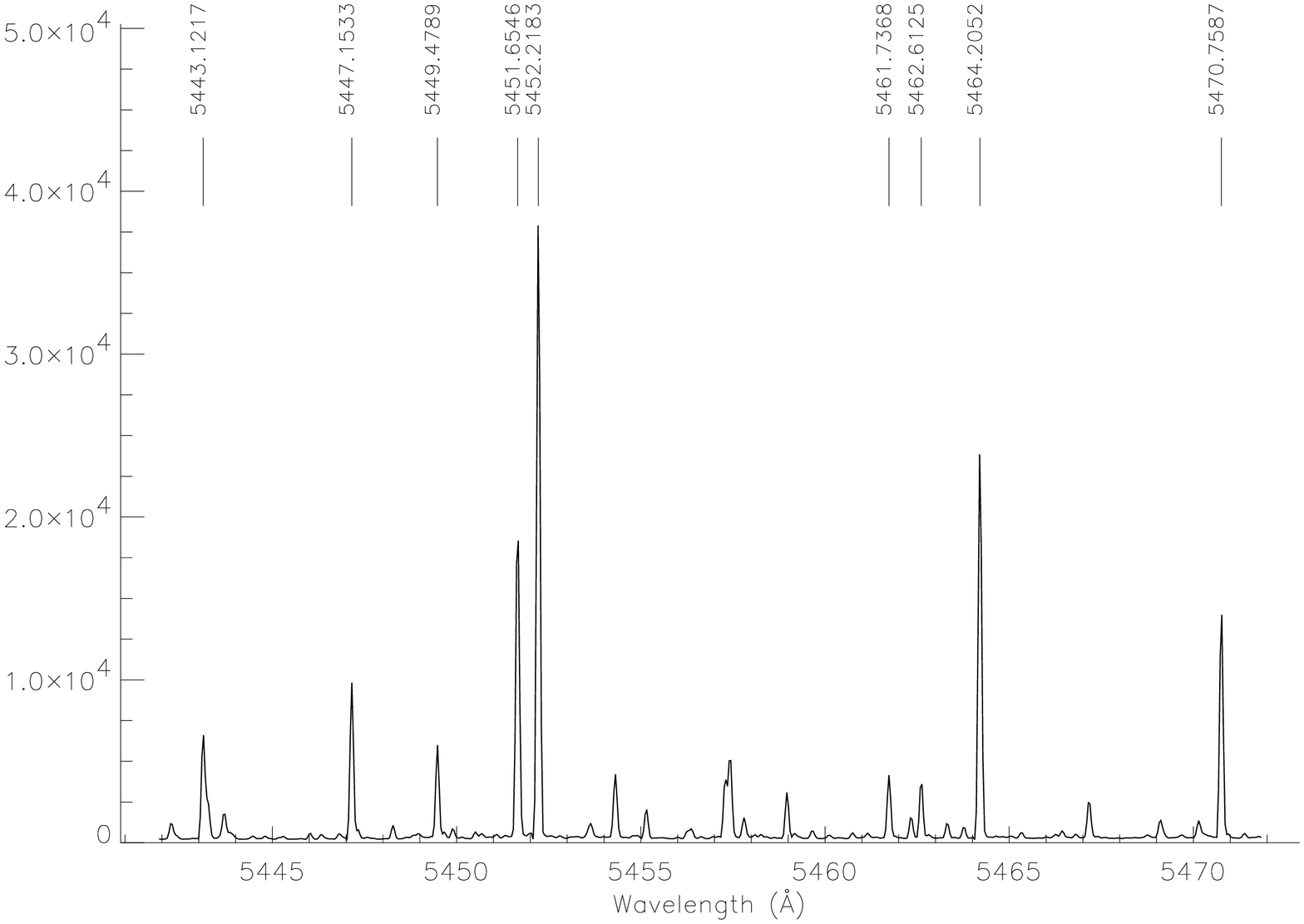}
\includegraphics[width=10cm,angle=90]{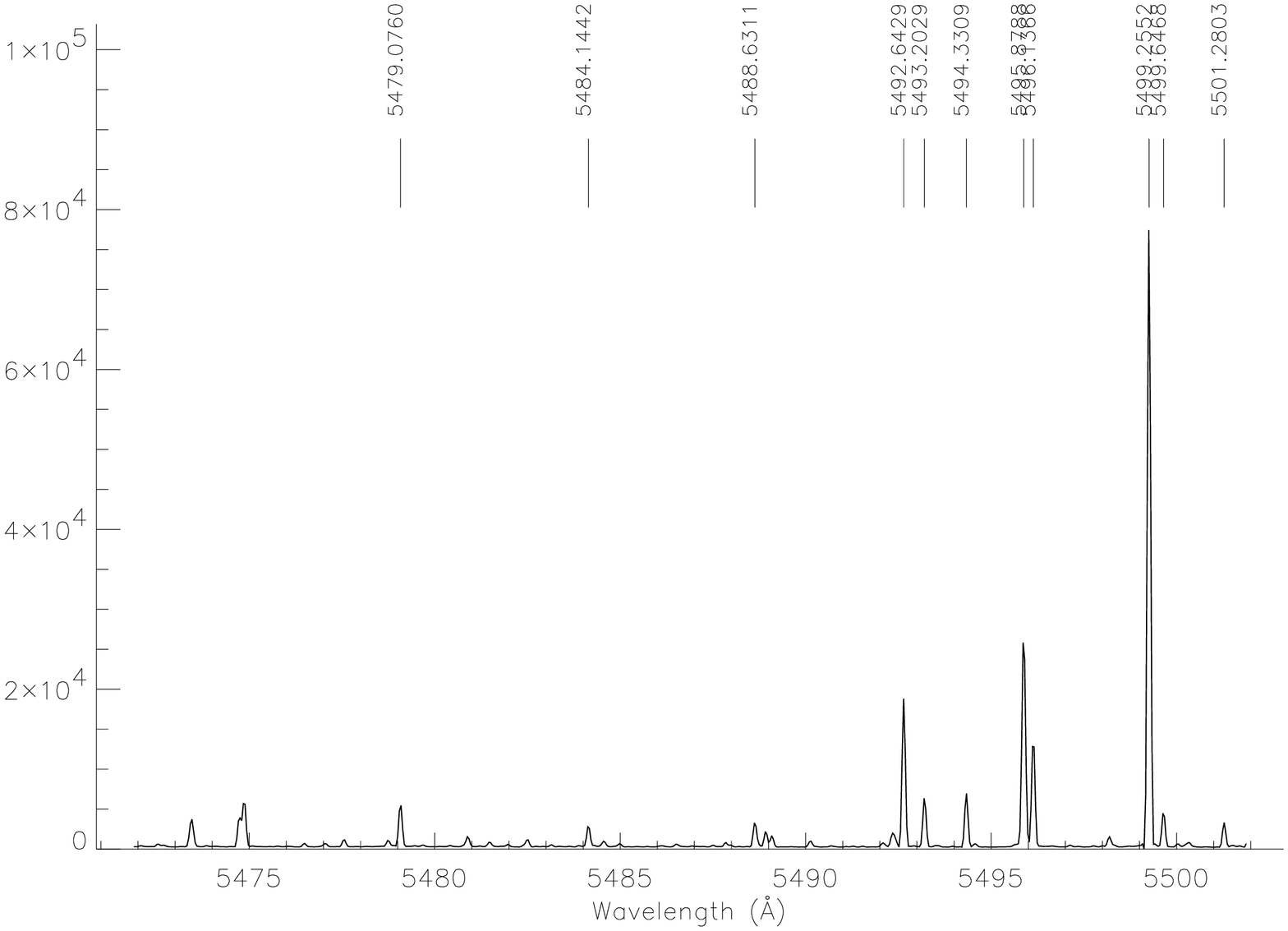}
\end{figure}
\clearpage
   
\begin{figure}
\centering
\includegraphics[width=10cm,angle=90]{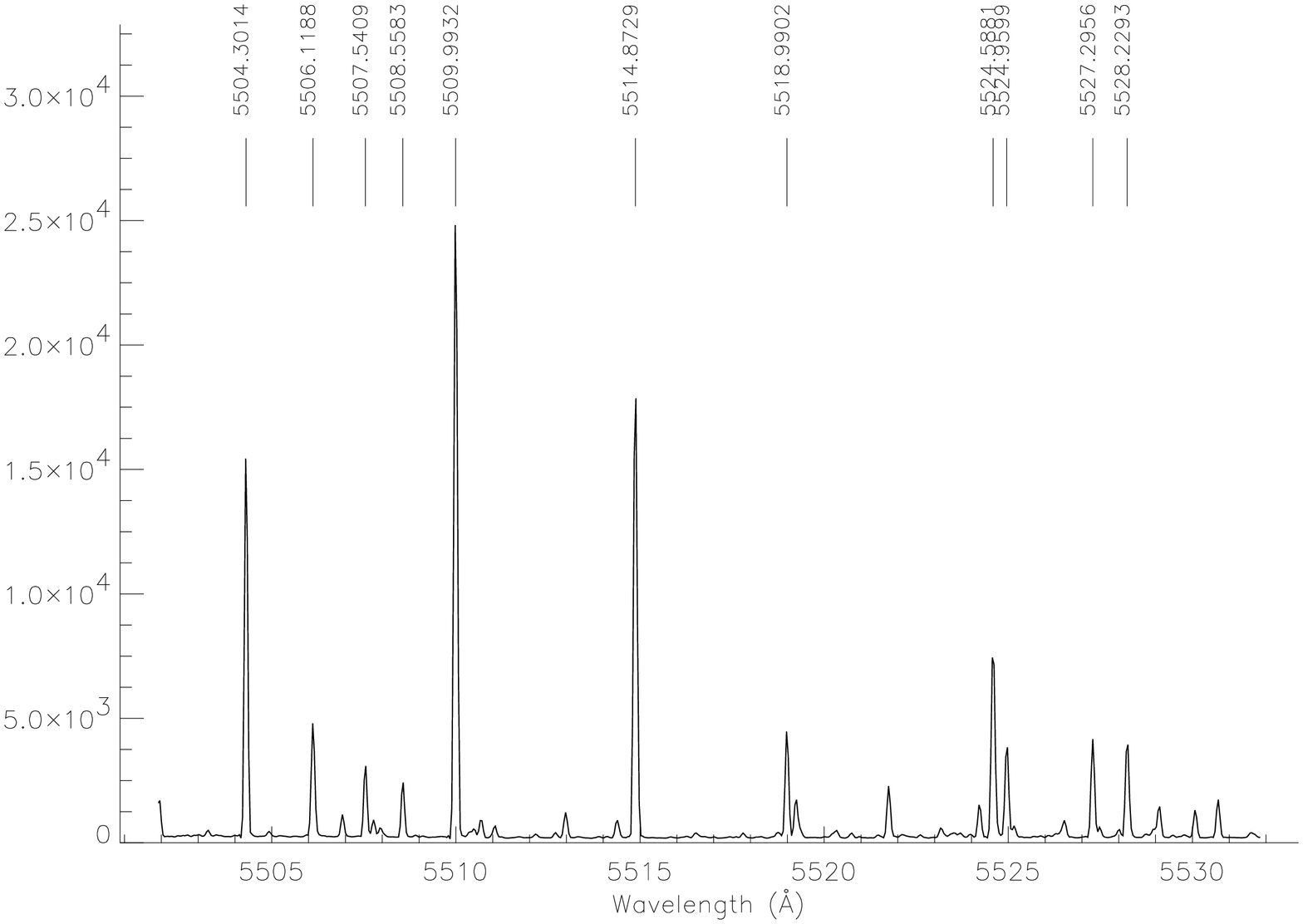}
\includegraphics[width=10cm,angle=90]{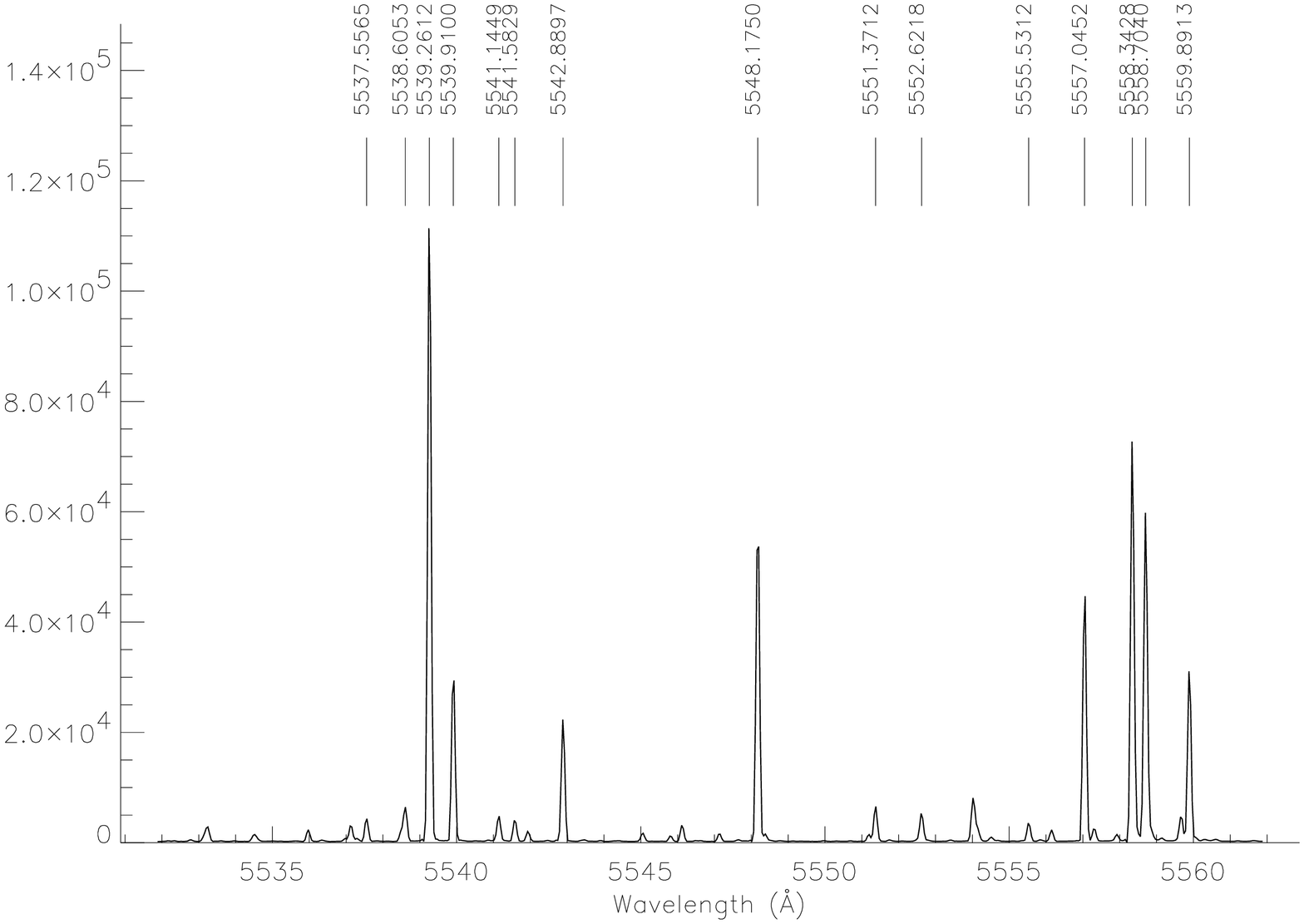}
\end{figure}
\clearpage
   
\begin{figure}
\centering
\includegraphics[width=10cm,angle=90]{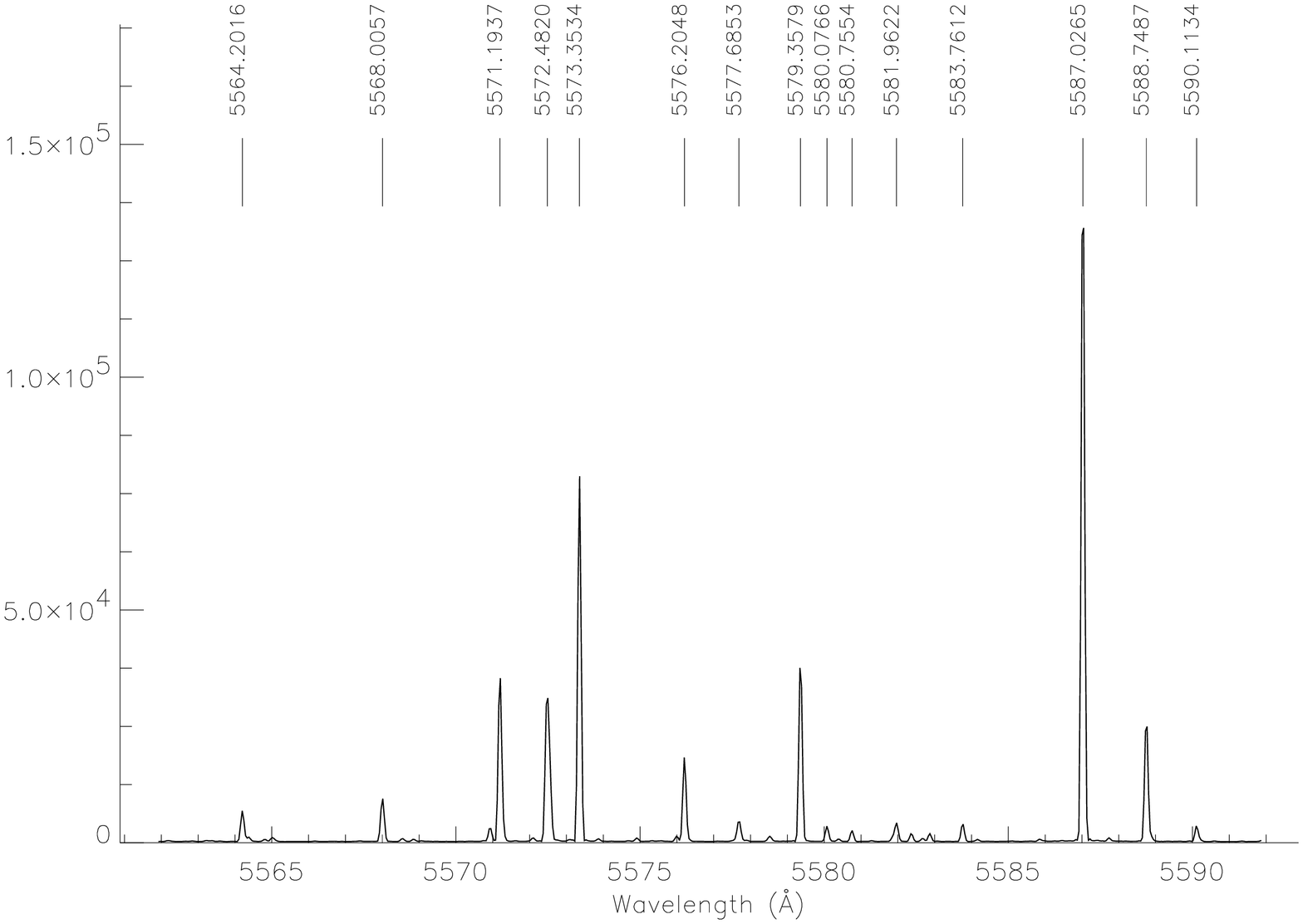}
\includegraphics[width=10cm,angle=90]{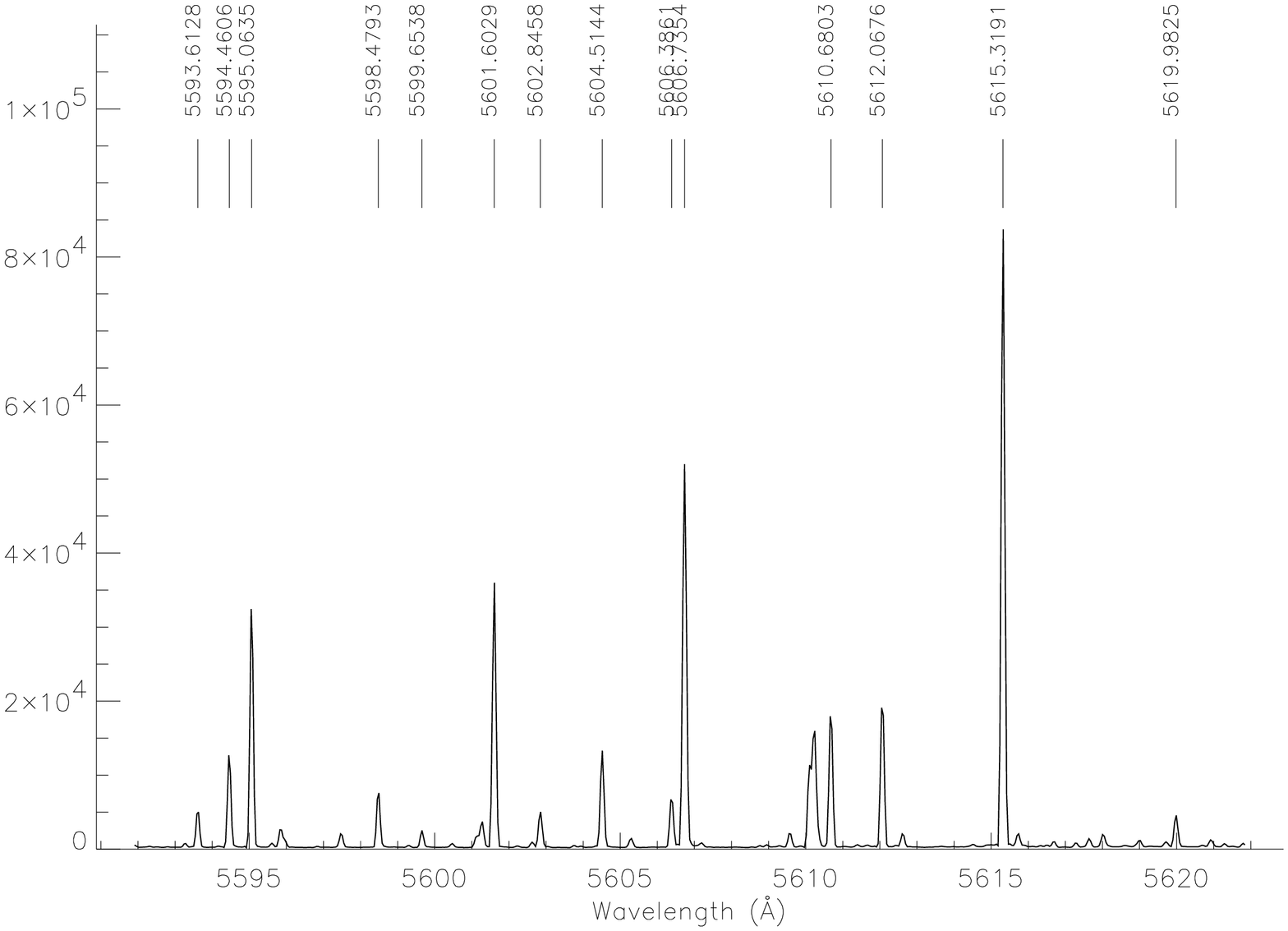}
\end{figure}
\clearpage
   
\begin{figure}
\centering
\includegraphics[width=10cm,angle=90]{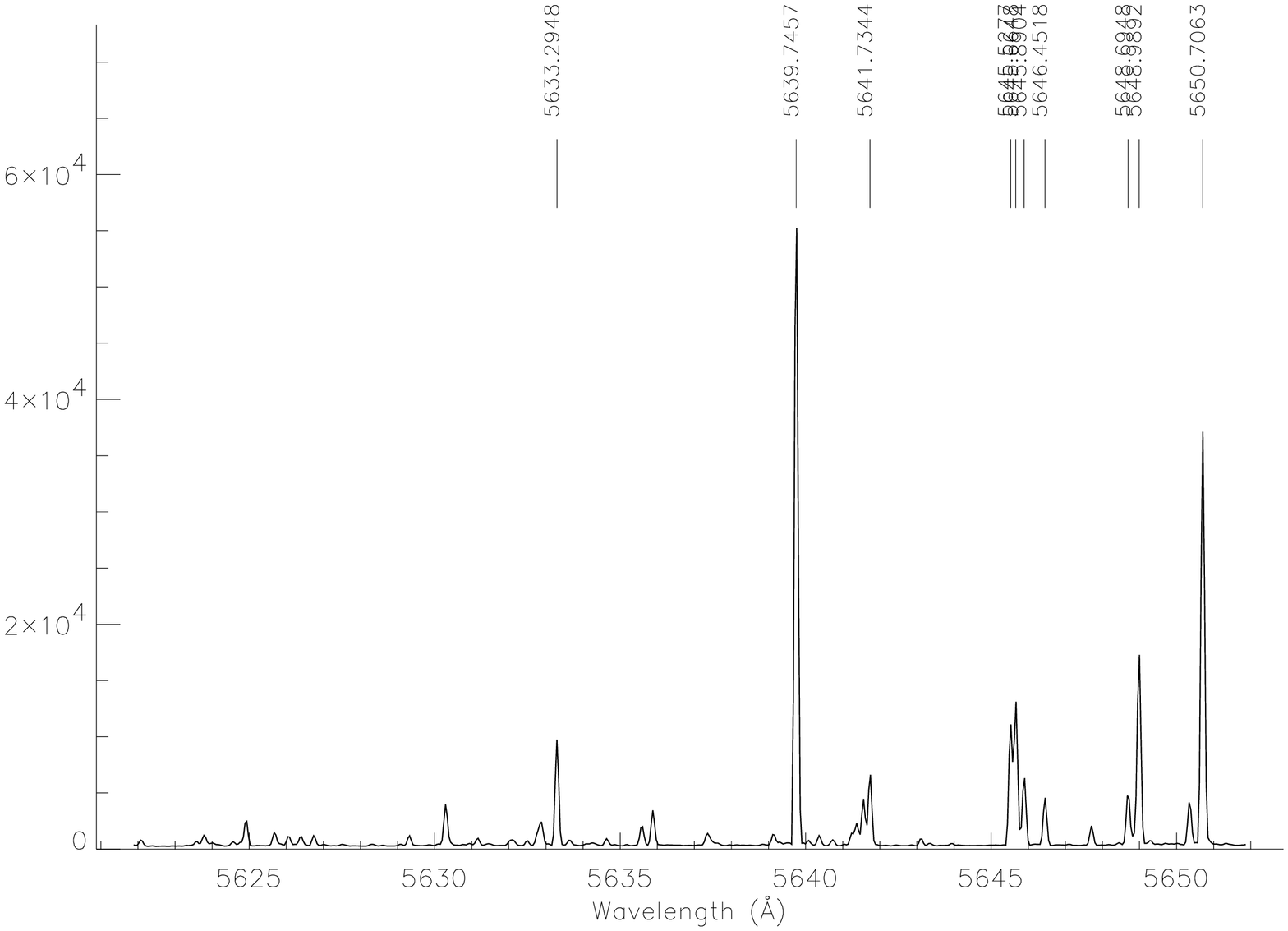}
\includegraphics[width=10cm,angle=90]{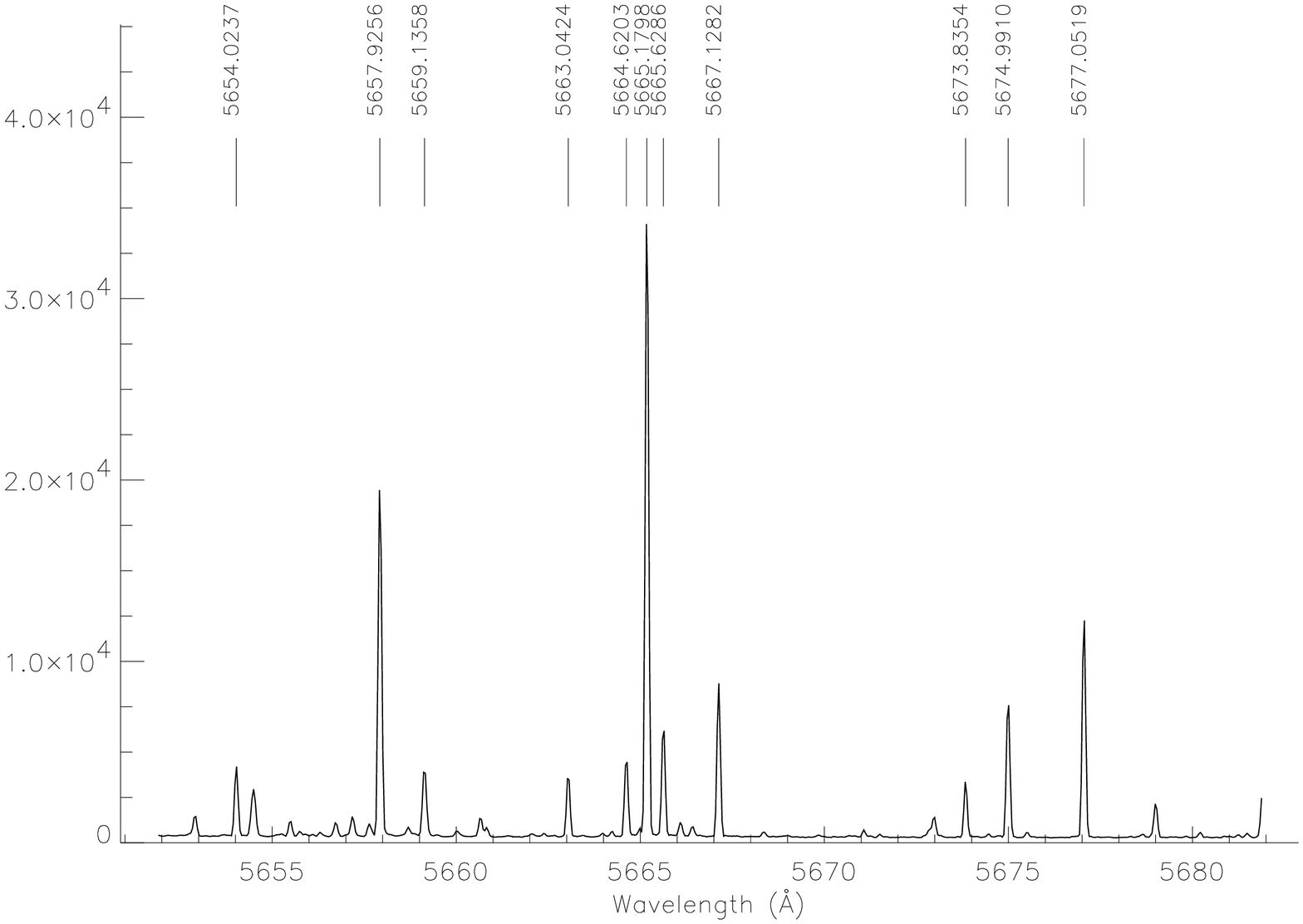}
\end{figure}
\clearpage
   
\begin{figure}
\centering
\includegraphics[width=10cm,angle=90]{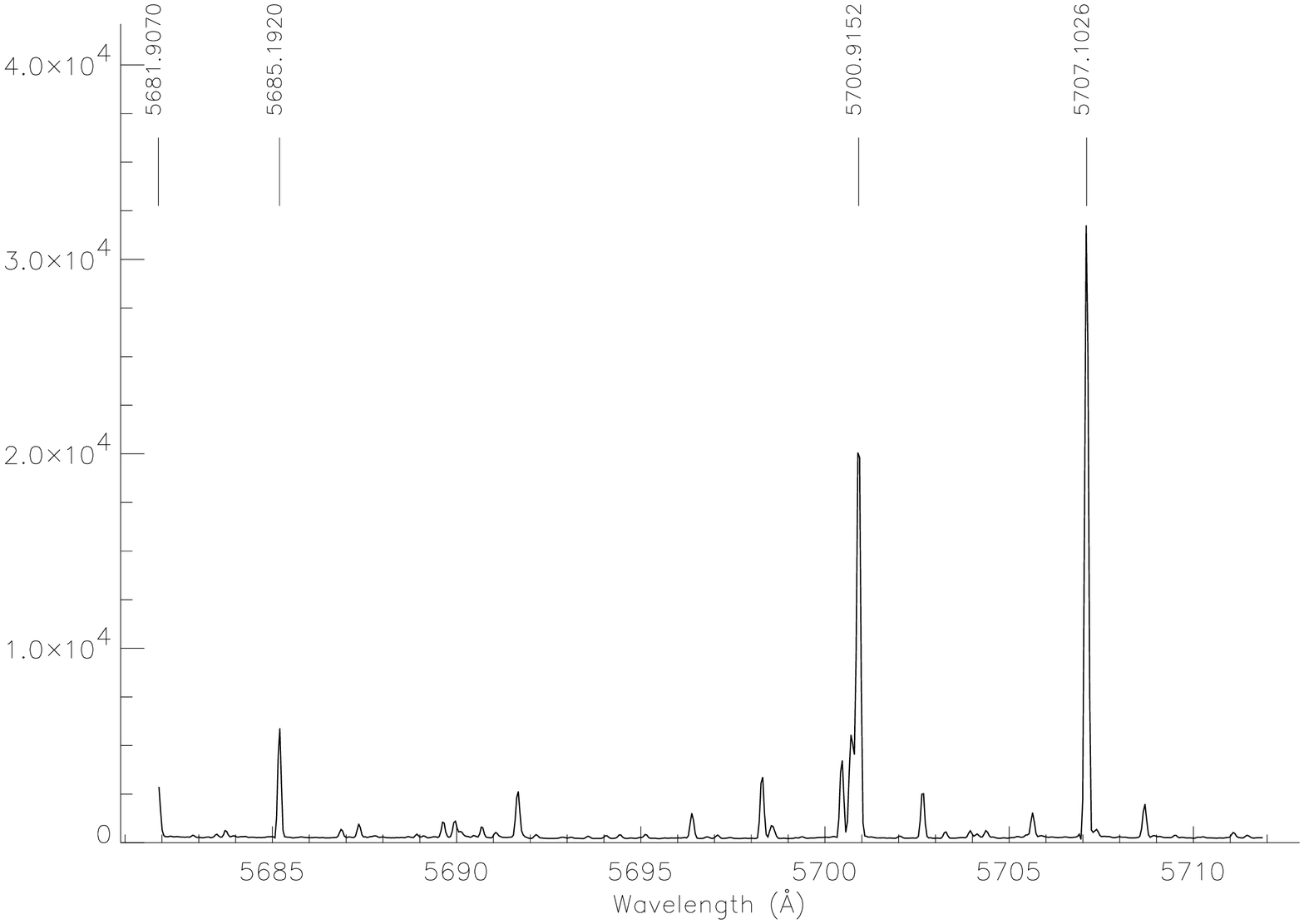}
\includegraphics[width=10cm,angle=90]{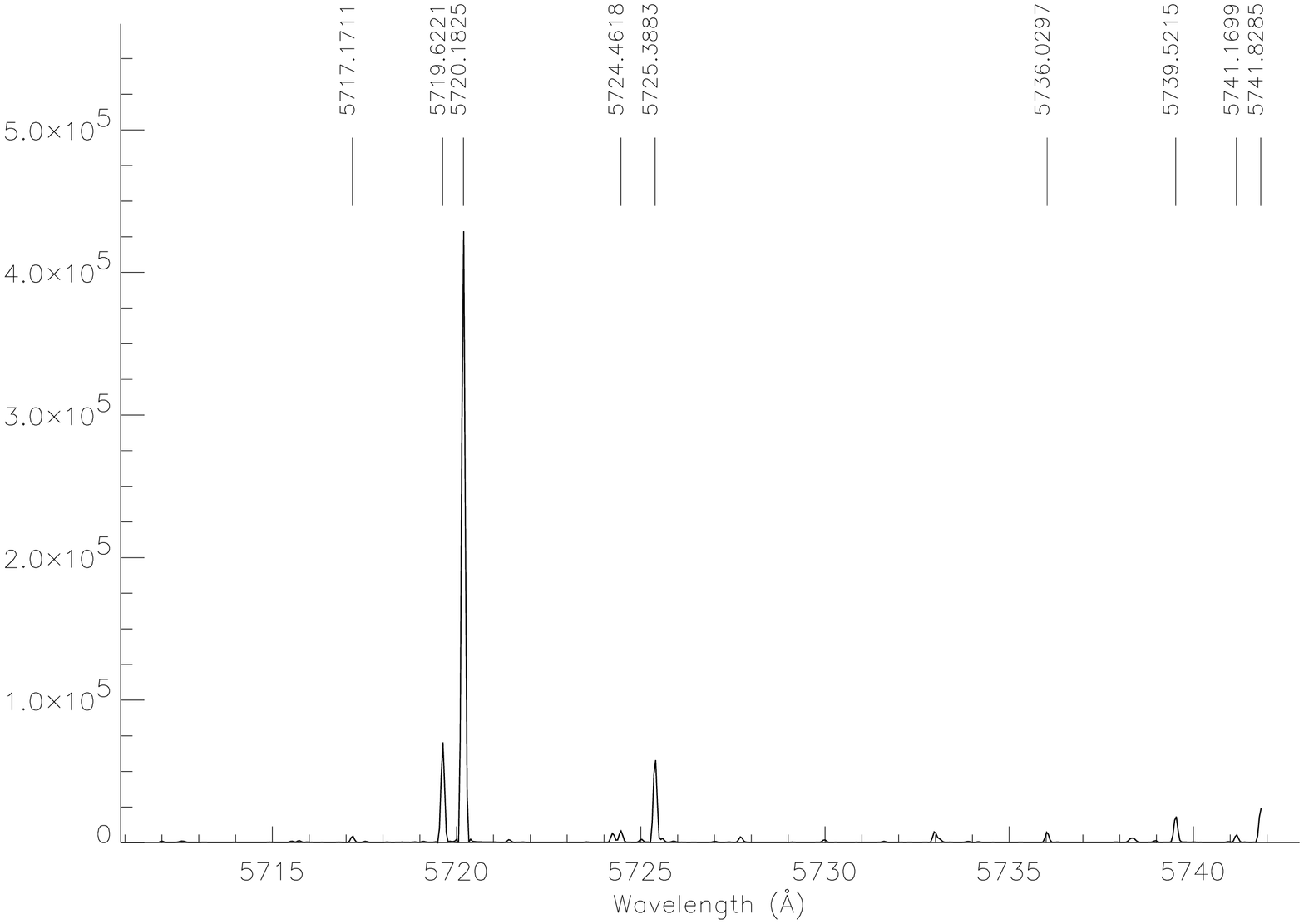}
\end{figure}
\clearpage
   
\begin{figure}
\centering
\includegraphics[width=10cm,angle=90]{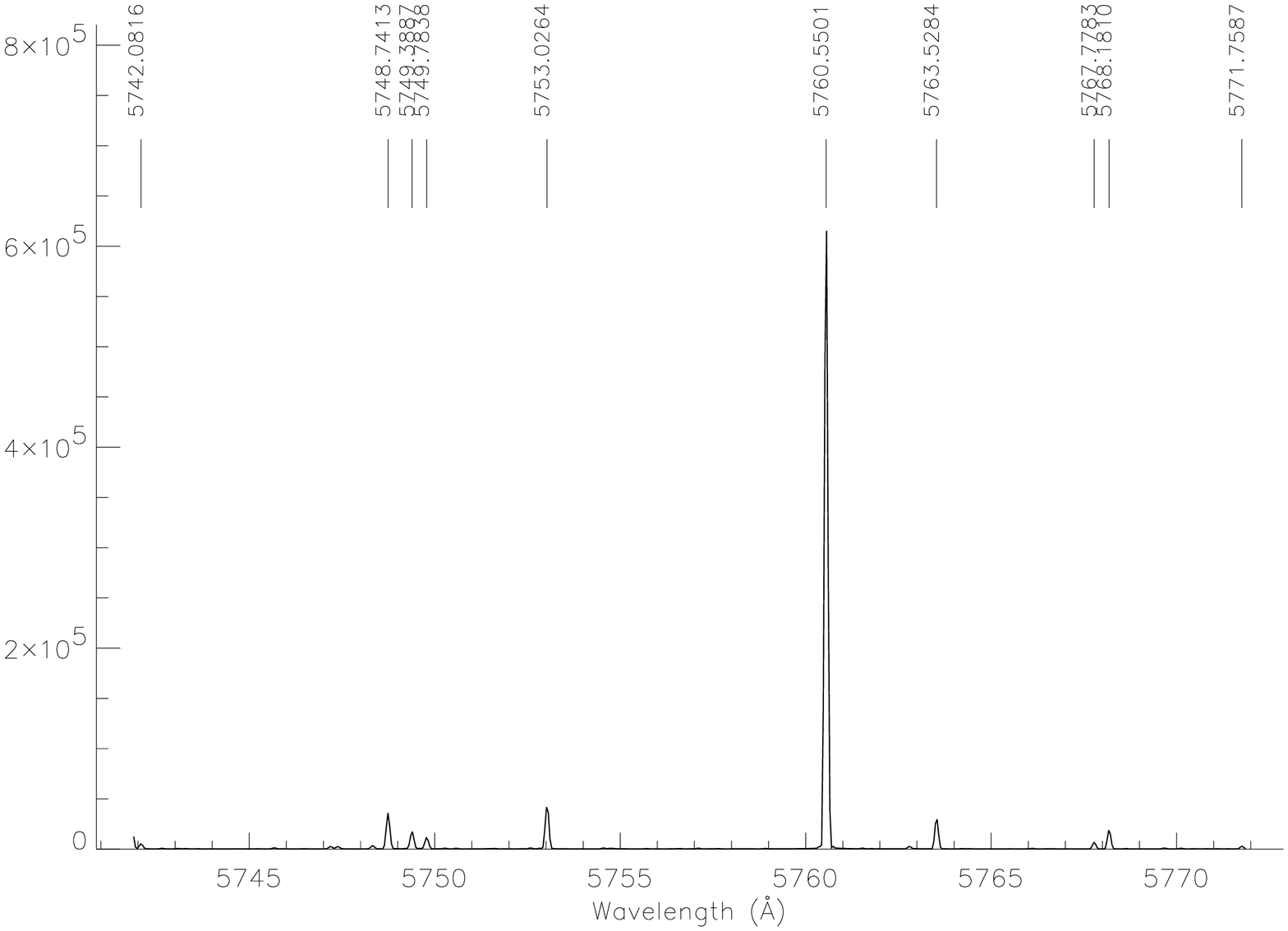}
\includegraphics[width=10cm,angle=90]{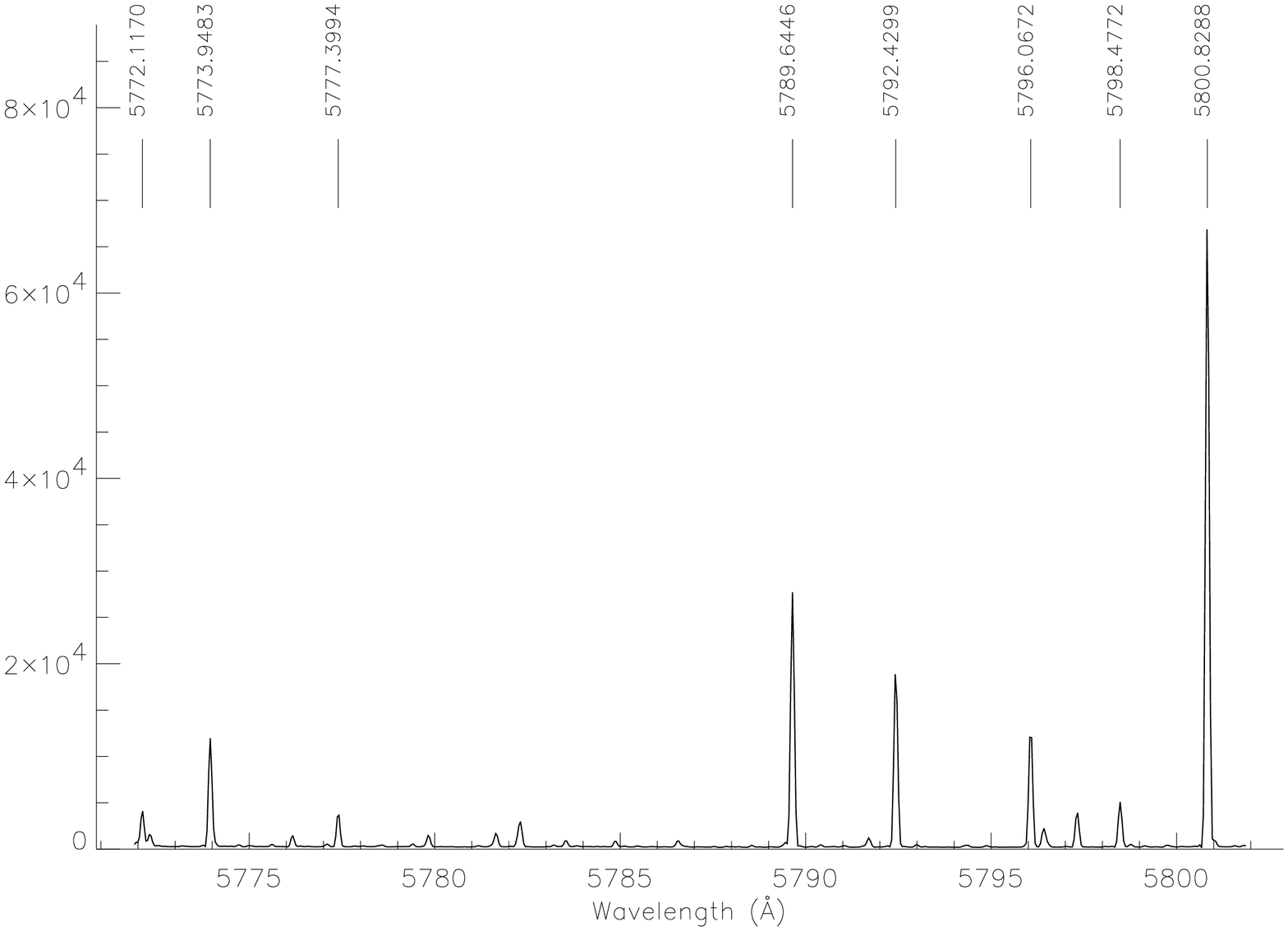}
\end{figure}
\clearpage
   
\begin{figure}
\centering
\includegraphics[width=10cm,angle=90]{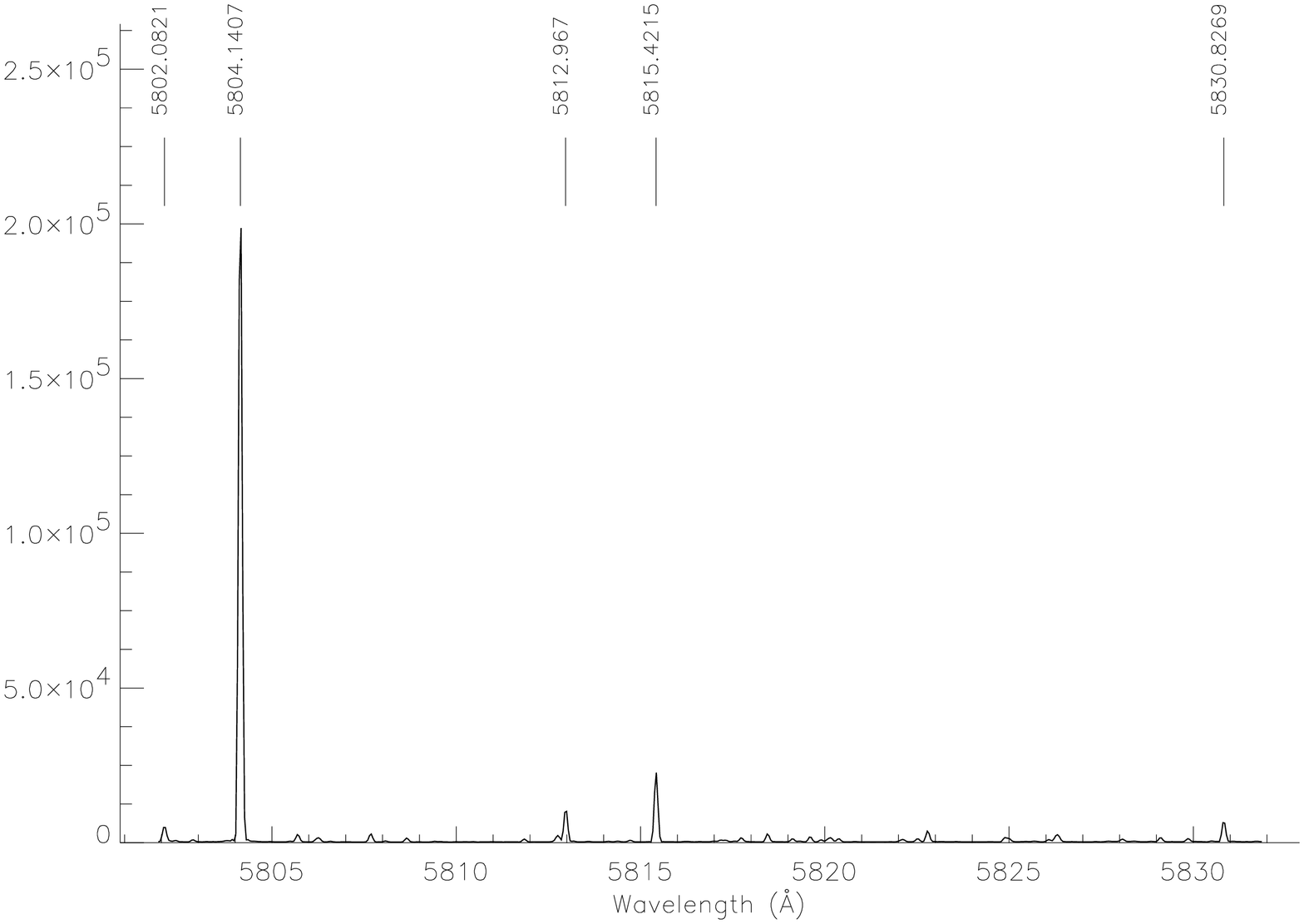}
\includegraphics[width=10cm,angle=90]{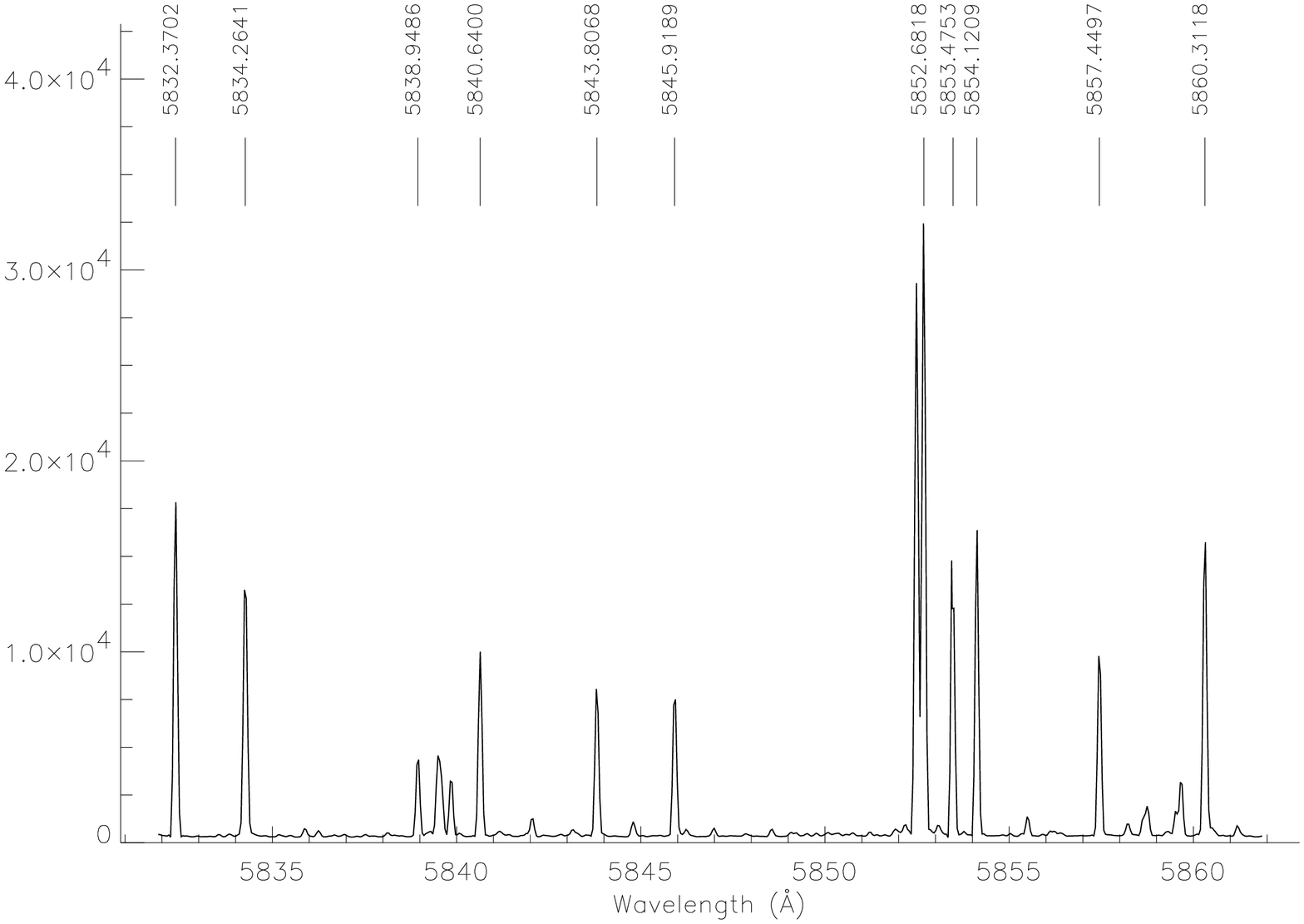}
\end{figure}
\clearpage
   
\begin{figure}
\centering
\includegraphics[width=10cm,angle=90]{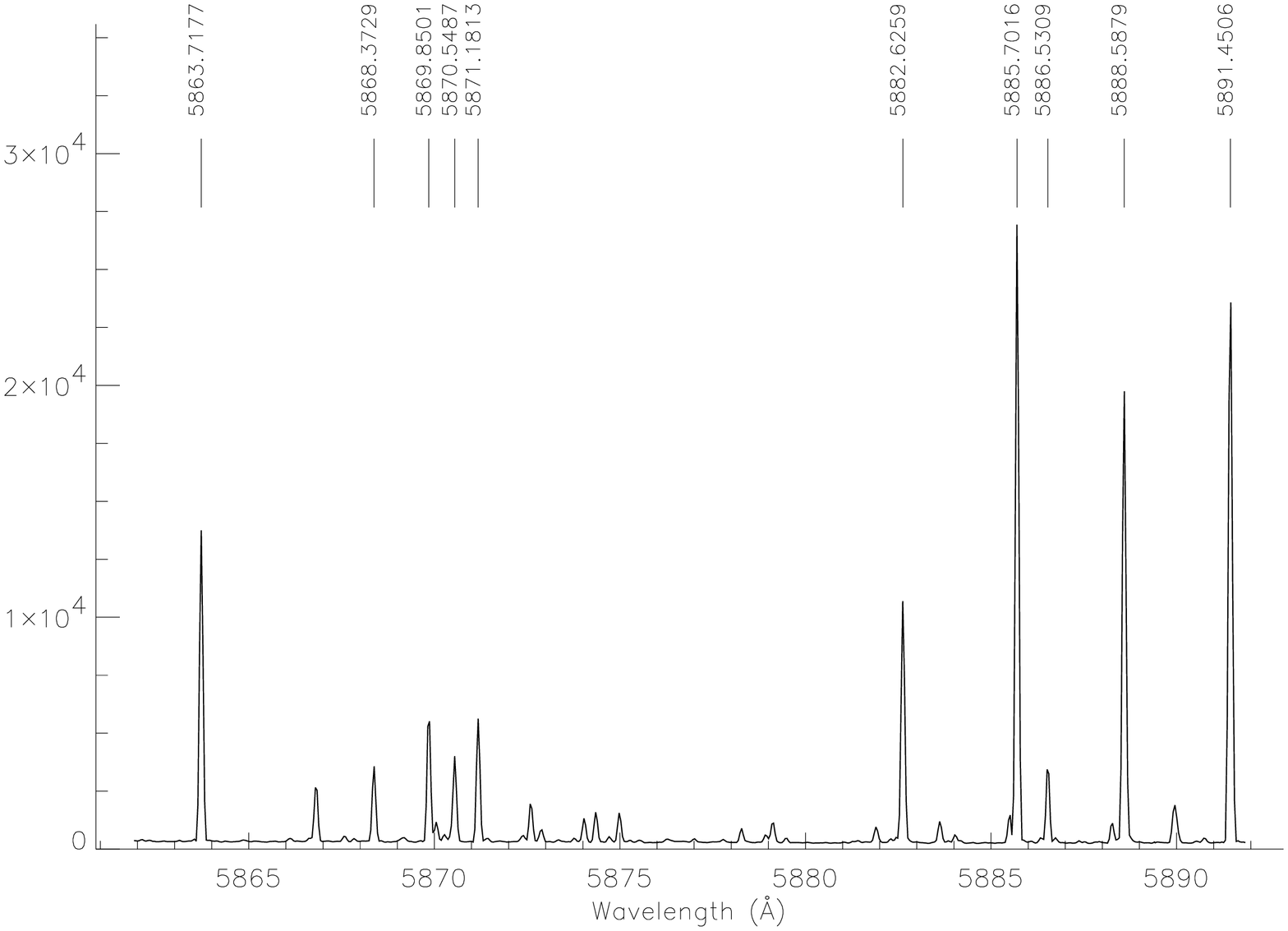}
\includegraphics[width=10cm,angle=90]{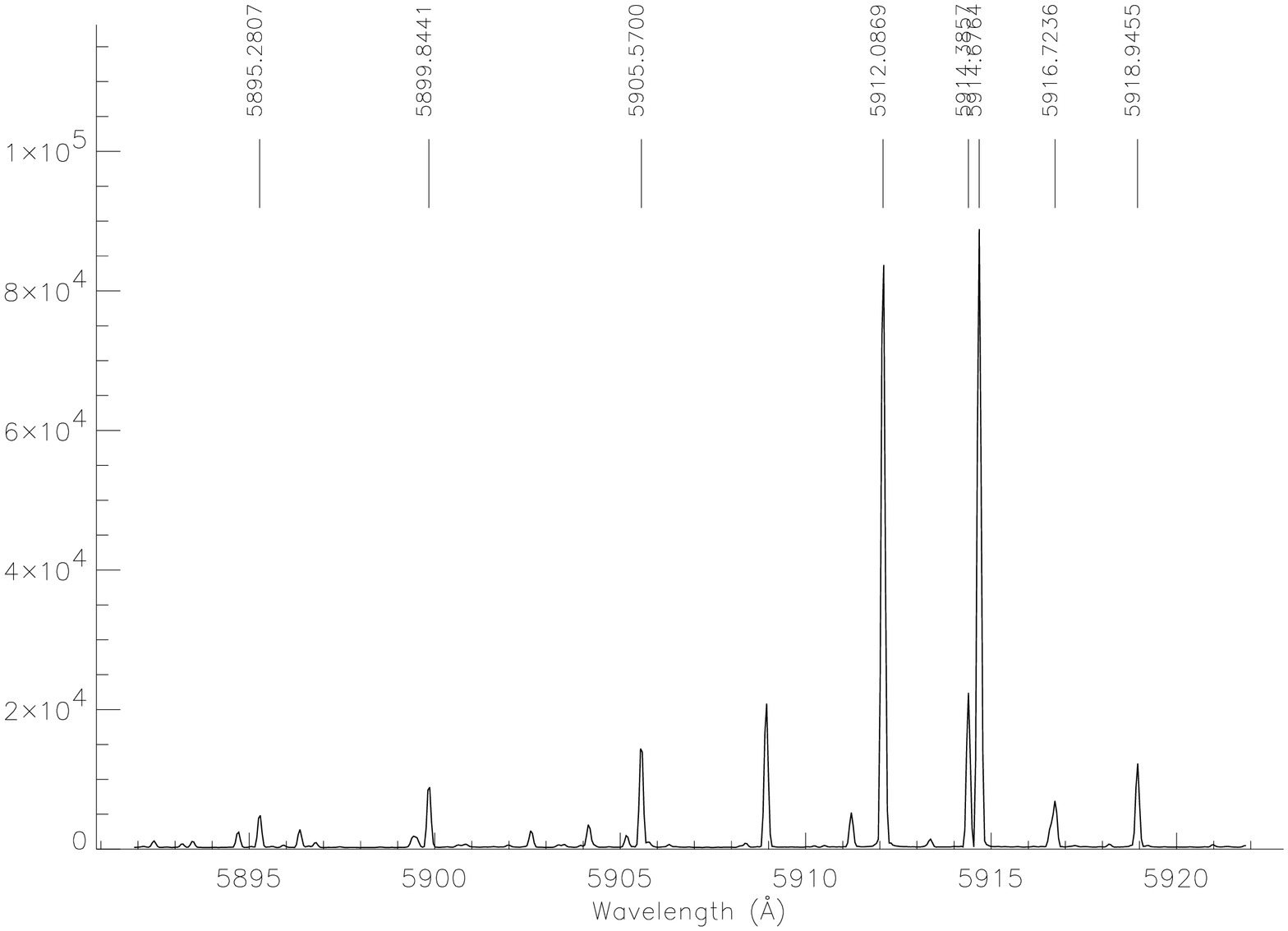}
\end{figure}
\clearpage
   
\begin{figure}
\centering
\includegraphics[width=10cm,angle=90]{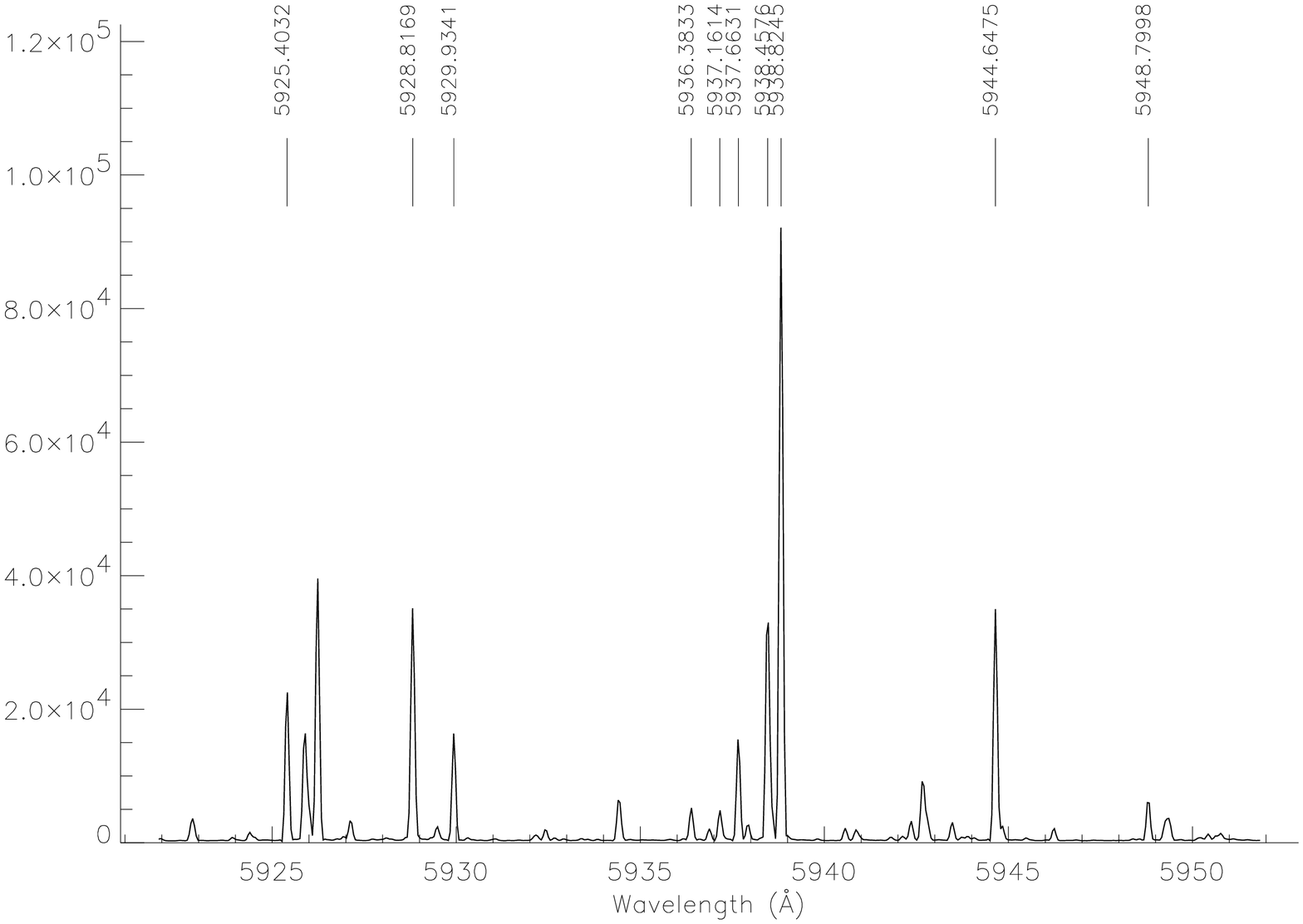}
\includegraphics[width=10cm,angle=90]{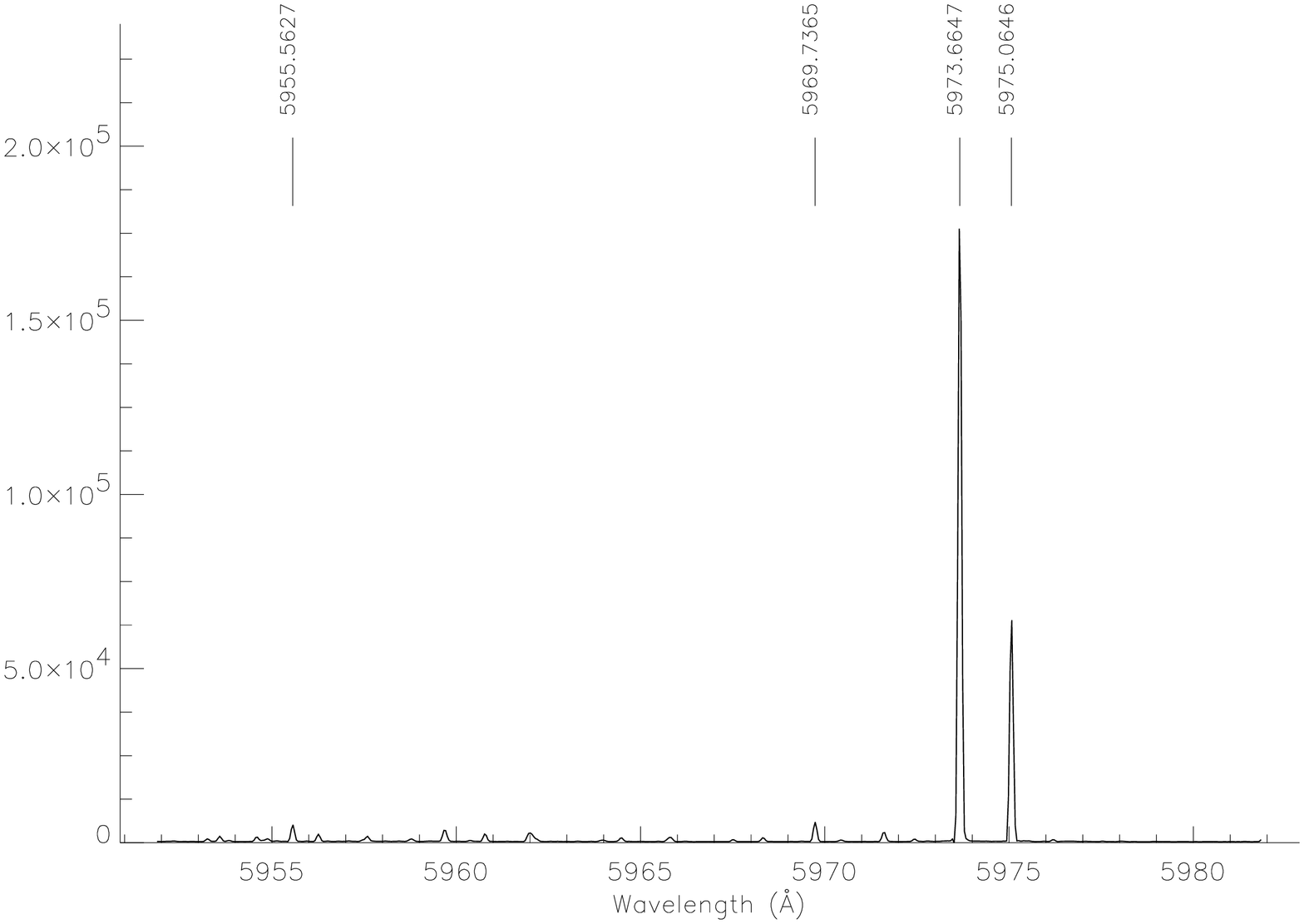}
\end{figure}
\clearpage
   
\begin{figure}
\centering
\includegraphics[width=10cm,angle=90]{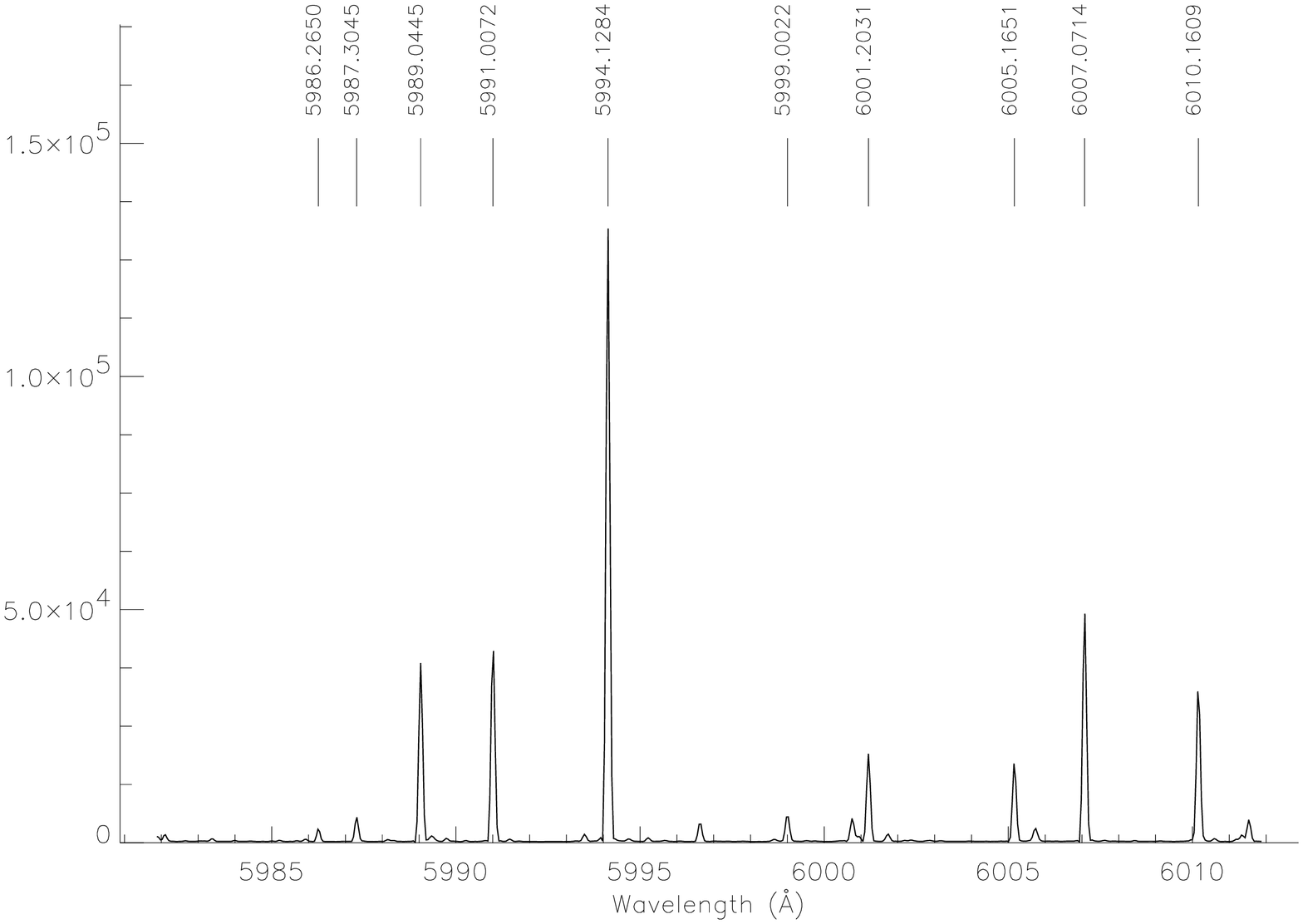}
\includegraphics[width=10cm,angle=90]{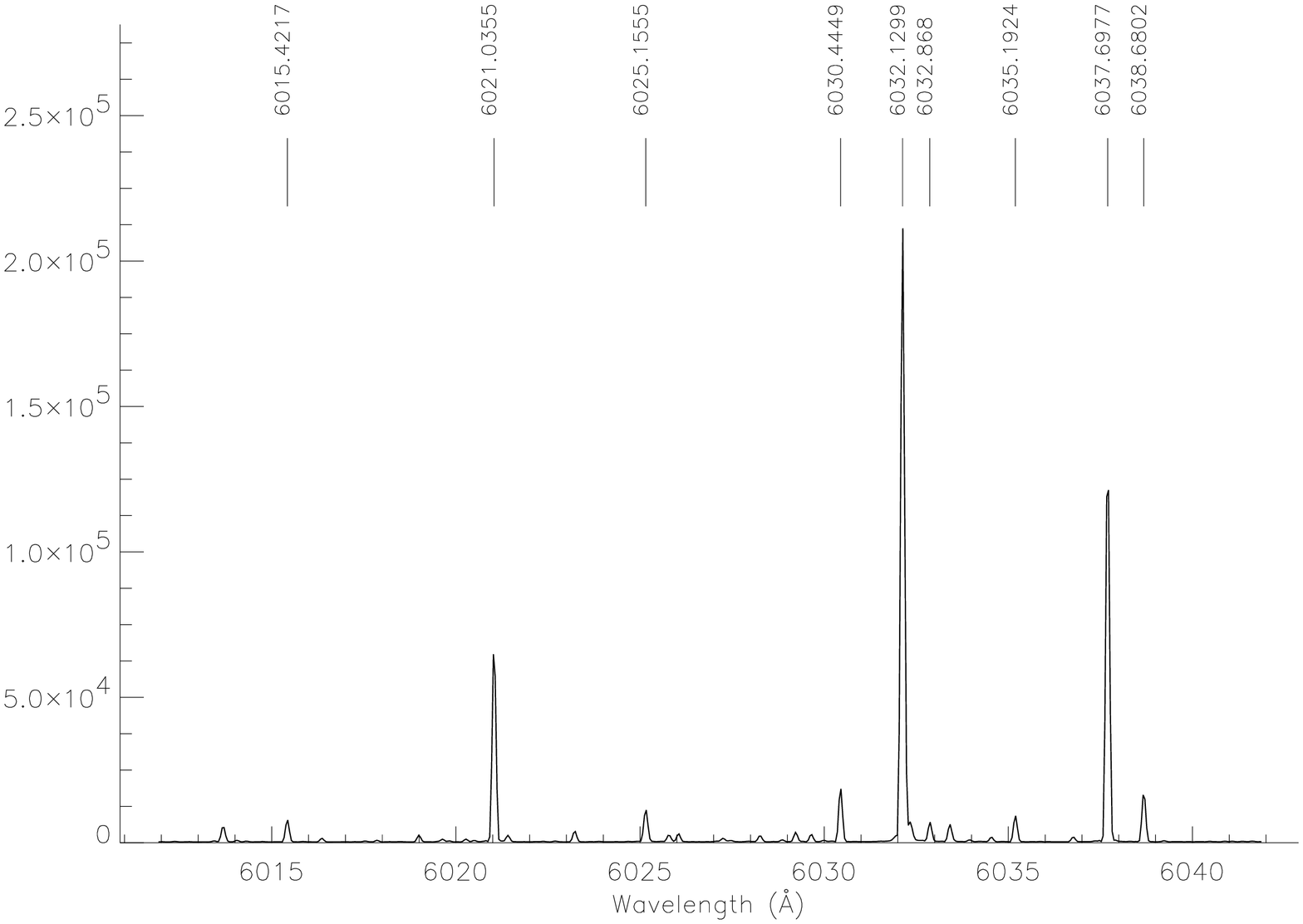}
\end{figure}
\clearpage
   
\begin{figure}
\centering
\includegraphics[width=10cm,angle=90]{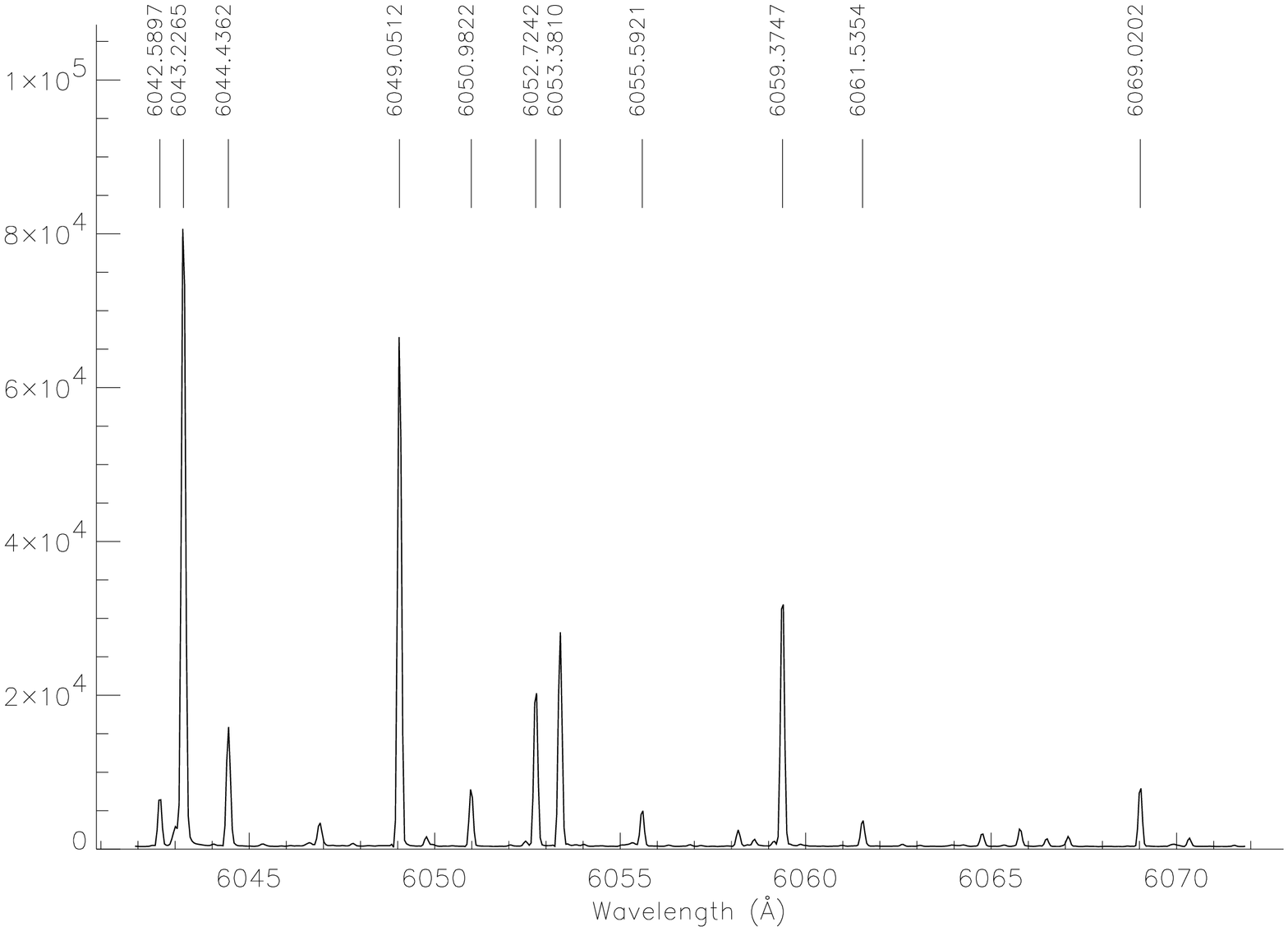}
\includegraphics[width=10cm,angle=90]{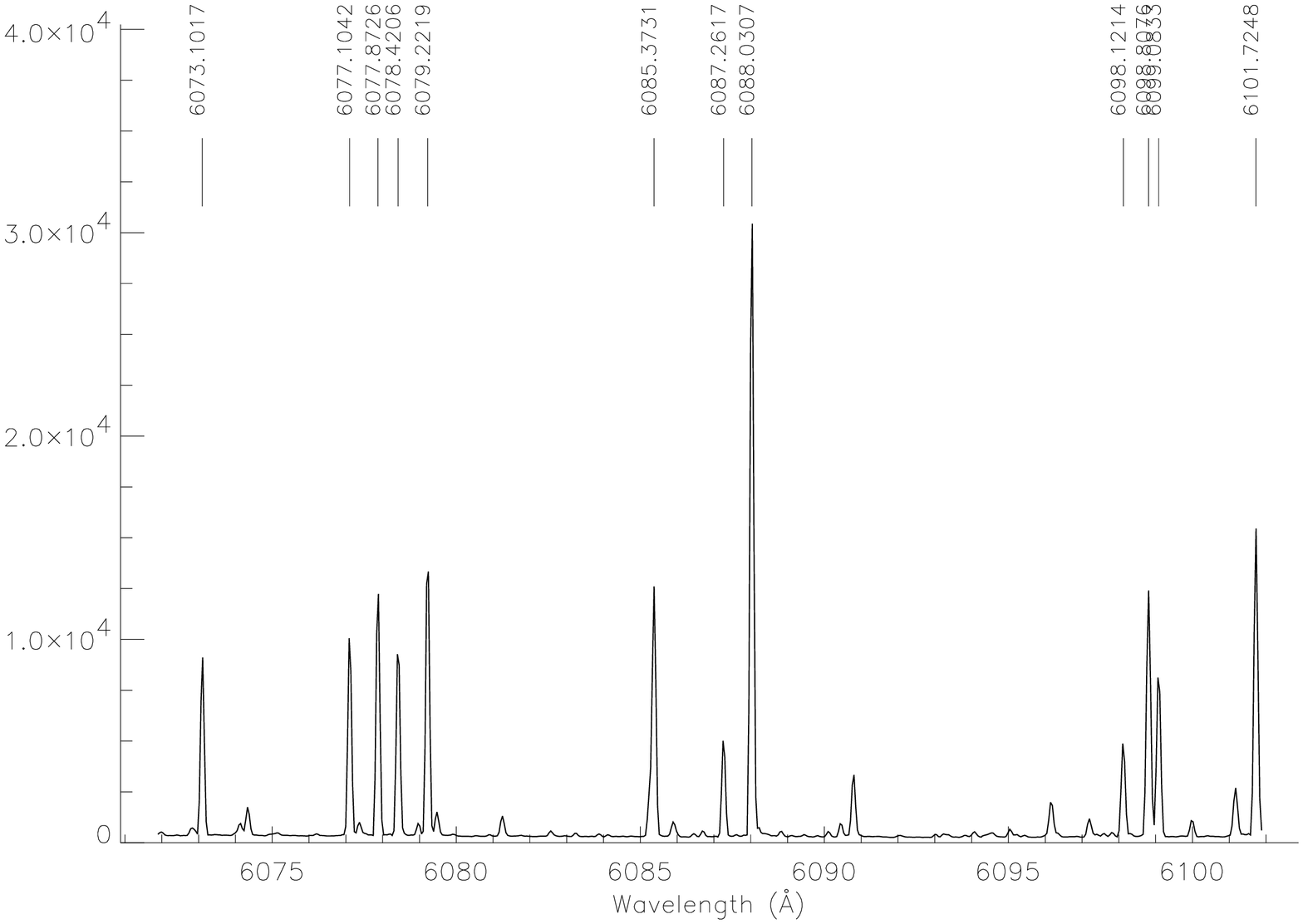}
\end{figure}
\clearpage
   
\begin{figure}
\centering
\includegraphics[width=10cm,angle=90]{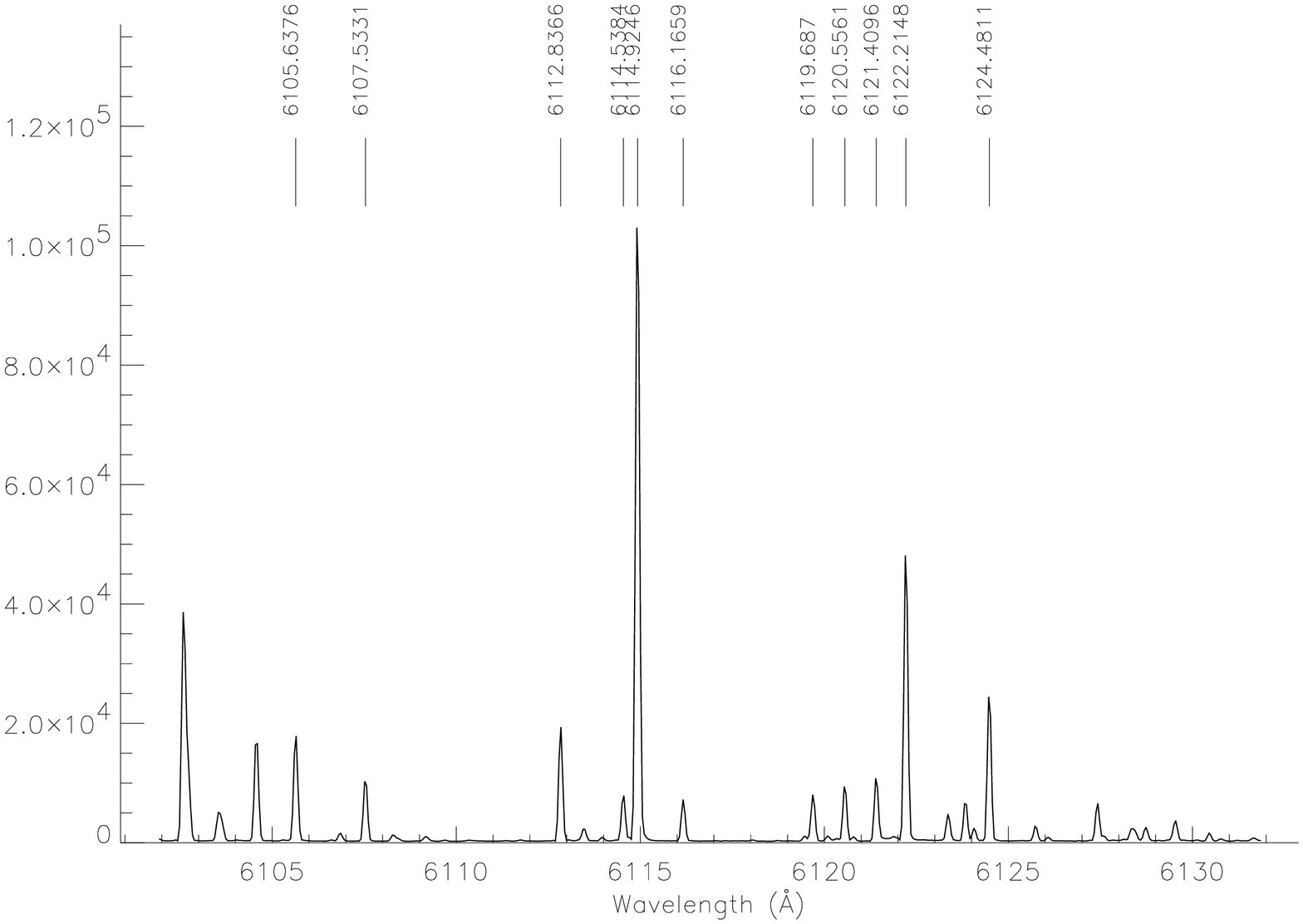}
\includegraphics[width=10cm,angle=90]{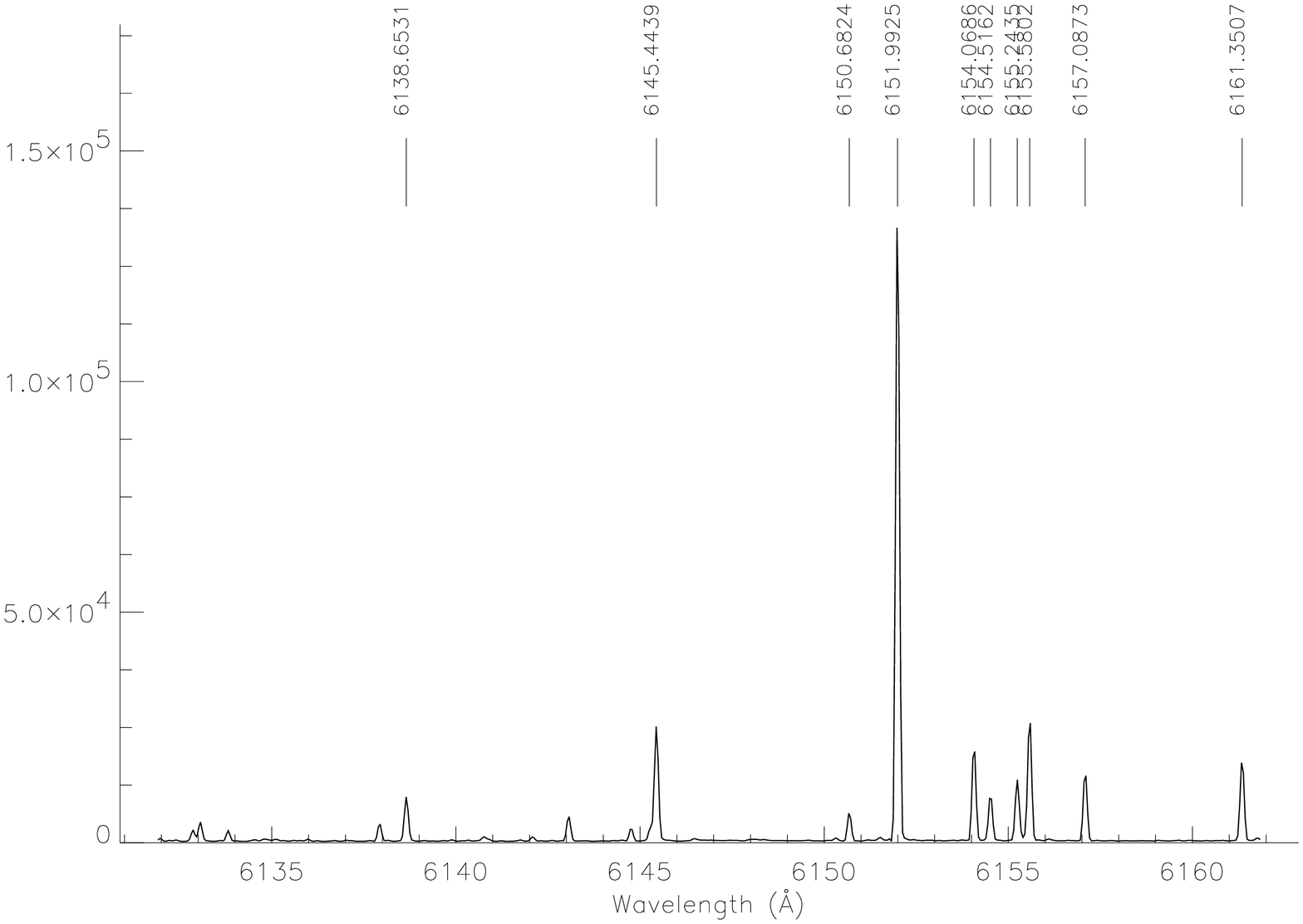}
\end{figure}
\clearpage
   
\begin{figure}
\centering
\includegraphics[width=10cm,angle=90]{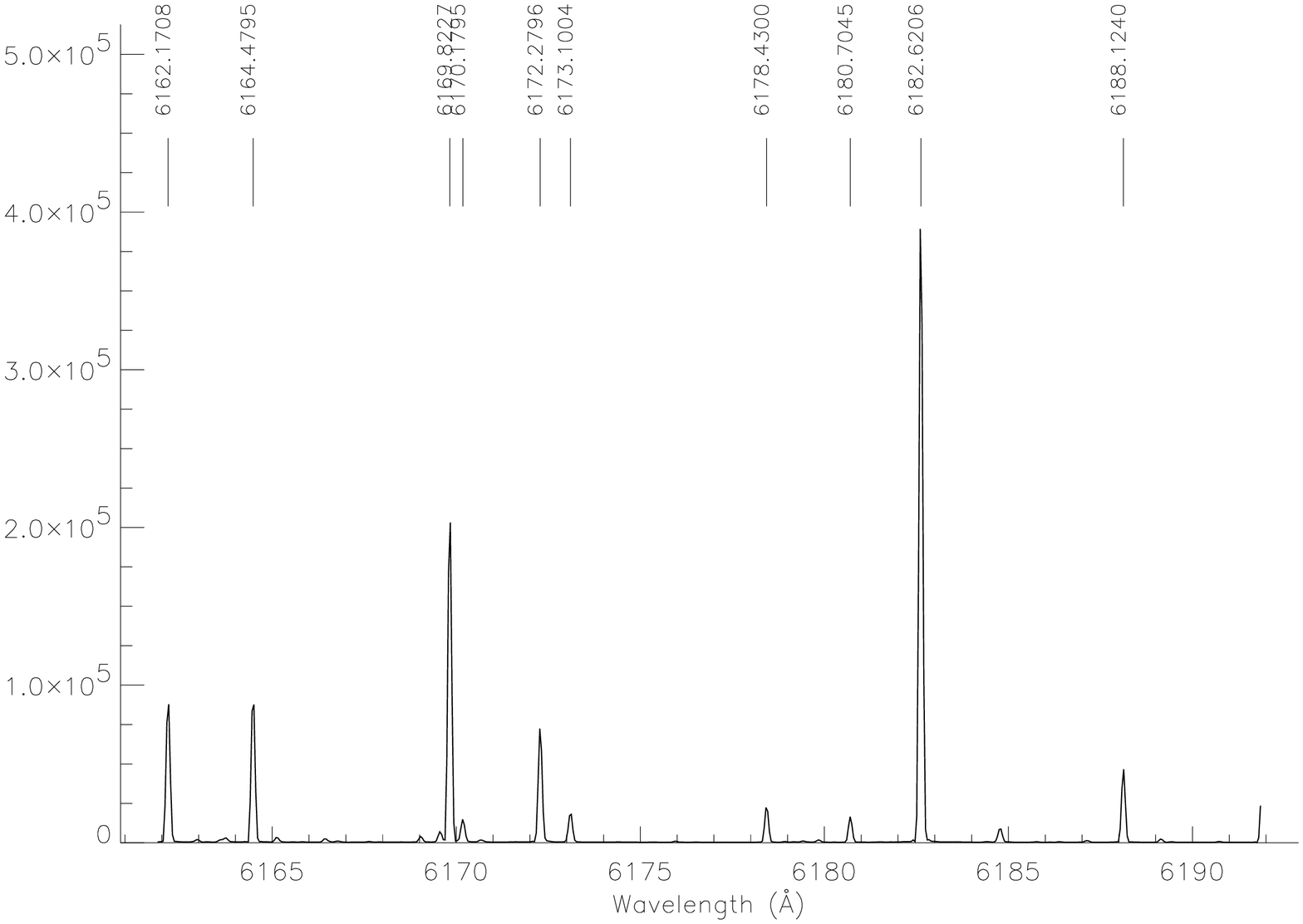}
\includegraphics[width=10cm,angle=90]{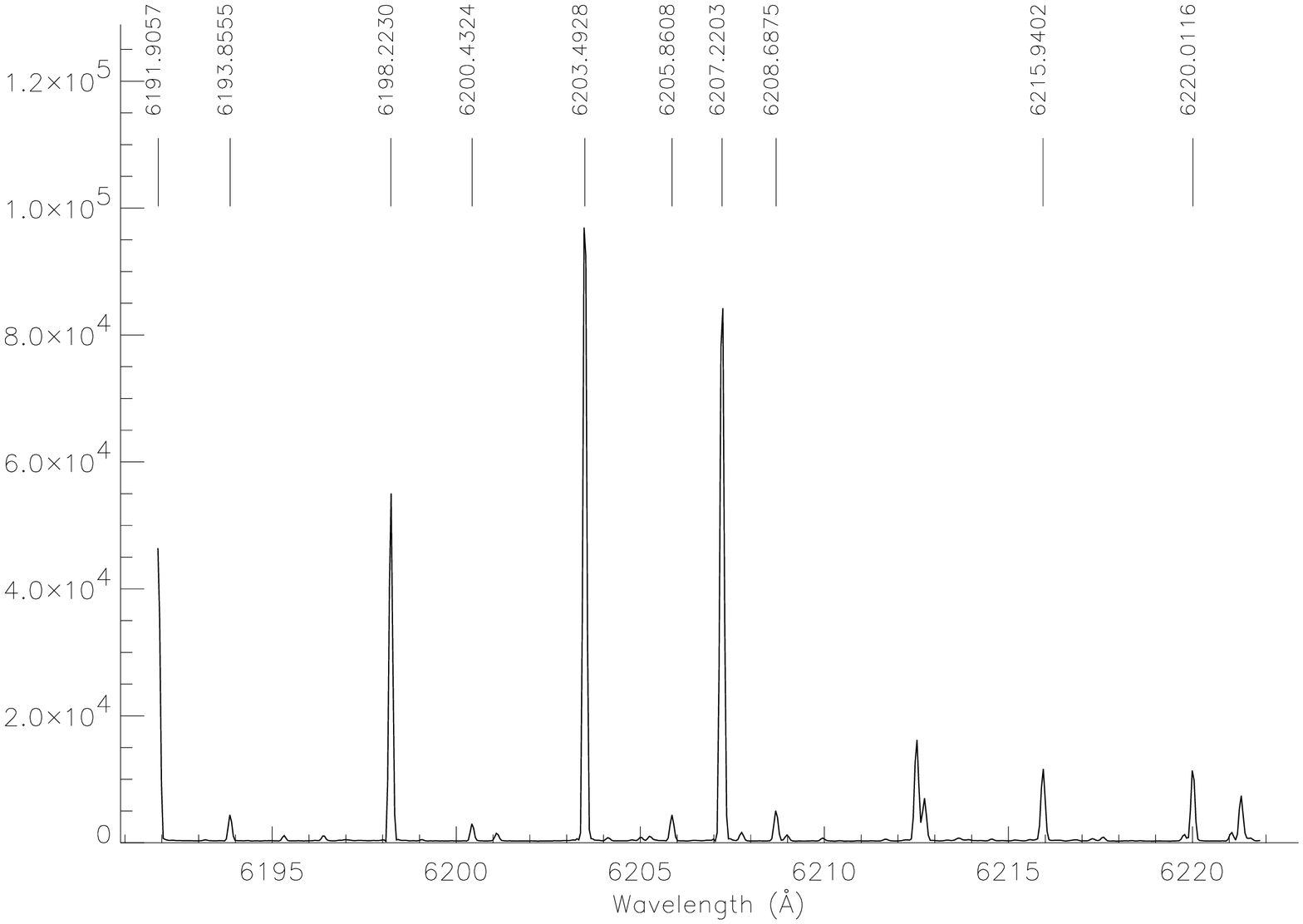}
\end{figure}
\clearpage
   
\begin{figure}
\centering
\includegraphics[width=10cm,angle=90]{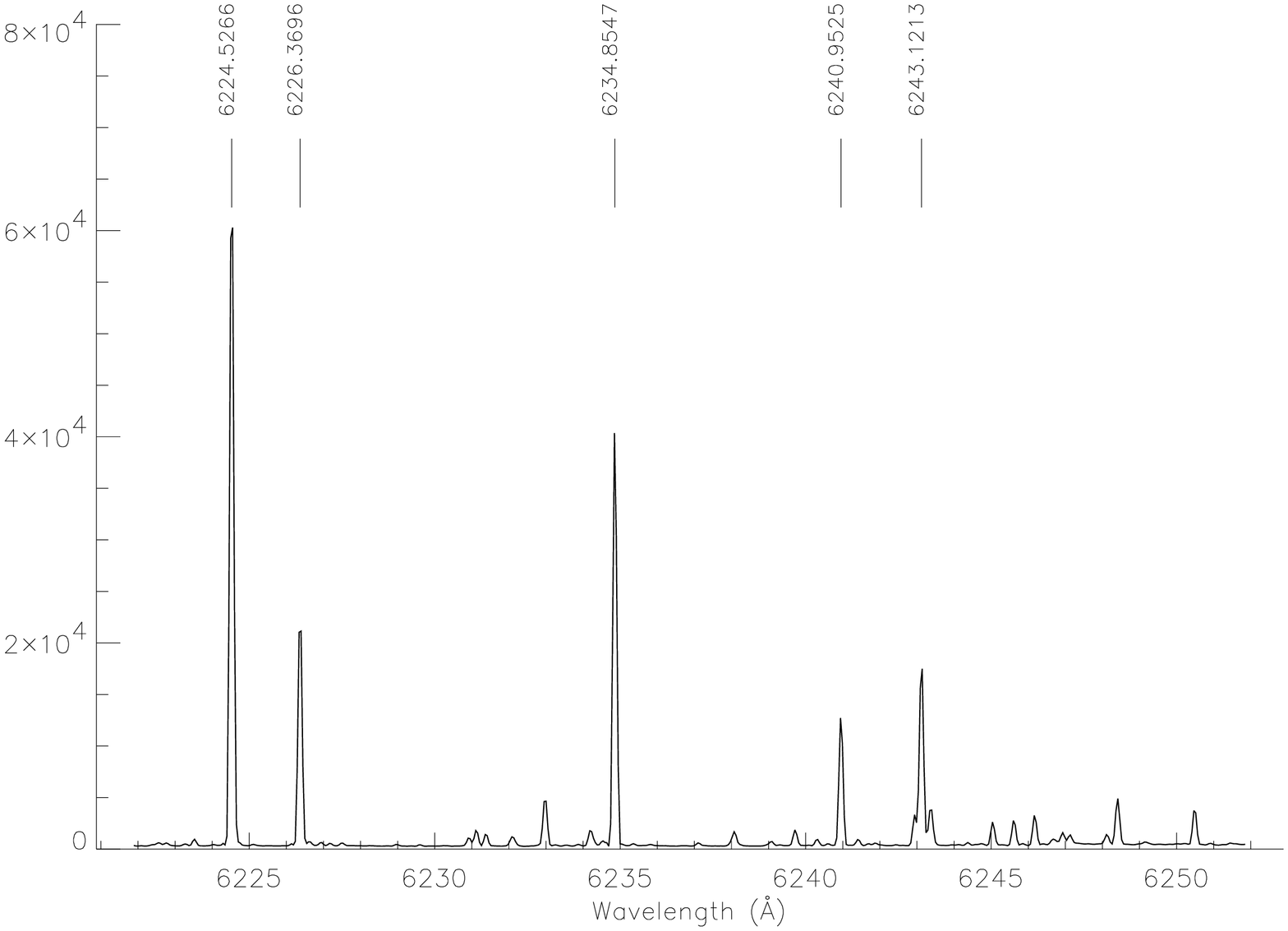}
\includegraphics[width=10cm,angle=90]{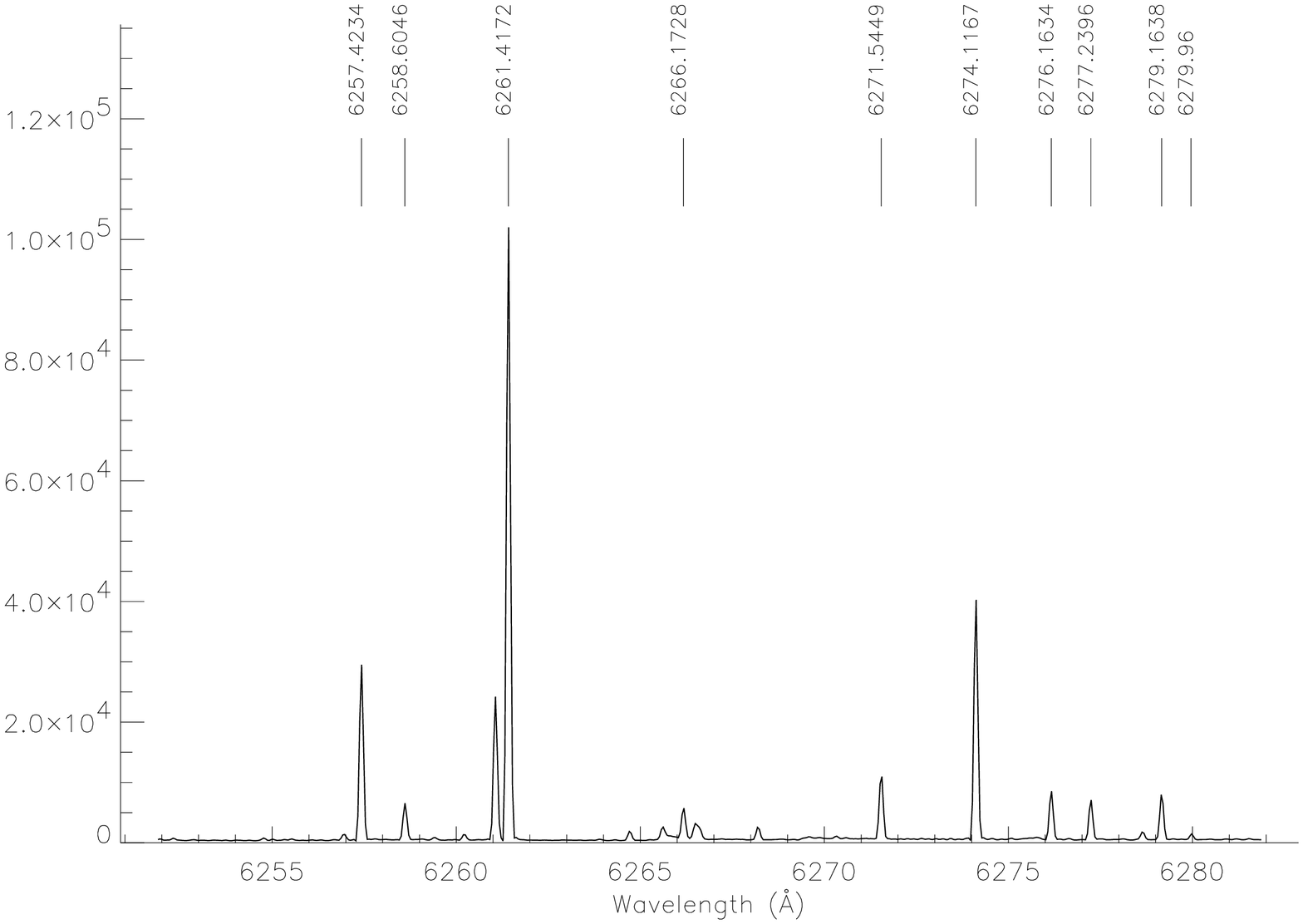}
\end{figure}
\clearpage
   
\begin{figure}
\centering
\includegraphics[width=10cm,angle=90]{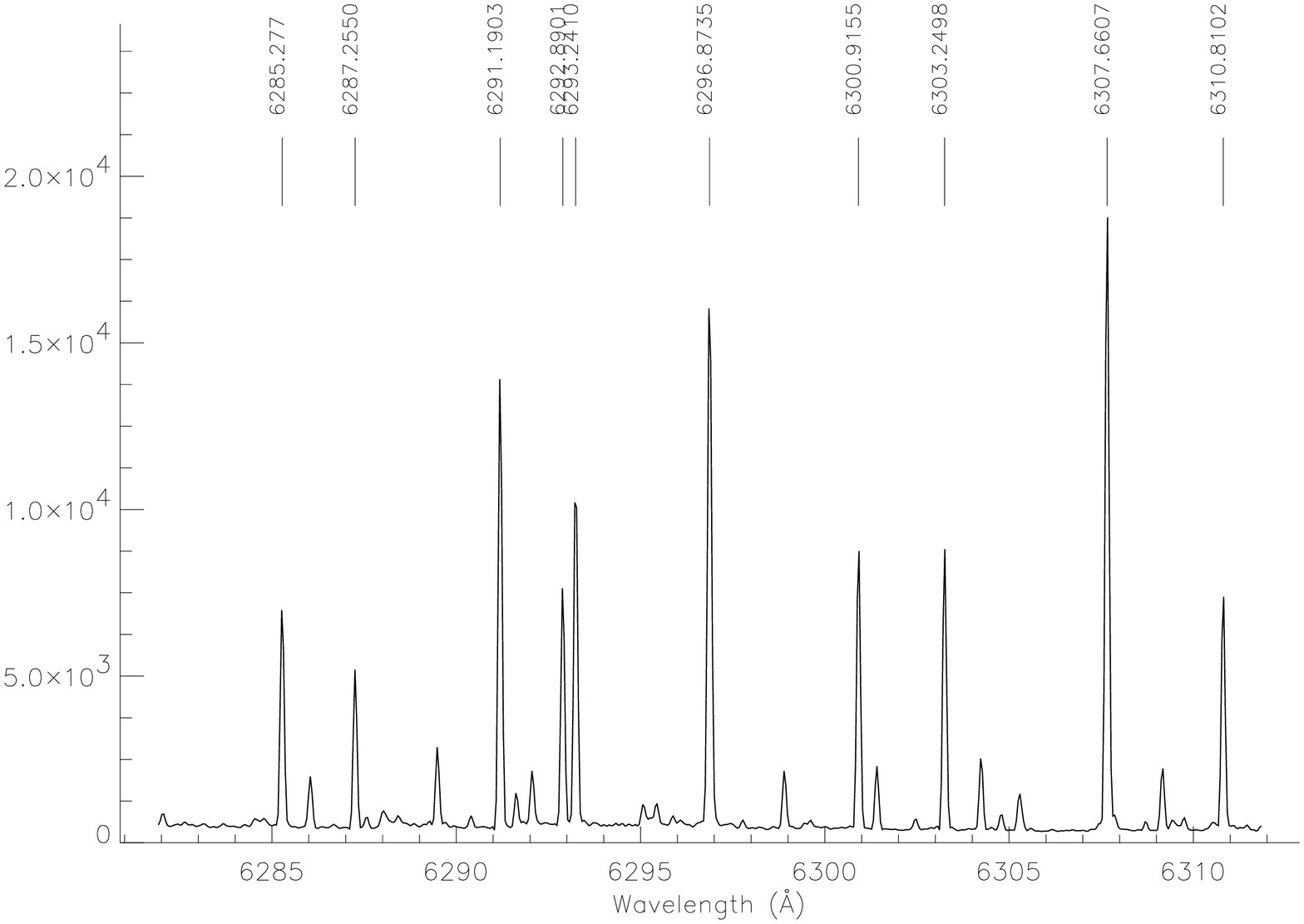}
\includegraphics[width=10cm,angle=90]{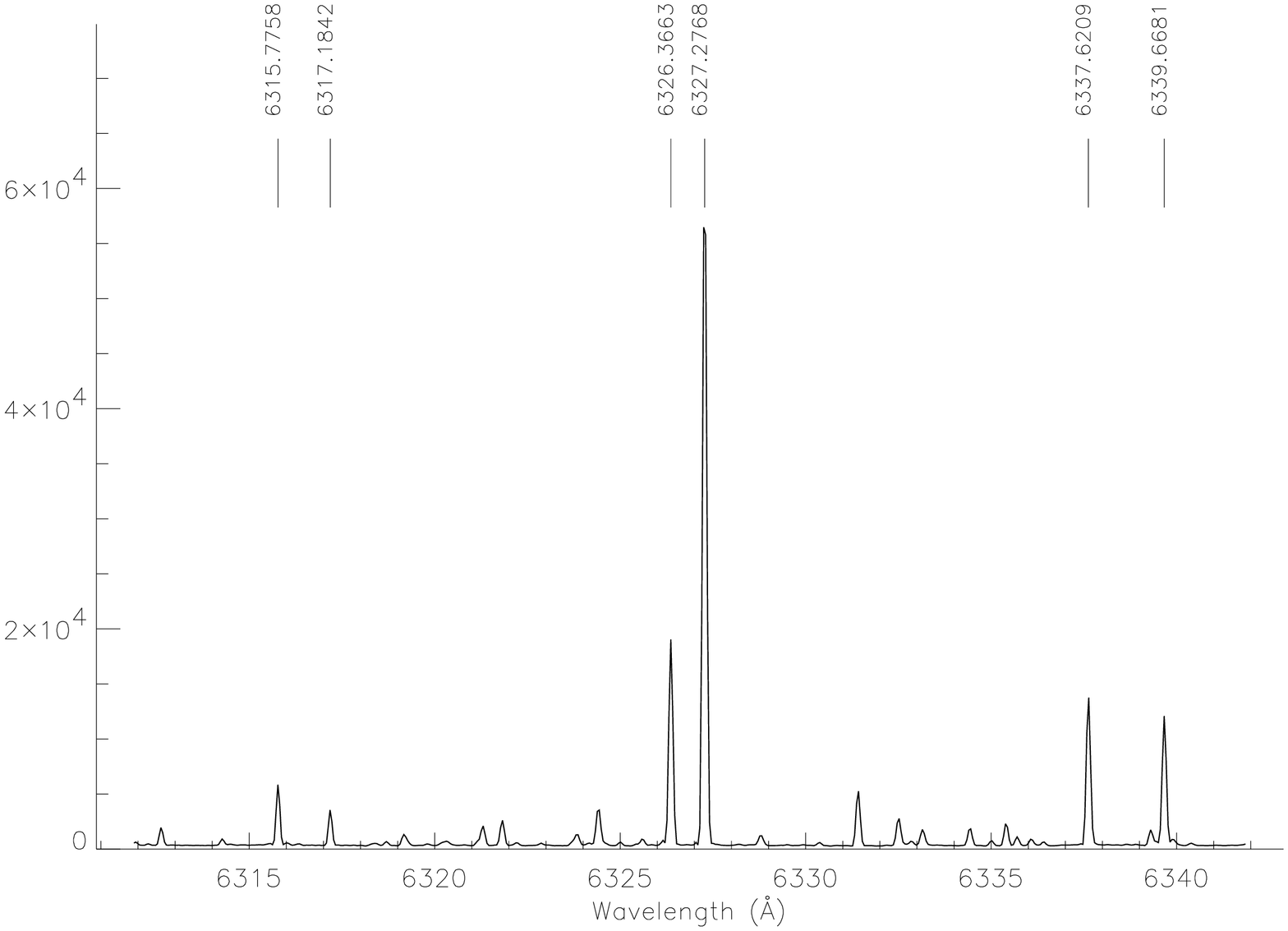}
\end{figure}
\clearpage
   
\begin{figure}
\centering
\includegraphics[width=10cm,angle=90]{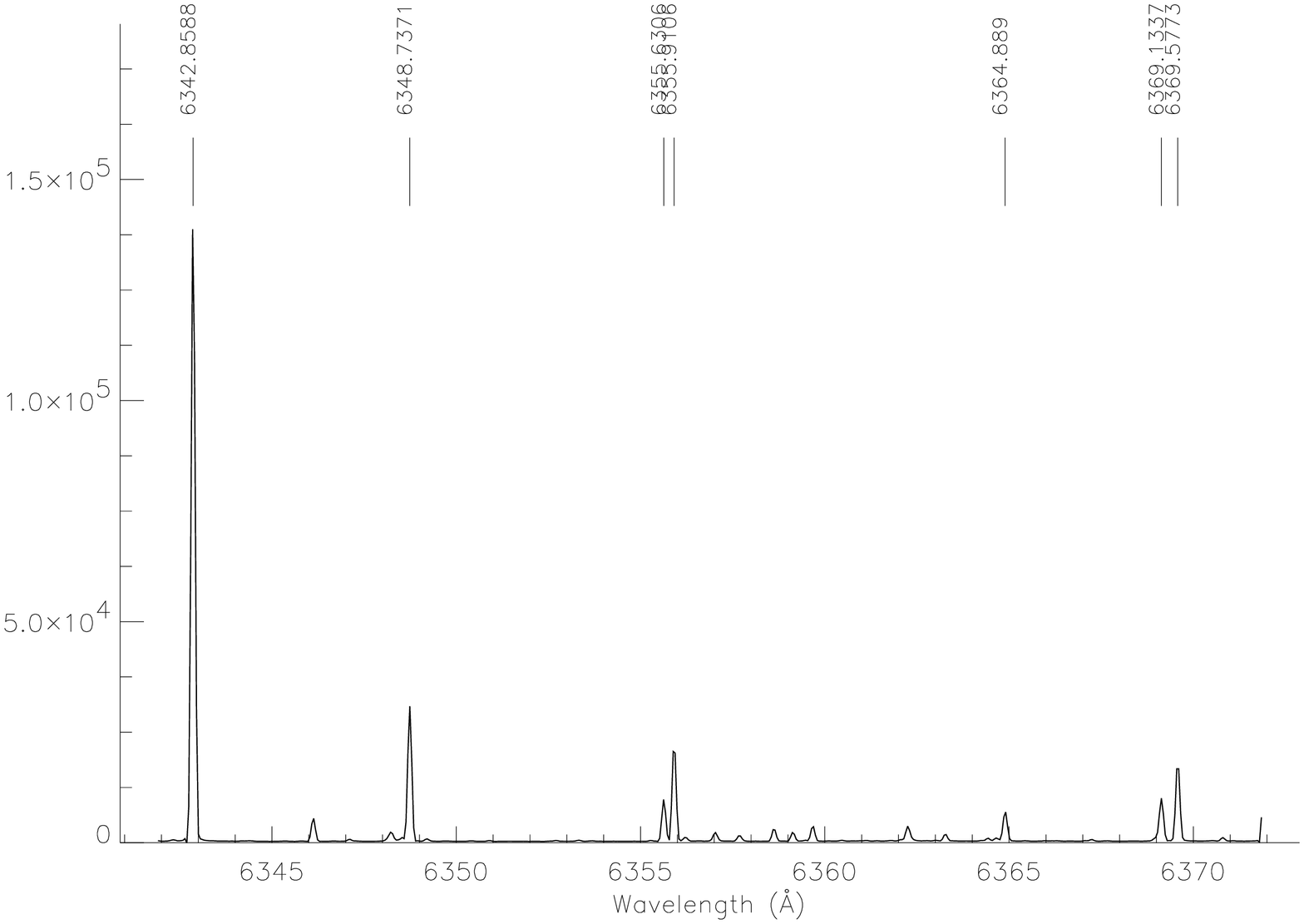}
\includegraphics[width=10cm,angle=90]{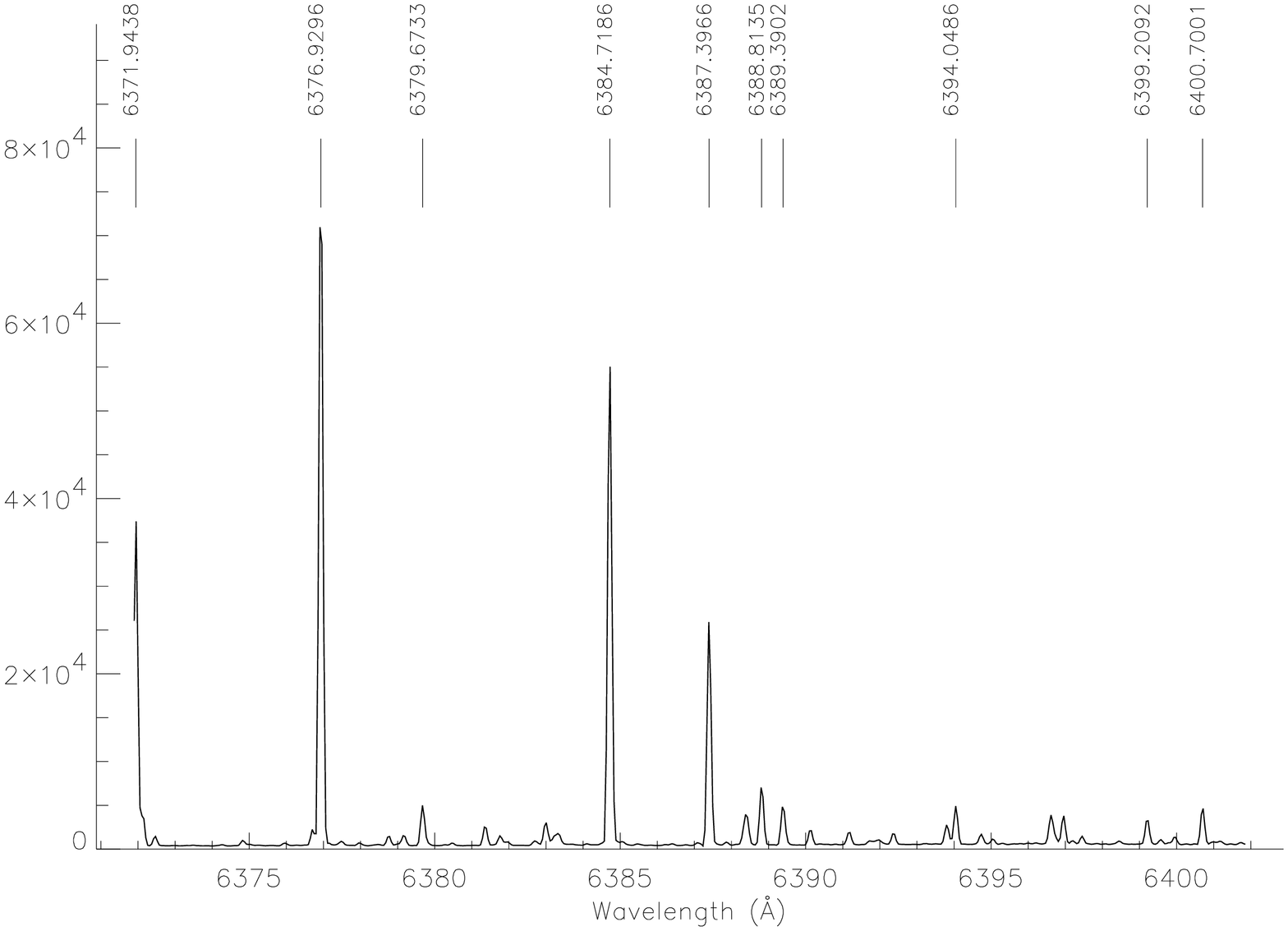}
\end{figure}
\clearpage
   
\begin{figure}
\centering
\includegraphics[width=10cm,angle=90]{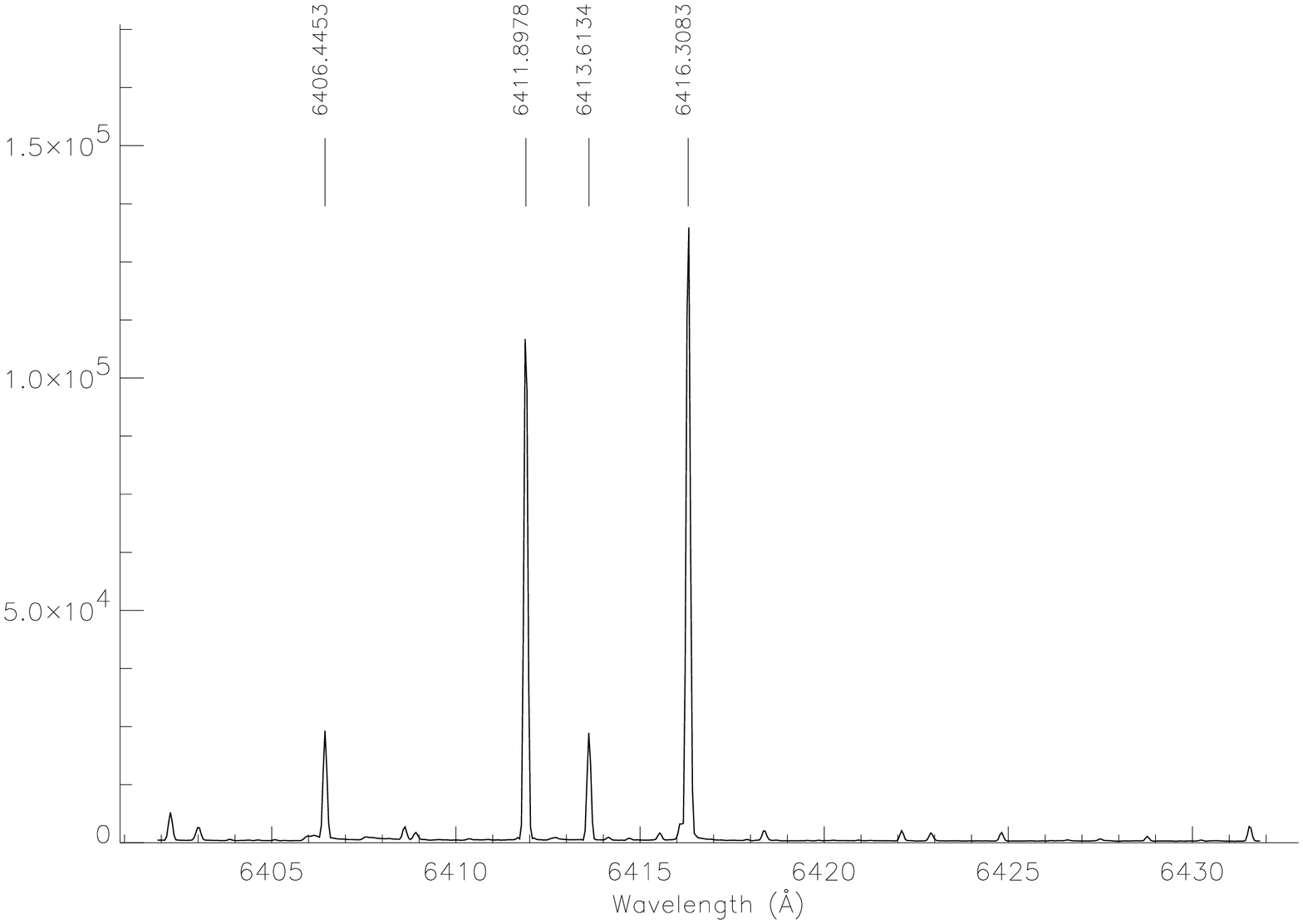}
\includegraphics[width=10cm,angle=90]{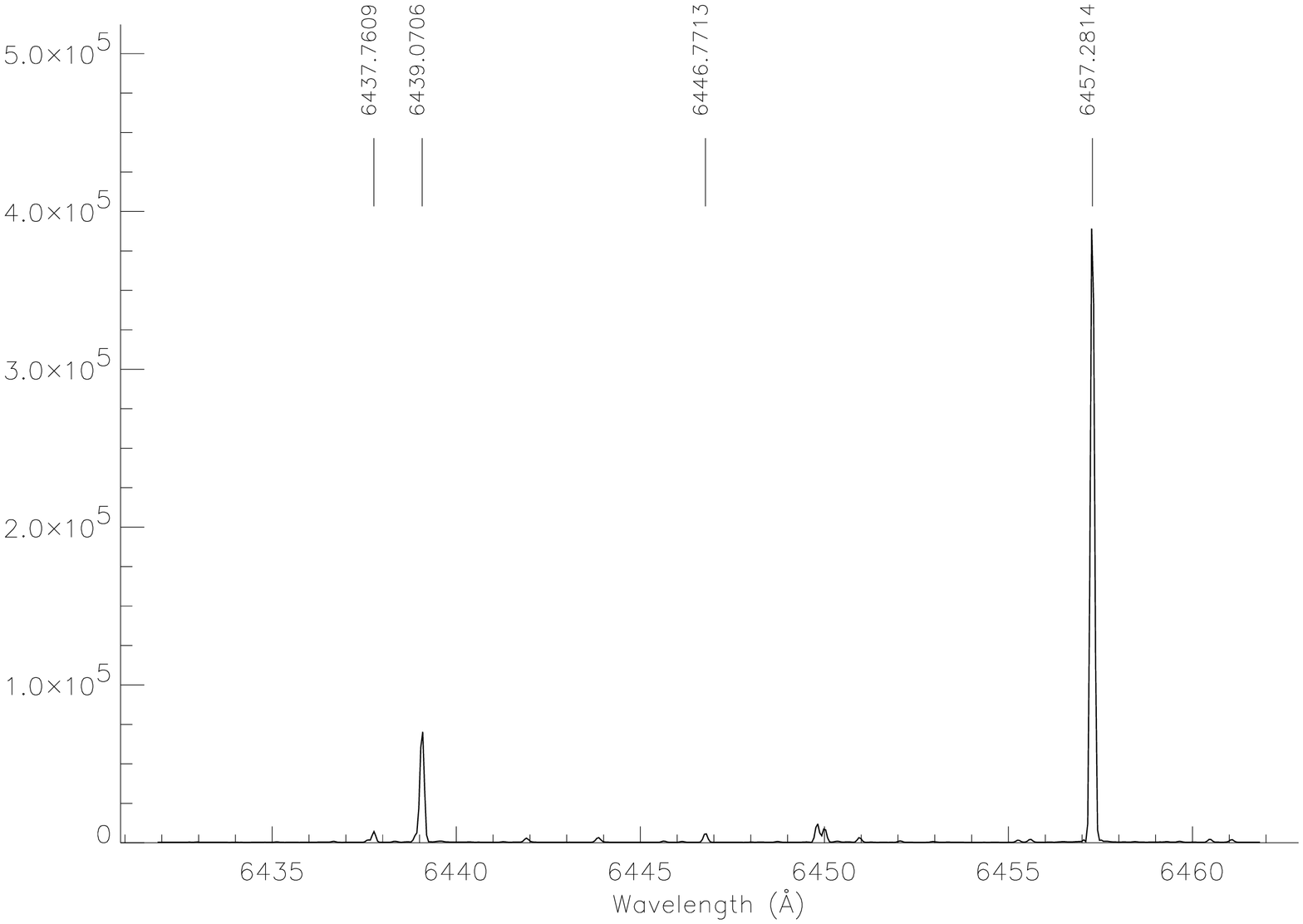}
\end{figure}
\clearpage
   
\begin{figure}
\centering
\includegraphics[width=10cm,angle=90]{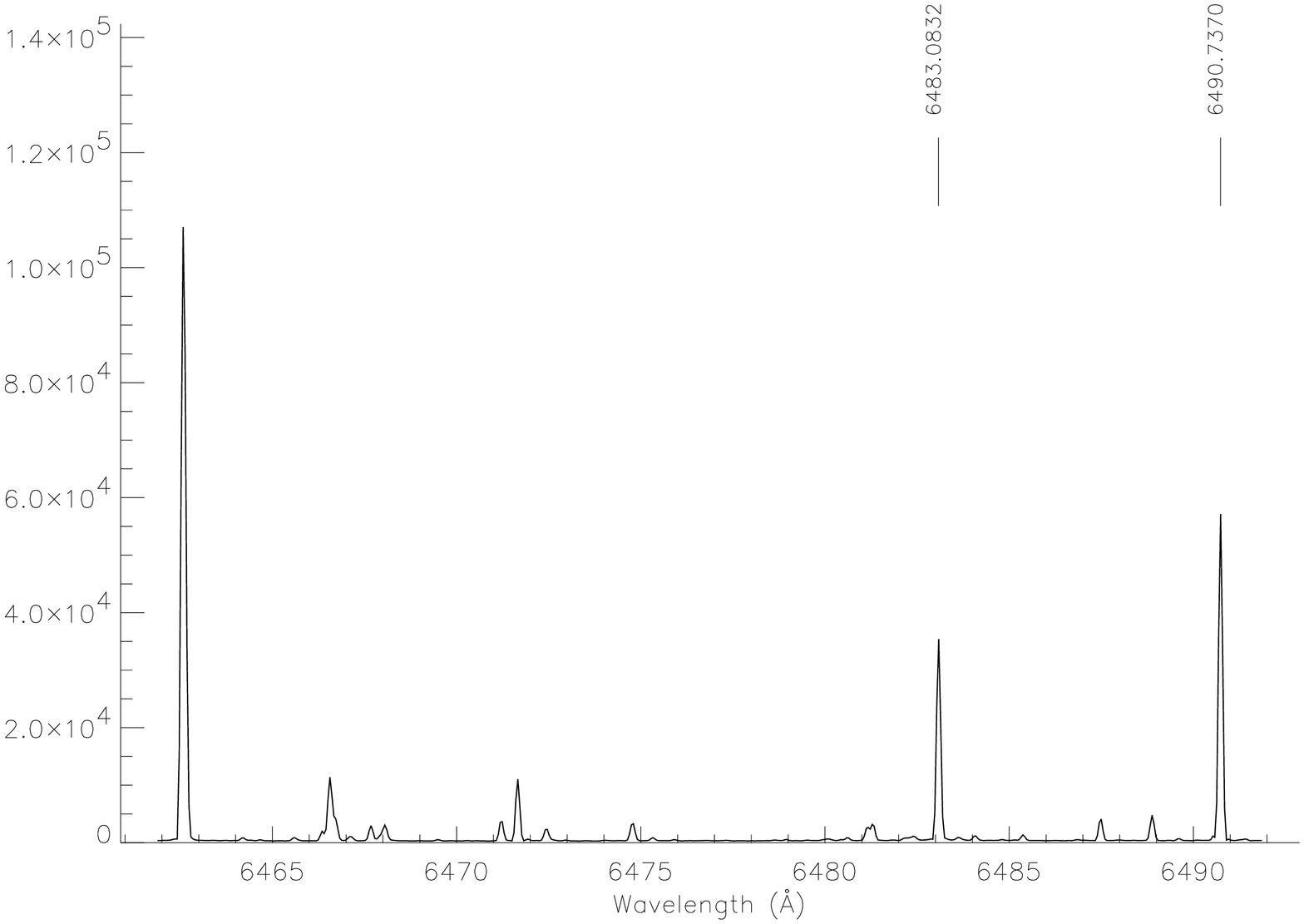}
\includegraphics[width=10cm,angle=90]{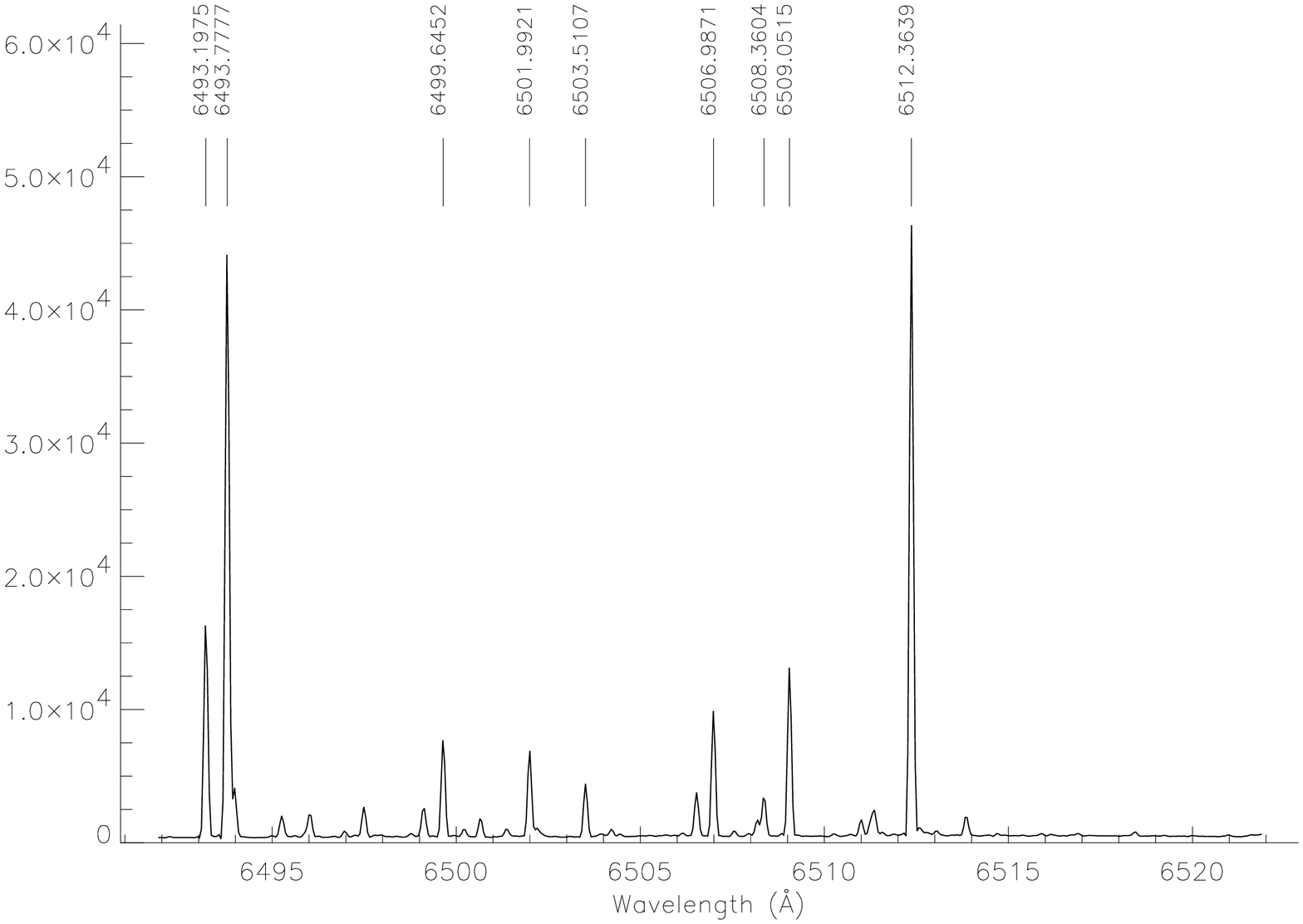}
\end{figure}
\clearpage
   
\begin{figure}
\centering
\includegraphics[width=10cm,angle=90]{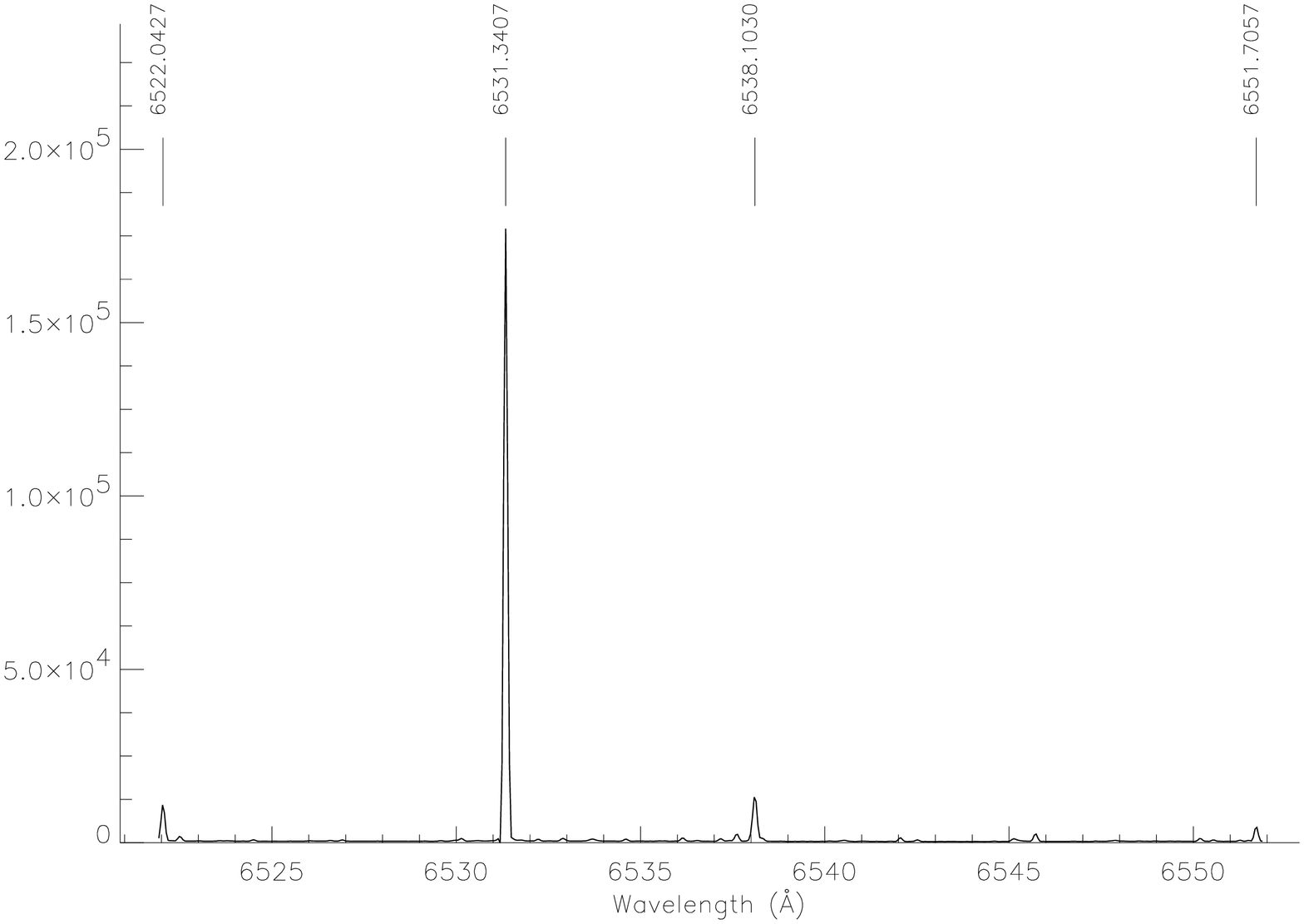}
\includegraphics[width=10cm,angle=90]{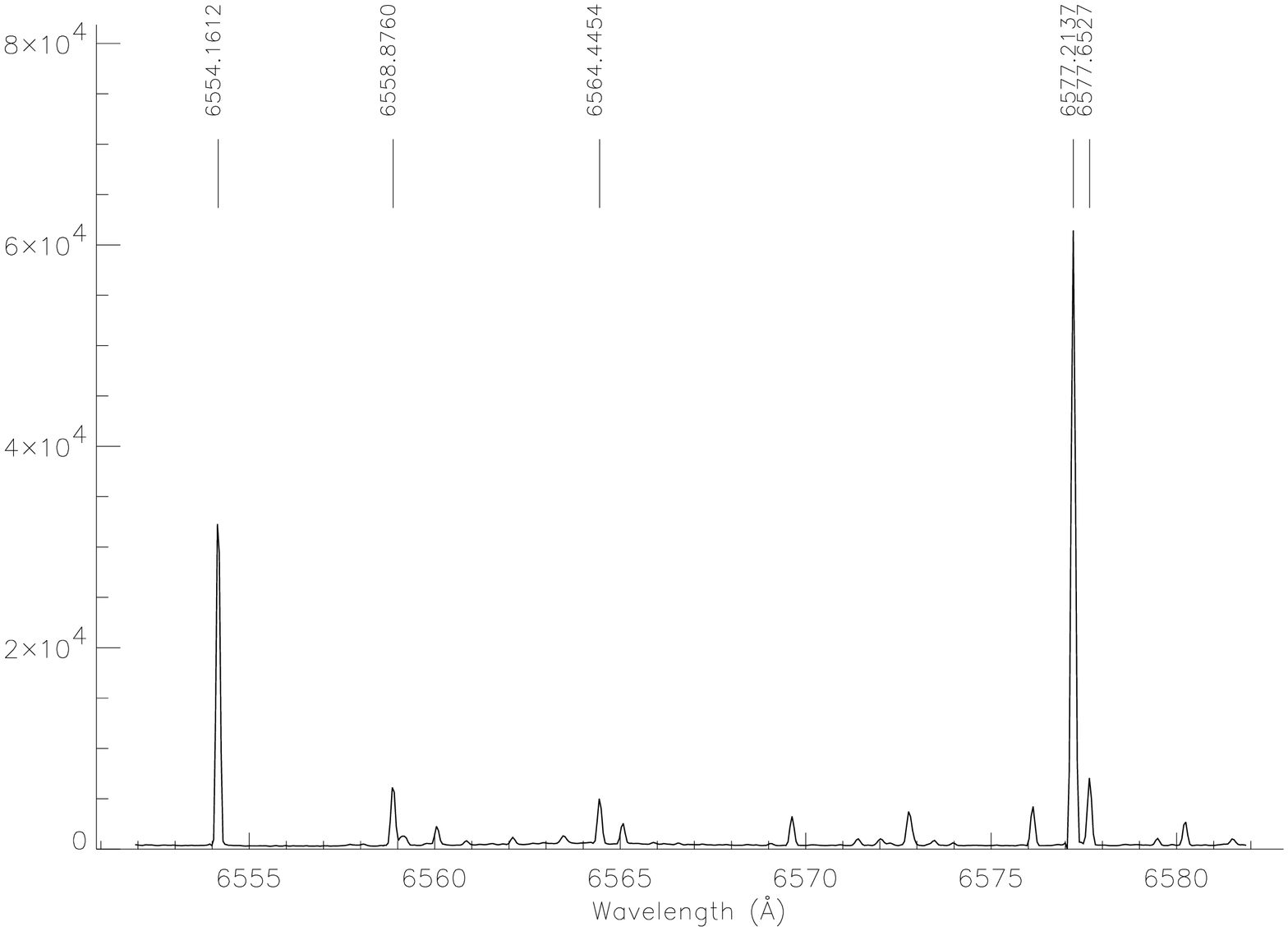}
\end{figure}
\clearpage
   
\begin{figure}
\centering
\includegraphics[width=10cm,angle=90]{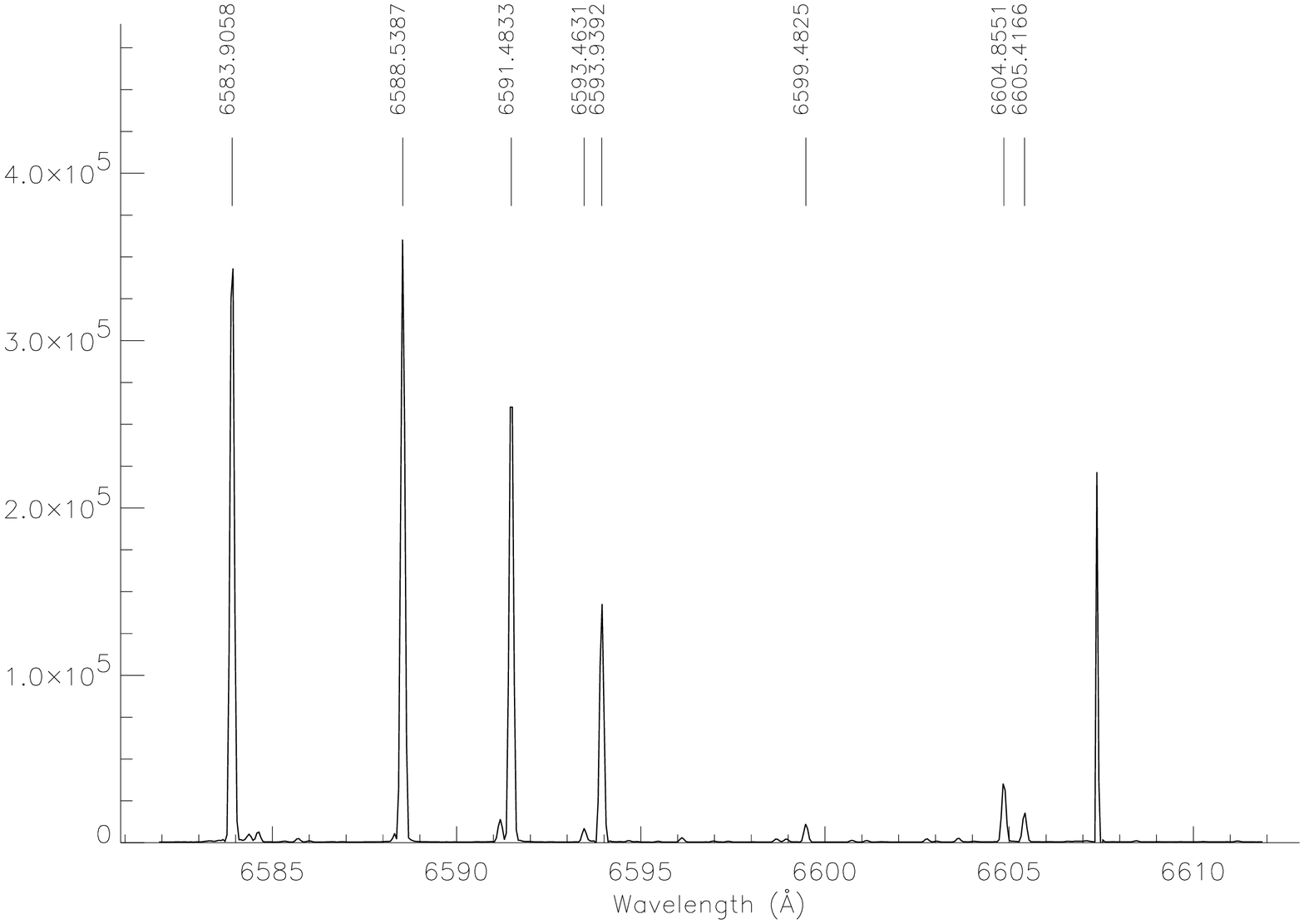}
\includegraphics[width=10cm,angle=90]{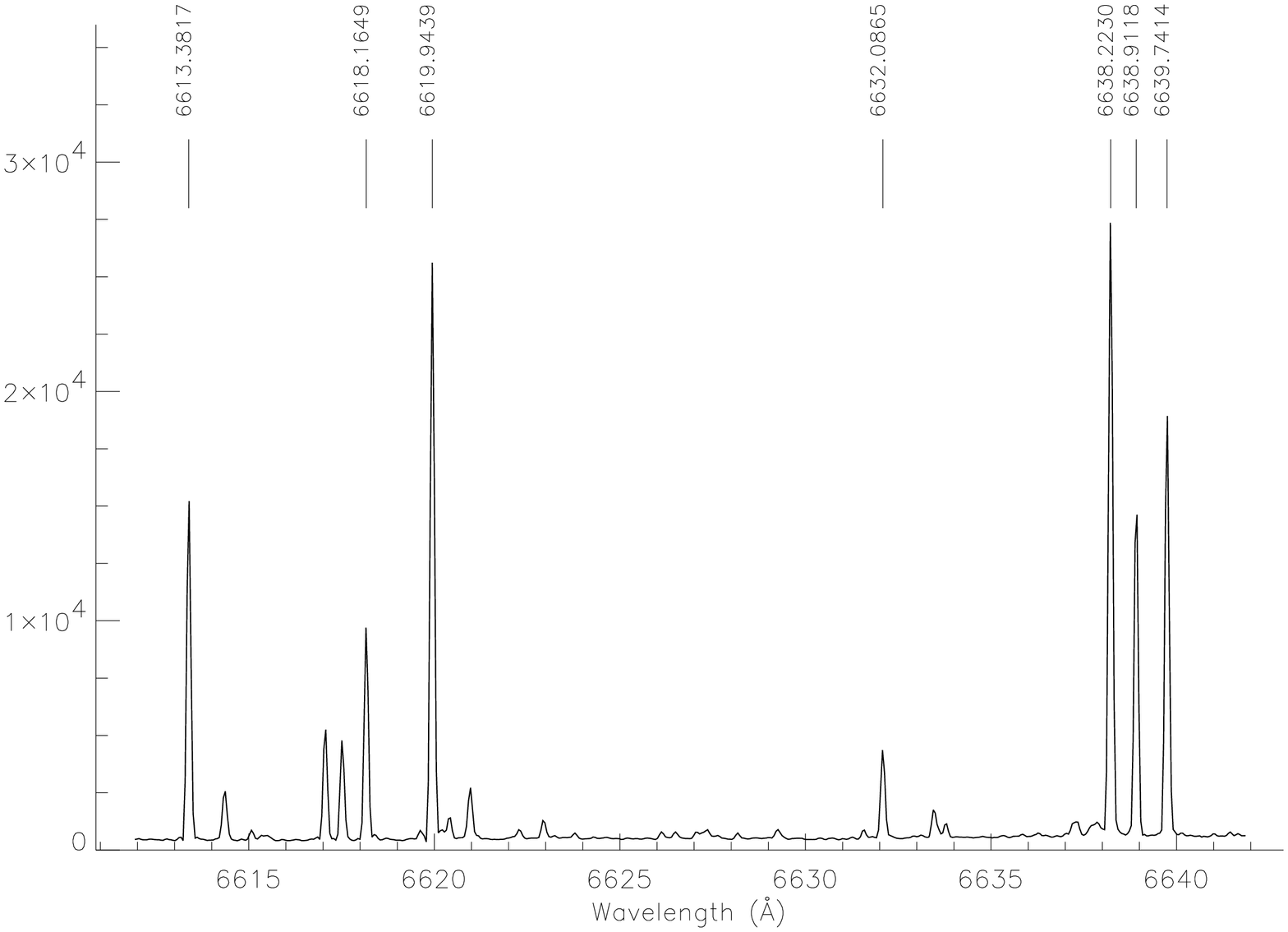}
\end{figure}
\clearpage
   
\begin{figure}
\centering
\includegraphics[width=10cm,angle=90]{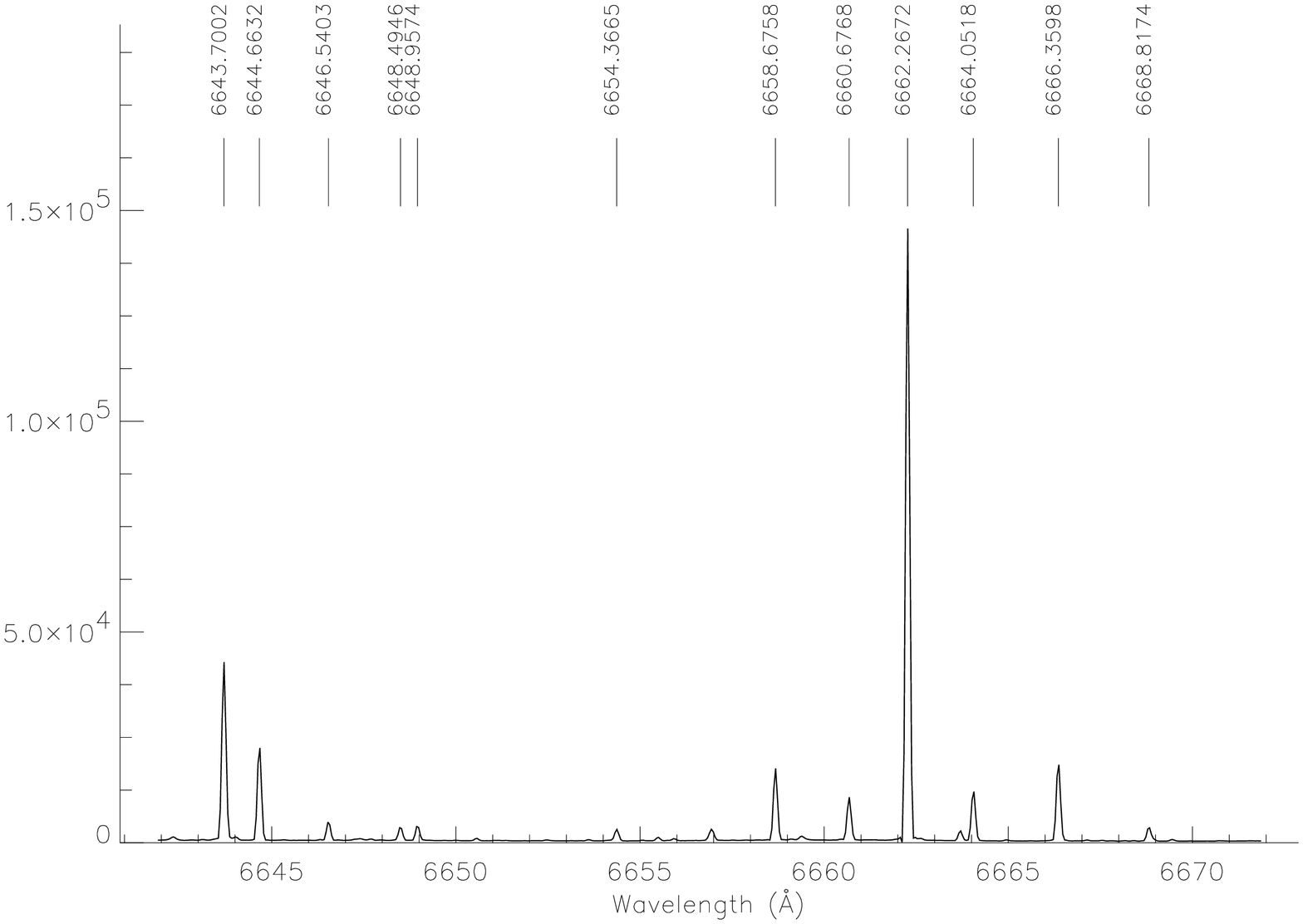}
\includegraphics[width=10cm,angle=90]{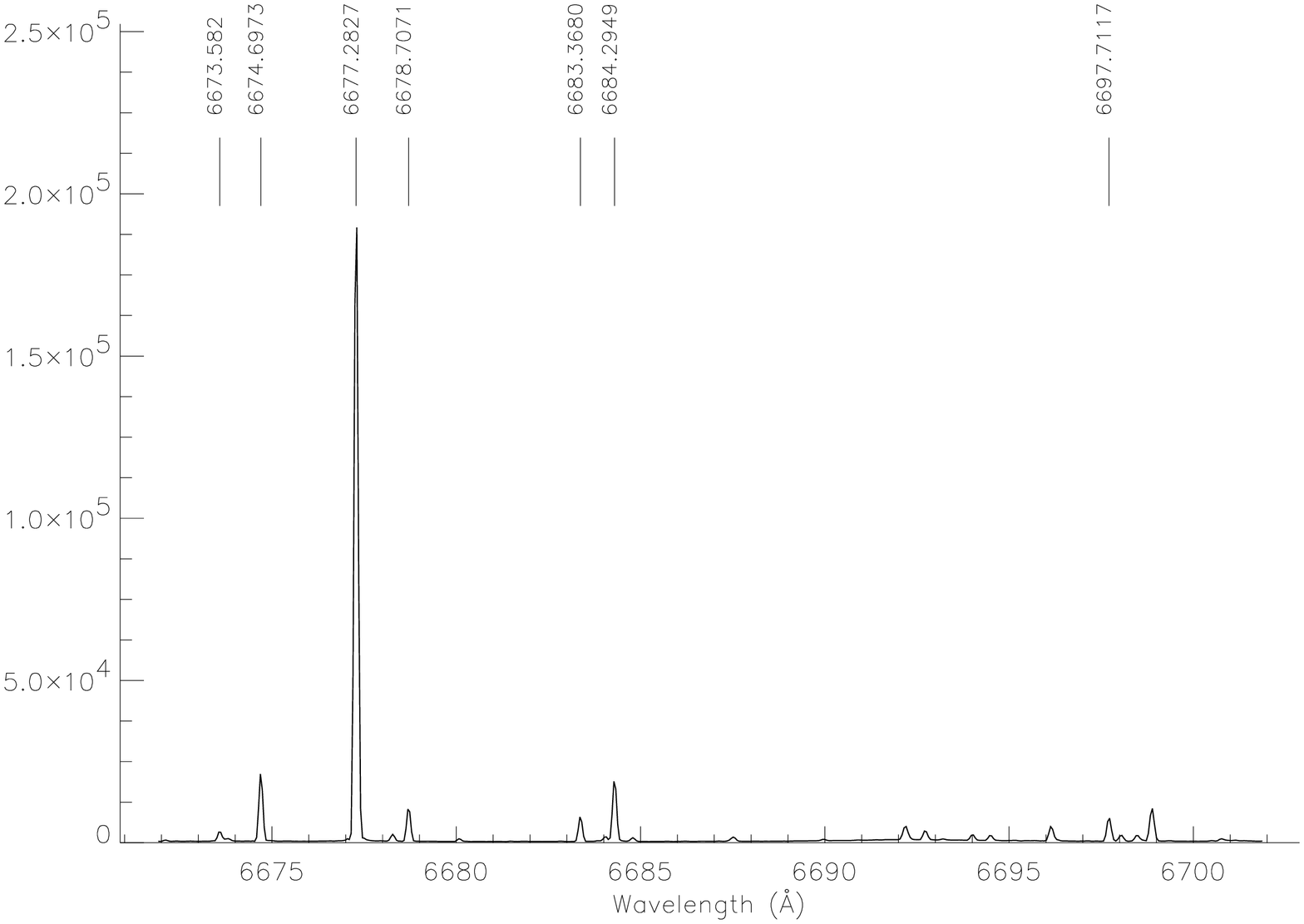}
\end{figure}
\clearpage
   
\begin{figure}
\centering
\includegraphics[width=10cm,angle=90]{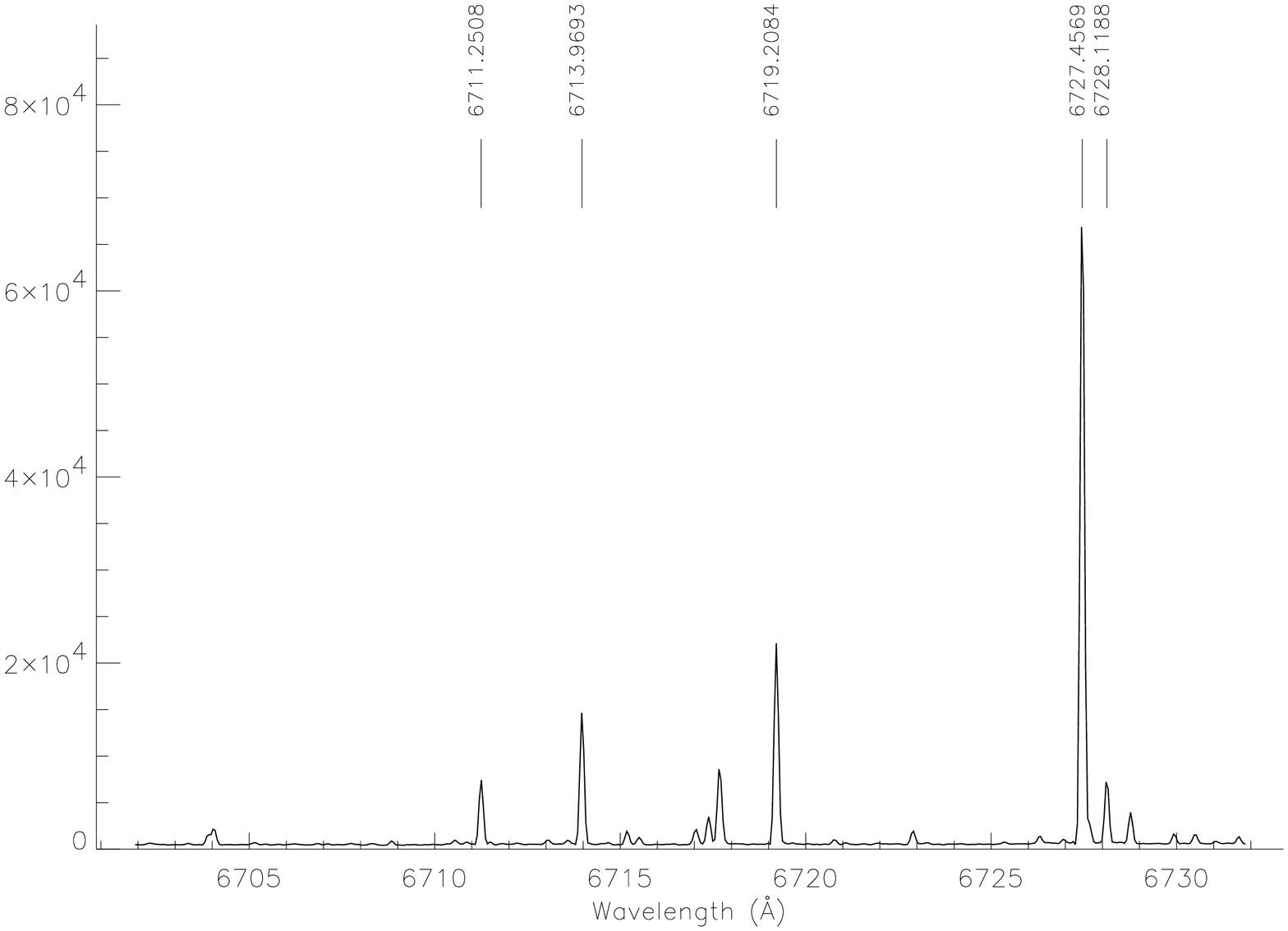}
\includegraphics[width=10cm,angle=90]{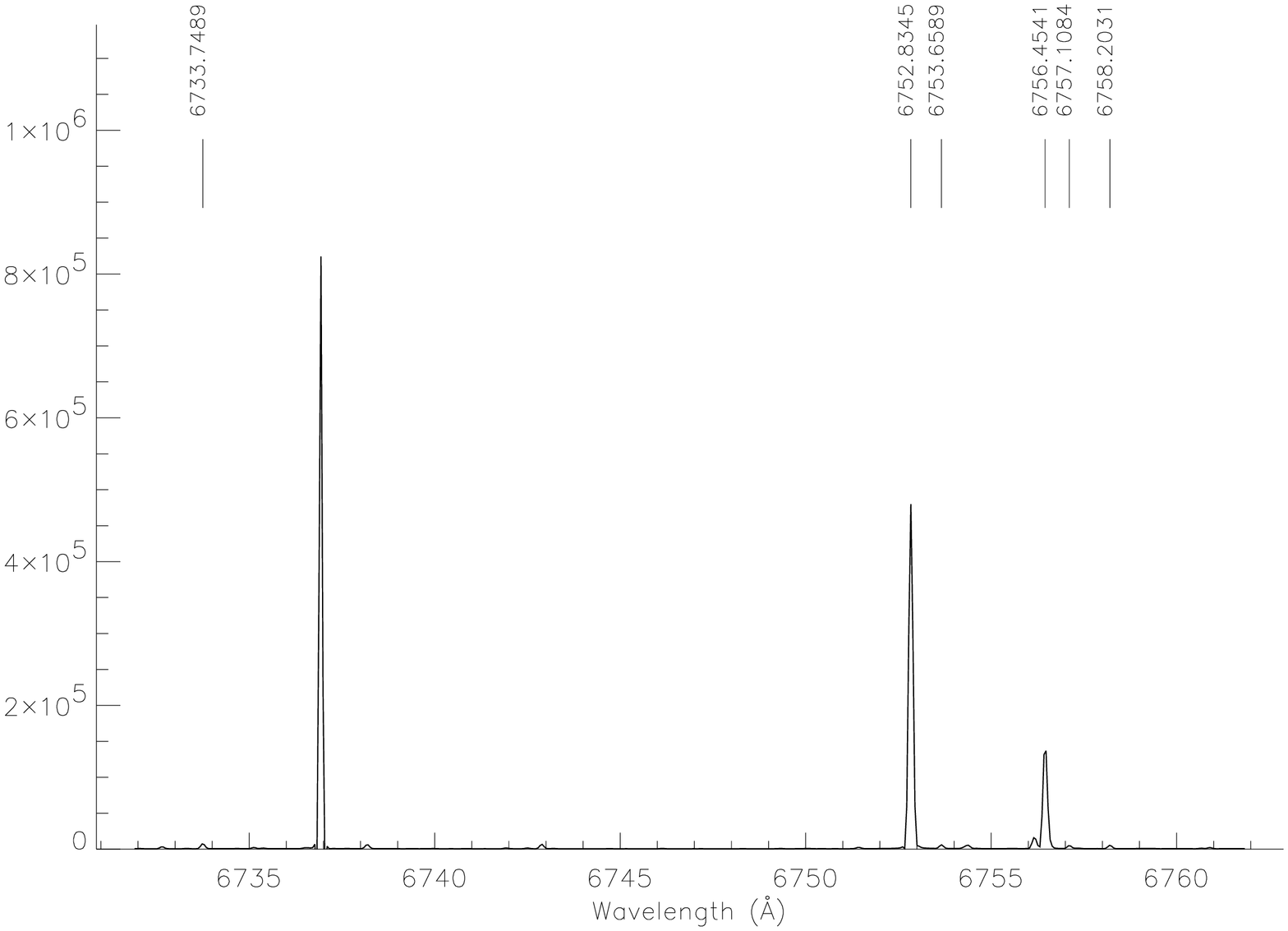}
\end{figure}
\clearpage
   
\begin{figure}
\centering
\includegraphics[width=10cm,angle=90]{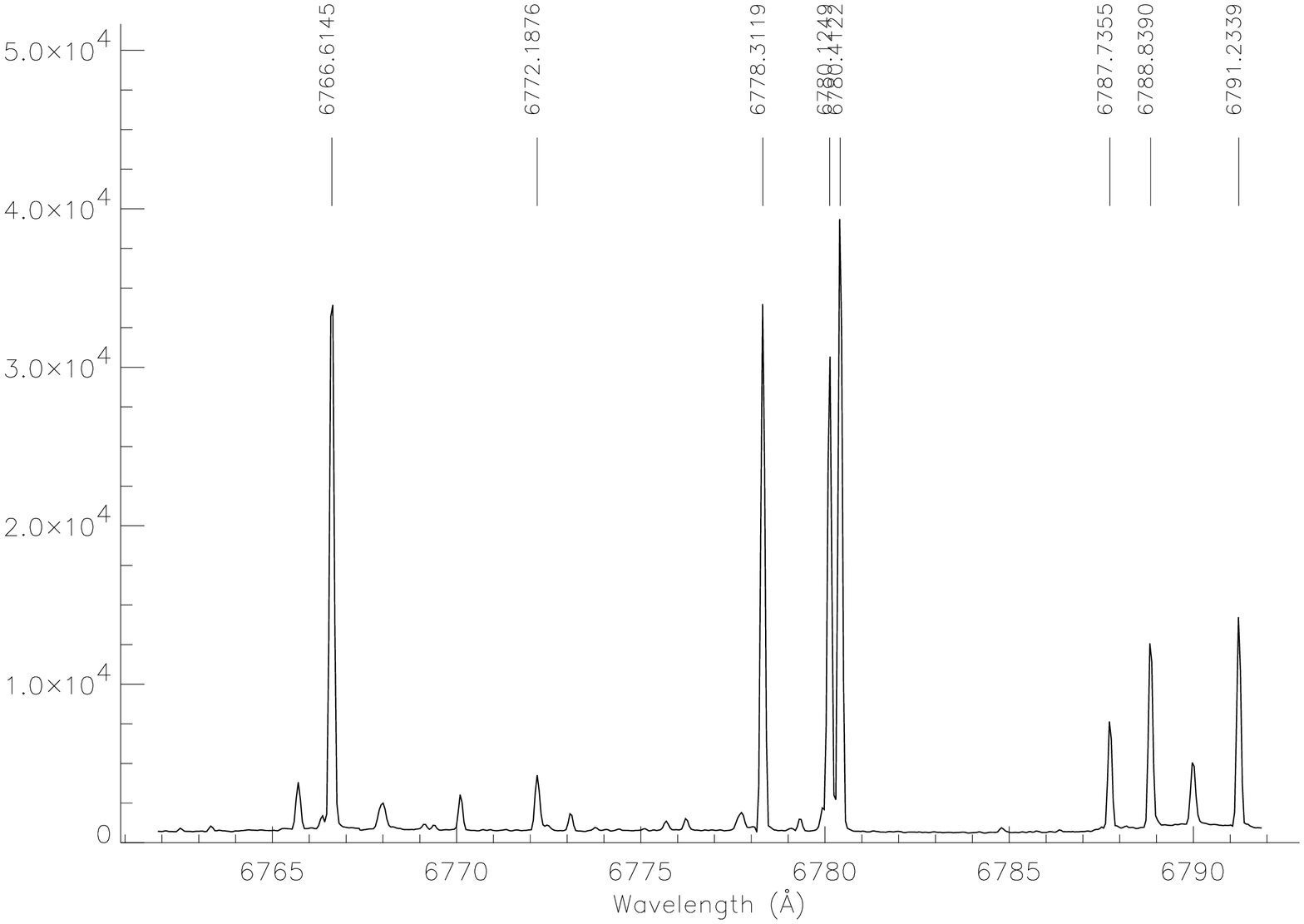}
\includegraphics[width=10cm,angle=90]{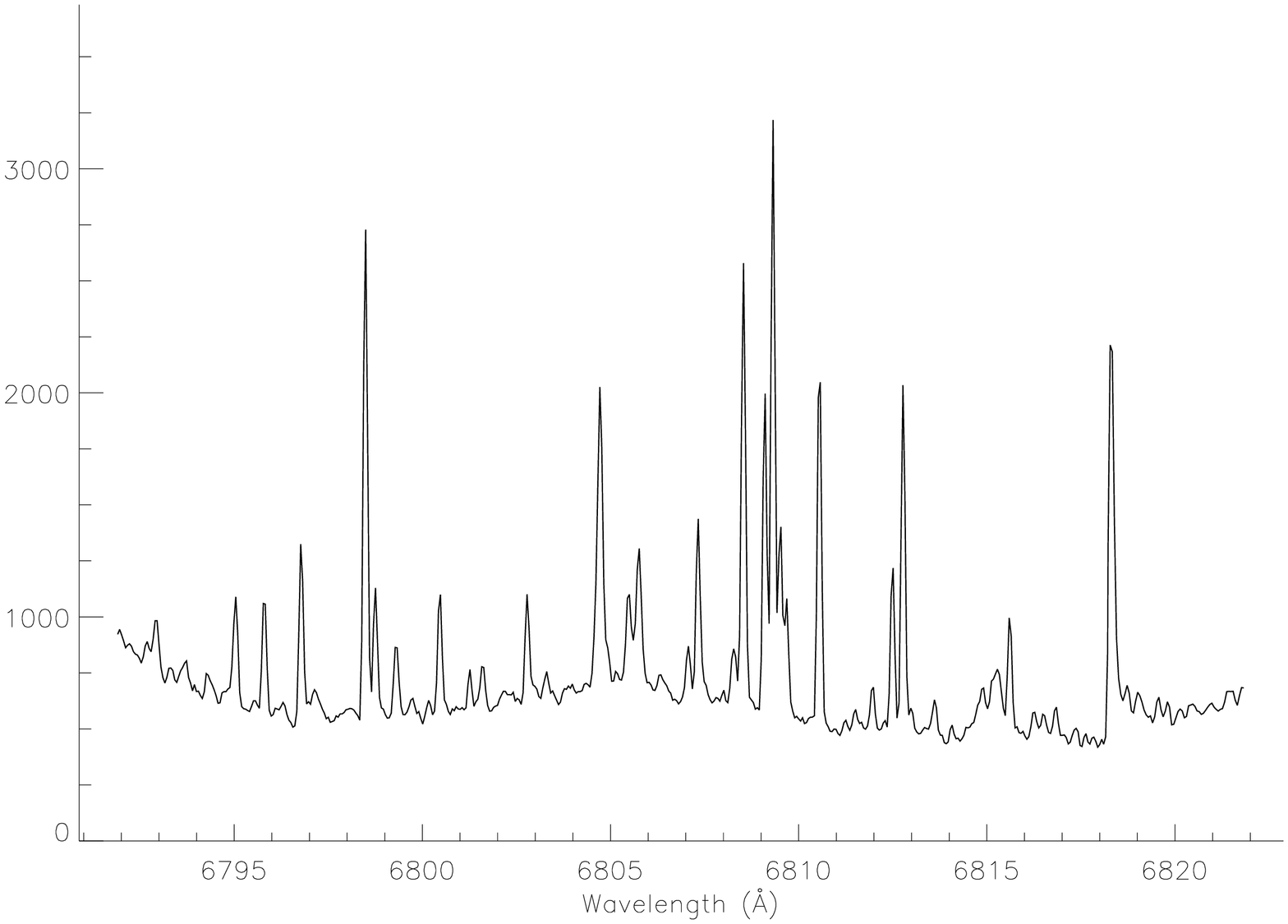}
\end{figure}
\clearpage
   
\begin{figure}
\centering
\includegraphics[width=10cm,angle=90]{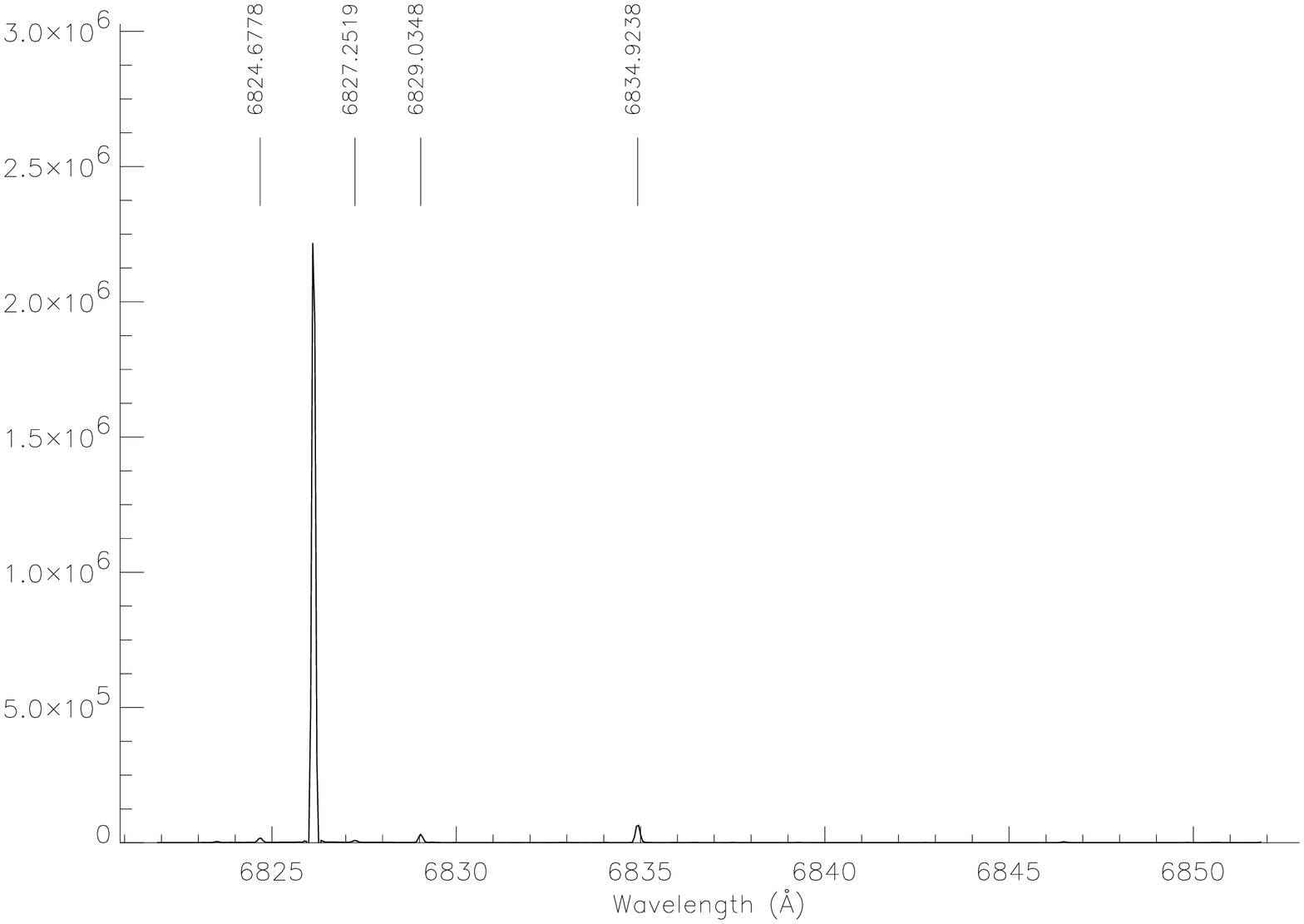}
\includegraphics[width=10cm,angle=90]{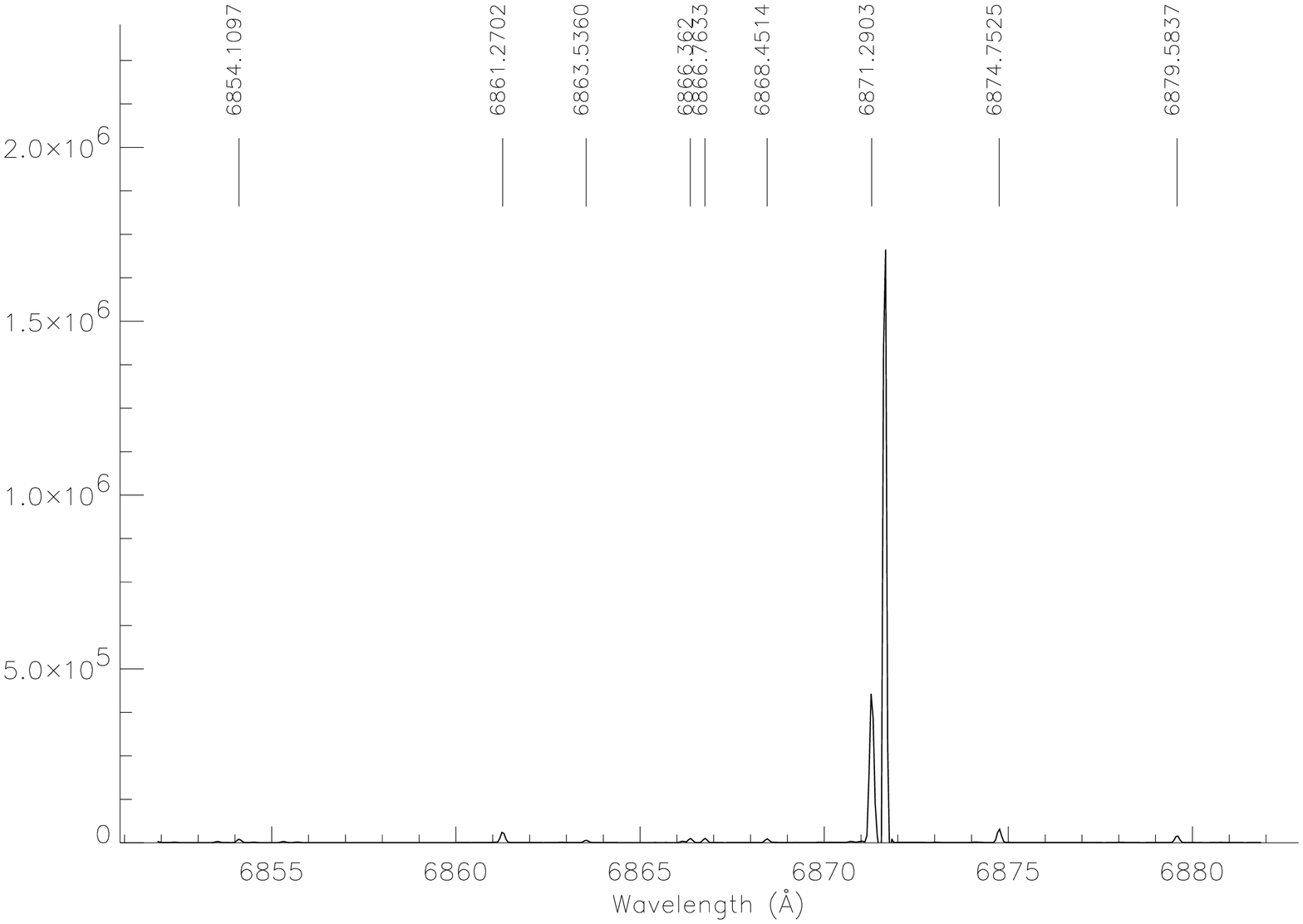}
\end{figure}
\clearpage
   
\begin{figure}
\centering
\includegraphics[width=10cm,angle=90]{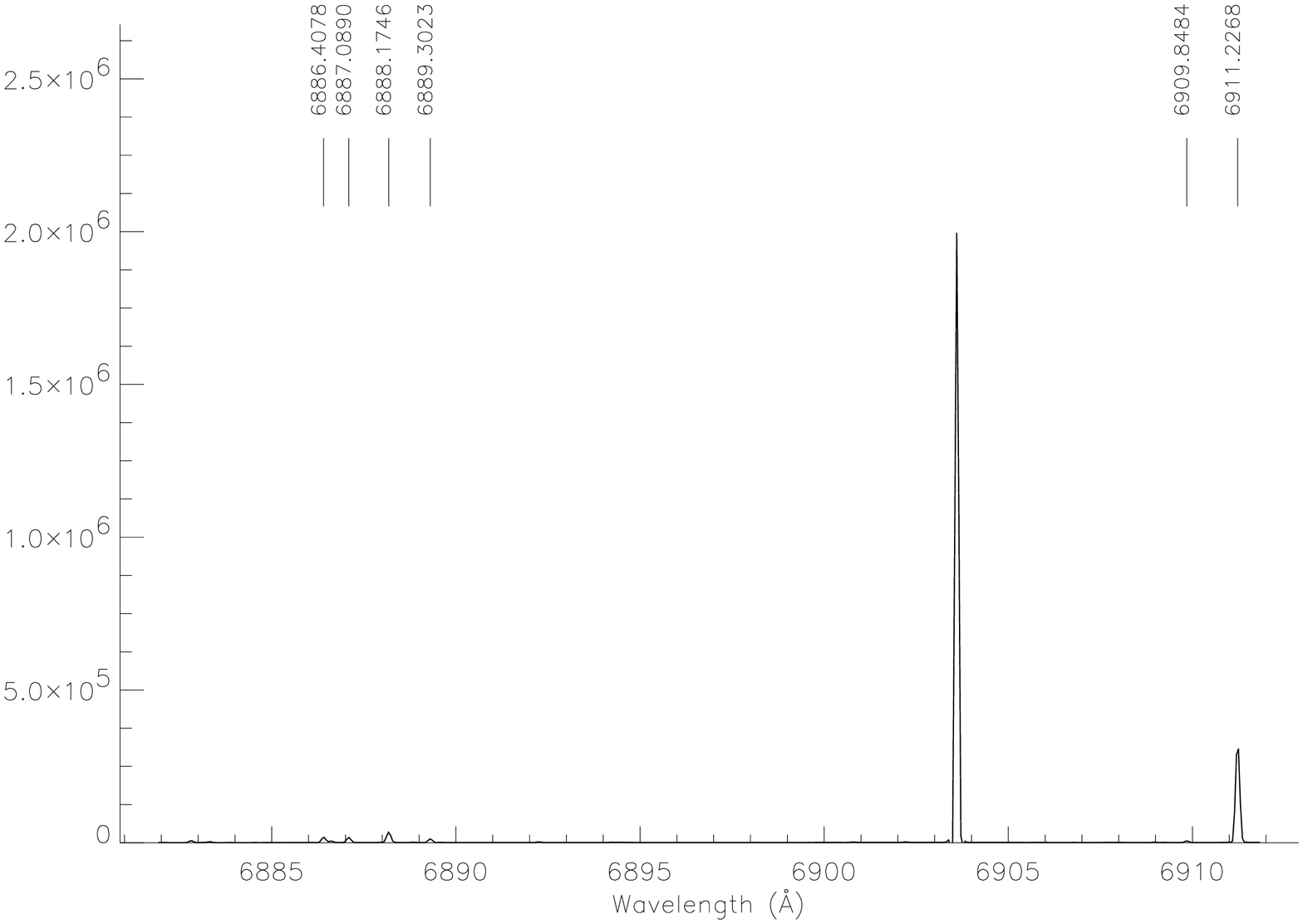}
\includegraphics[width=10cm,angle=90]{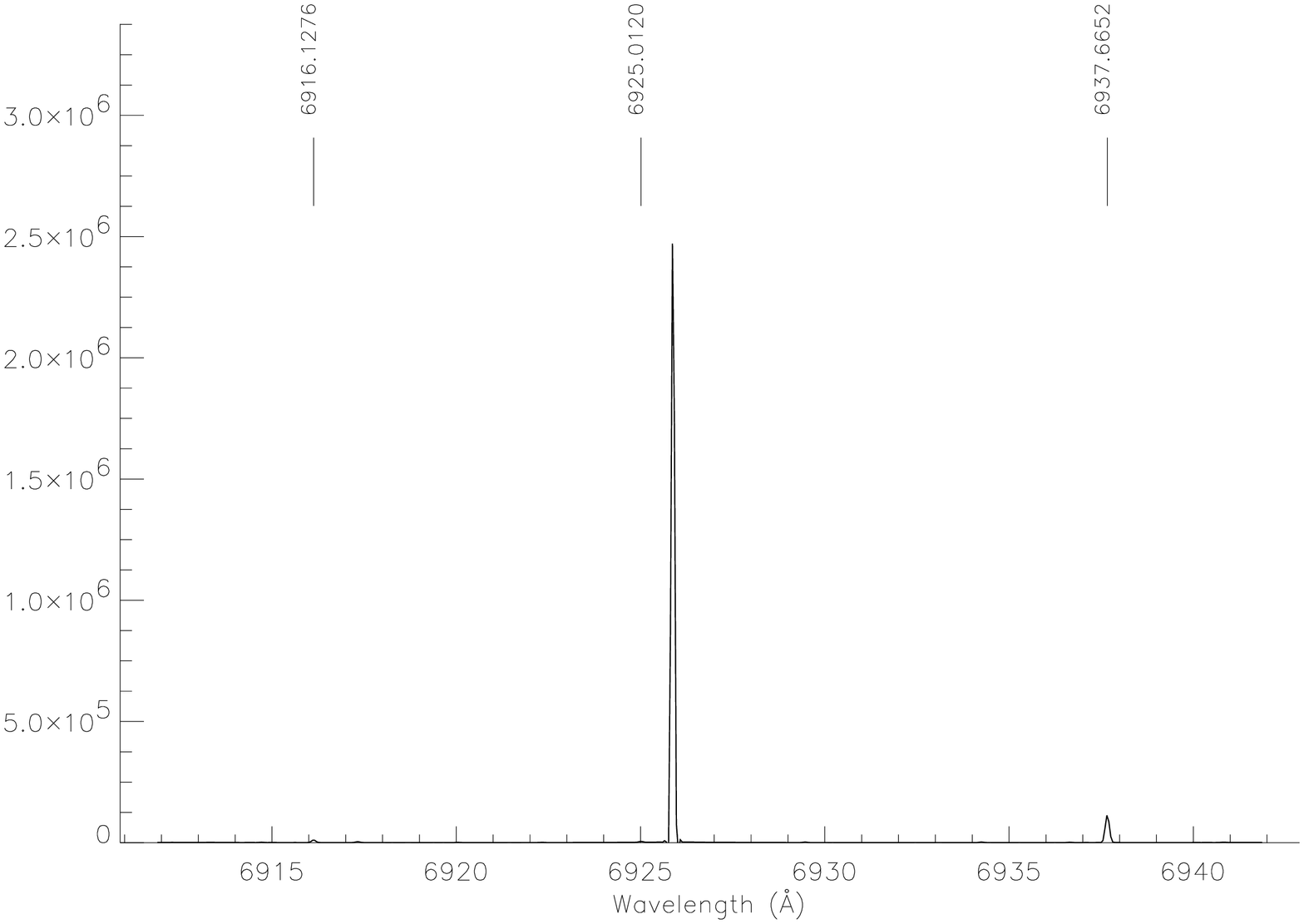}
\end{figure}
\clearpage
   
\begin{figure}
\centering
\includegraphics[width=10cm,angle=90]{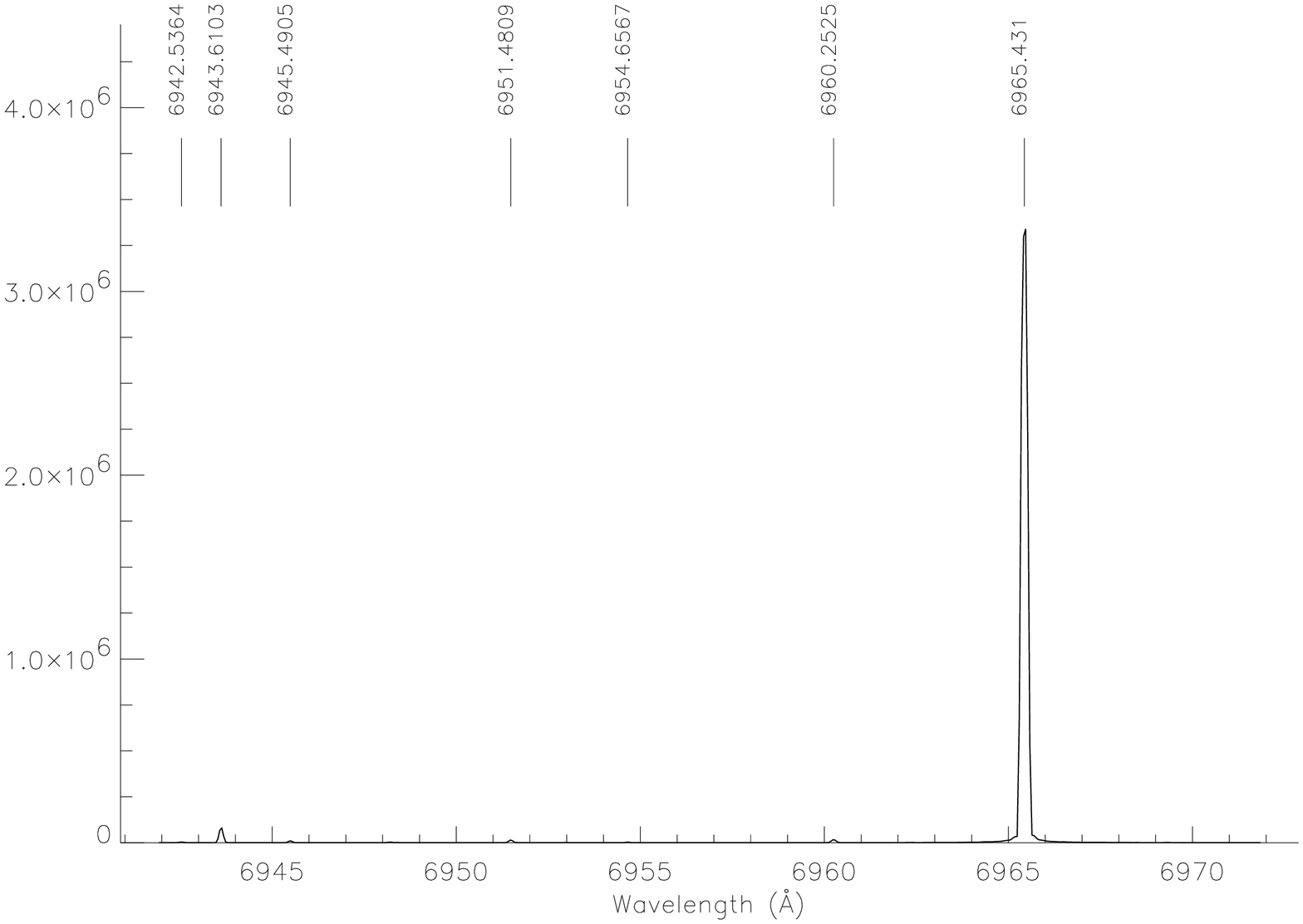}
\includegraphics[width=10cm,angle=90]{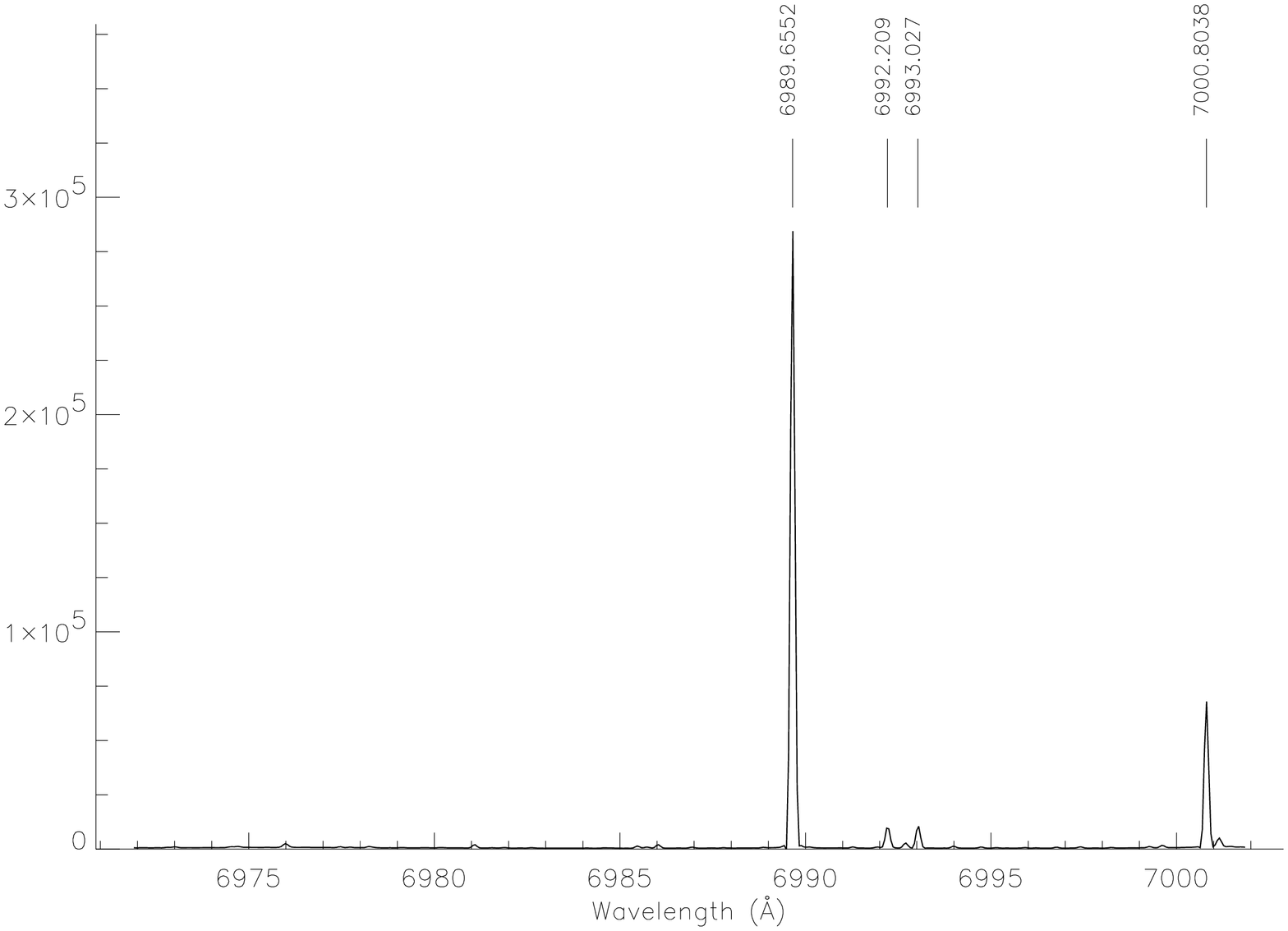}
\end{figure}
\clearpage
   
\begin{figure}
\centering
\includegraphics[width=10cm,angle=90]{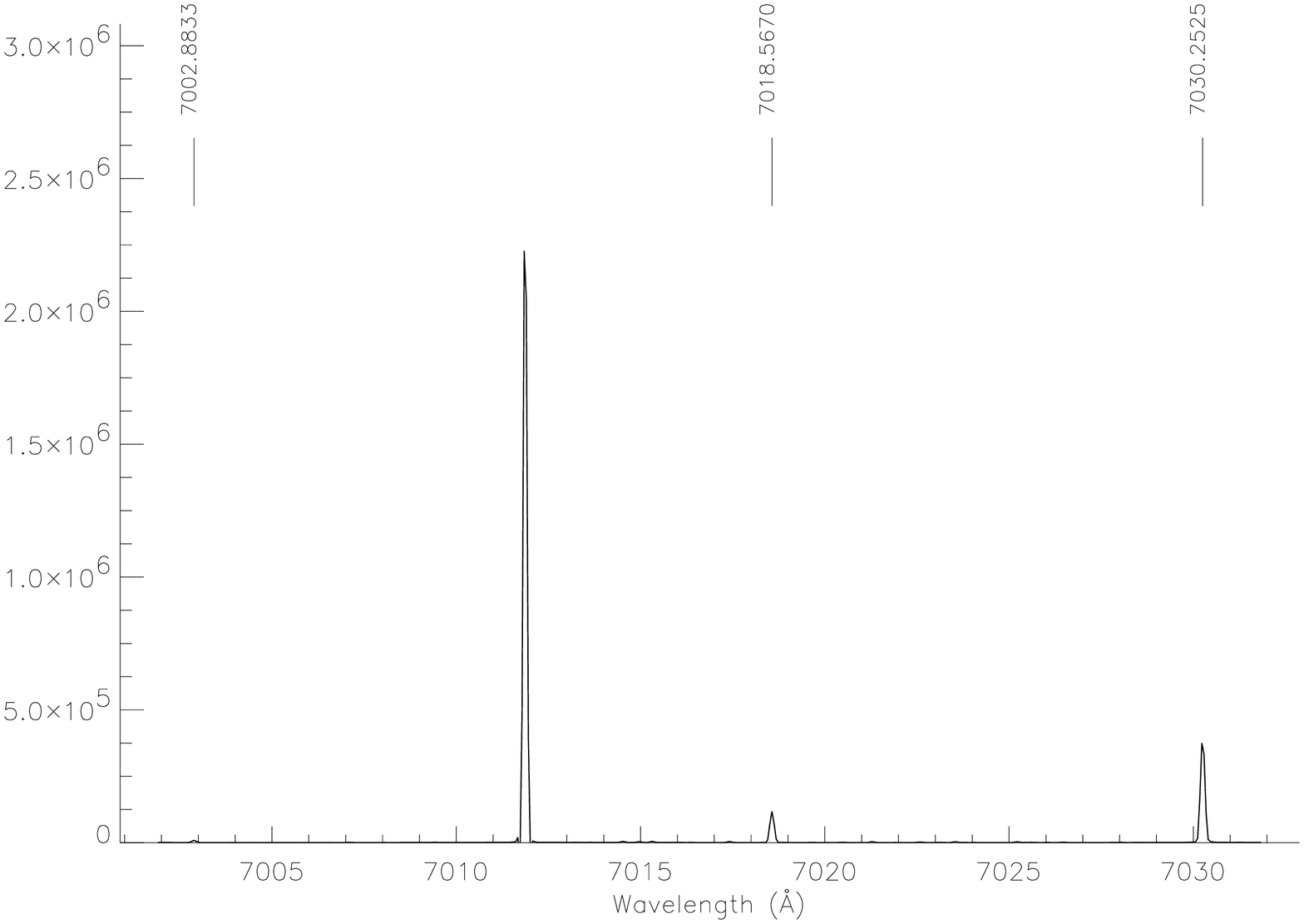}
\includegraphics[width=10cm,angle=90]{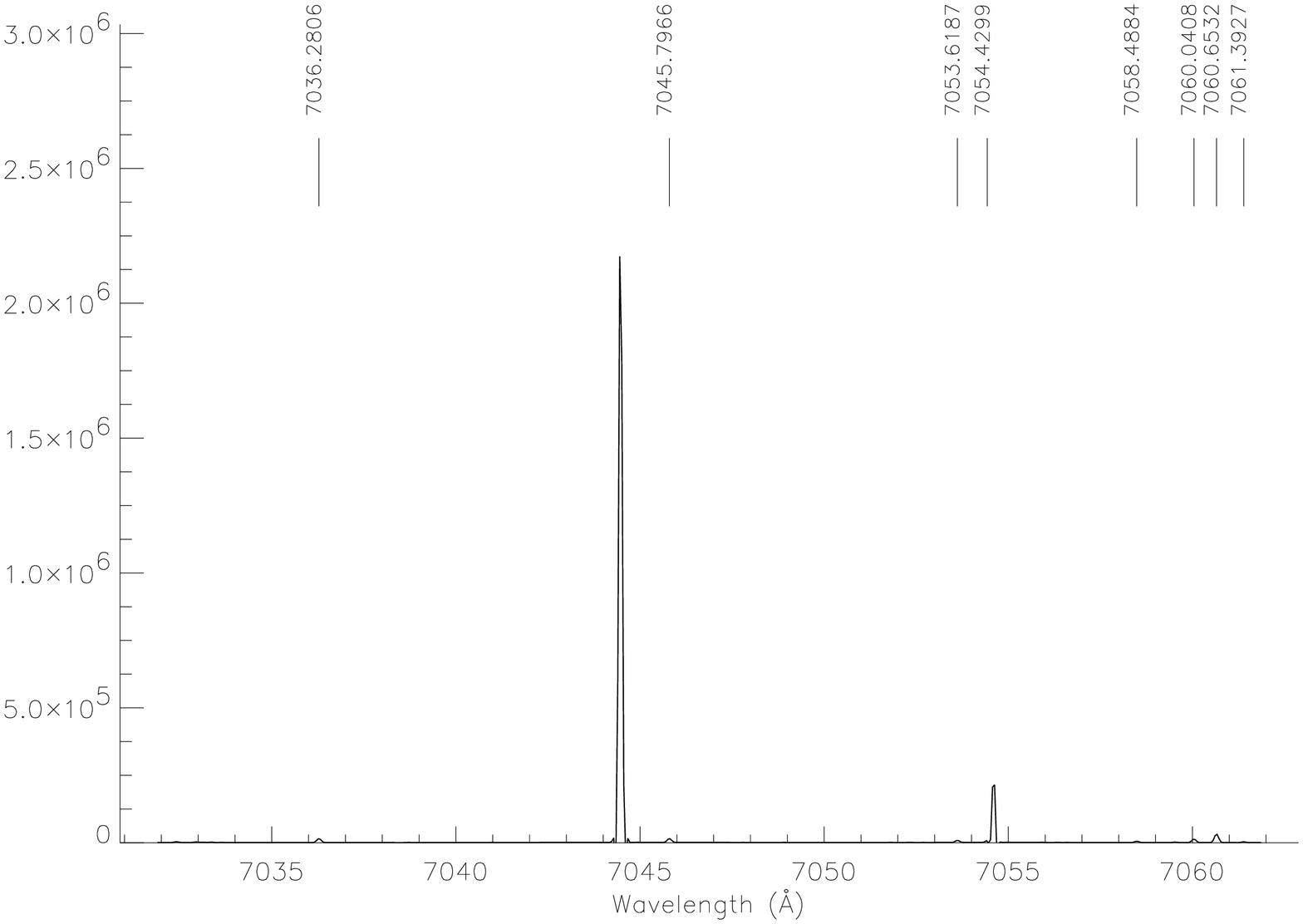}
\end{figure}
\clearpage
   
\begin{figure}
\centering
\includegraphics[width=10cm,angle=90]{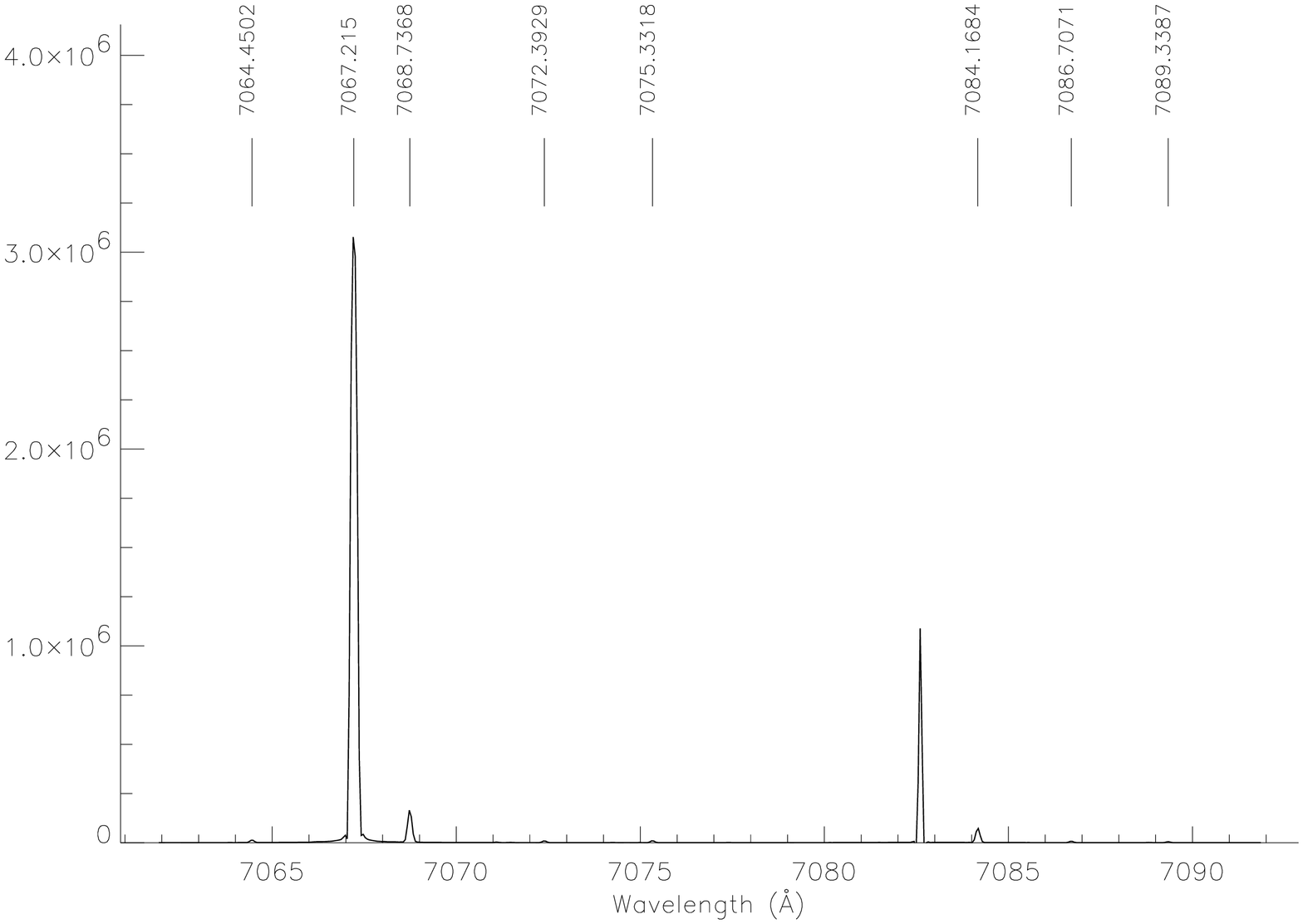}
\includegraphics[width=10cm,angle=90]{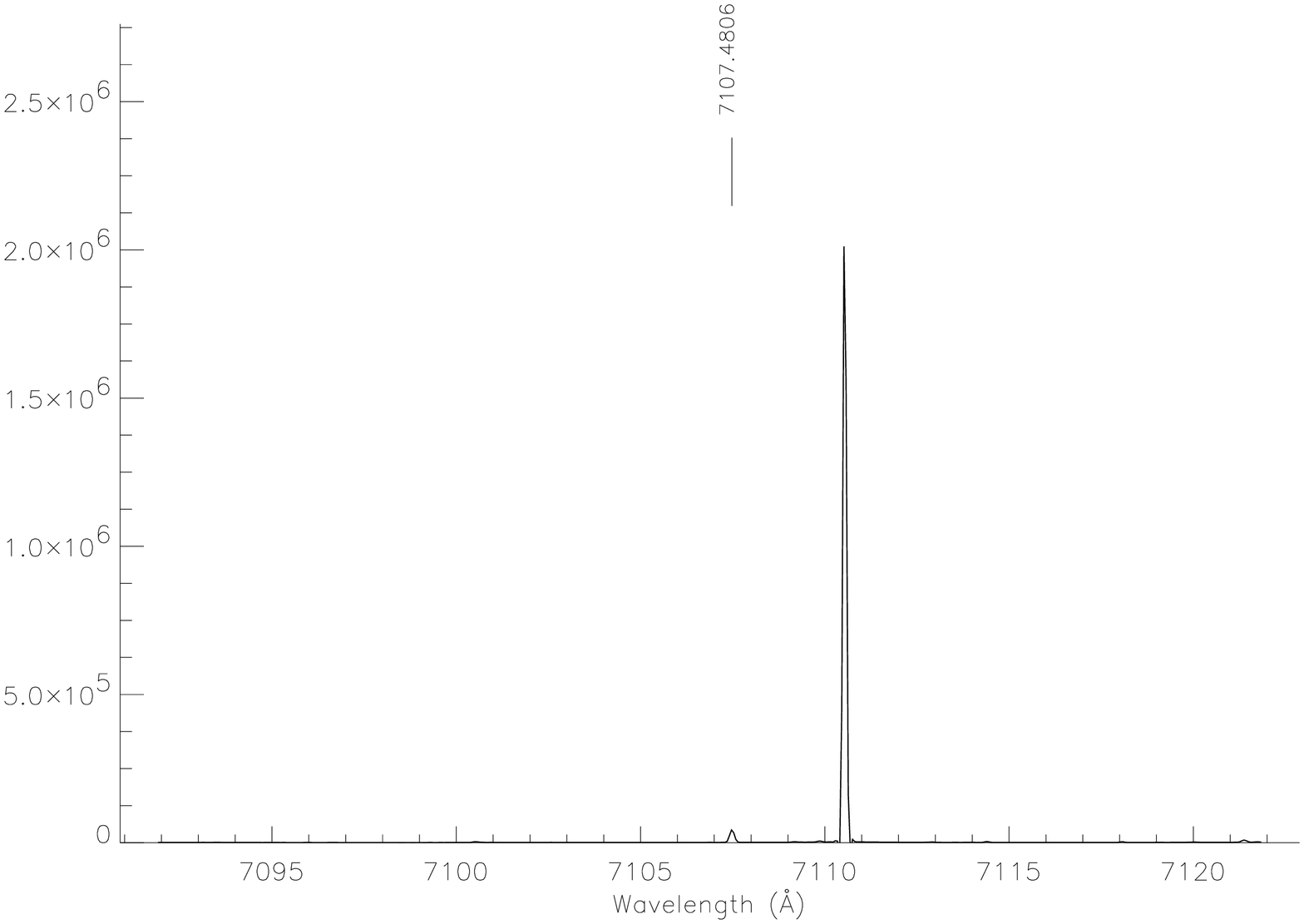}
\end{figure}
\clearpage
   
\begin{figure}
\centering
\includegraphics[width=10cm,angle=90]{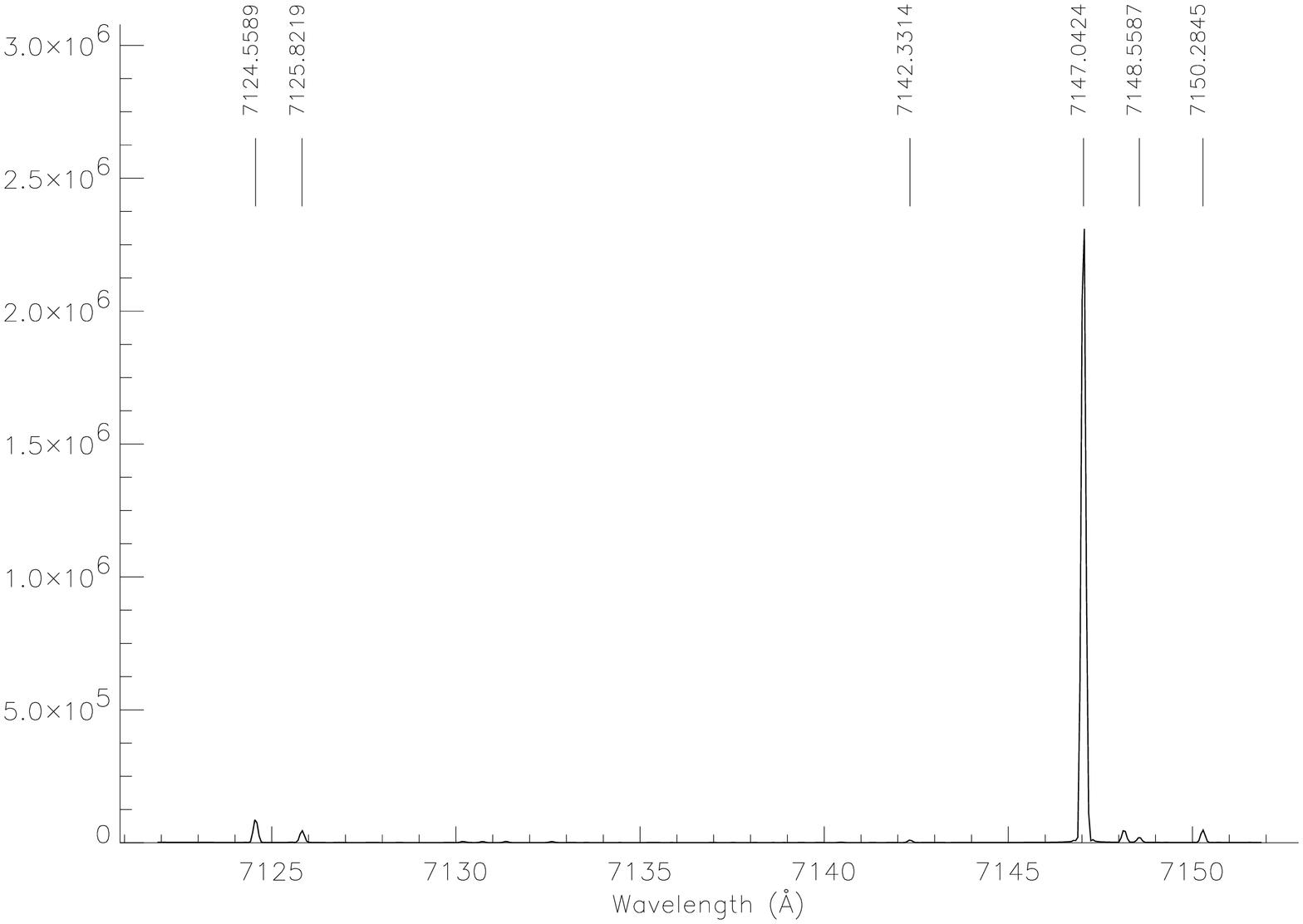}
\includegraphics[width=10cm,angle=90]{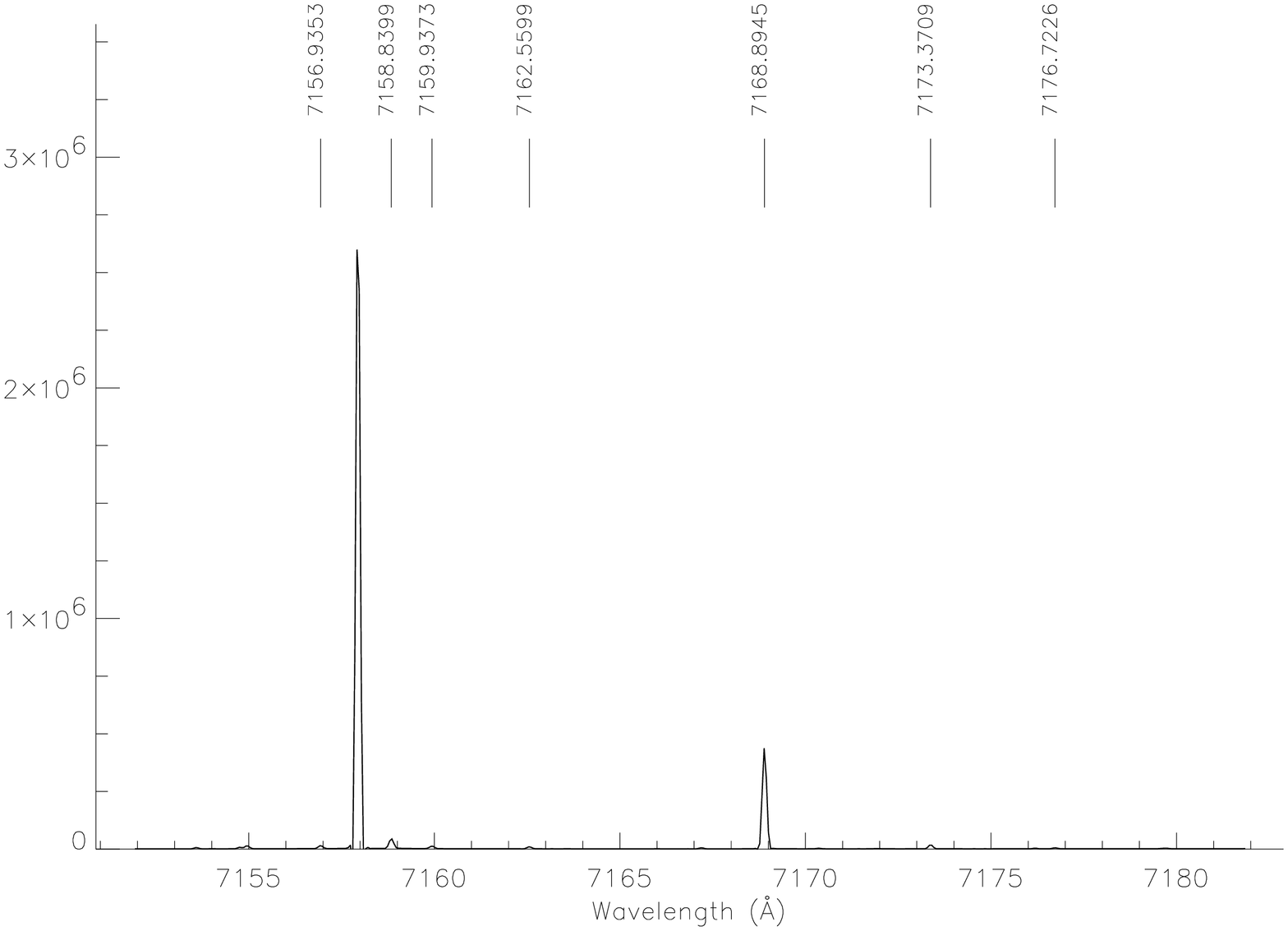}
\end{figure}
\clearpage
   
\begin{figure}
\centering
\includegraphics[width=10cm,angle=90]{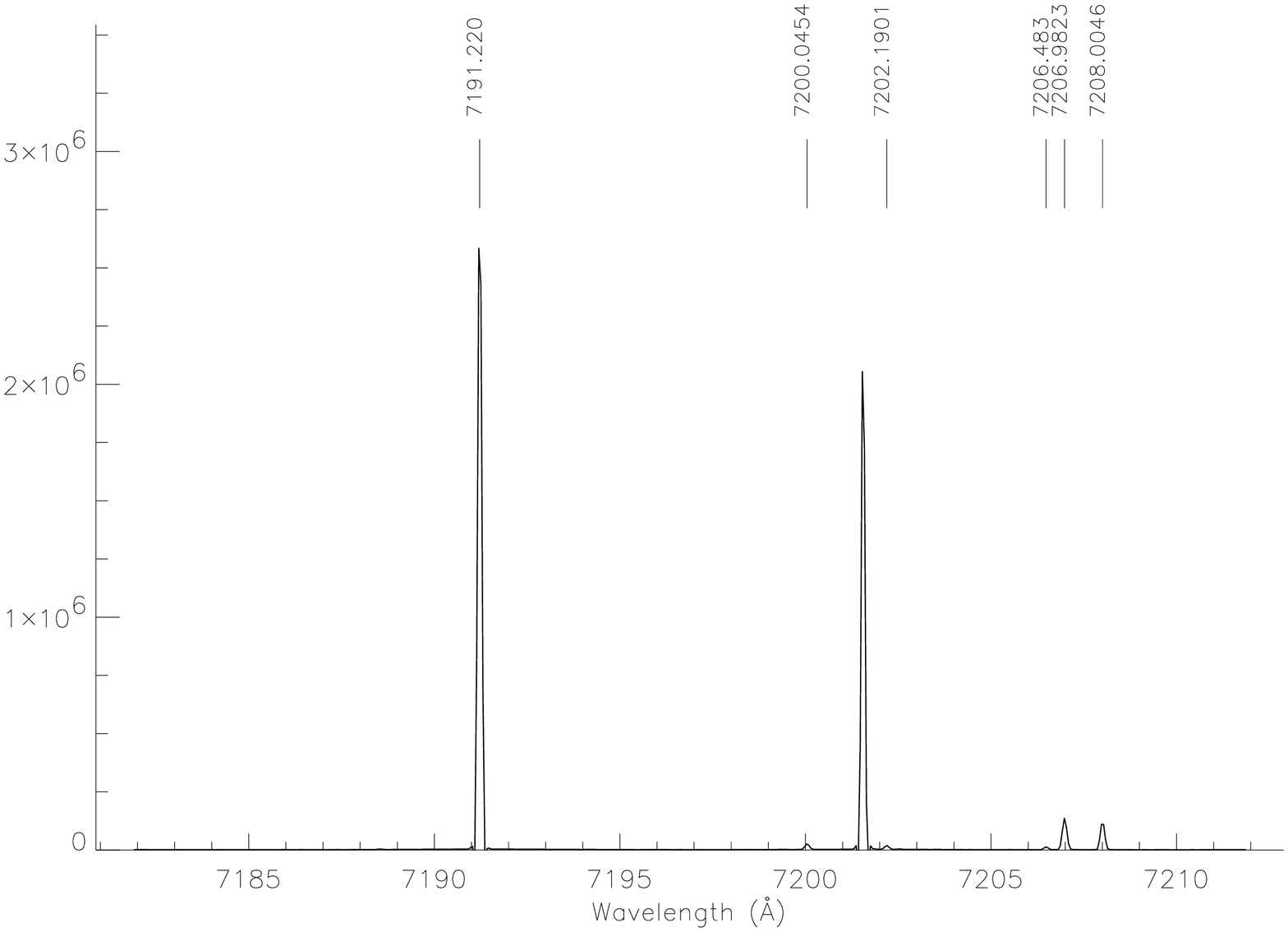}
\includegraphics[width=10cm,angle=90]{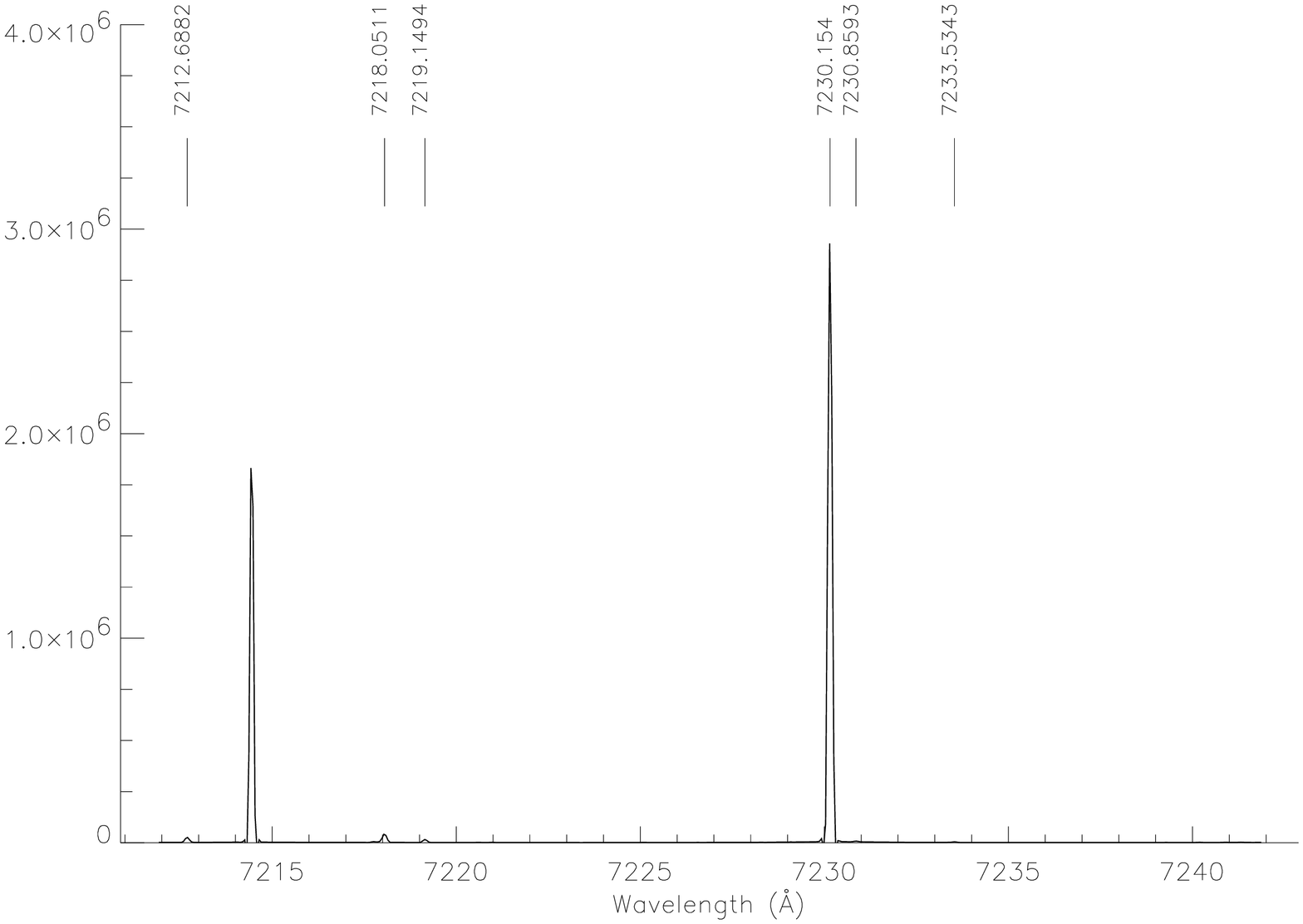}
\end{figure}
\clearpage
   
\begin{figure}
\centering
\includegraphics[width=10cm,angle=90]{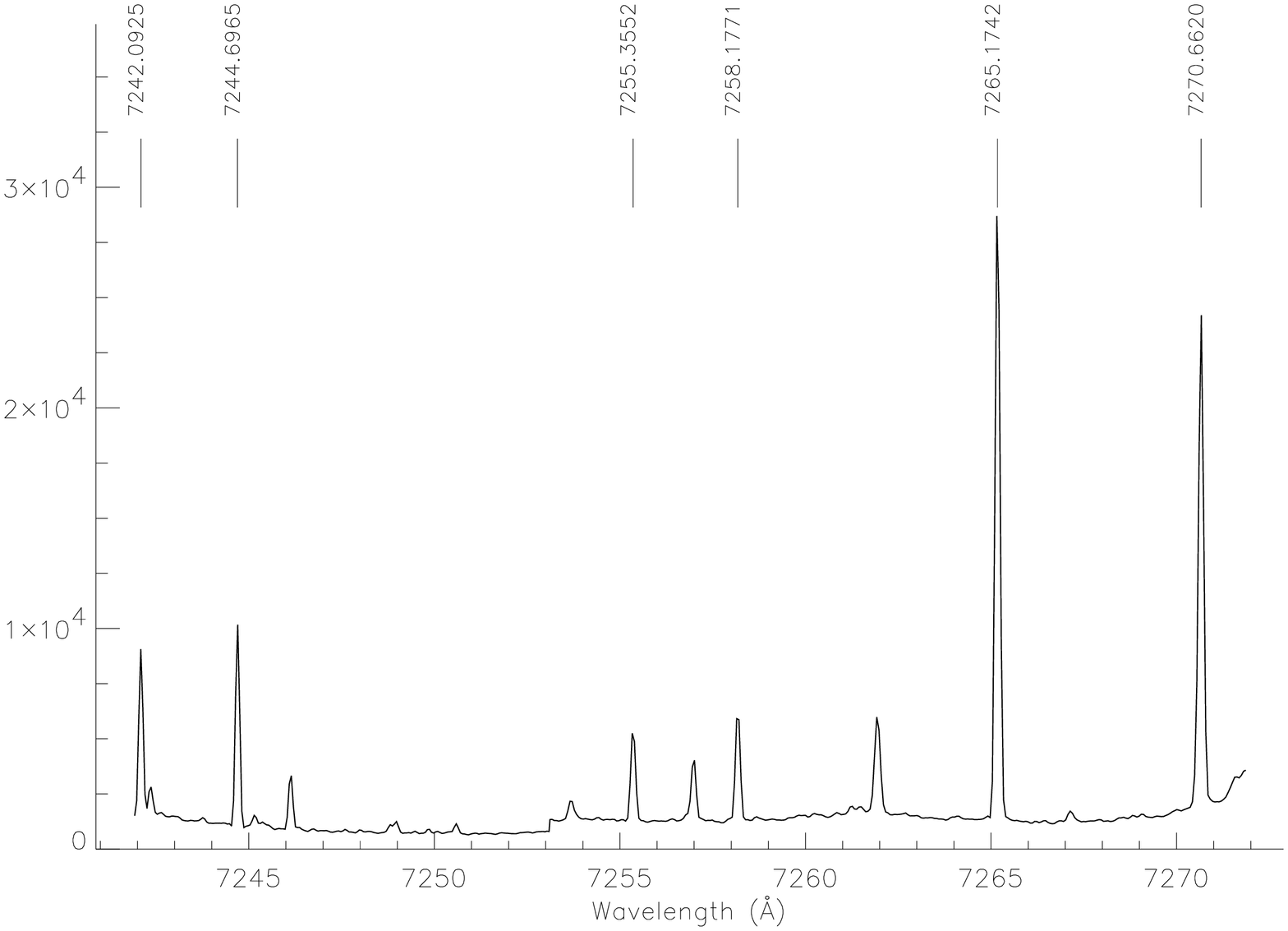}
\includegraphics[width=10cm,angle=90]{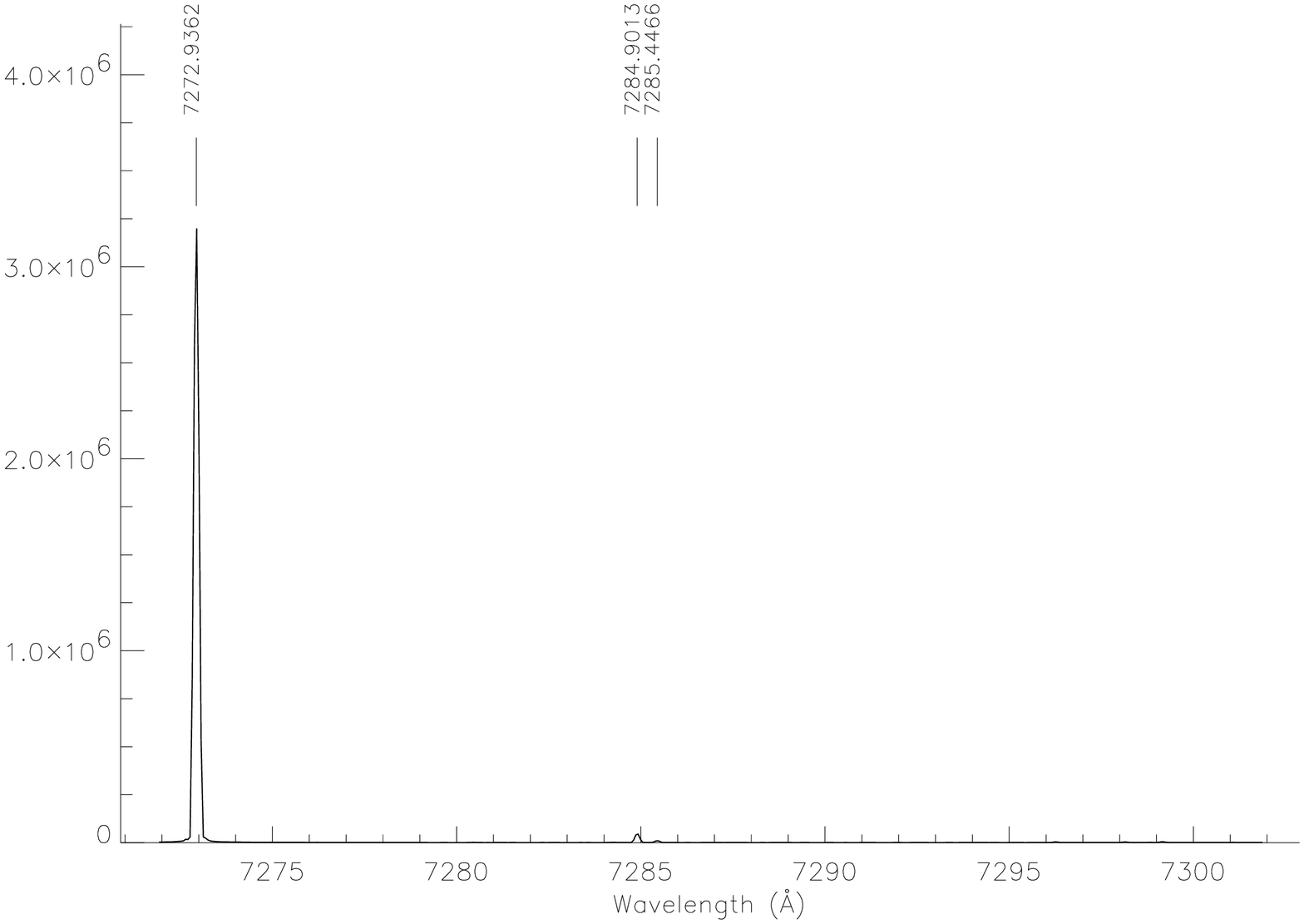}
\end{figure}
\clearpage
   
\begin{figure}
\centering
\includegraphics[width=10cm,angle=90]{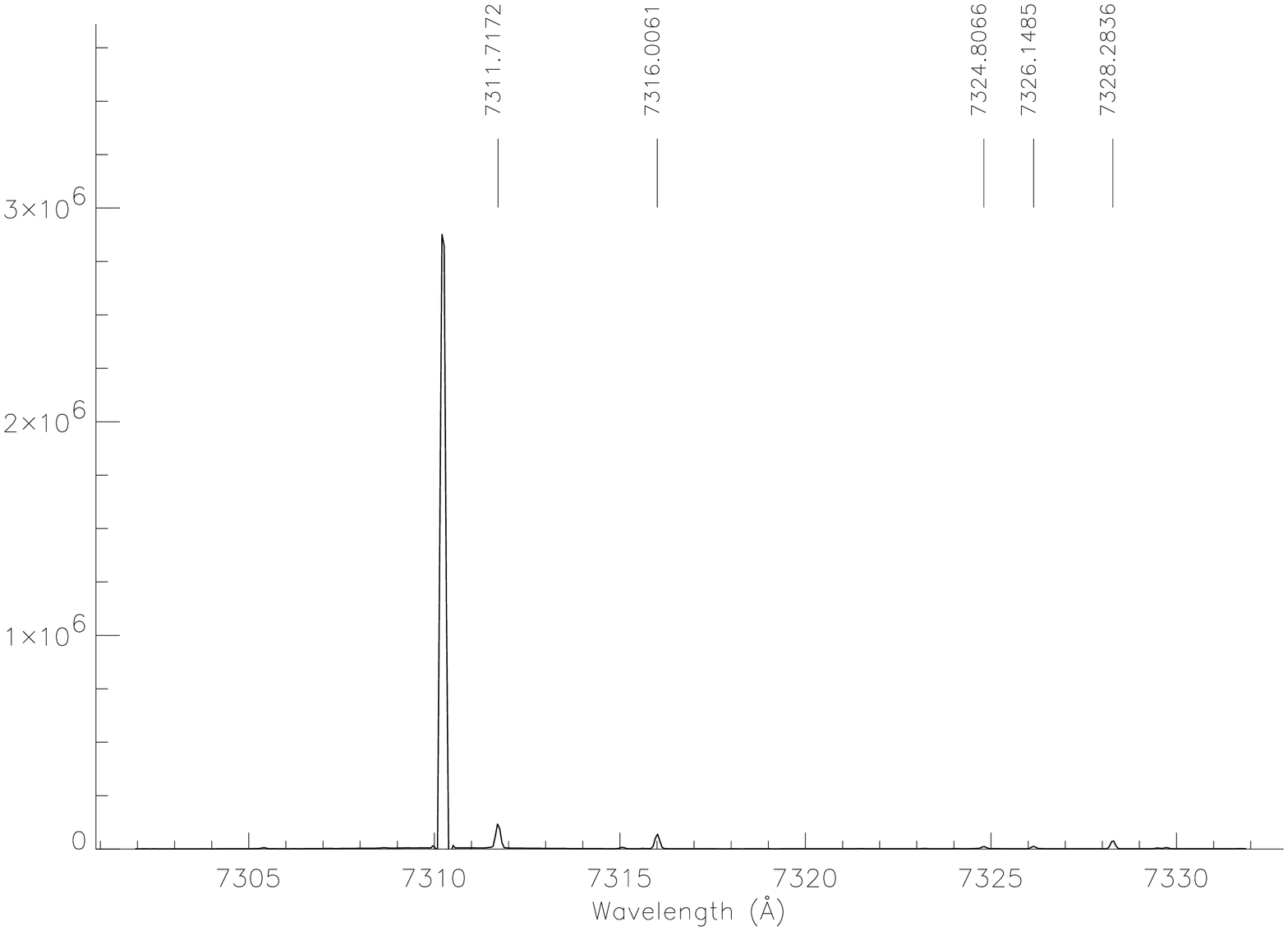}
\includegraphics[width=10cm,angle=90]{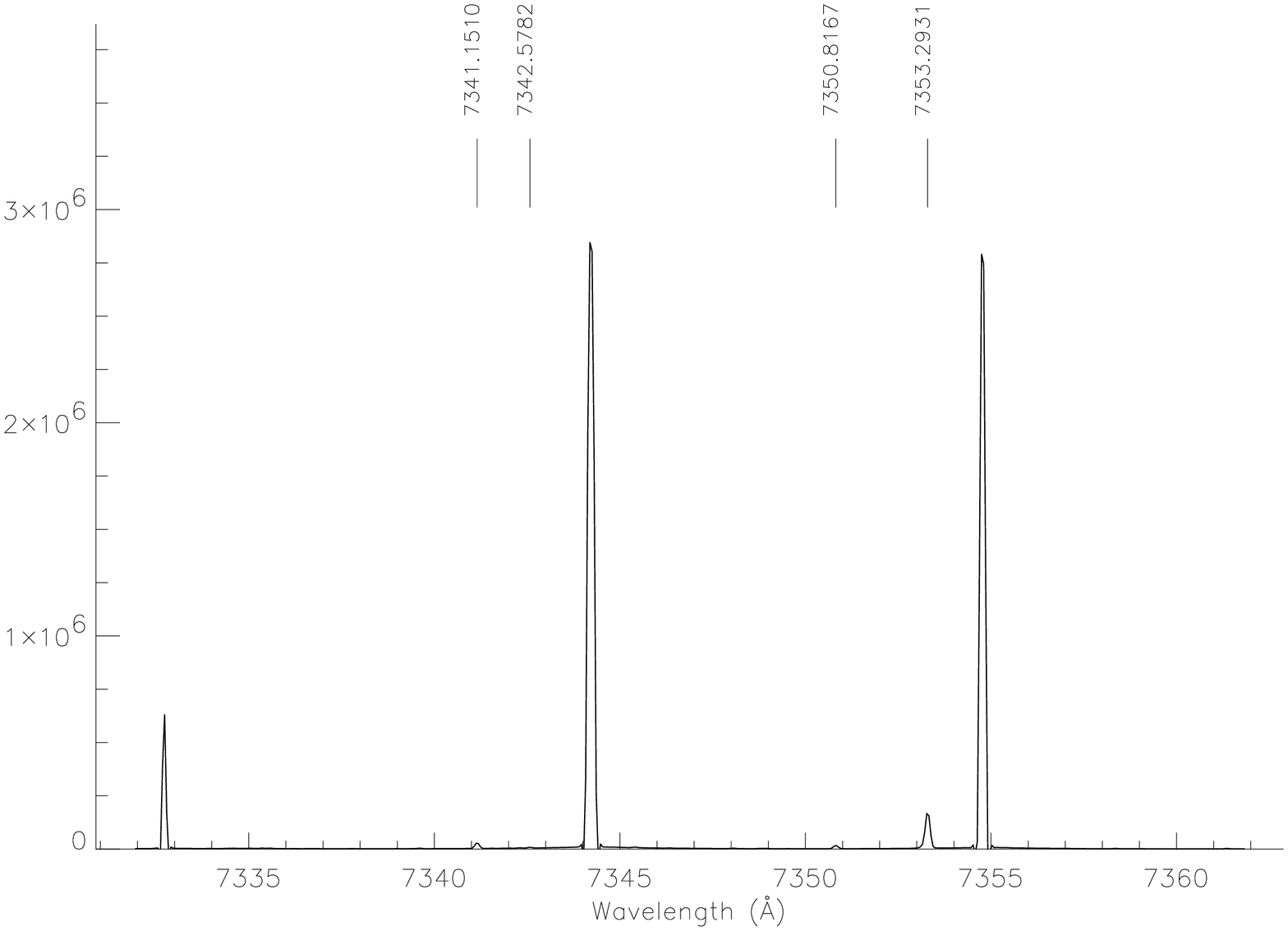}
\end{figure}
\clearpage
   
\begin{figure}
\centering
\includegraphics[width=10cm,angle=90]{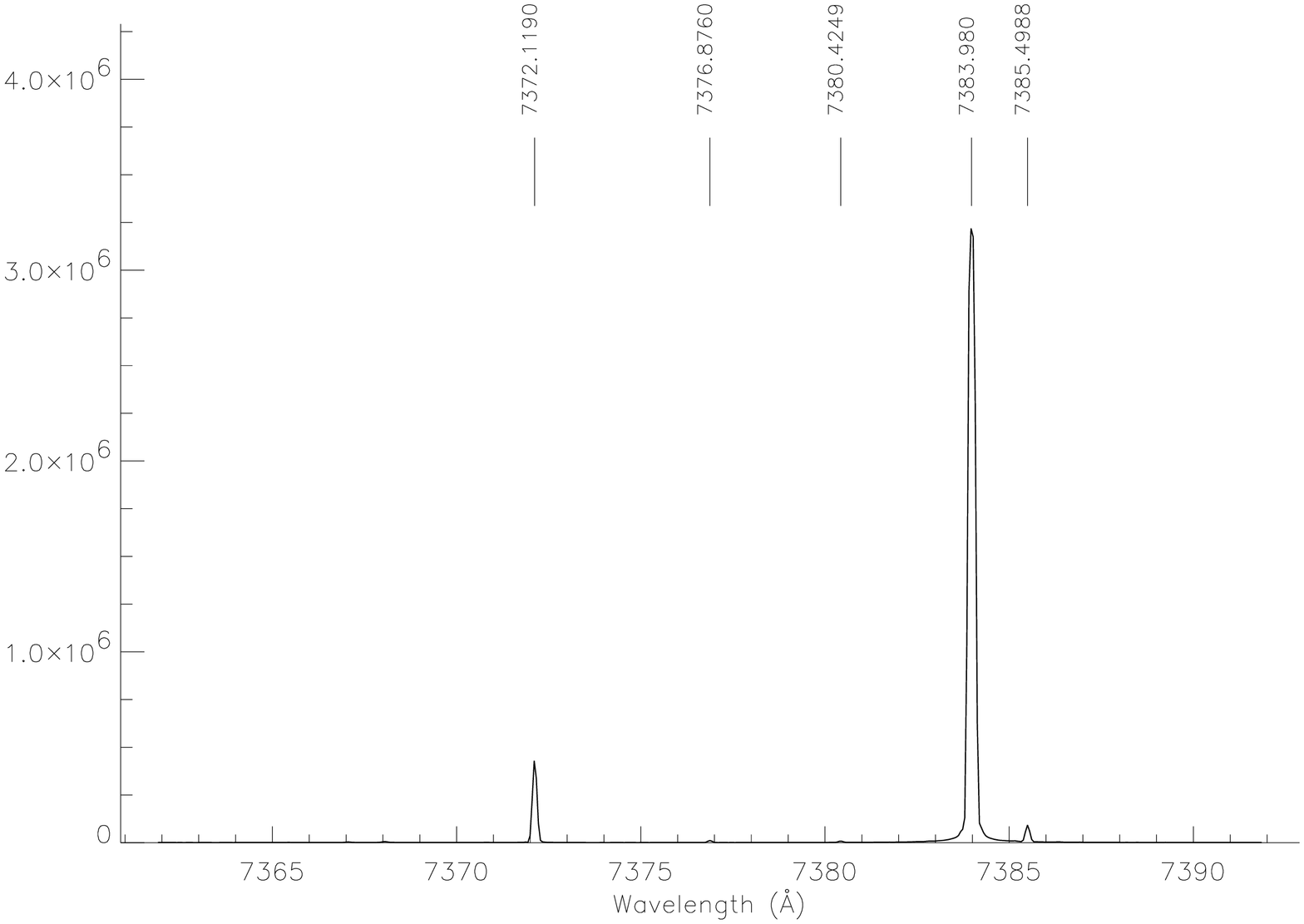}
\includegraphics[width=10cm,angle=90]{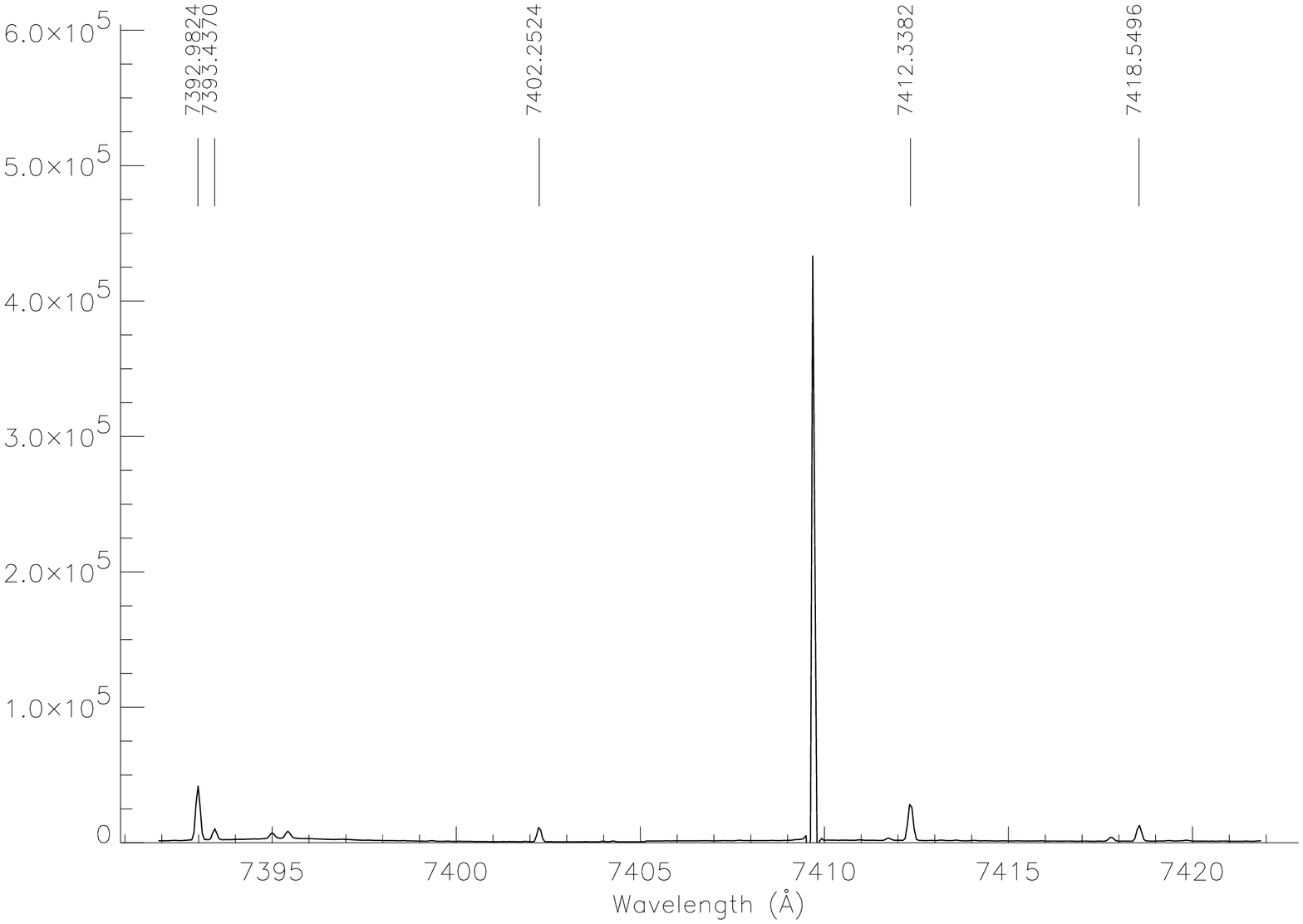}
\end{figure}
\clearpage
   
\begin{figure}
\centering
\includegraphics[width=10cm,angle=90]{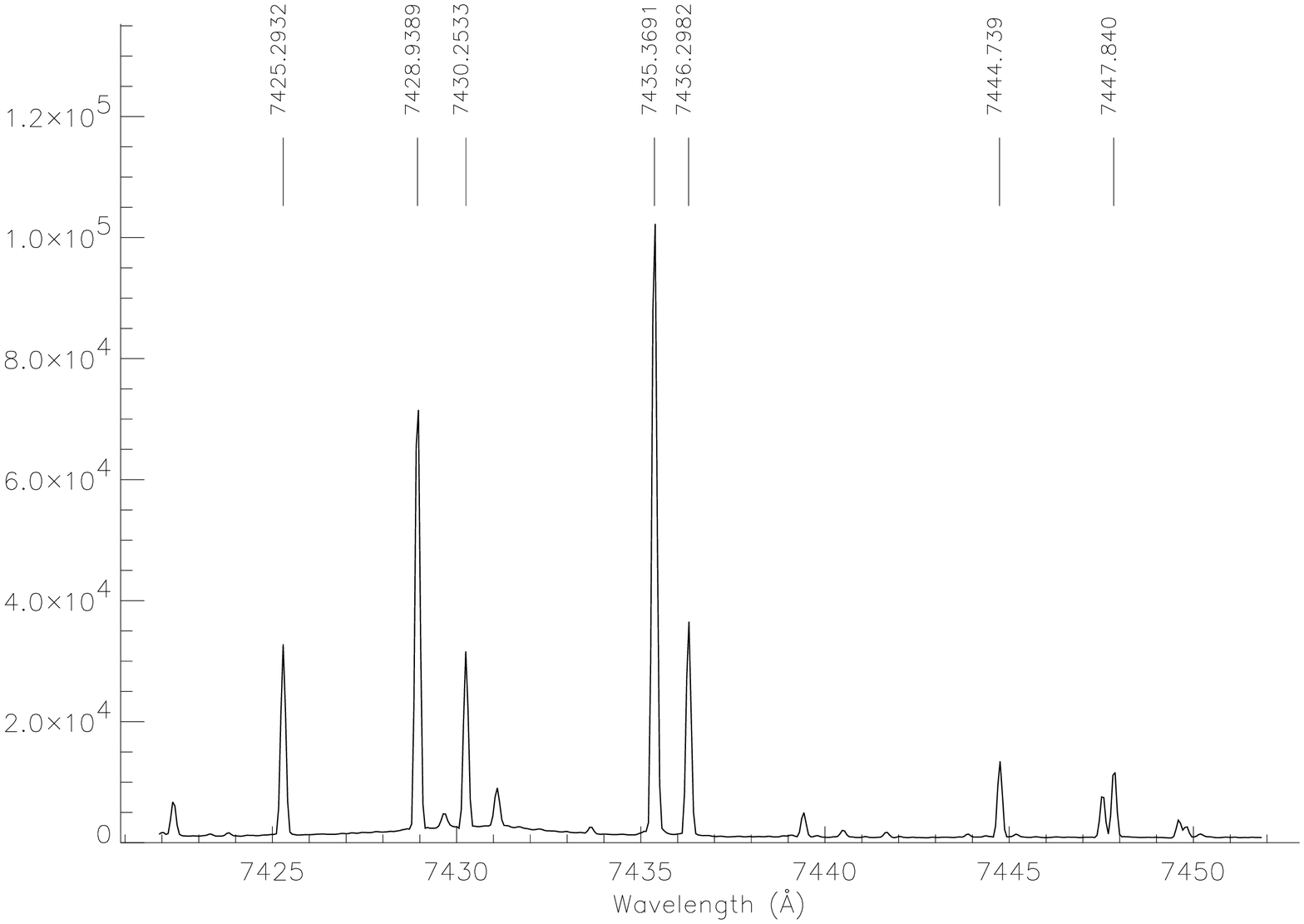}
\includegraphics[width=10cm,angle=90]{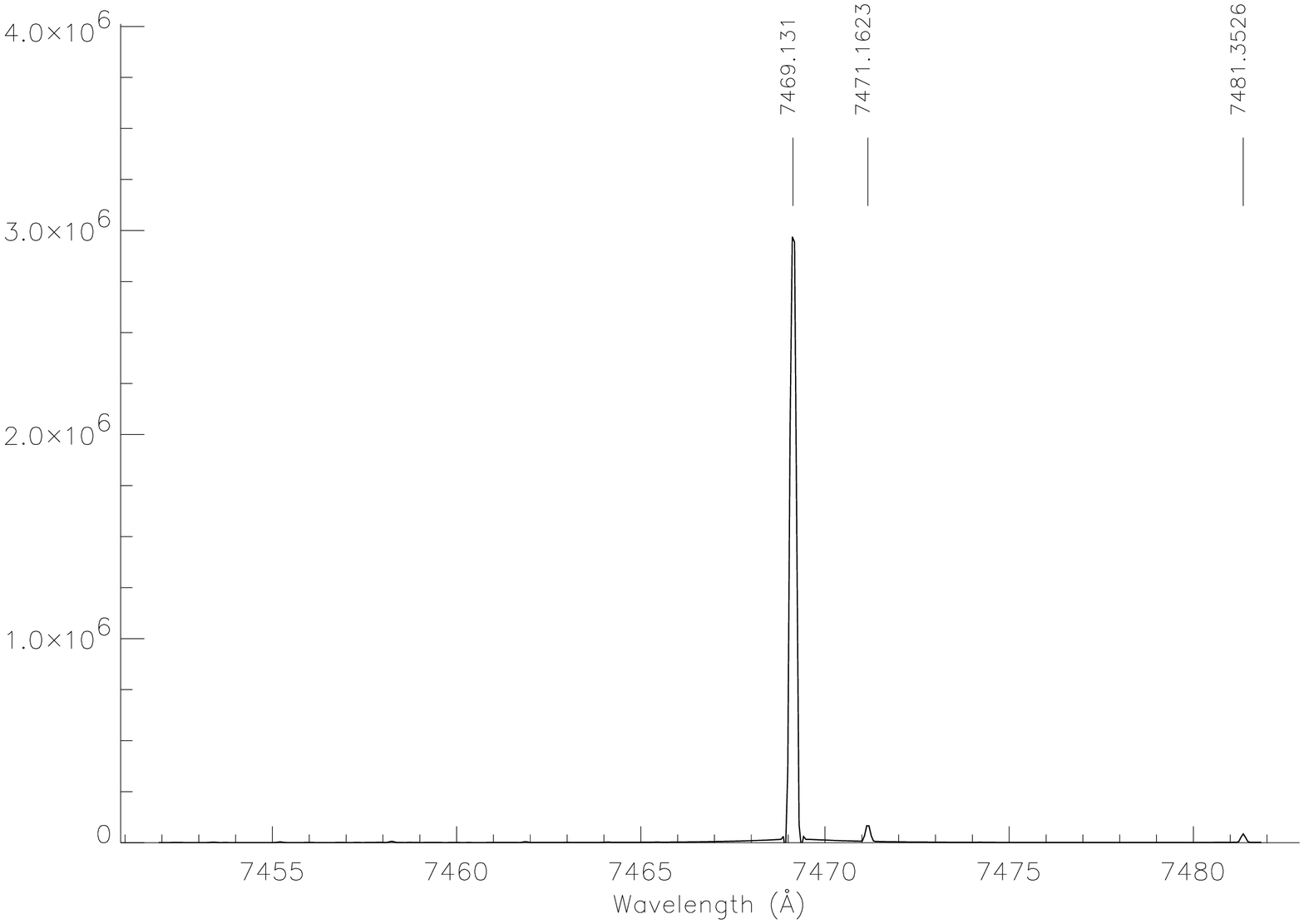}
\end{figure}
\clearpage
   
\begin{figure}
\centering
\includegraphics[width=10cm,angle=90]{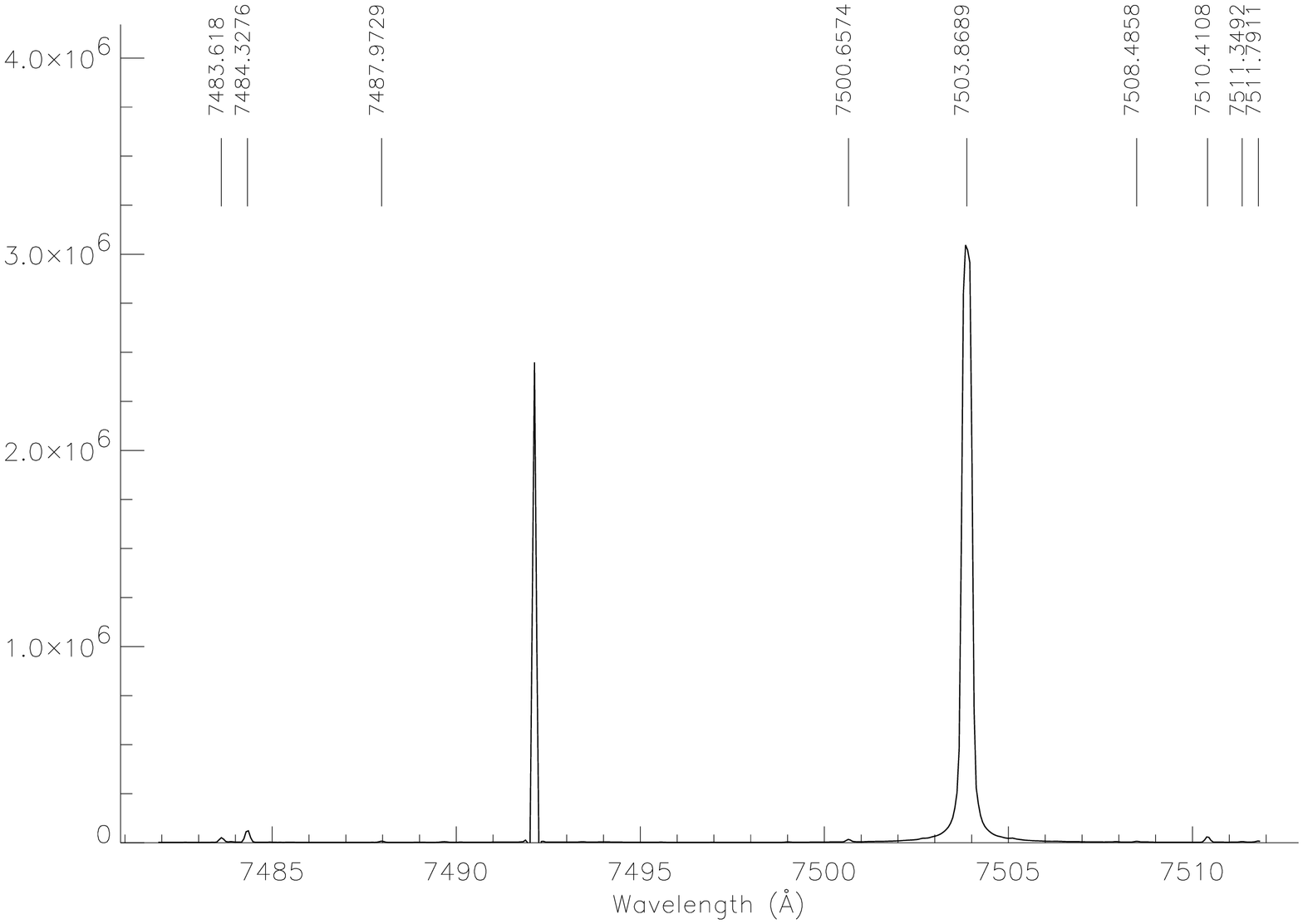}
\includegraphics[width=10cm,angle=90]{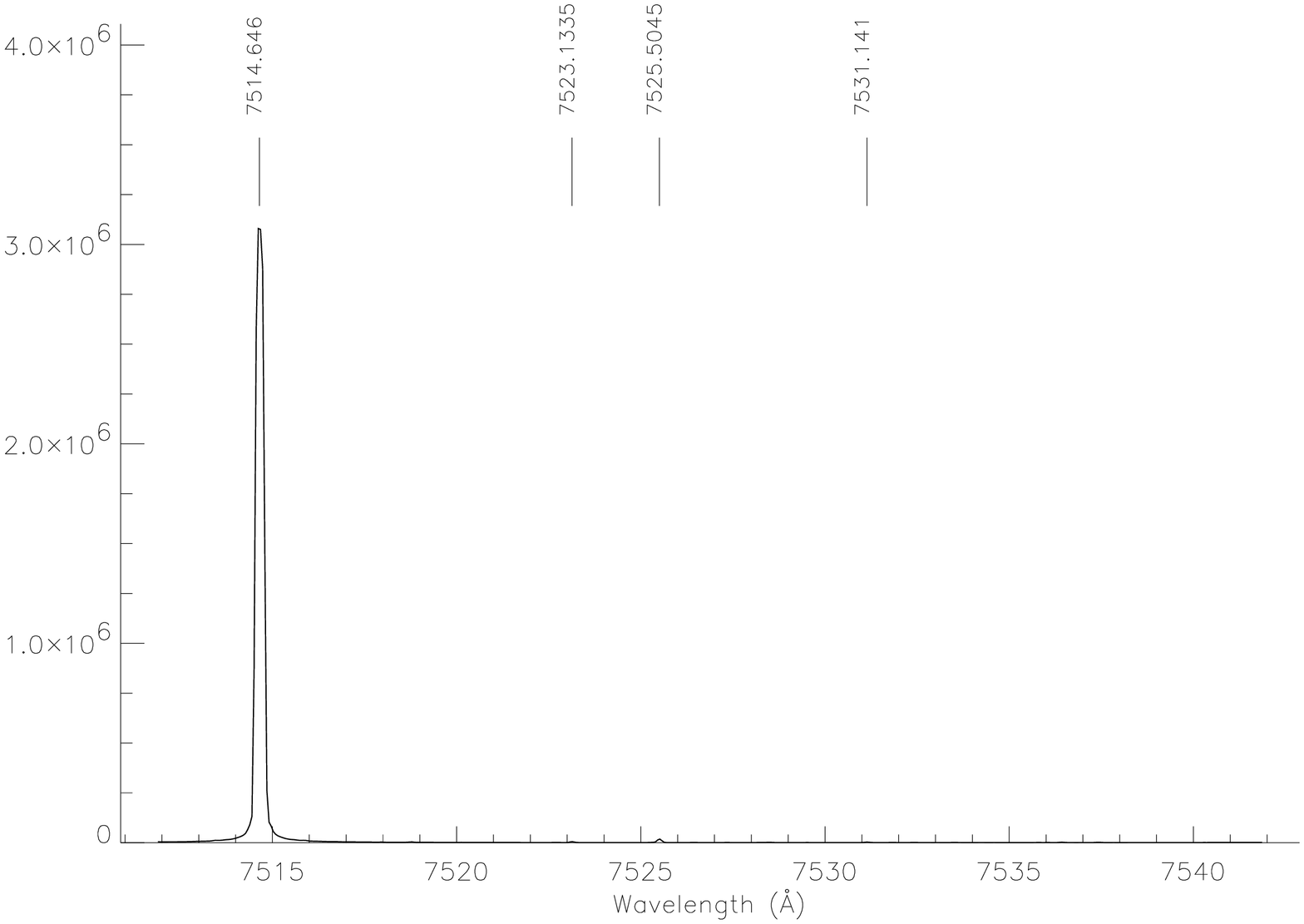}
\end{figure}
\clearpage
   
\begin{figure}
\centering
\includegraphics[width=10cm,angle=90]{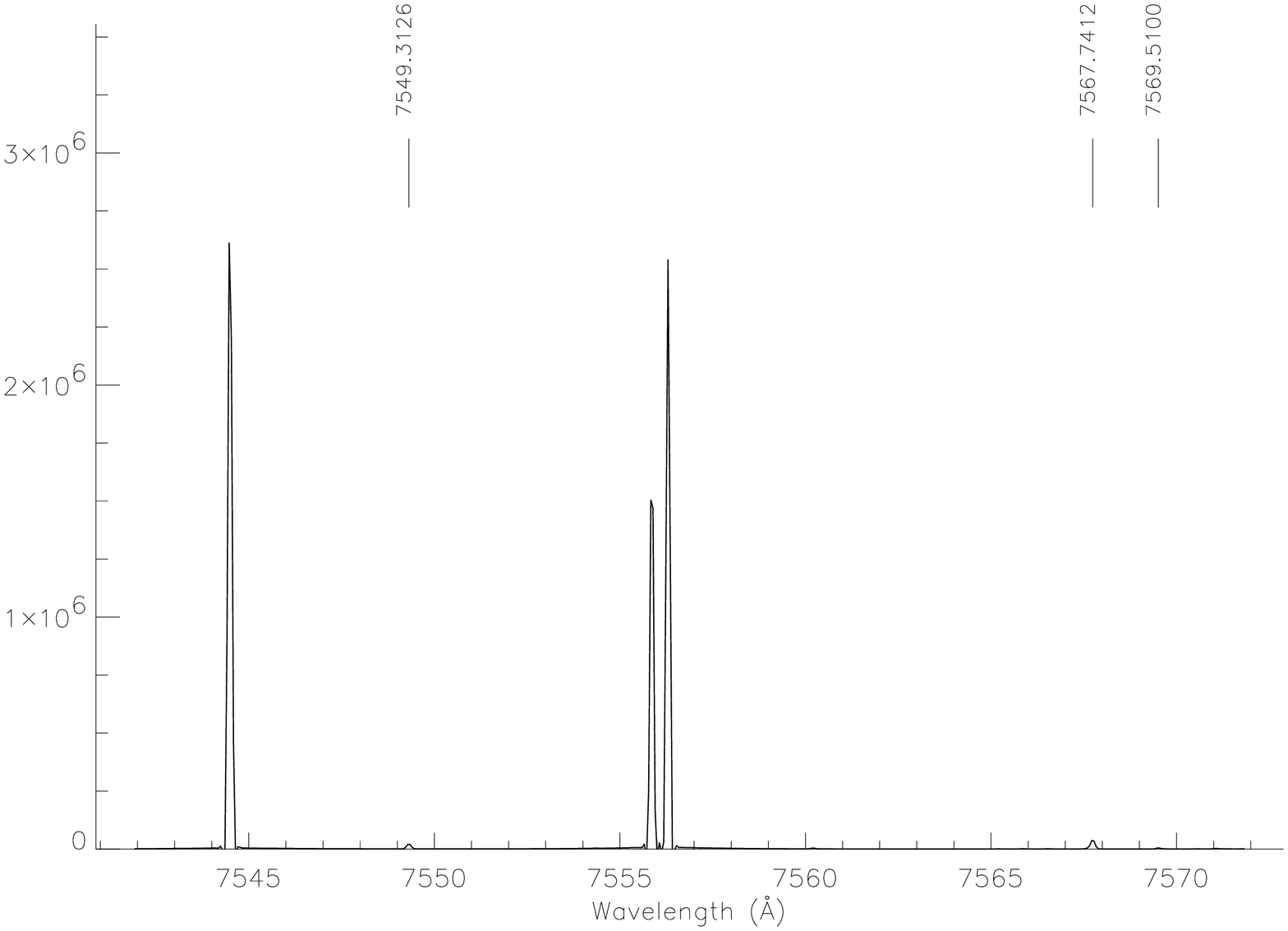}
\includegraphics[width=10cm,angle=90]{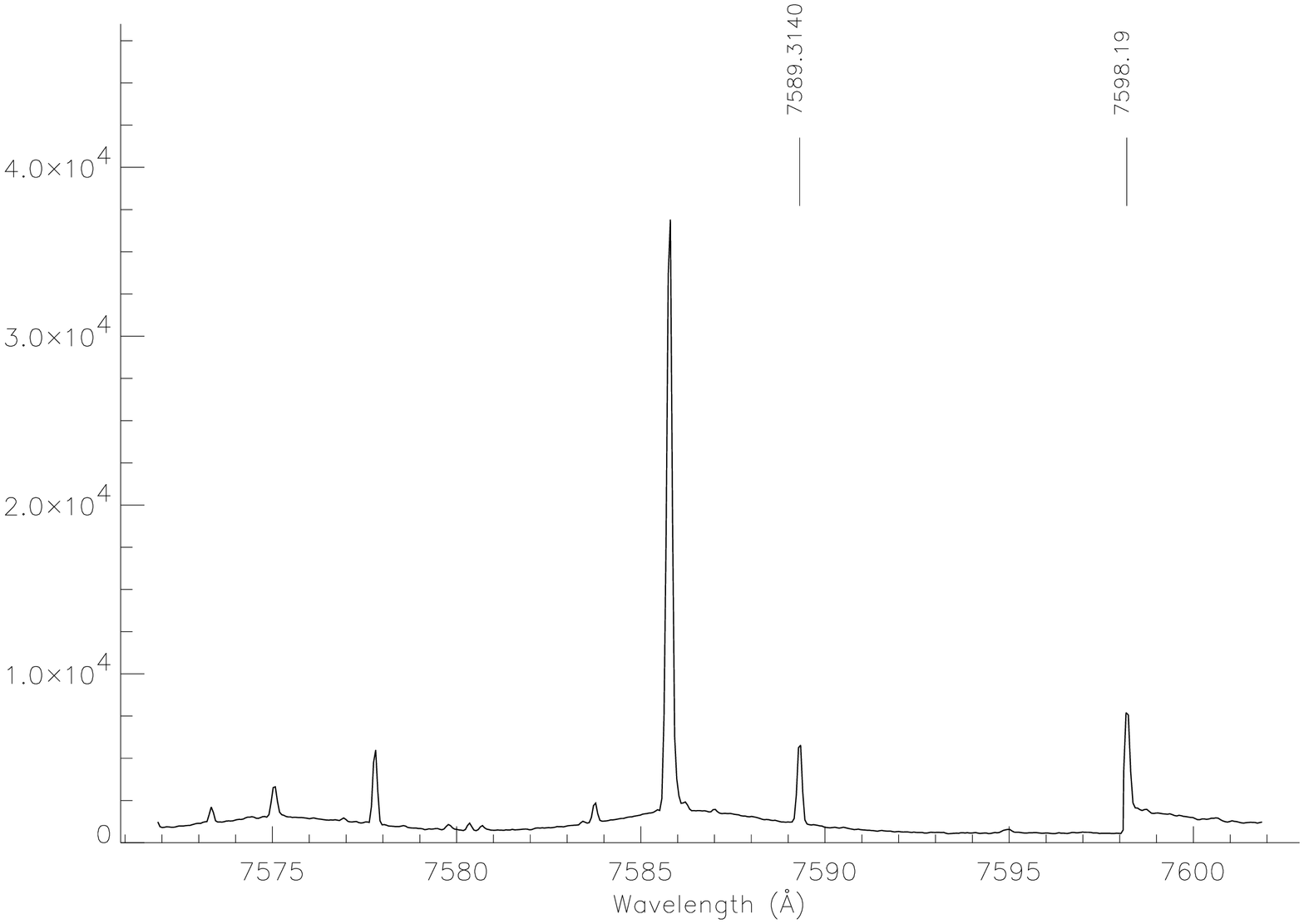}
\end{figure}
\clearpage
   
\begin{figure}
\centering
\includegraphics[width=10cm,angle=90]{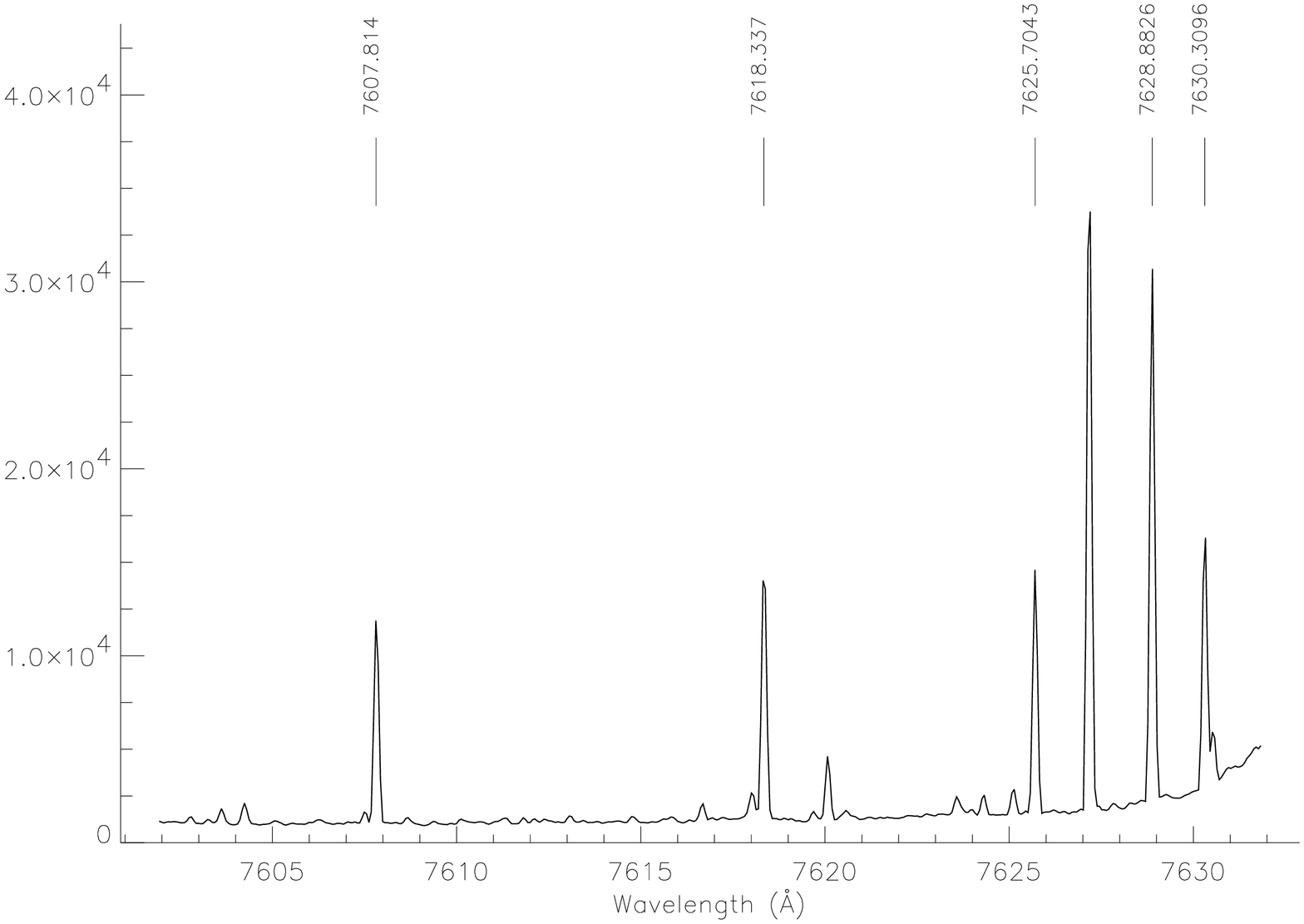}
\includegraphics[width=10cm,angle=90]{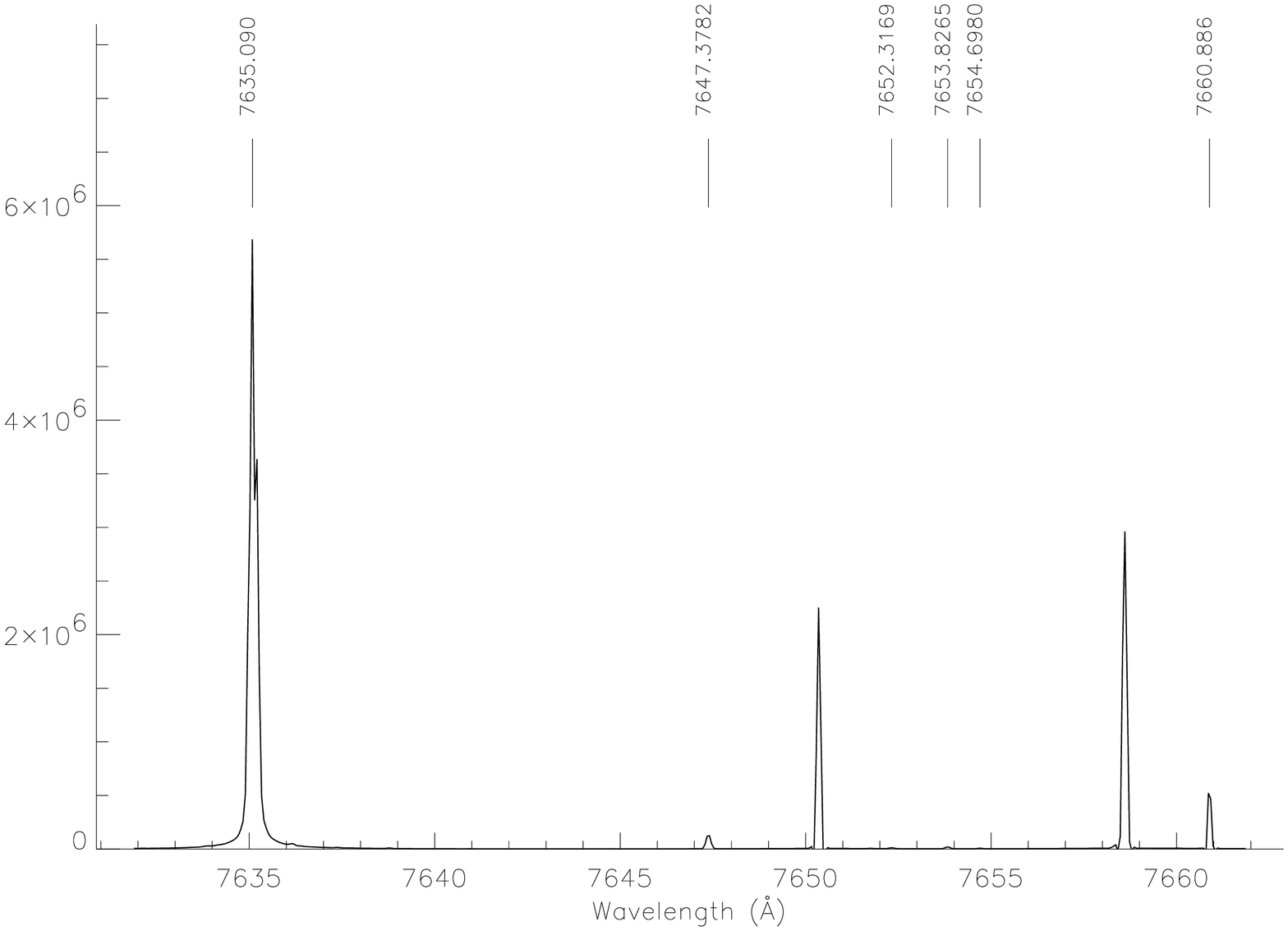}
\end{figure}
\clearpage
   
\begin{figure}
\centering
\includegraphics[width=10cm,angle=90]{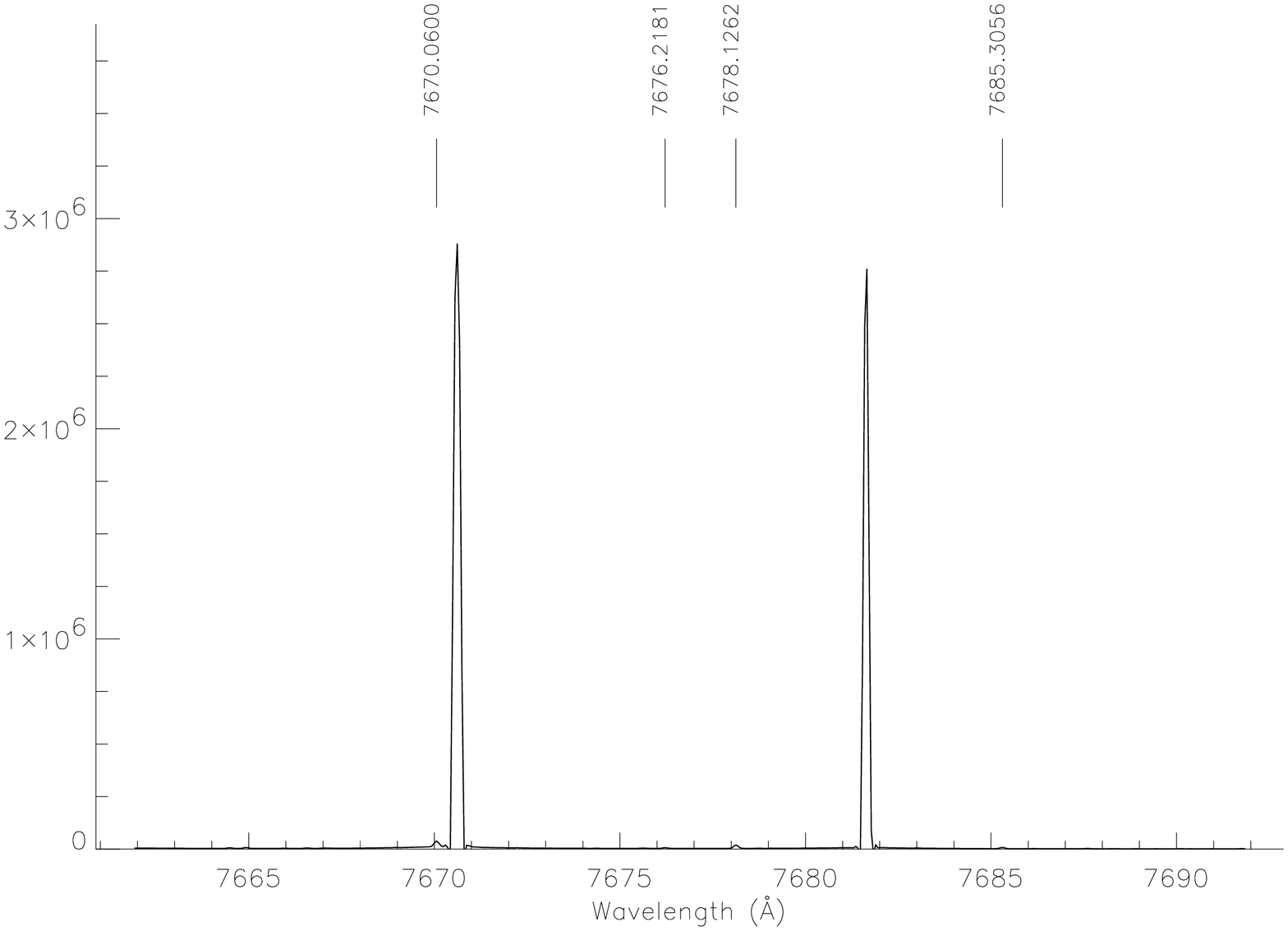}
\includegraphics[width=10cm,angle=90]{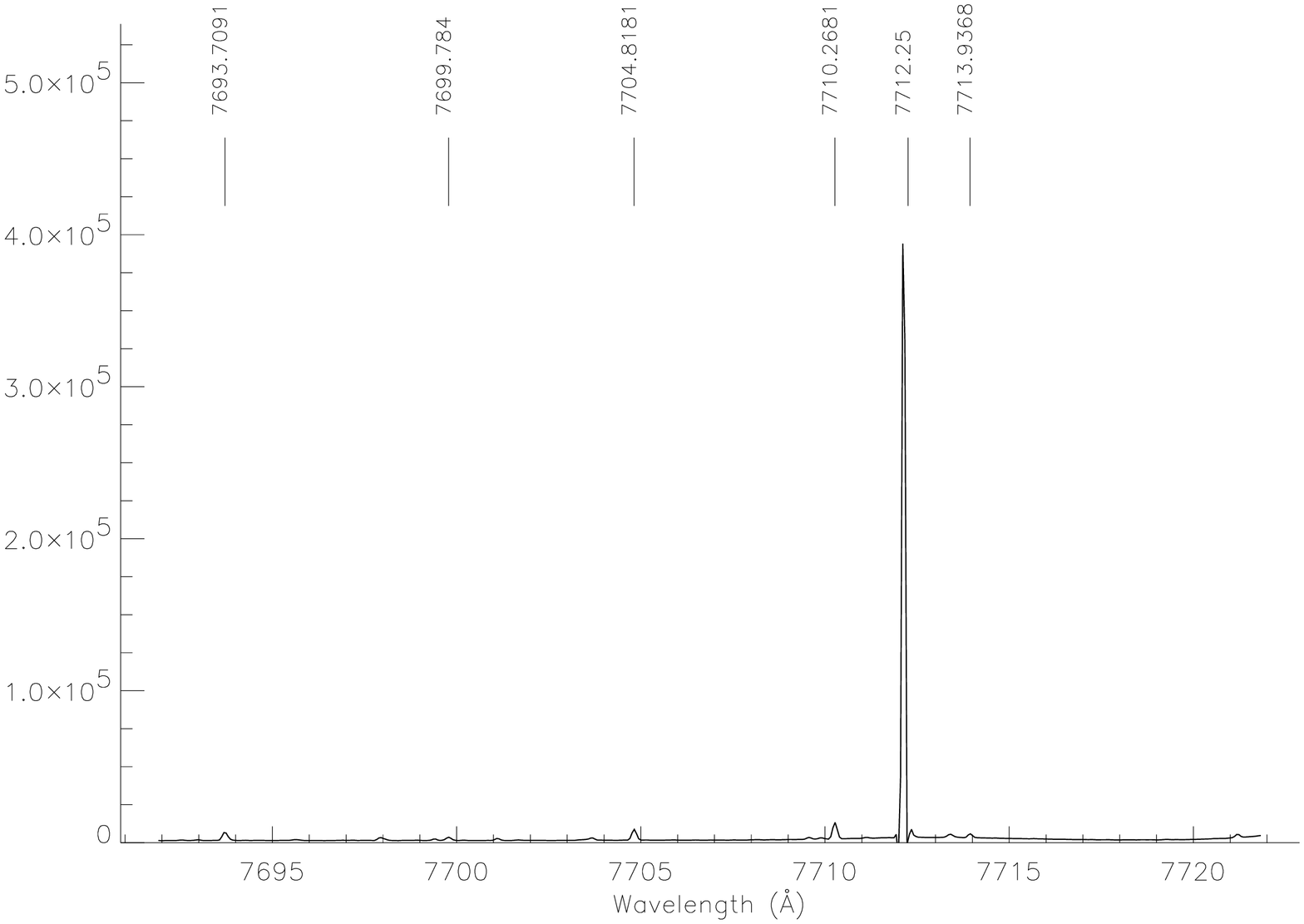}
\end{figure}
\clearpage
   
\begin{figure}
\centering
\includegraphics[width=10cm,angle=90]{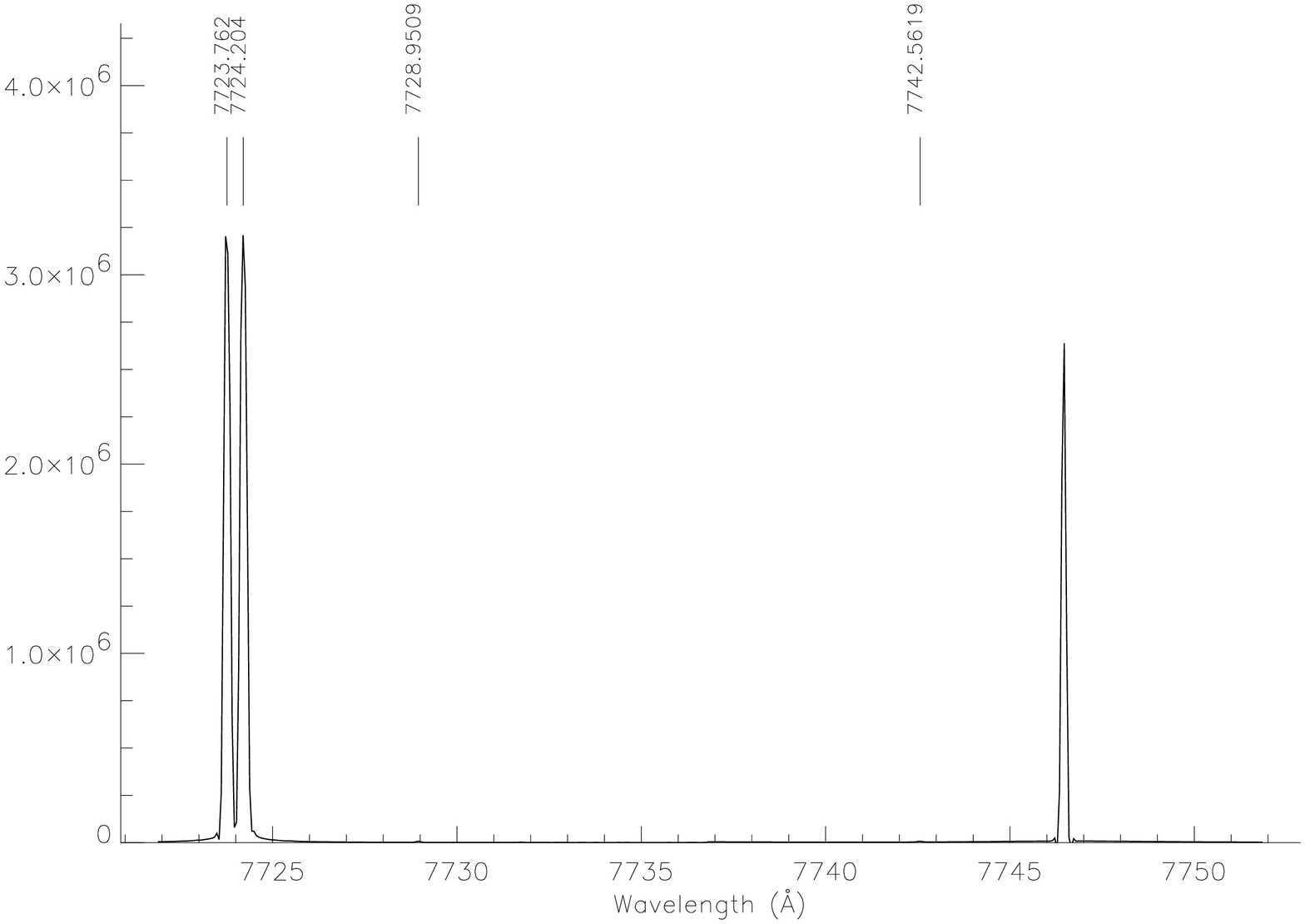}
\includegraphics[width=10cm,angle=90]{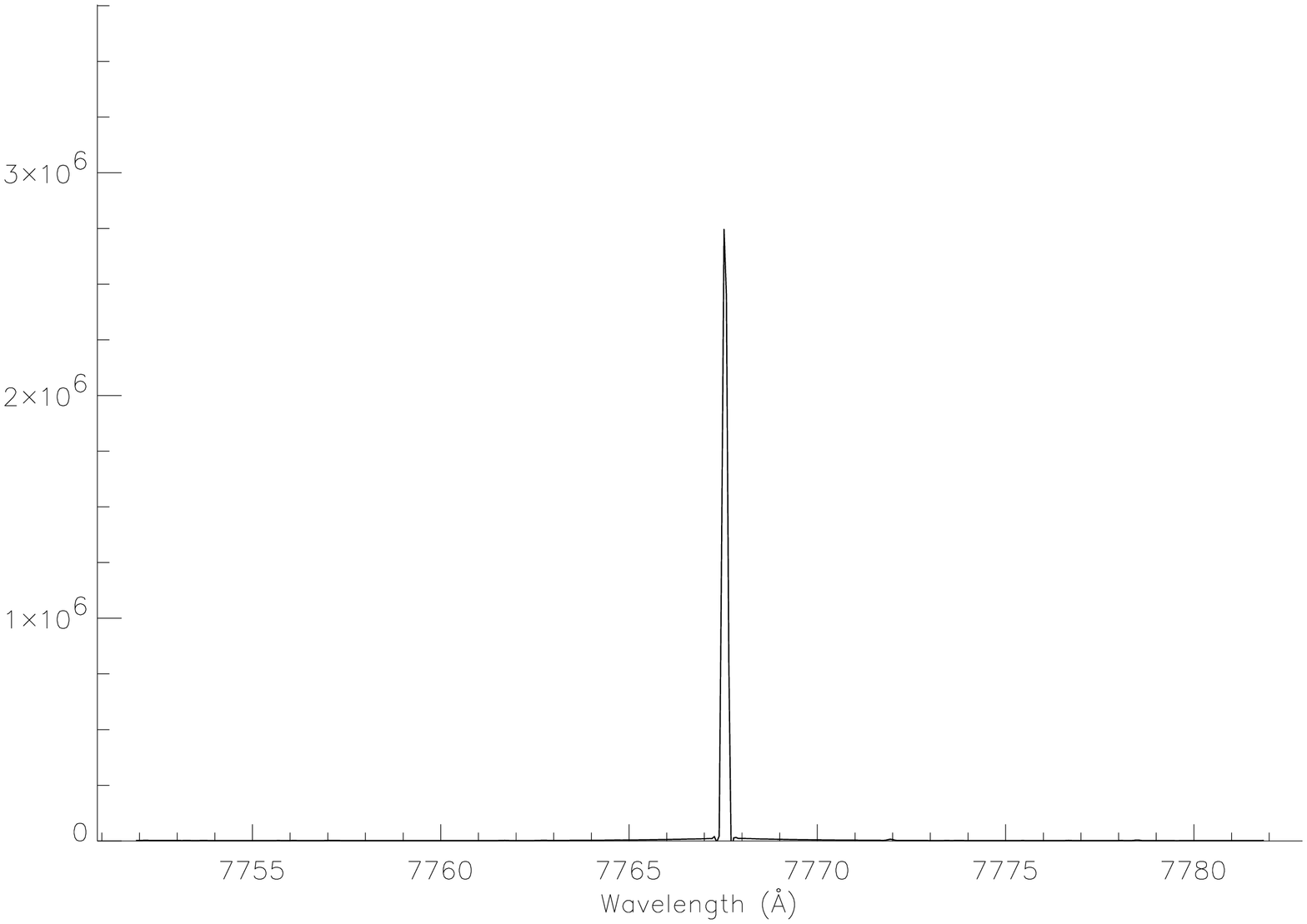}
\end{figure}
\clearpage
   
\begin{figure}
\centering
\includegraphics[width=10cm,angle=90]{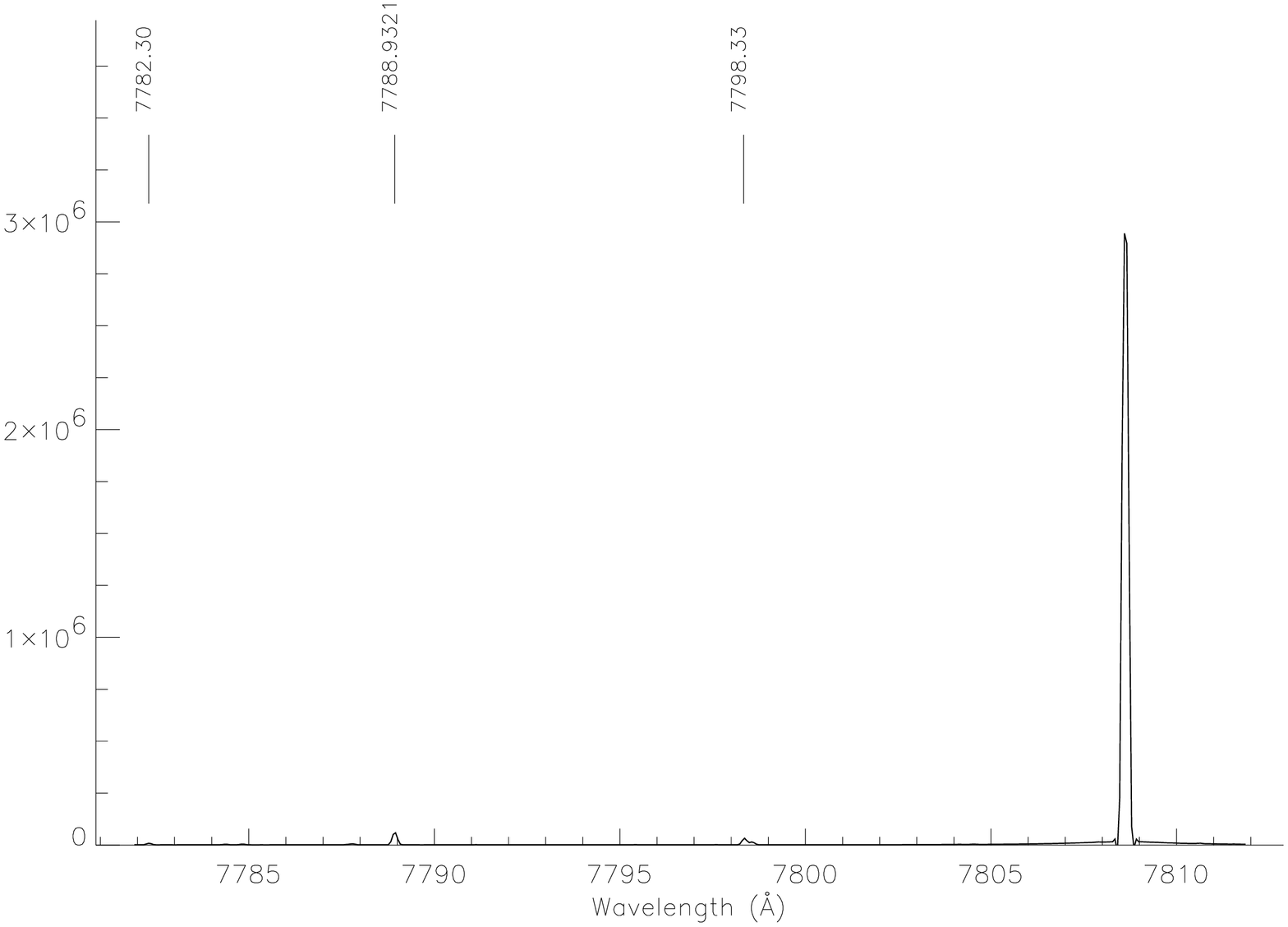}
\includegraphics[width=10cm,angle=90]{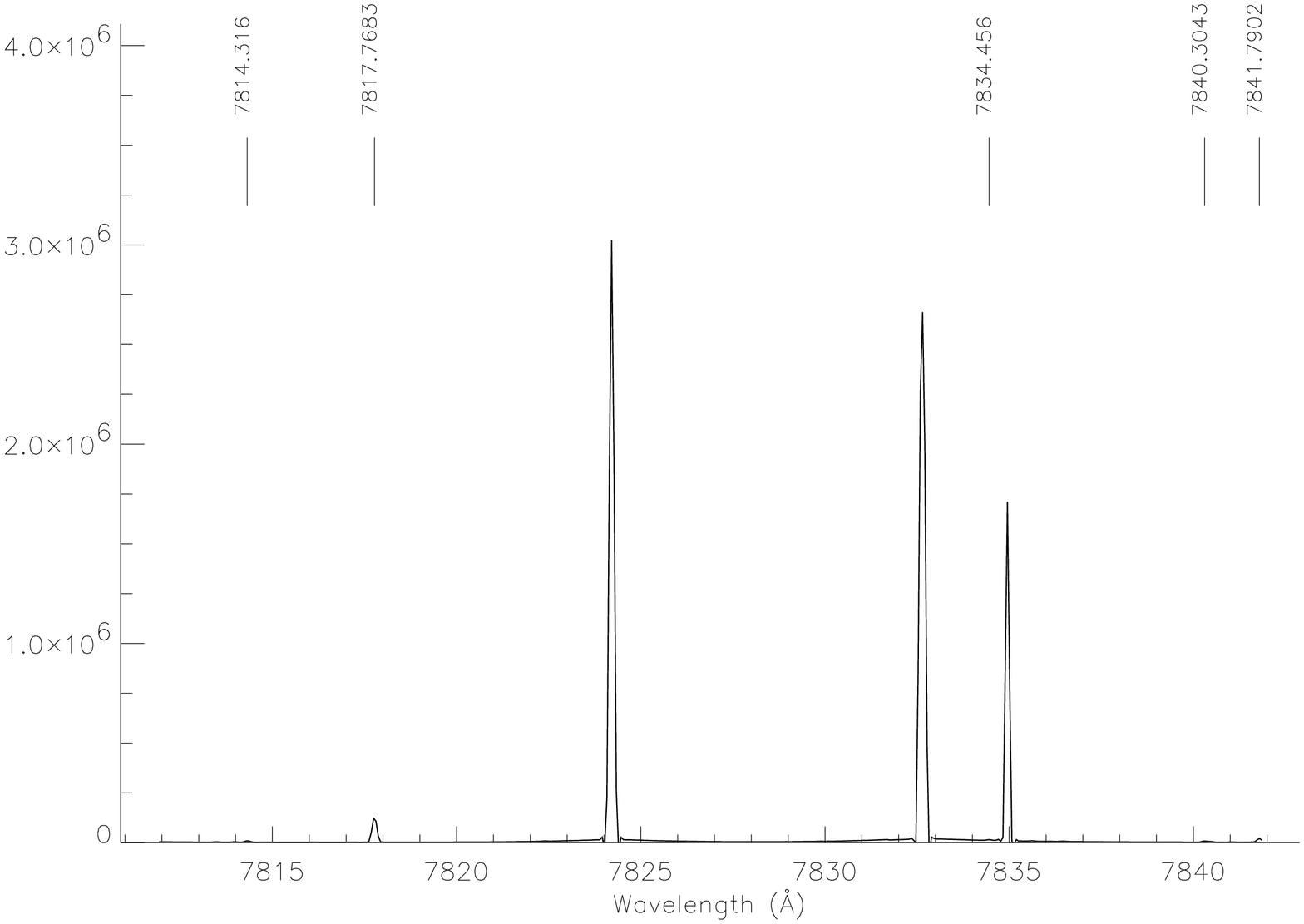}
\end{figure}
\clearpage
   
\begin{figure}
\centering
\includegraphics[width=10cm,angle=90]{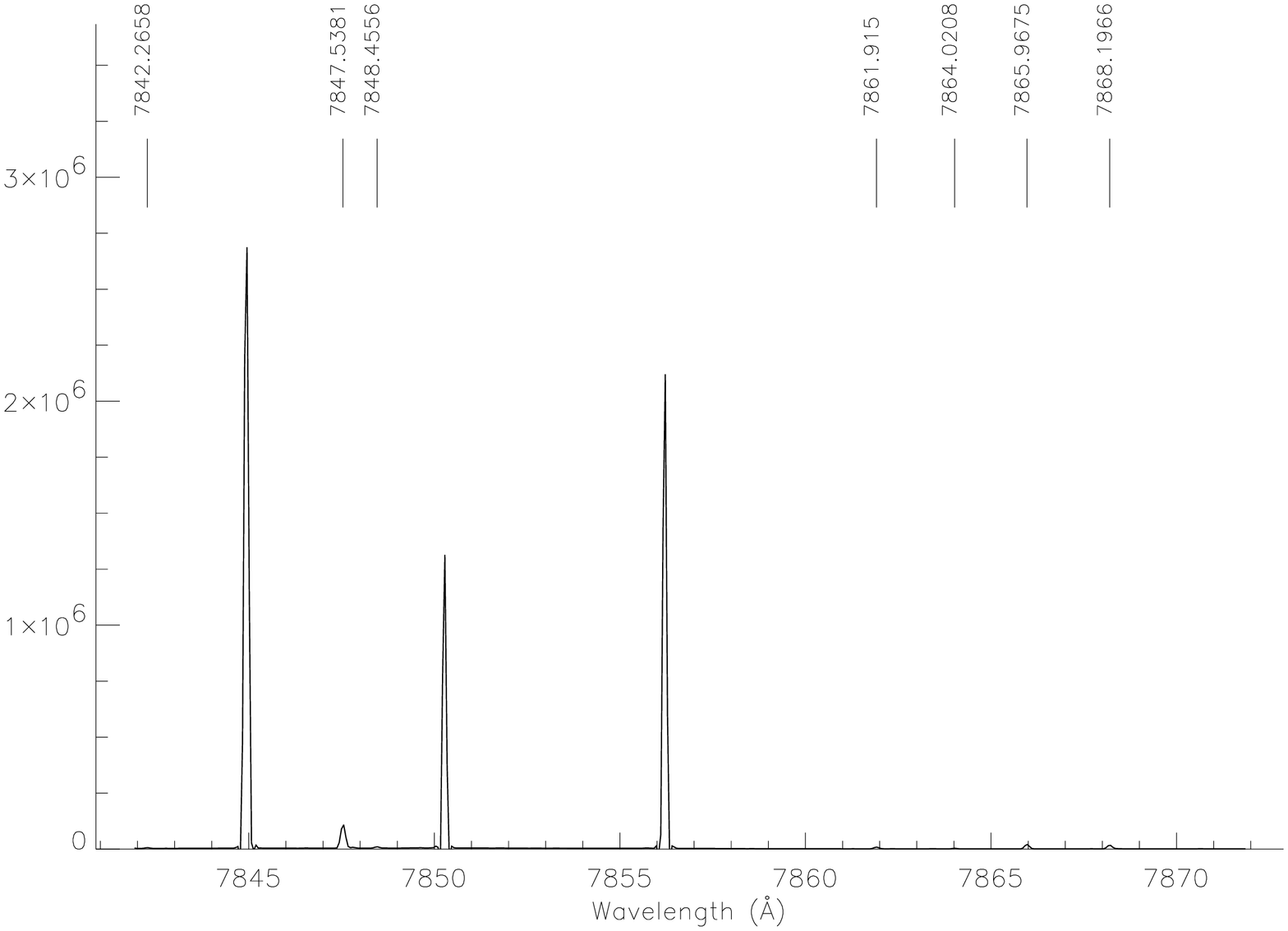}
\includegraphics[width=10cm,angle=90]{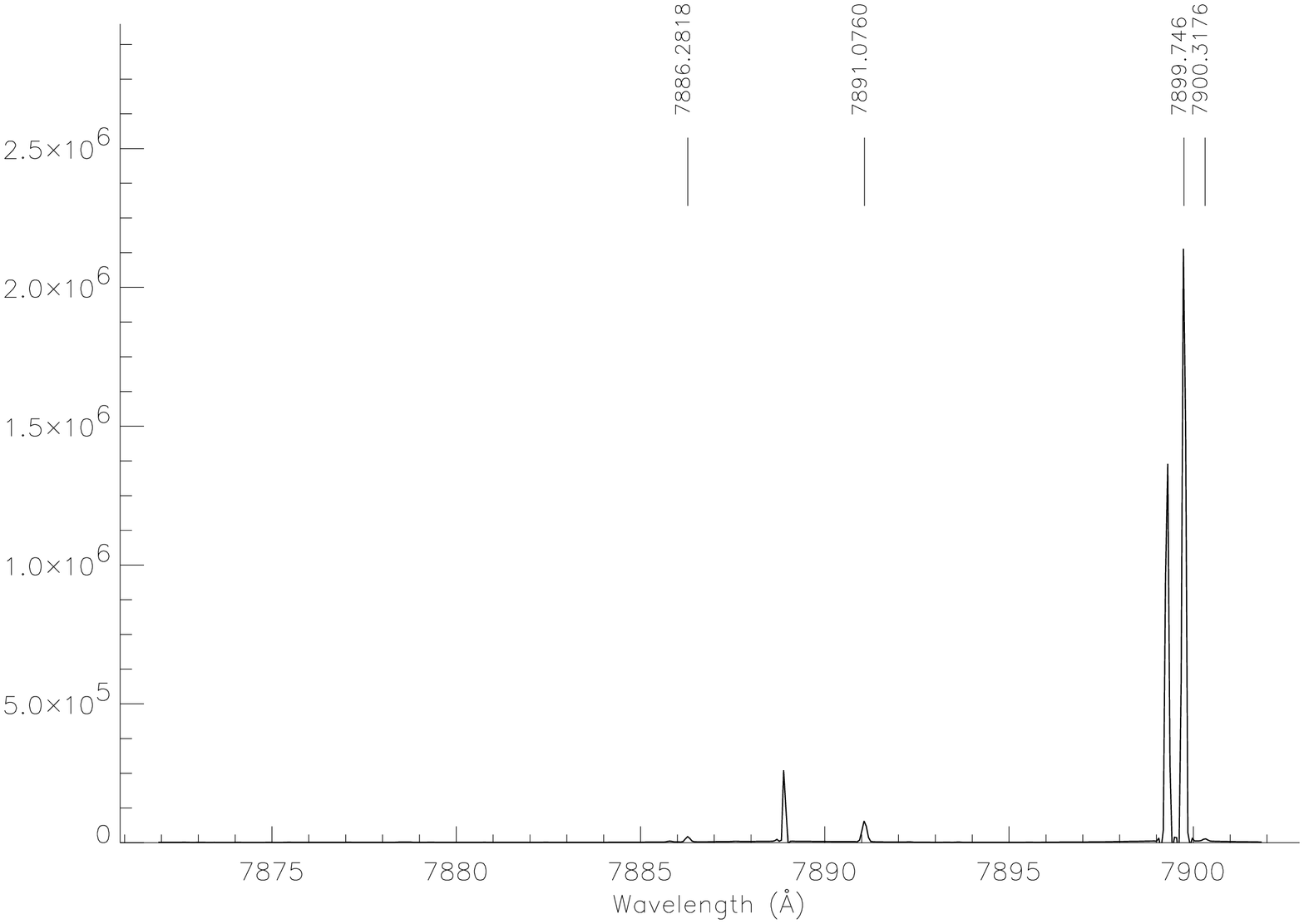}
\end{figure}
\clearpage
   
\begin{figure}
\centering
\includegraphics[width=10cm,angle=90]{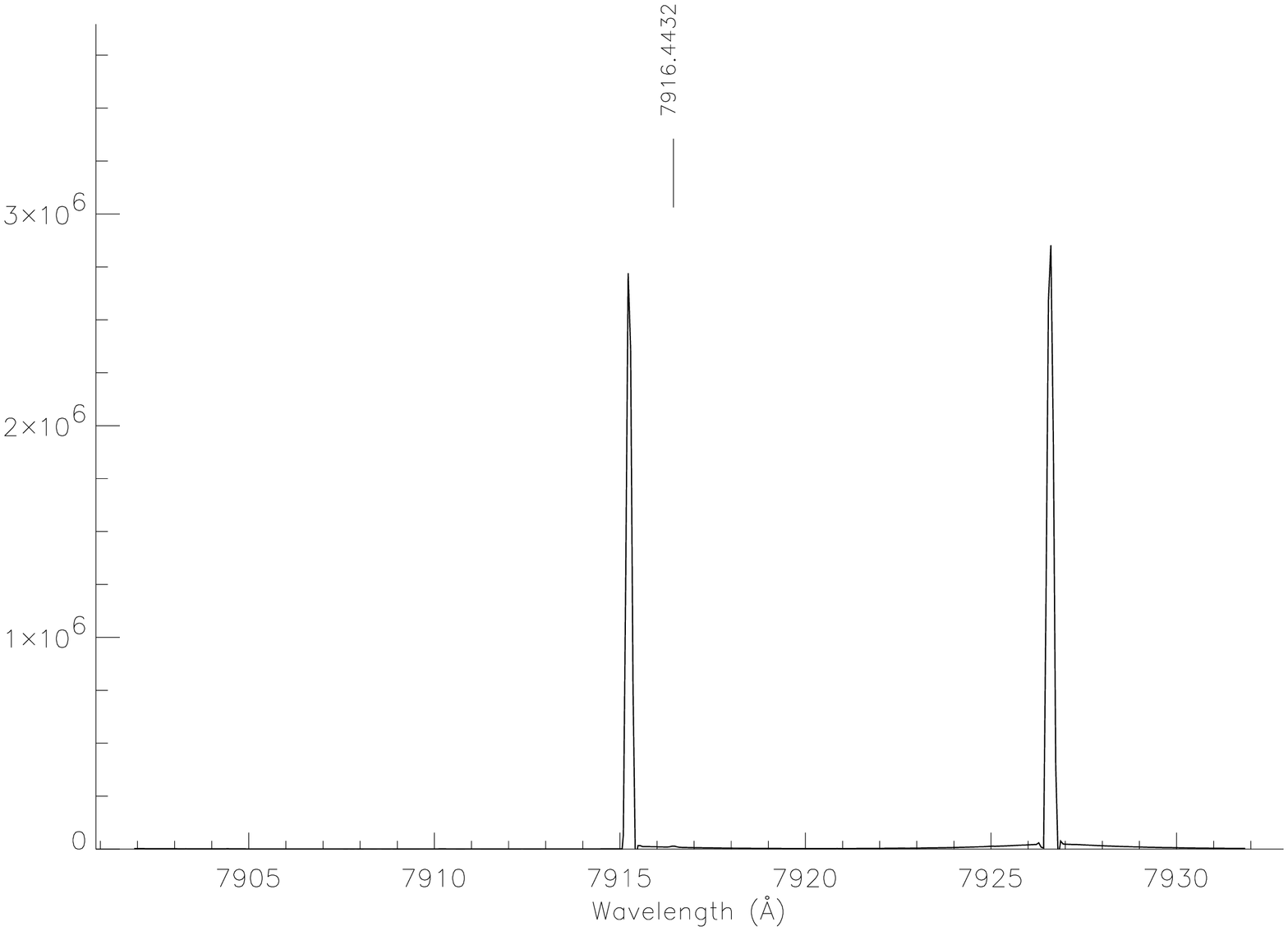}
\includegraphics[width=10cm,angle=90]{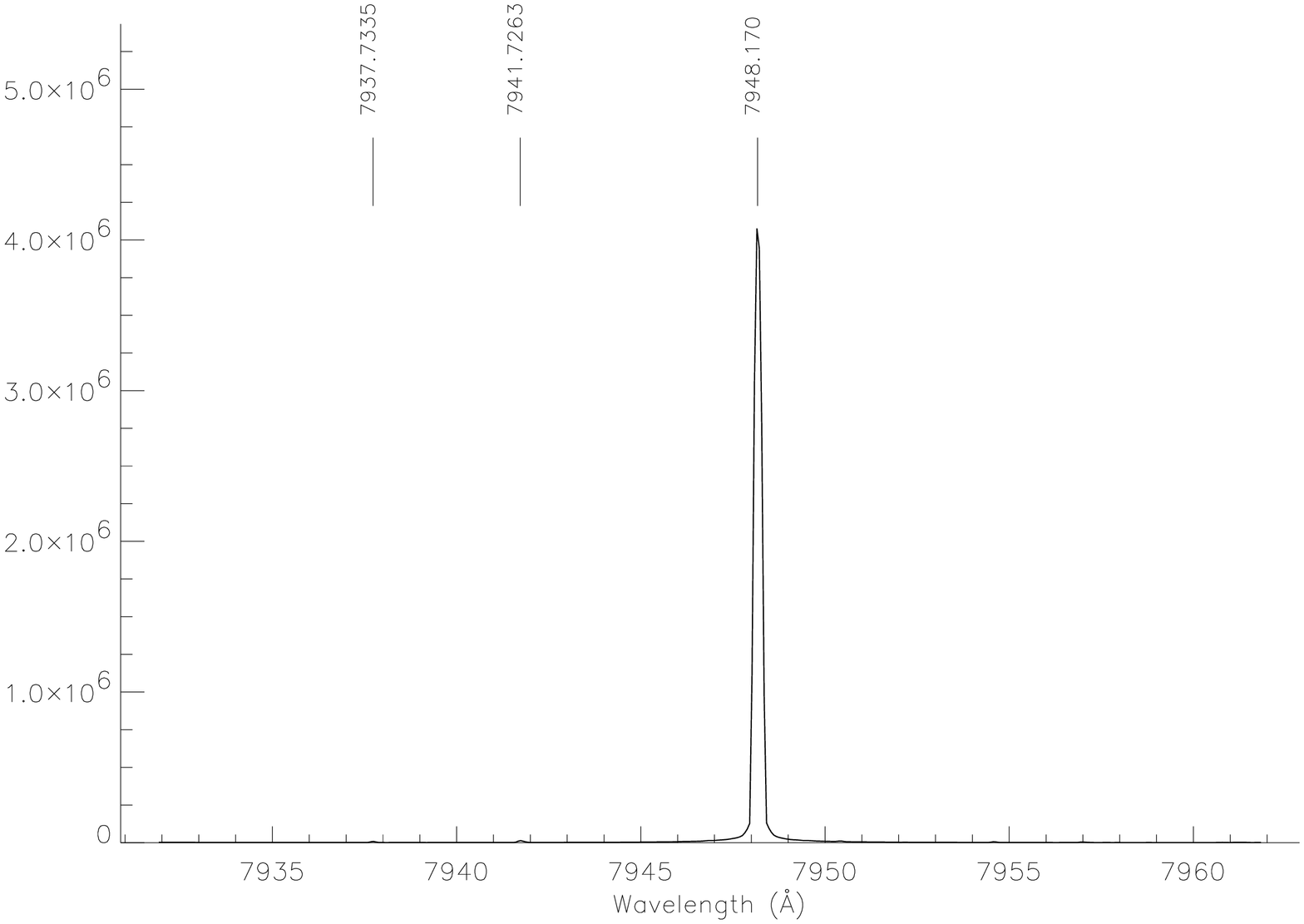}
\end{figure}
\clearpage
   
\begin{figure}
\centering
\includegraphics[width=10cm,angle=90]{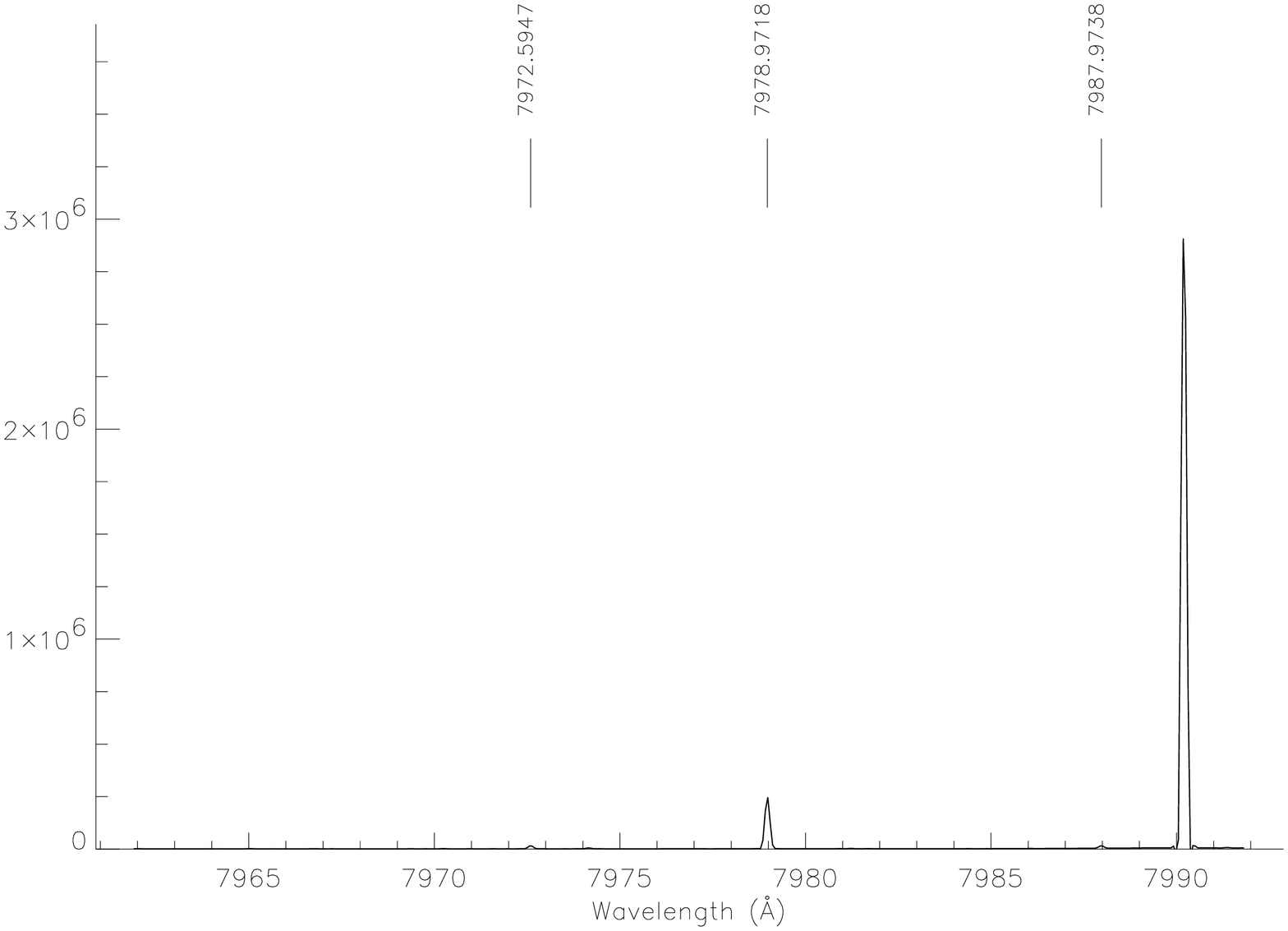}
\includegraphics[width=10cm,angle=90]{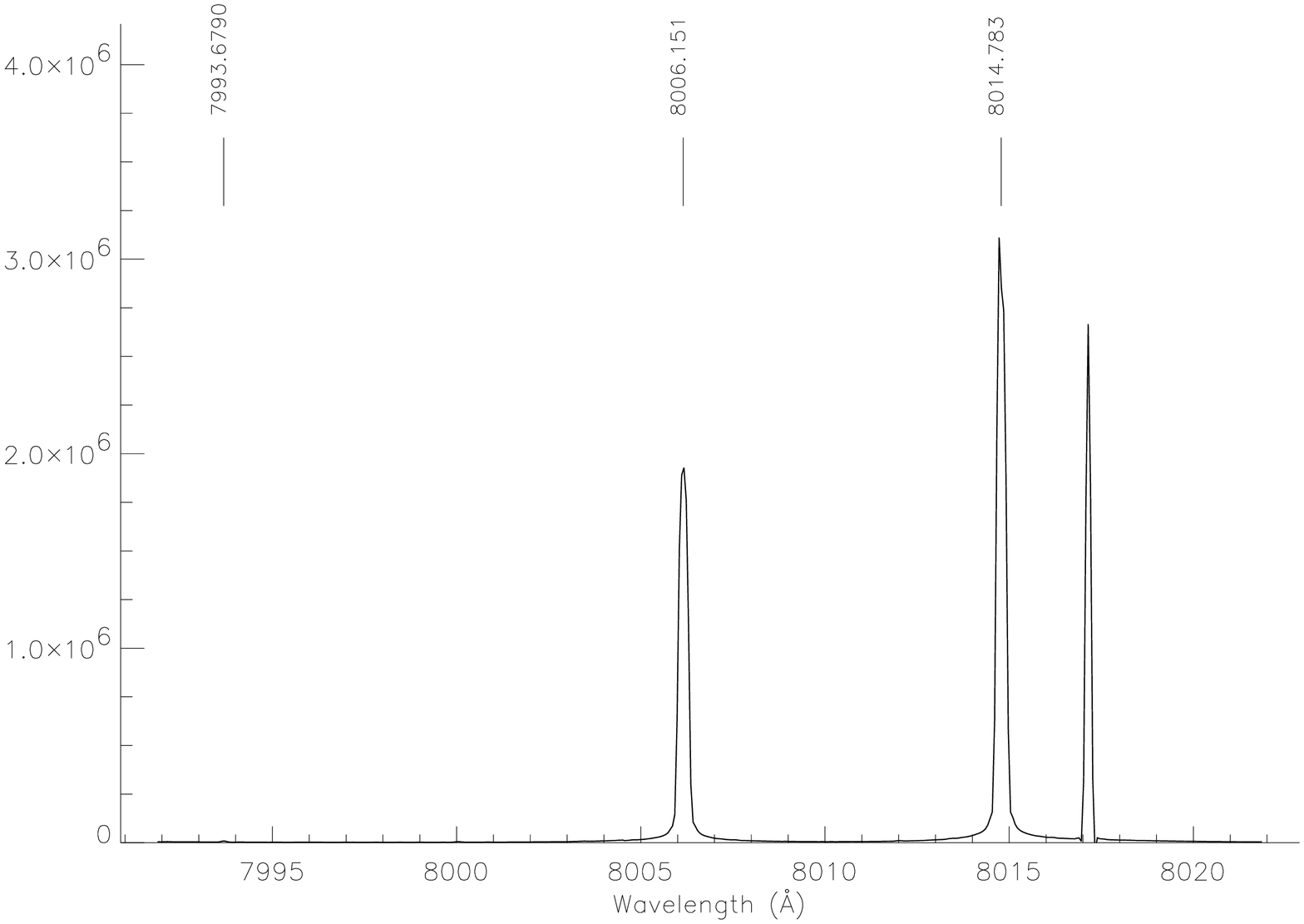}
\end{figure}
\clearpage
   
\begin{figure}
\centering
\includegraphics[width=10cm,angle=90]{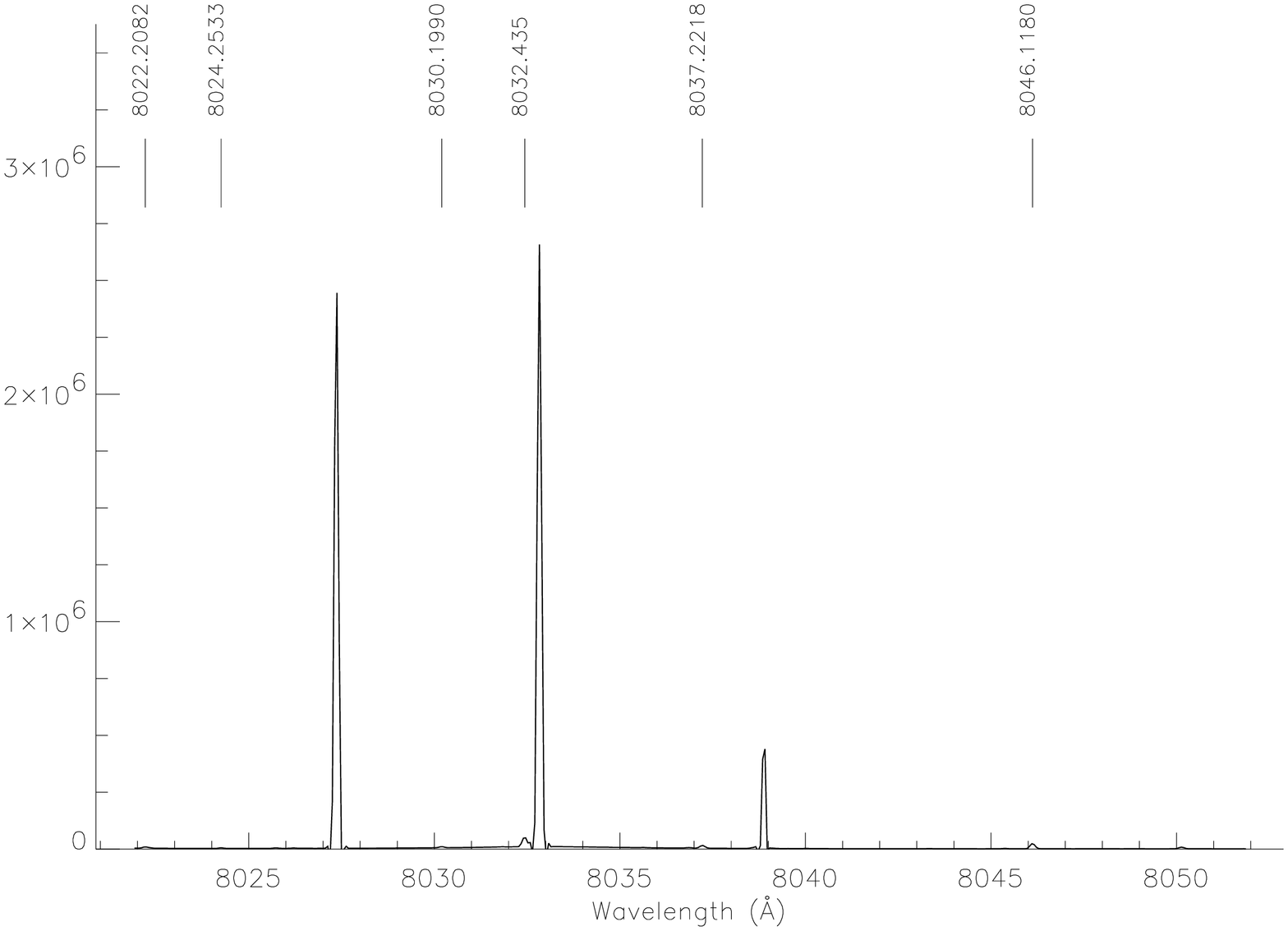}
\includegraphics[width=10cm,angle=90]{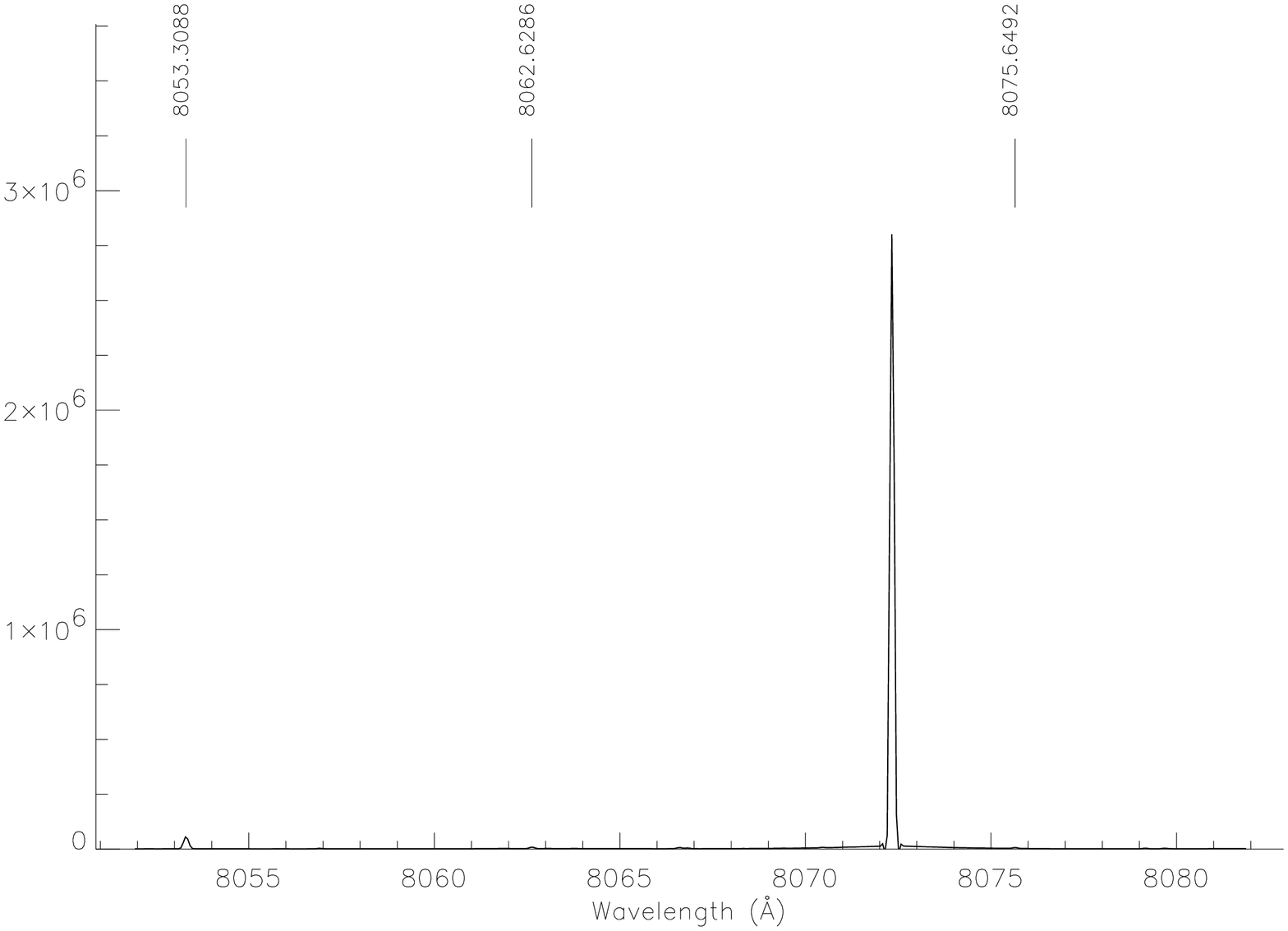}
\end{figure}
\clearpage
   
\begin{figure}
\centering
\includegraphics[width=10cm,angle=90]{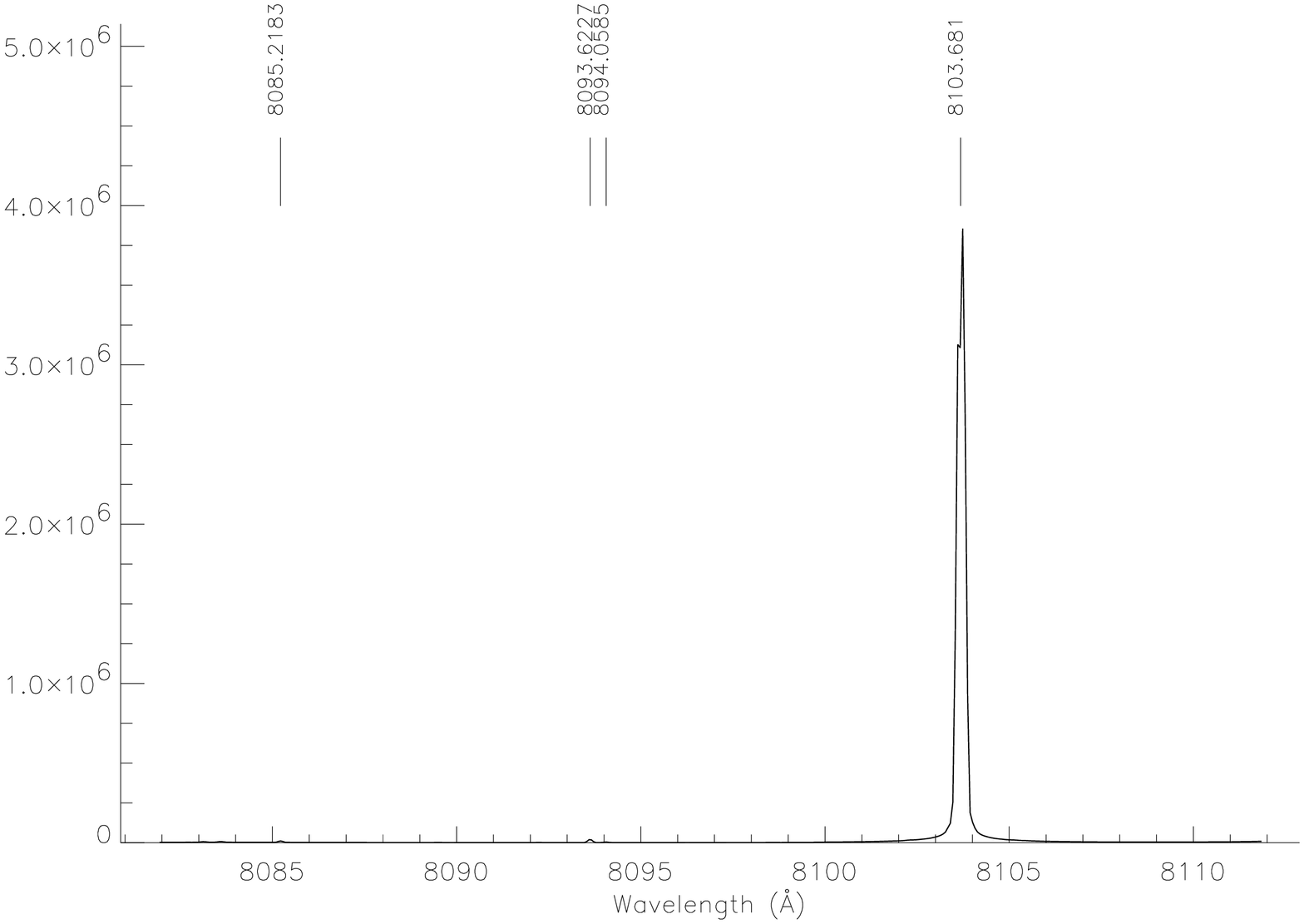}
\includegraphics[width=10cm,angle=90]{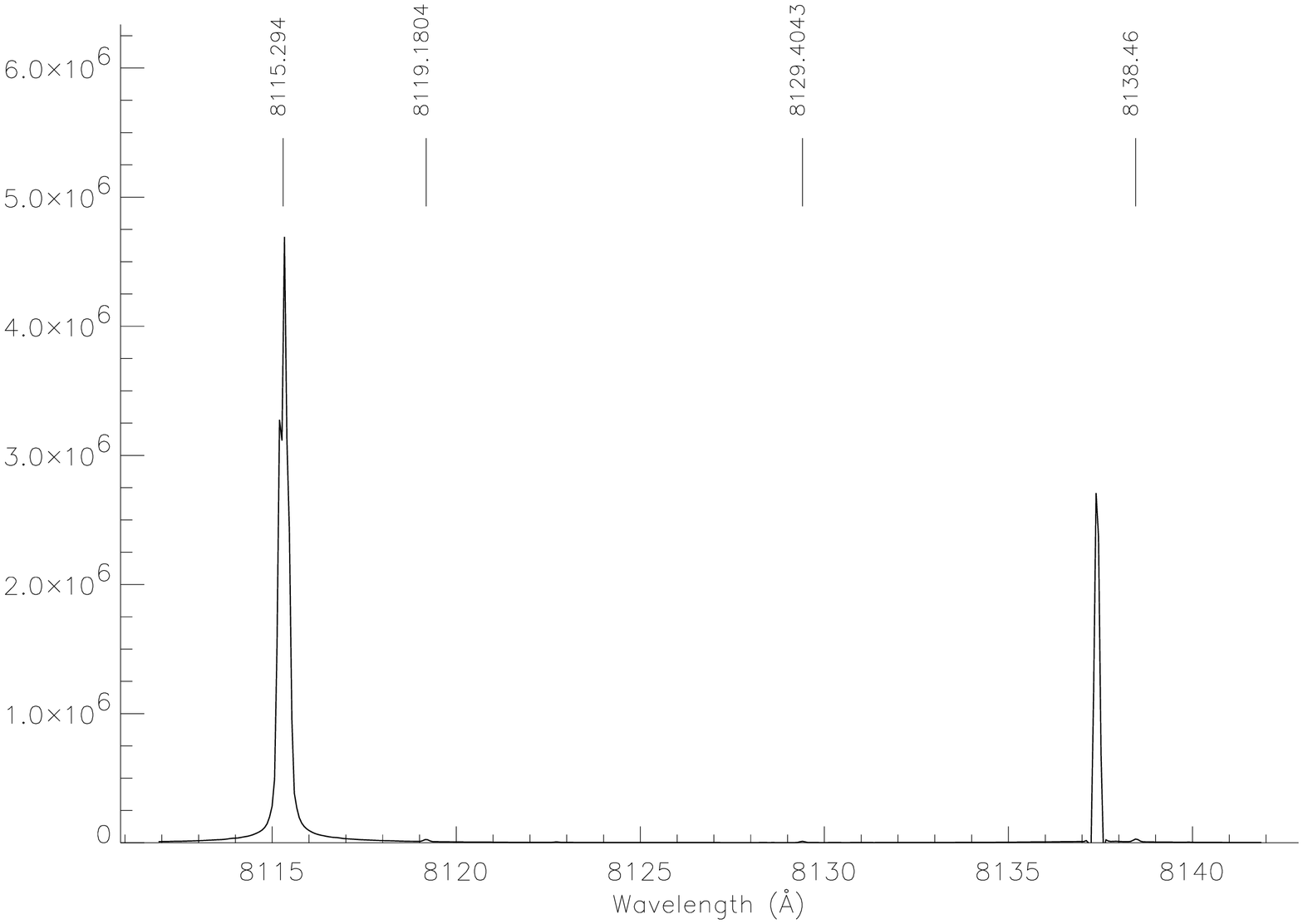}
\end{figure}
\clearpage
   
\begin{figure}
\centering
\includegraphics[width=10cm,angle=90]{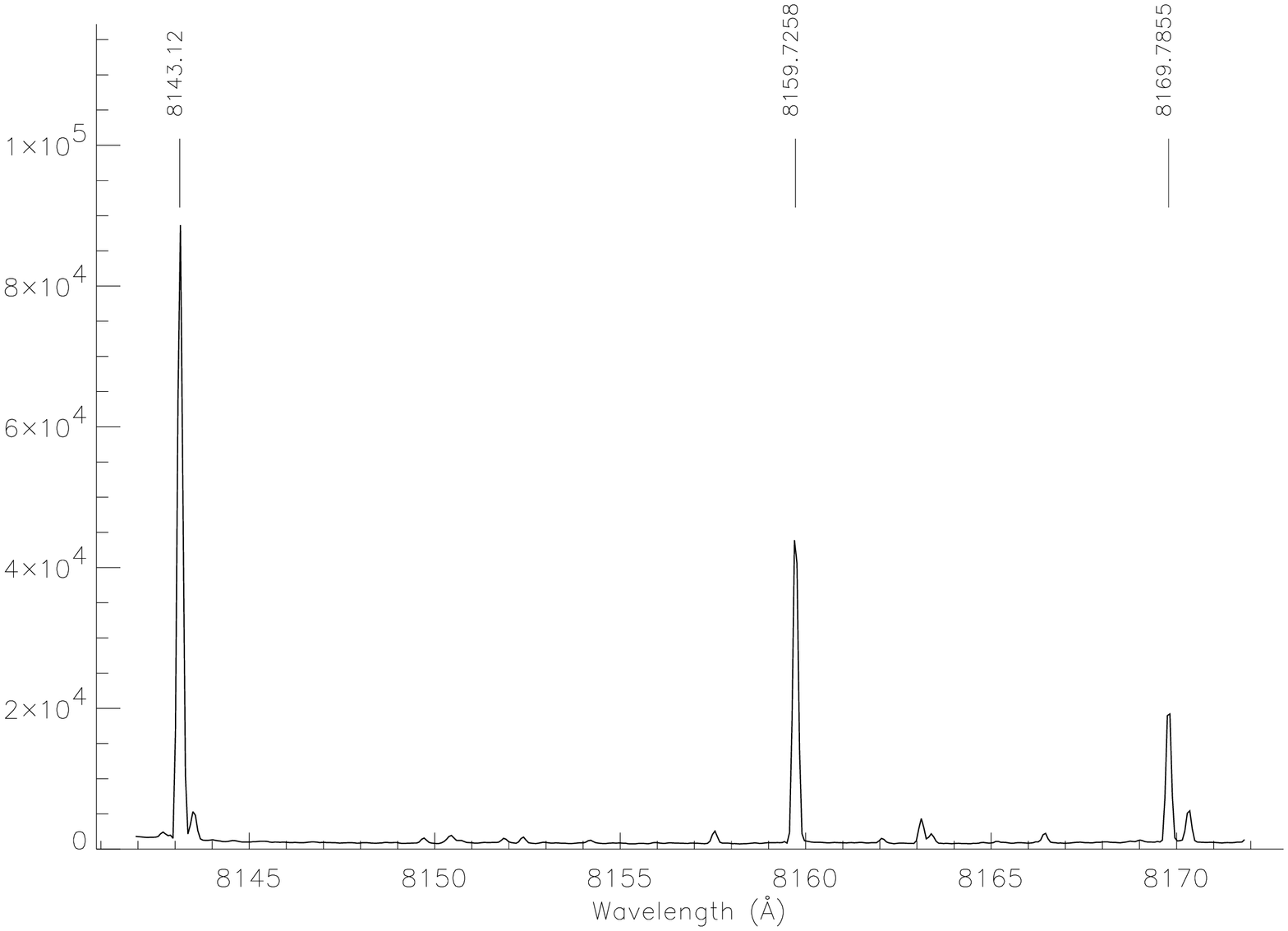}
\includegraphics[width=10cm,angle=90]{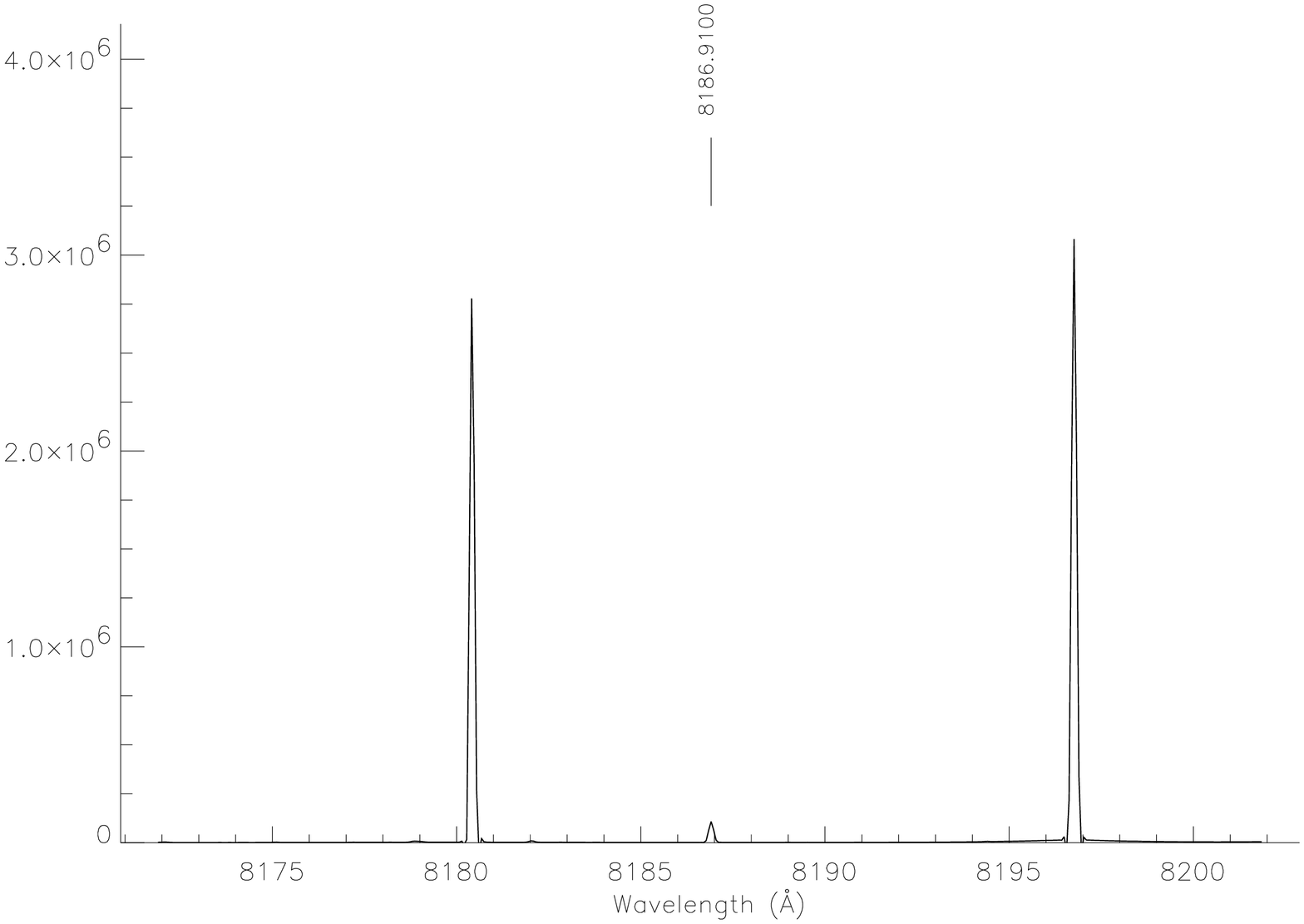}
\end{figure}
\clearpage
   
\begin{figure}
\centering
\includegraphics[width=10cm,angle=90]{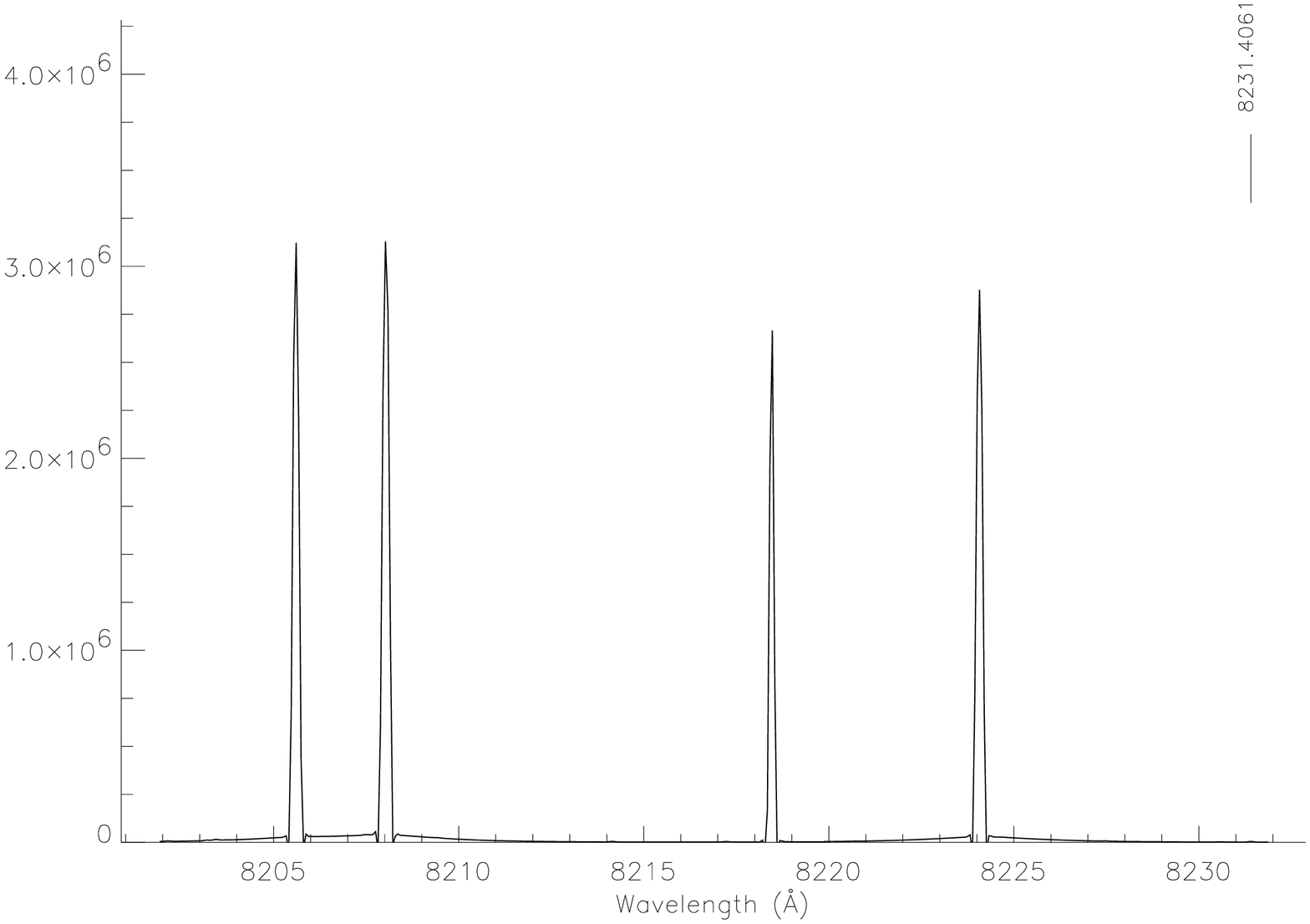}
\includegraphics[width=10cm,angle=90]{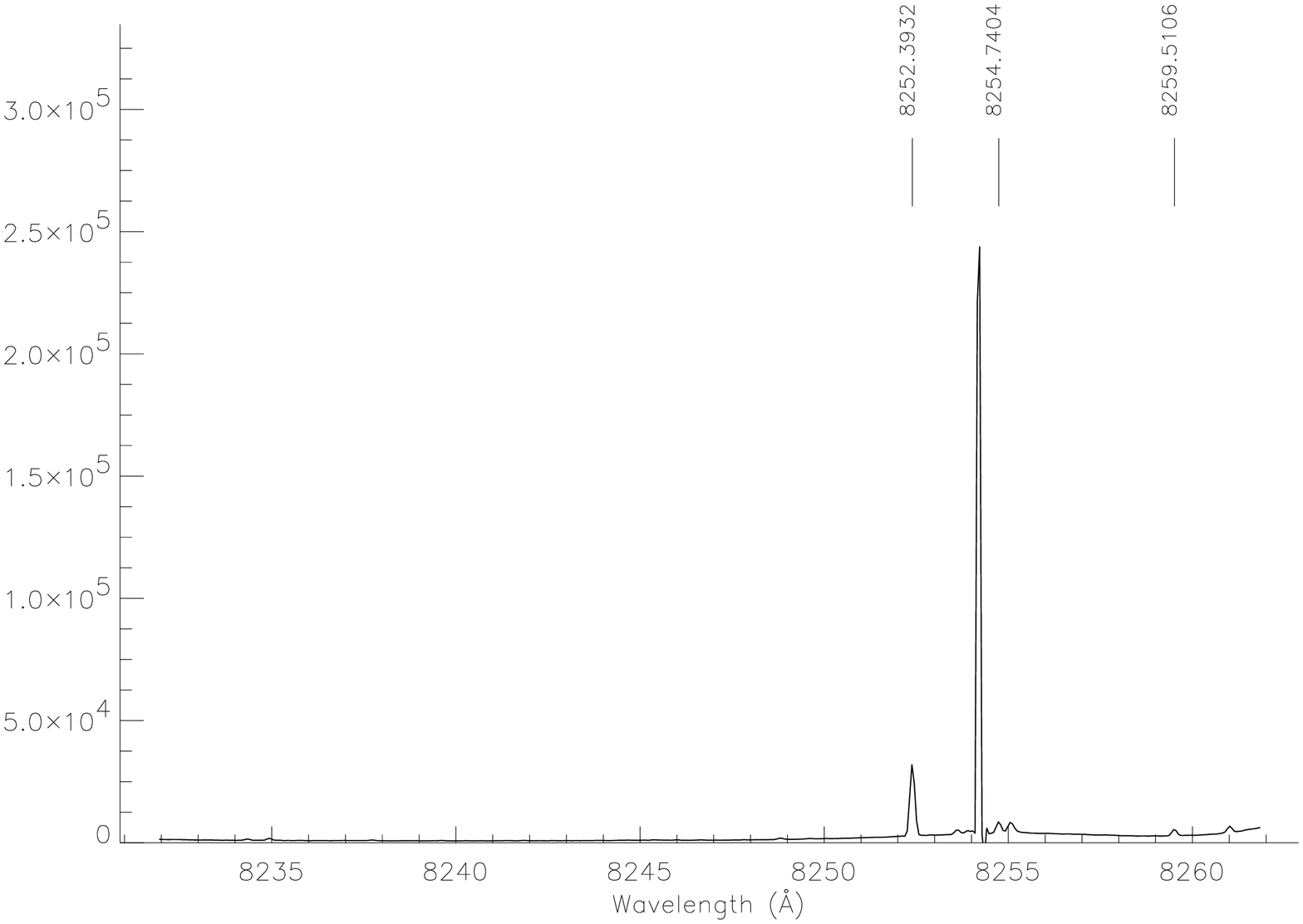}
\end{figure}
\clearpage
   
\begin{figure}
\centering
\includegraphics[width=10cm,angle=90]{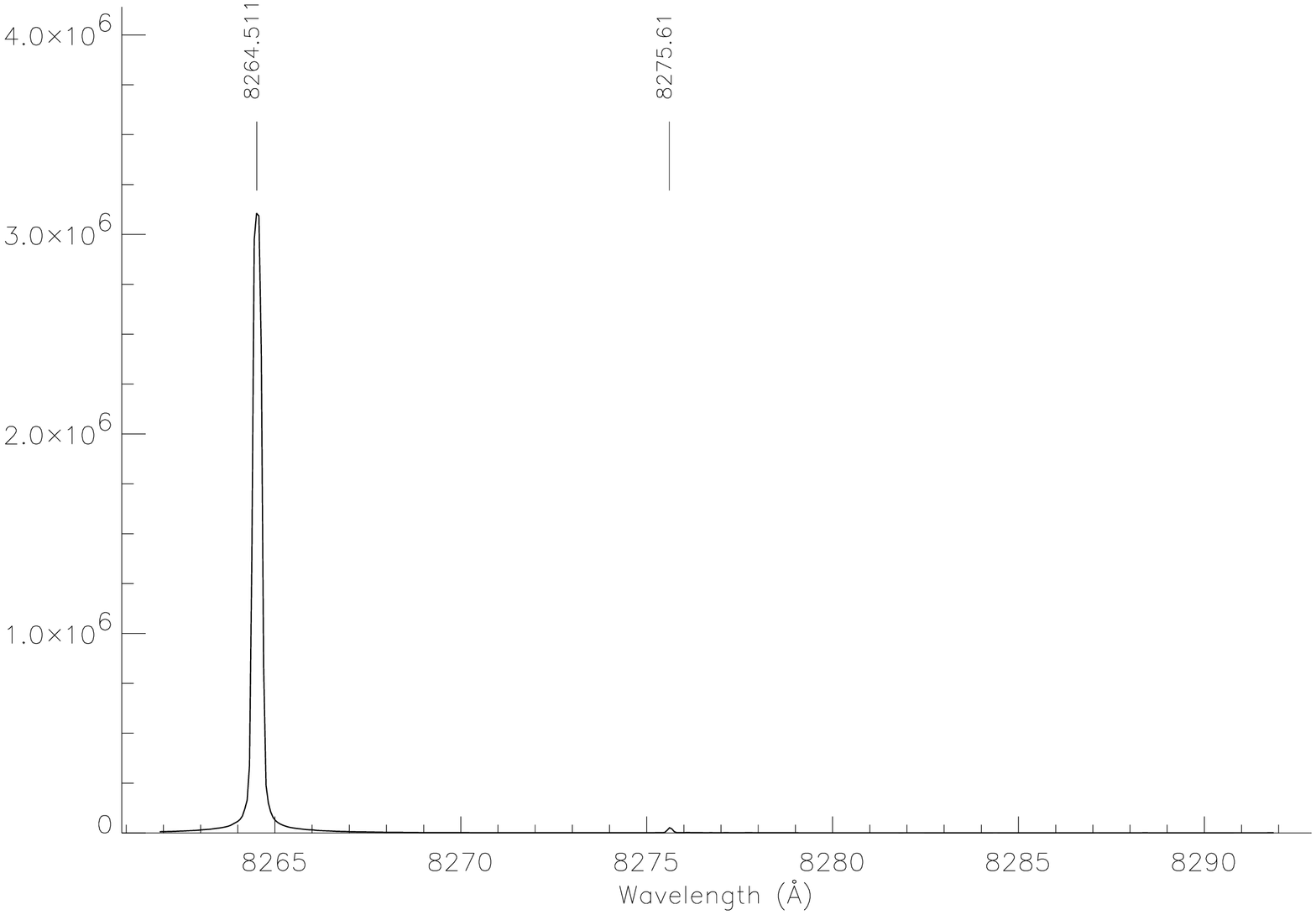}
\includegraphics[width=10cm,angle=90]{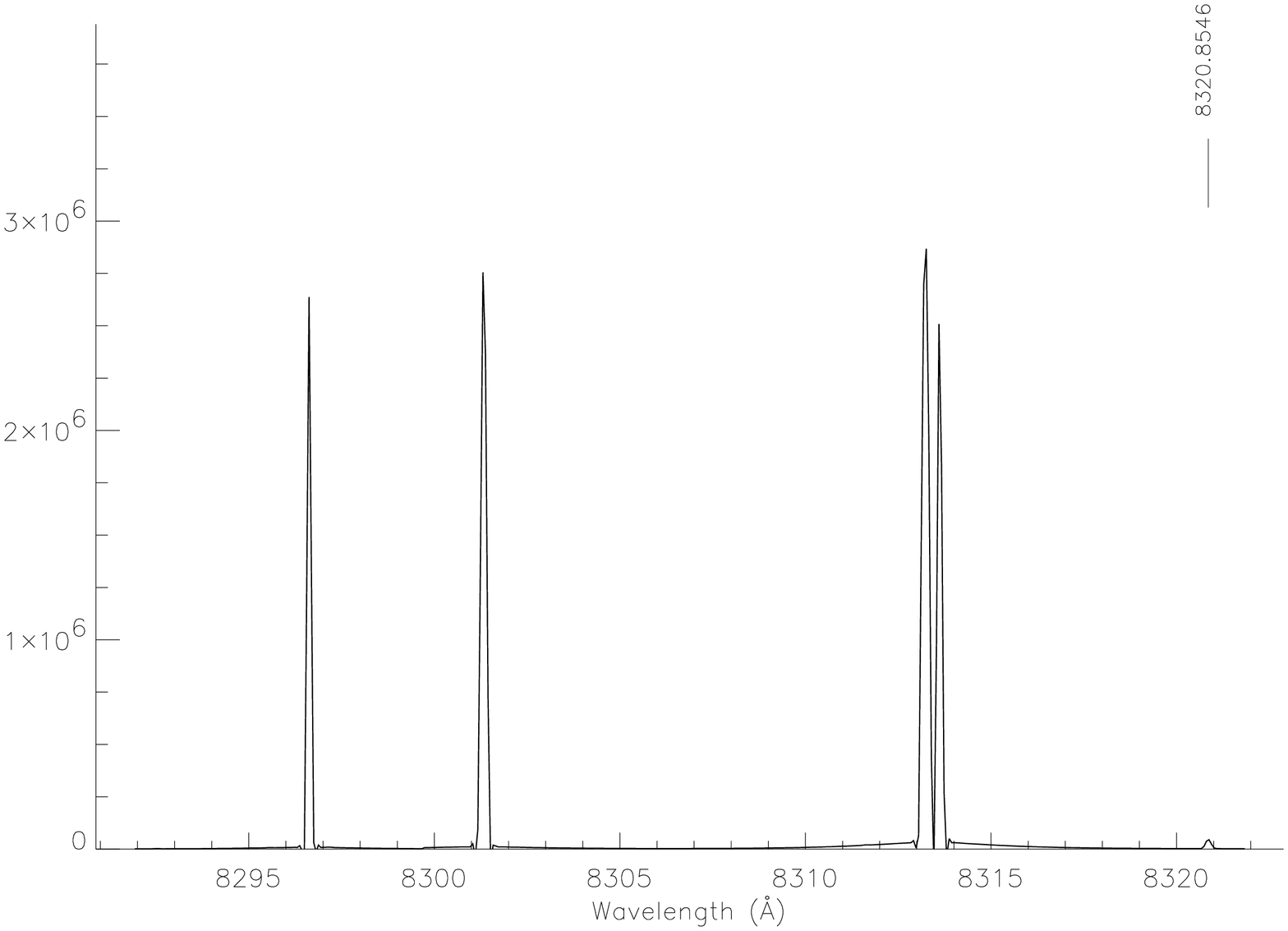}
\end{figure}
\clearpage
   
\begin{figure}
\centering
\includegraphics[width=10cm,angle=90]{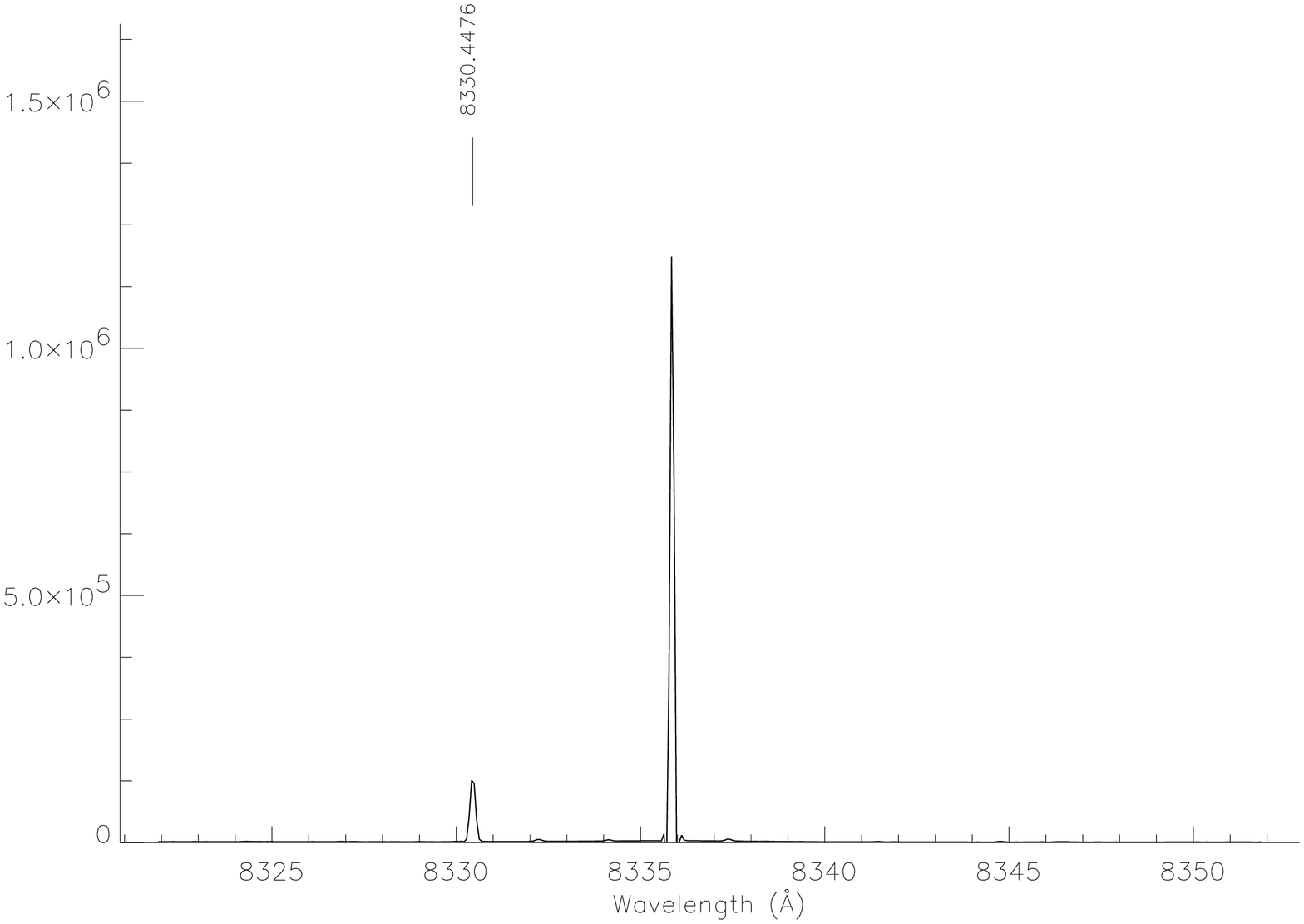}
\includegraphics[width=10cm,angle=90]{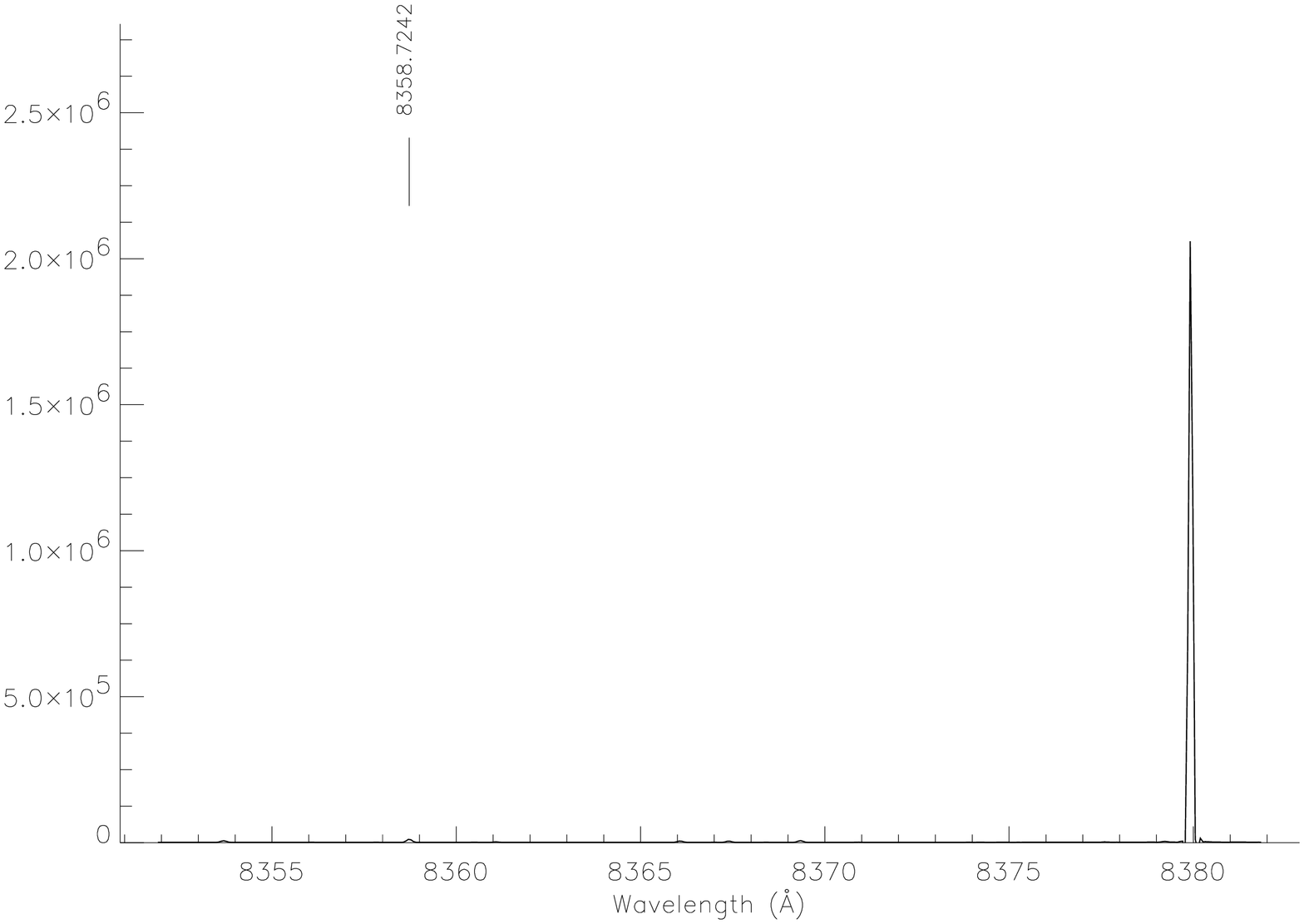}
\end{figure}
\clearpage
   
\begin{figure}
\centering
\includegraphics[width=10cm,angle=90]{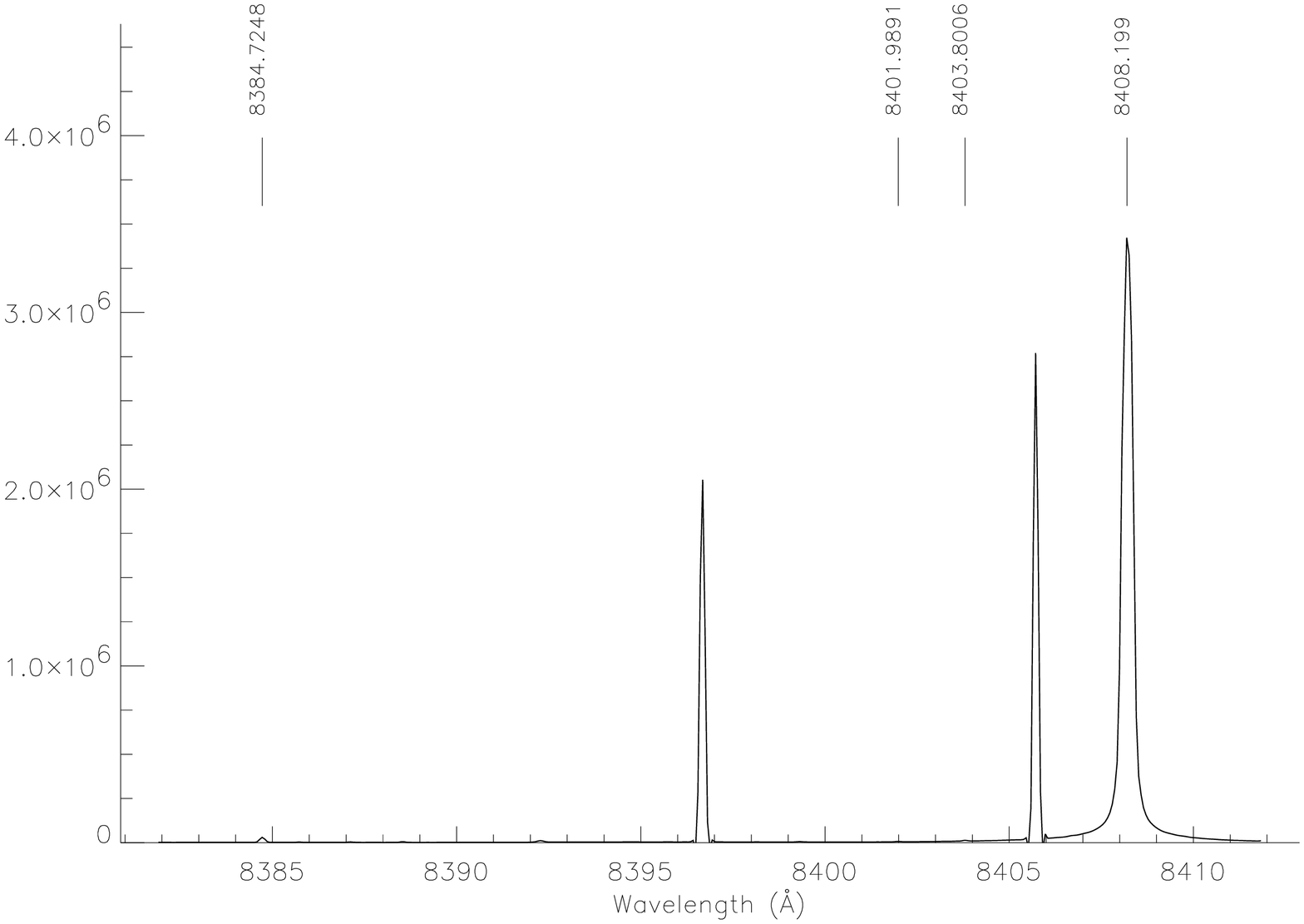}
\includegraphics[width=10cm,angle=90]{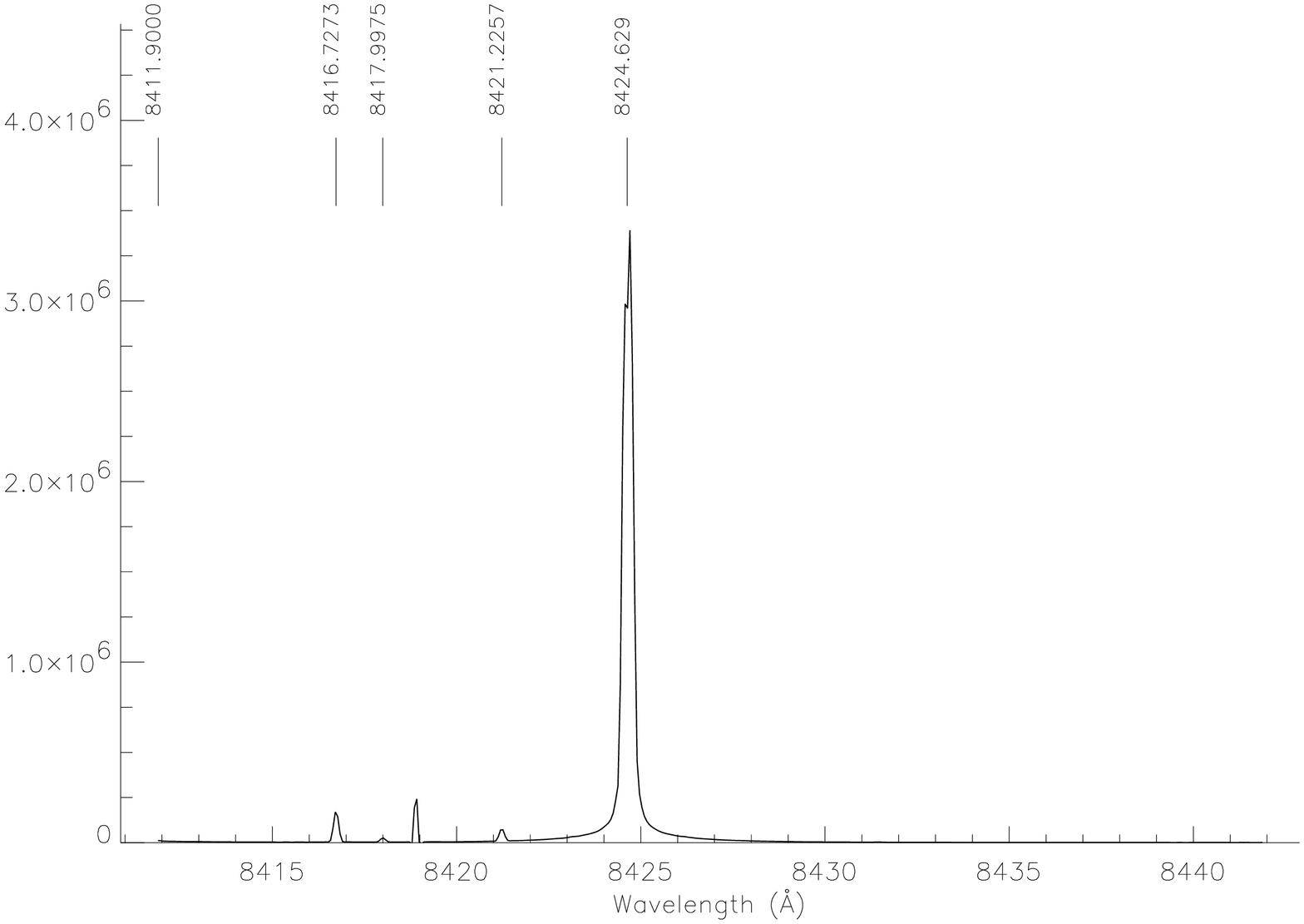}
\end{figure}
\clearpage
   
\begin{figure}
\centering
\includegraphics[width=10cm,angle=90]{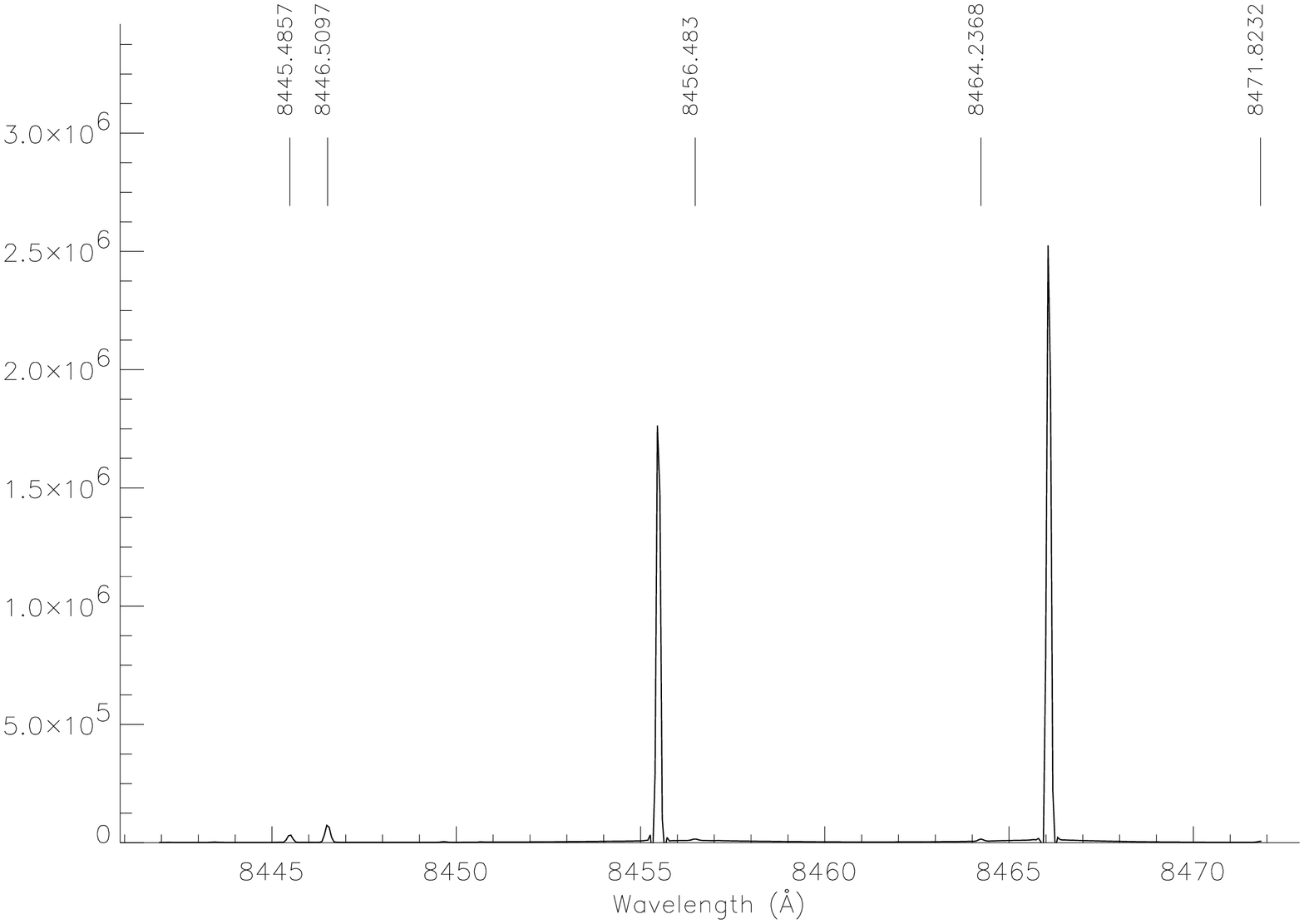}
\includegraphics[width=10cm,angle=90]{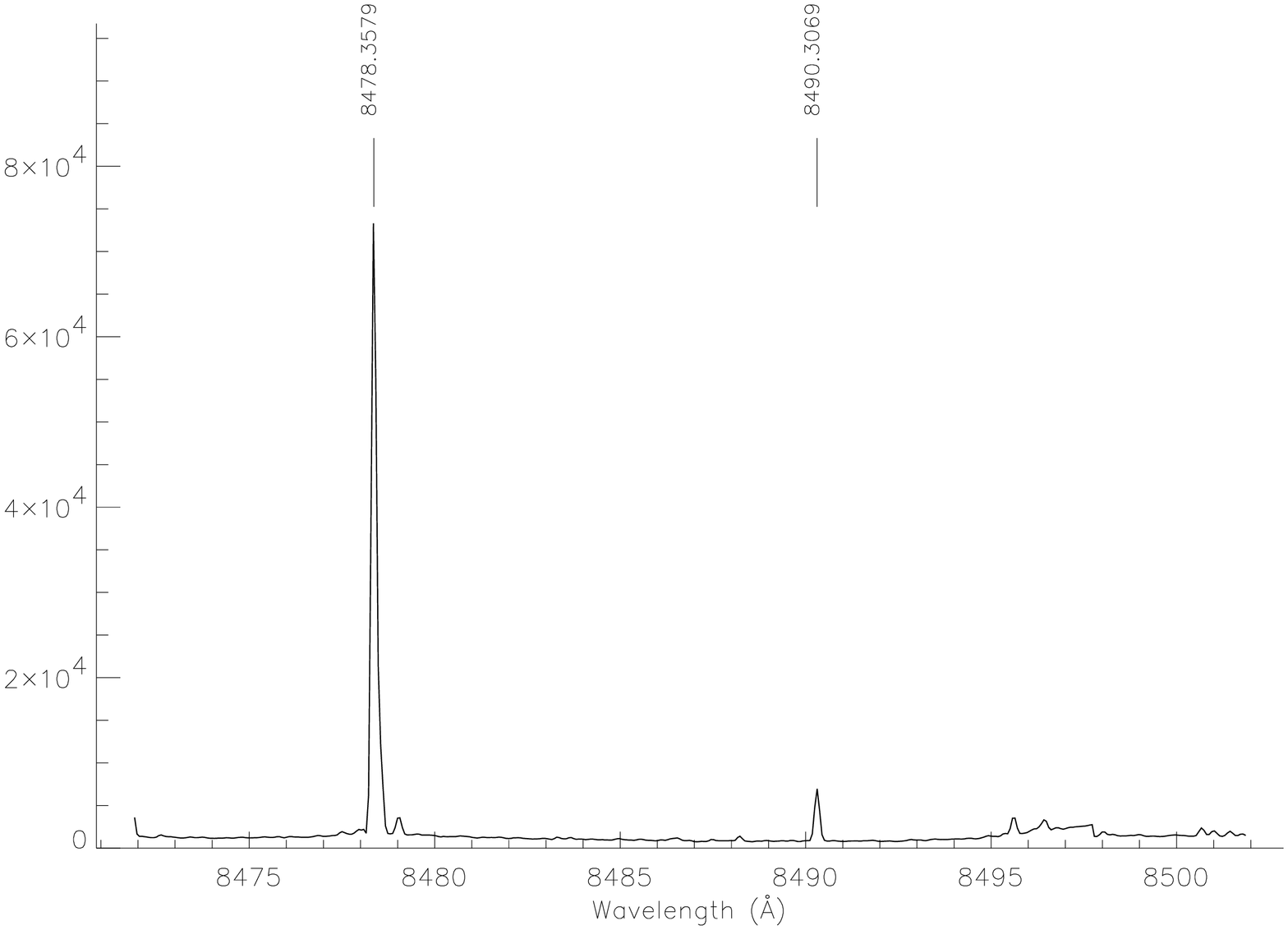}
\end{figure}
\clearpage
   
\begin{figure}
\centering
\includegraphics[width=10cm,angle=90]{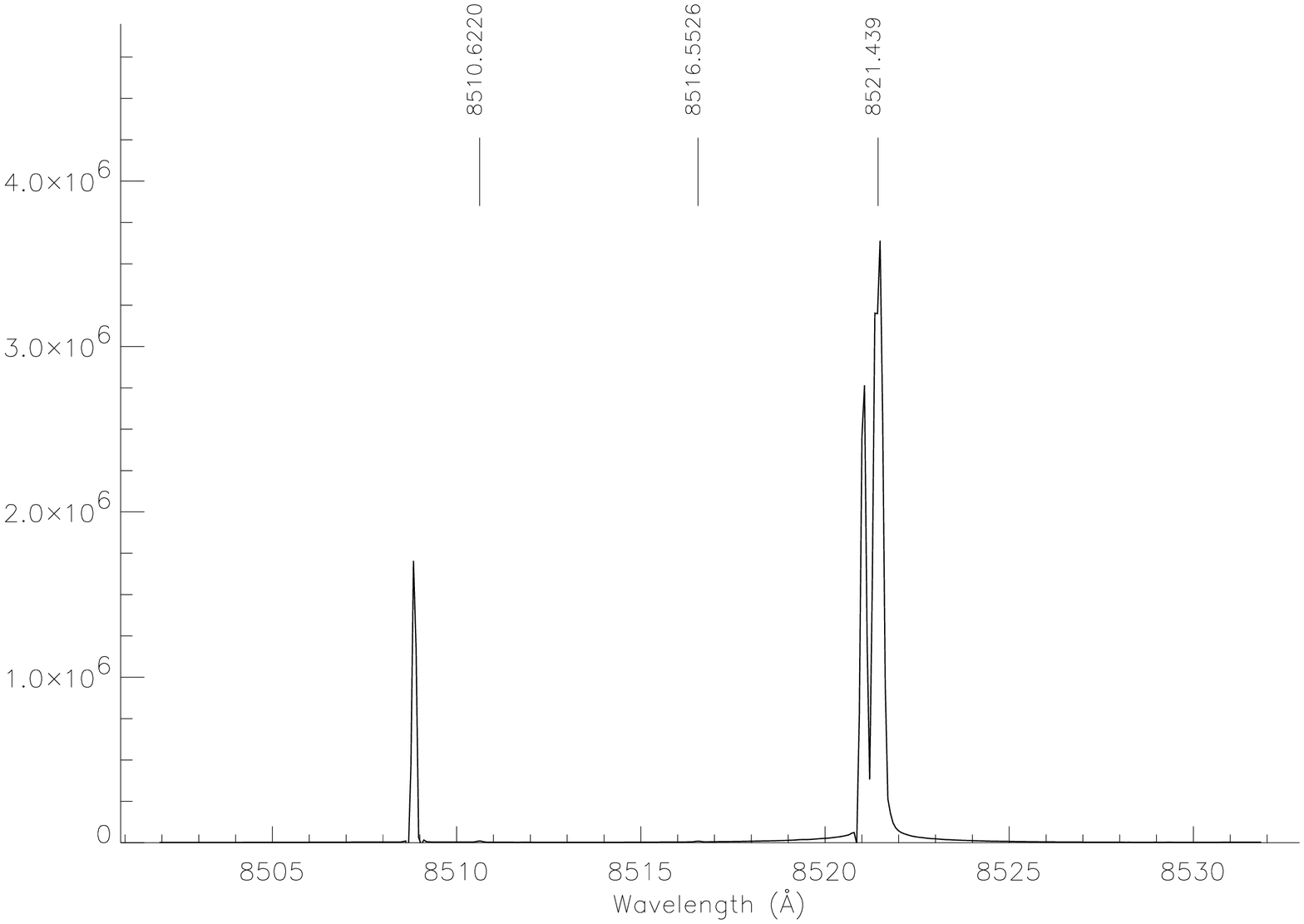}
\includegraphics[width=10cm,angle=90]{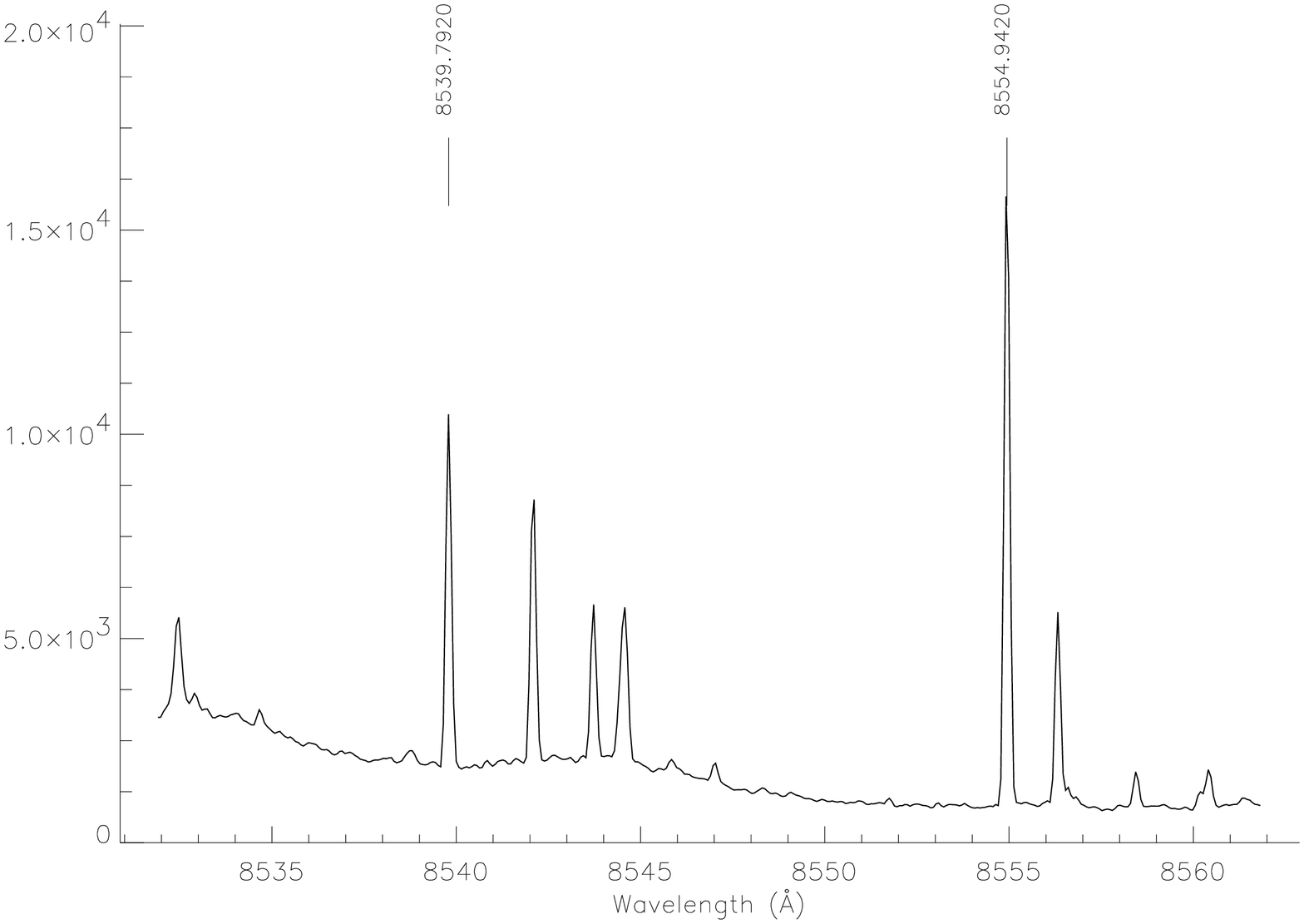}
\end{figure}
\clearpage
   
\begin{figure}
\centering
\includegraphics[width=10cm,angle=90]{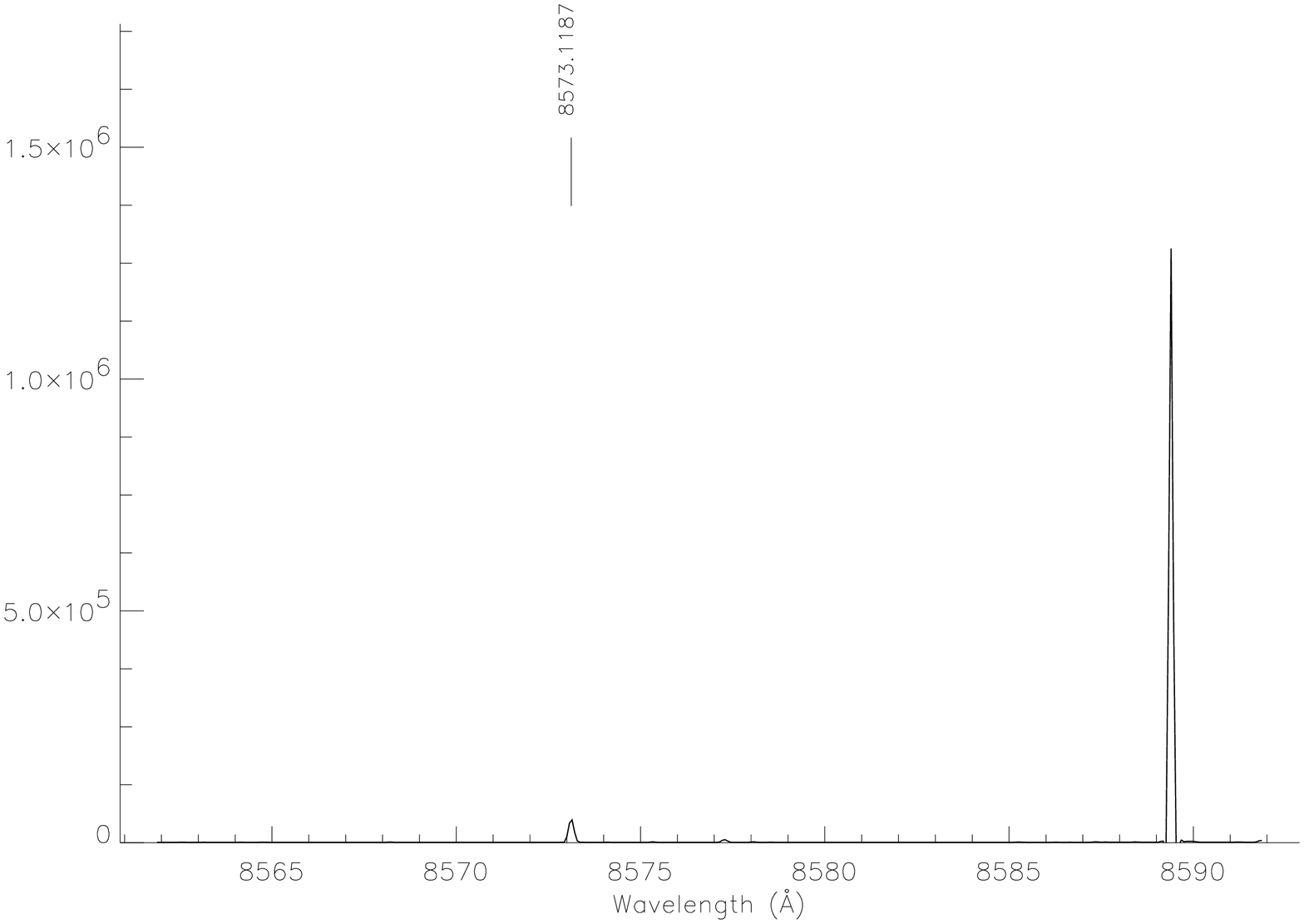}
\includegraphics[width=10cm,angle=90]{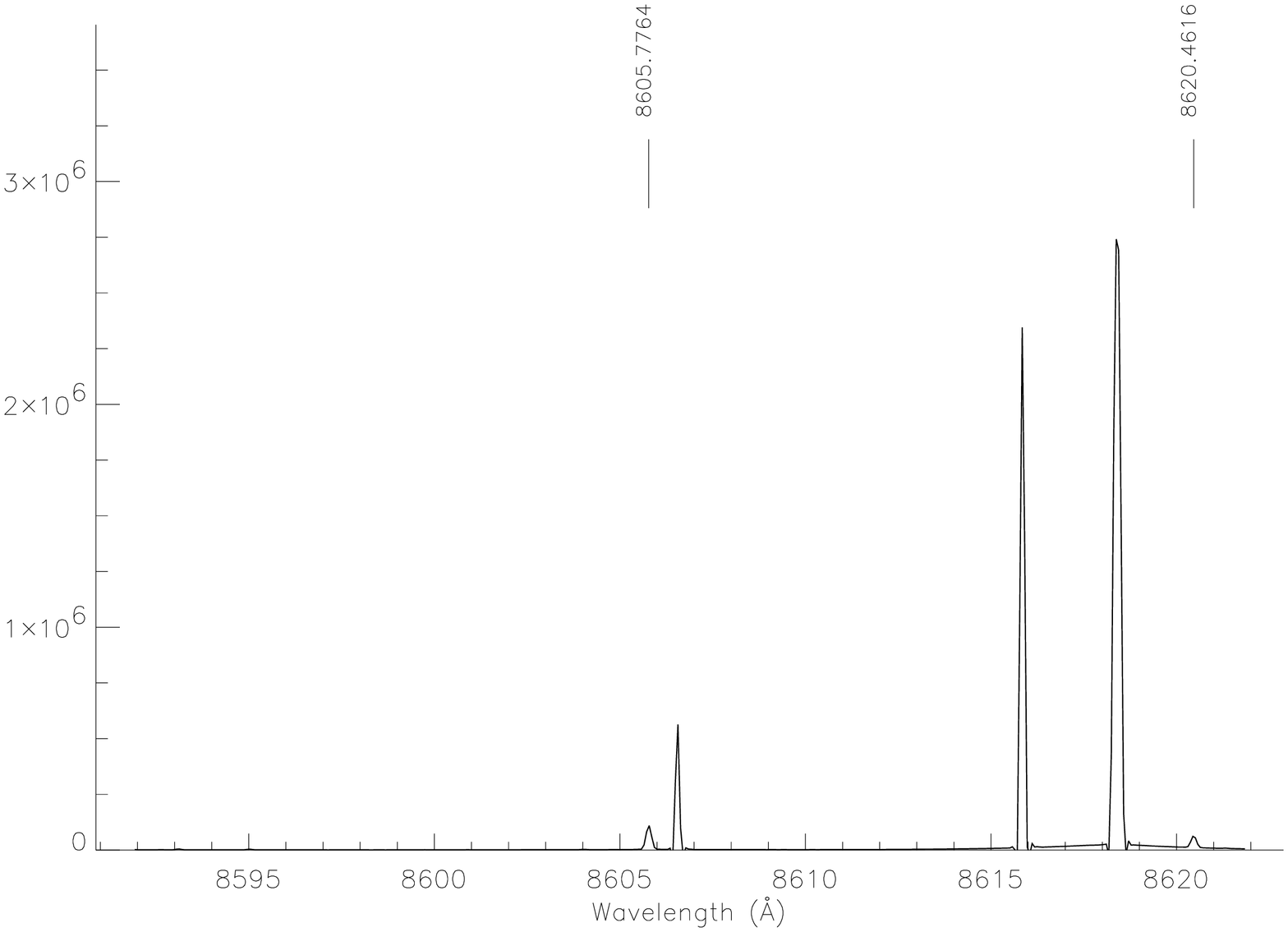}
\end{figure}
\clearpage
   
\begin{figure}
\centering
\includegraphics[width=10cm,angle=90]{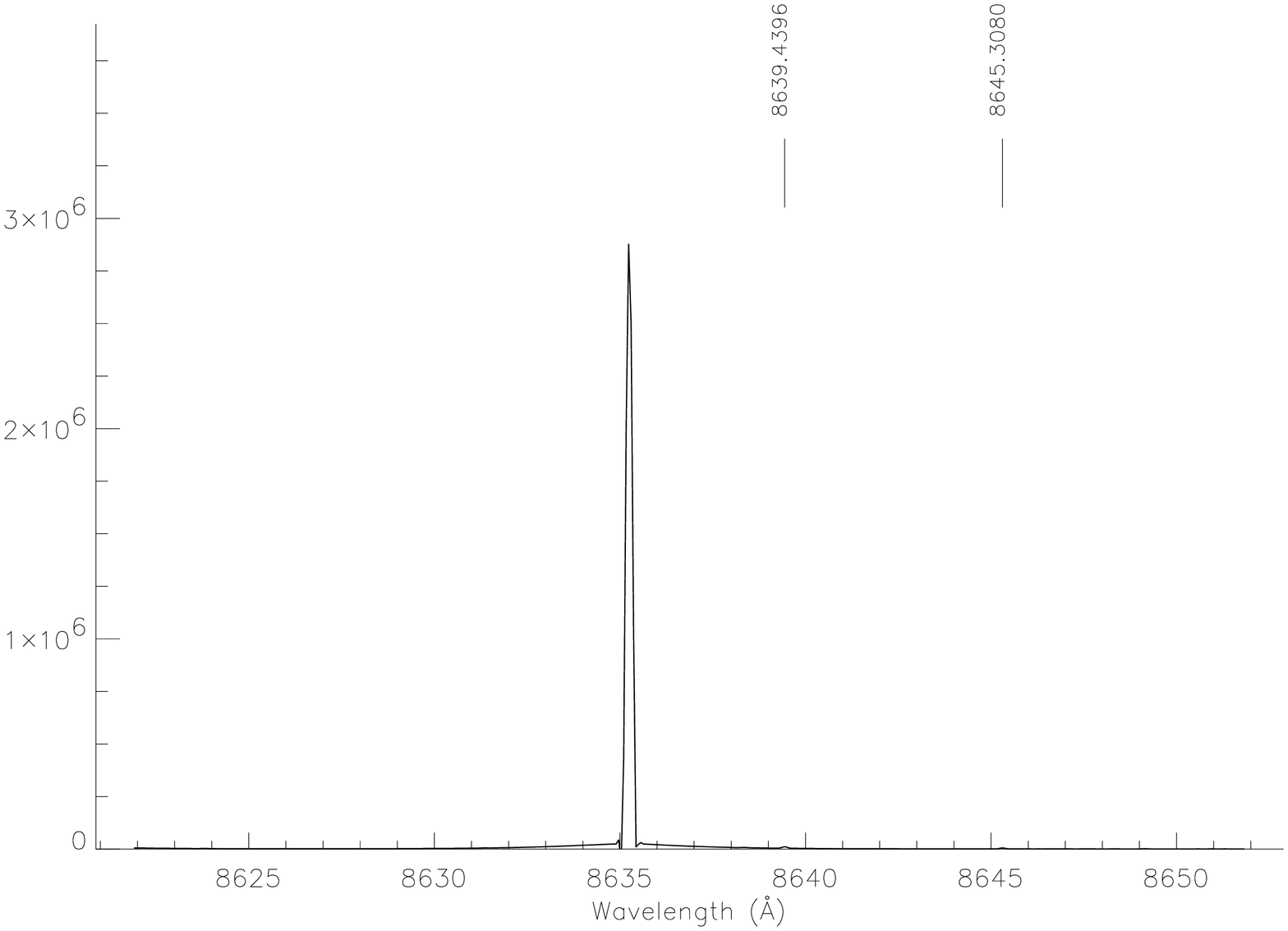}
\includegraphics[width=10cm,angle=90]{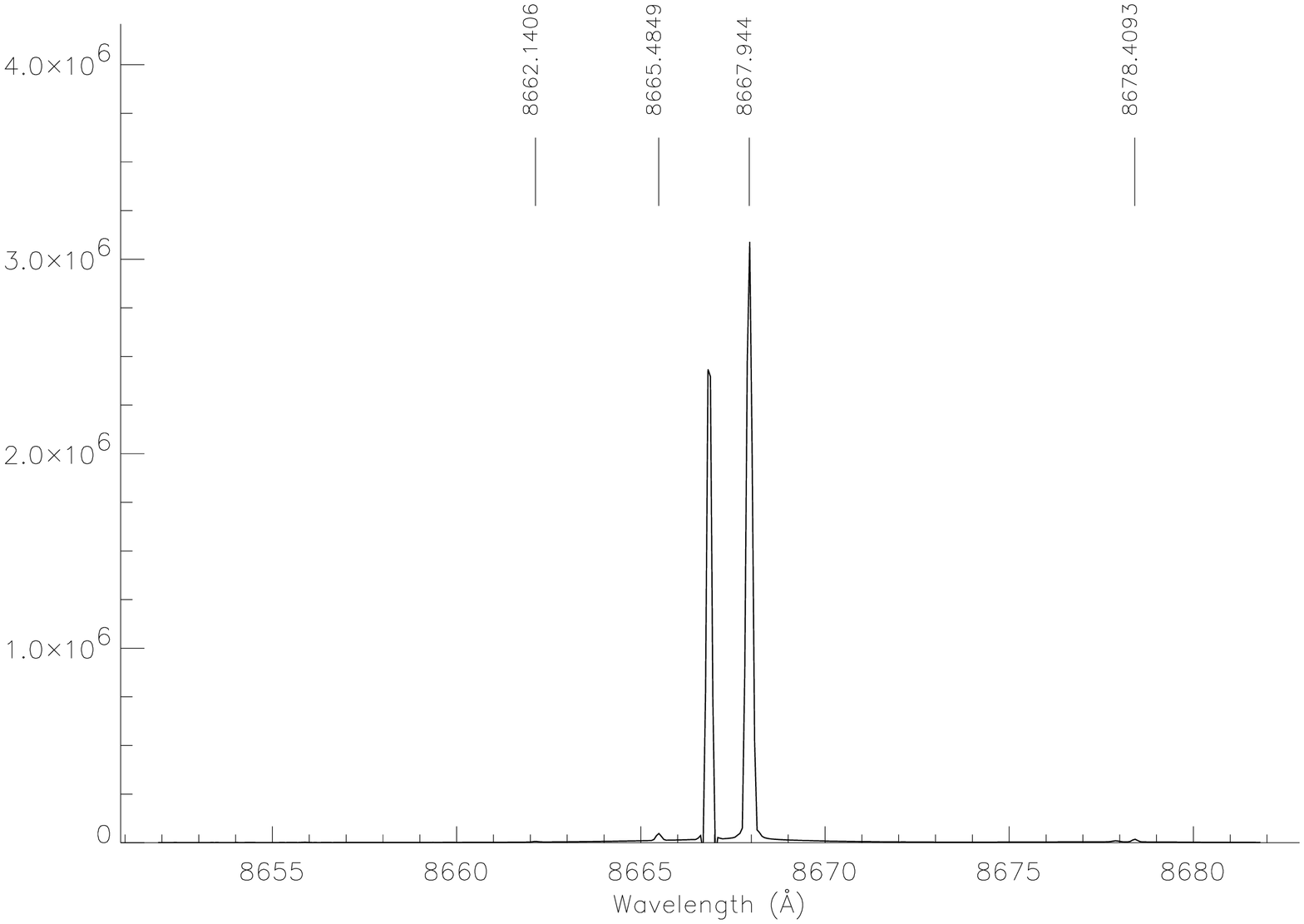}
\end{figure}
\clearpage
   
\begin{figure}
\centering
\includegraphics[width=10cm,angle=90]{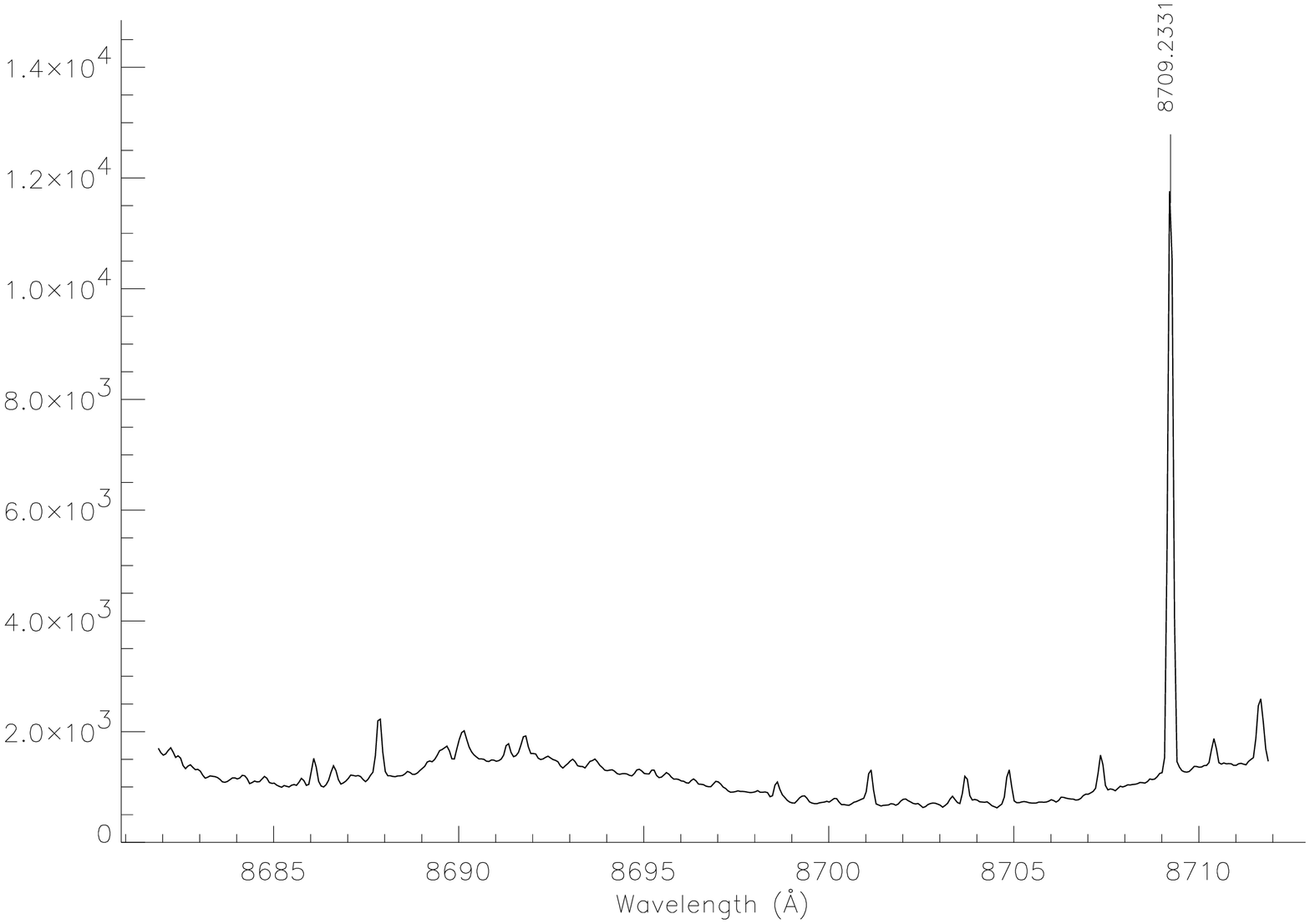}
\includegraphics[width=10cm,angle=90]{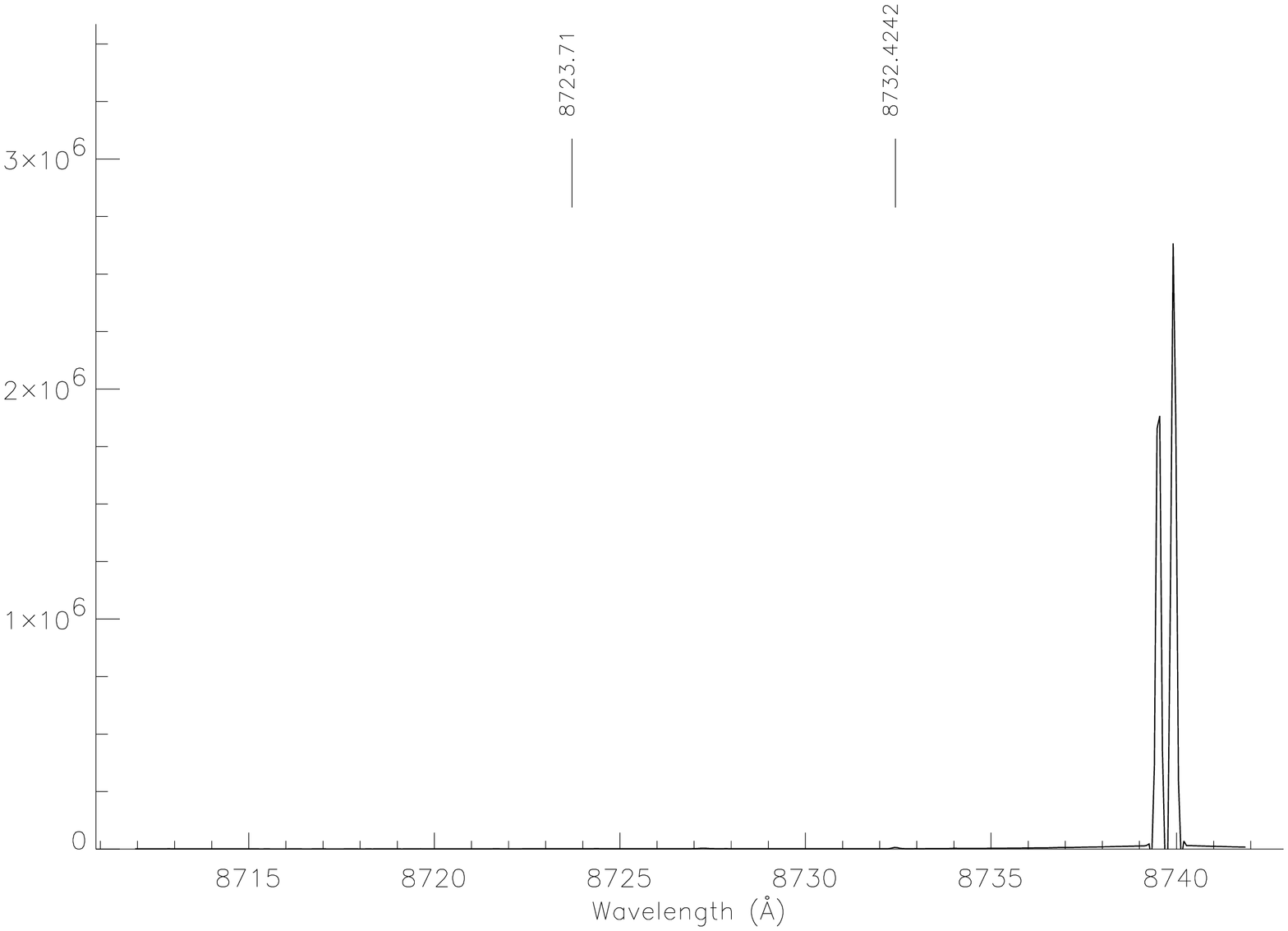}
\end{figure}
\clearpage
   
\begin{figure}
\centering
\includegraphics[width=10cm,angle=90]{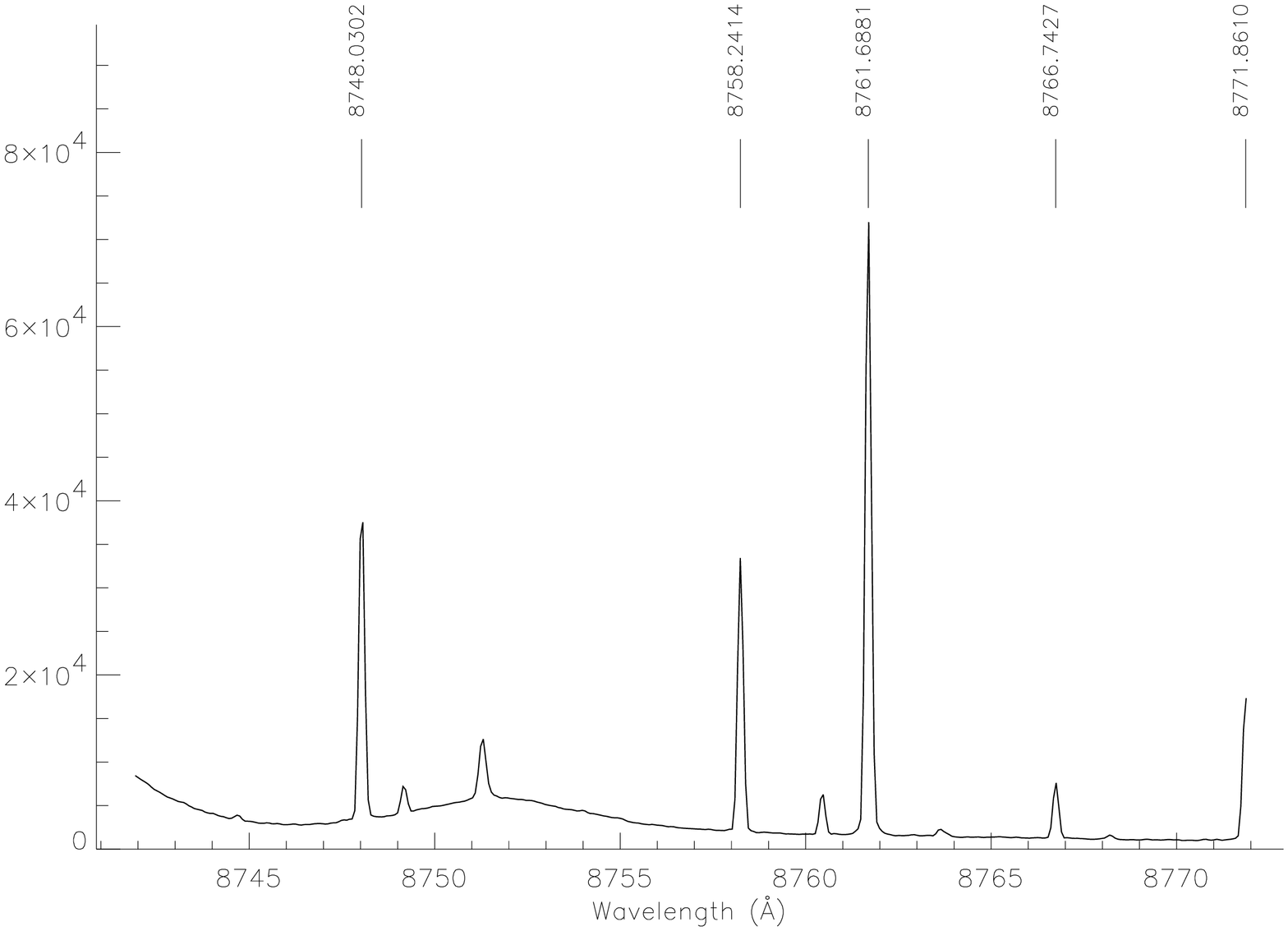}
\includegraphics[width=10cm,angle=90]{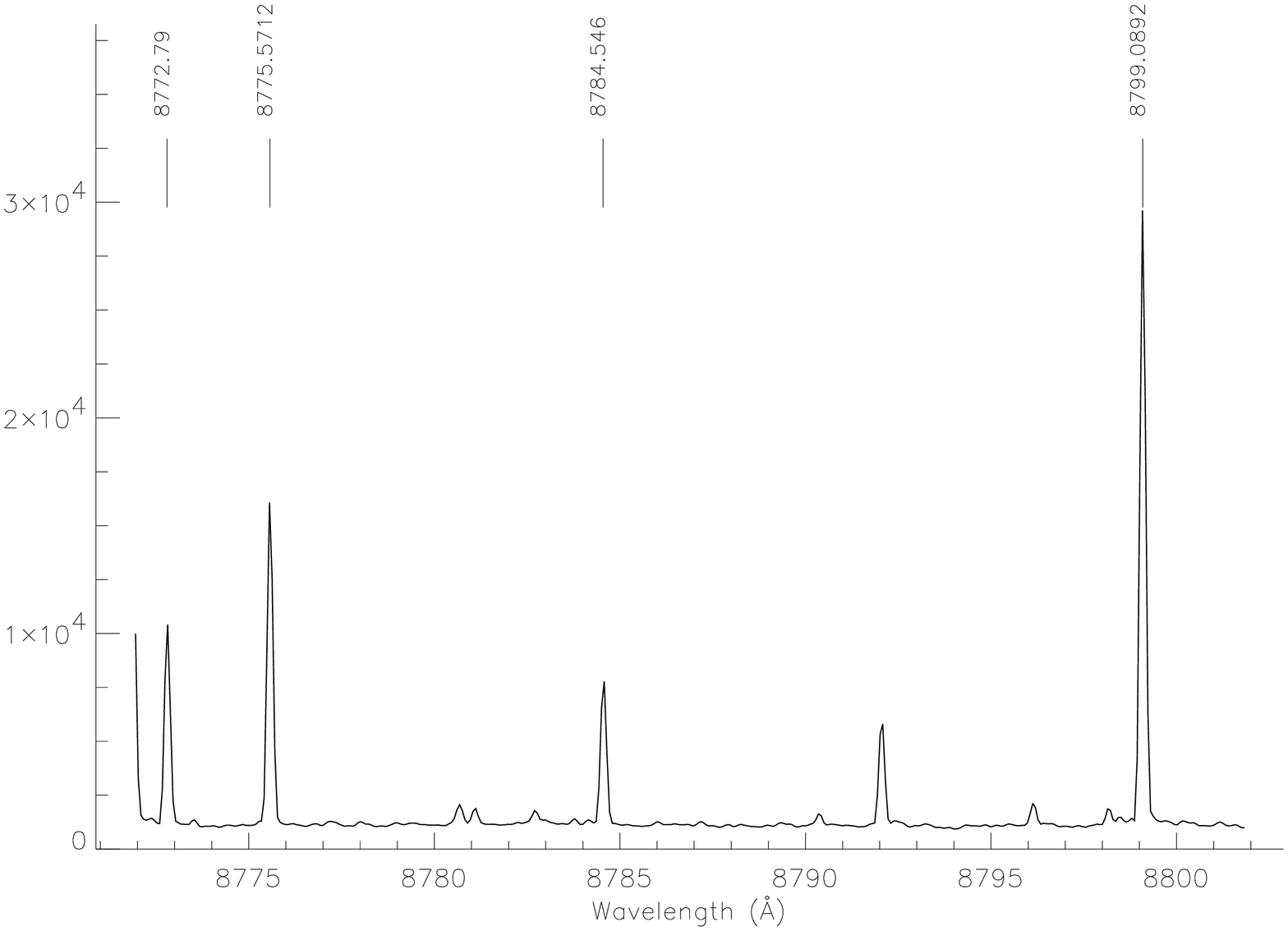}
\end{figure}
\clearpage
   
\begin{figure}
\centering
\includegraphics[width=10cm,angle=90]{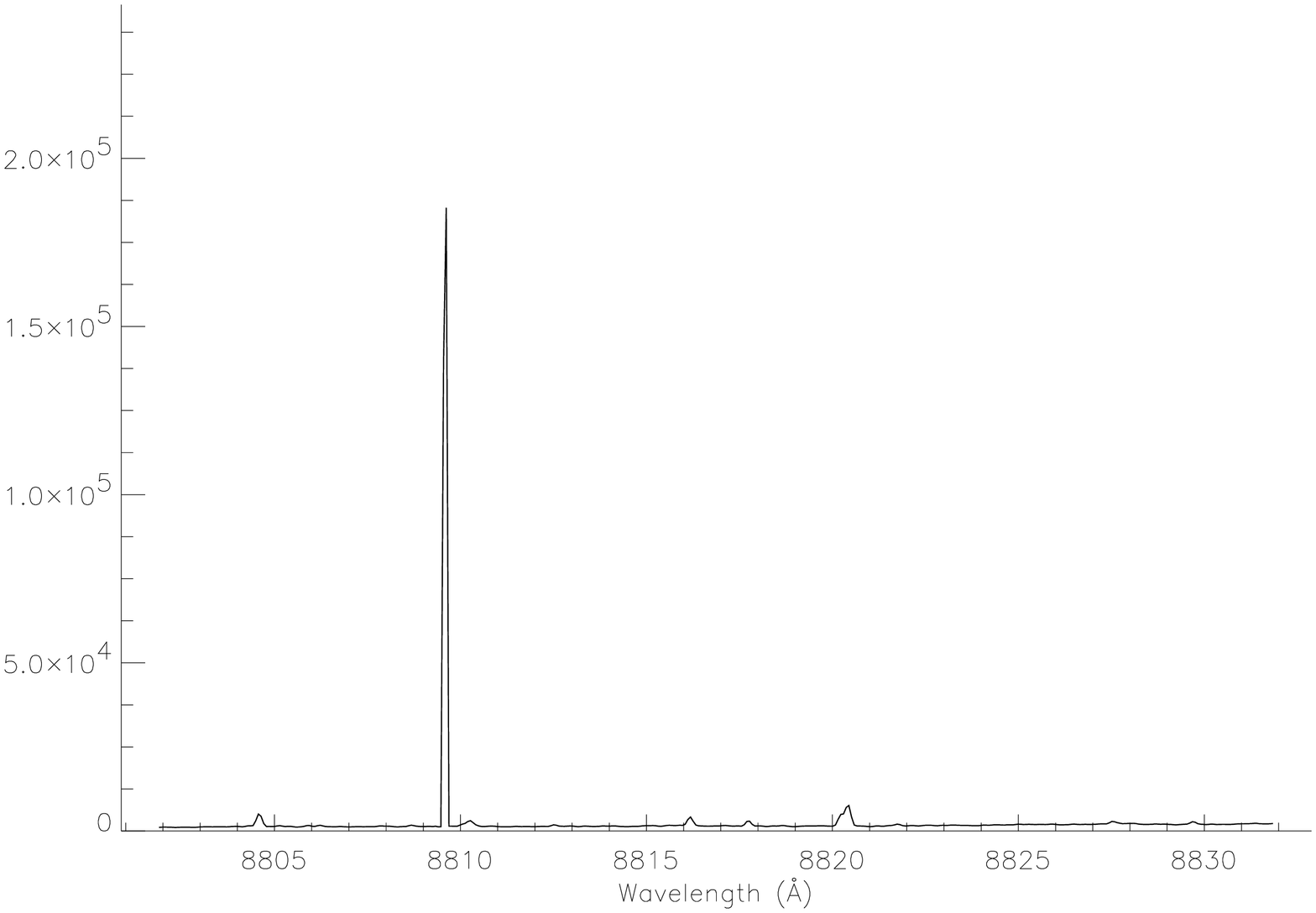}
\includegraphics[width=10cm,angle=90]{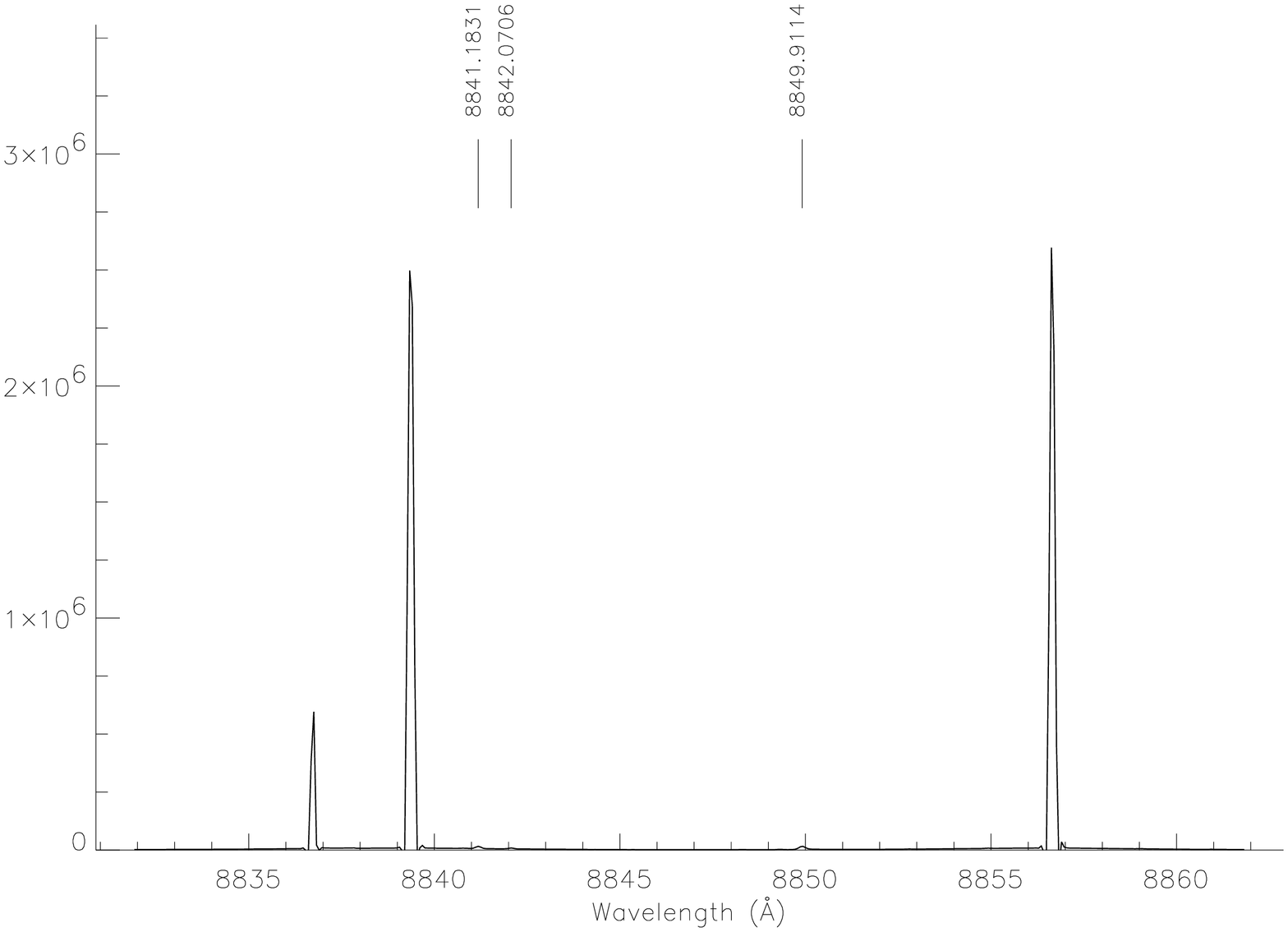}
\end{figure}
\clearpage
   
\begin{figure}
\centering
\includegraphics[width=10cm,angle=90]{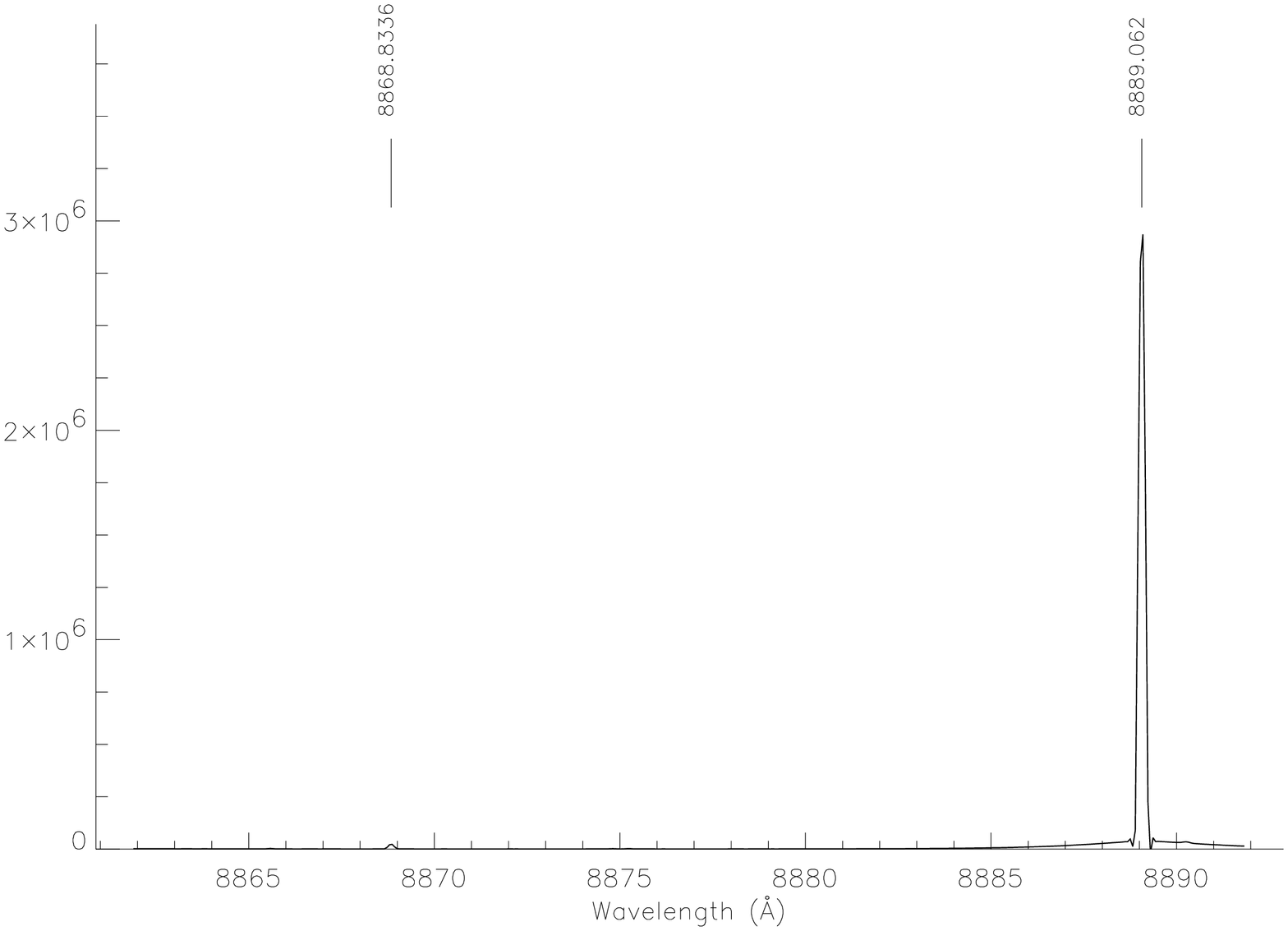}
\includegraphics[width=10cm,angle=90]{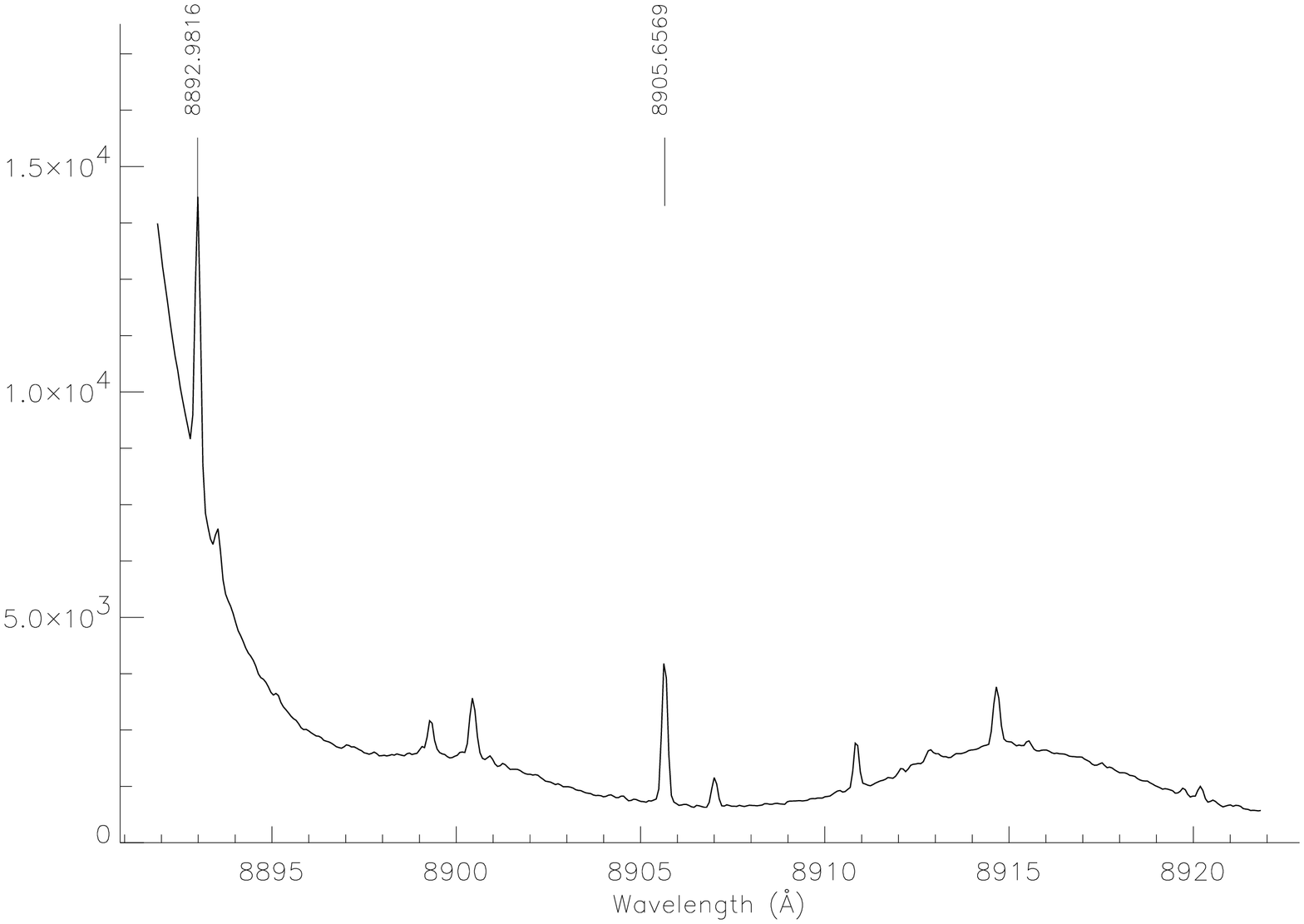}
\end{figure}
\clearpage
   
\begin{figure}
\centering
\includegraphics[width=10cm,angle=90]{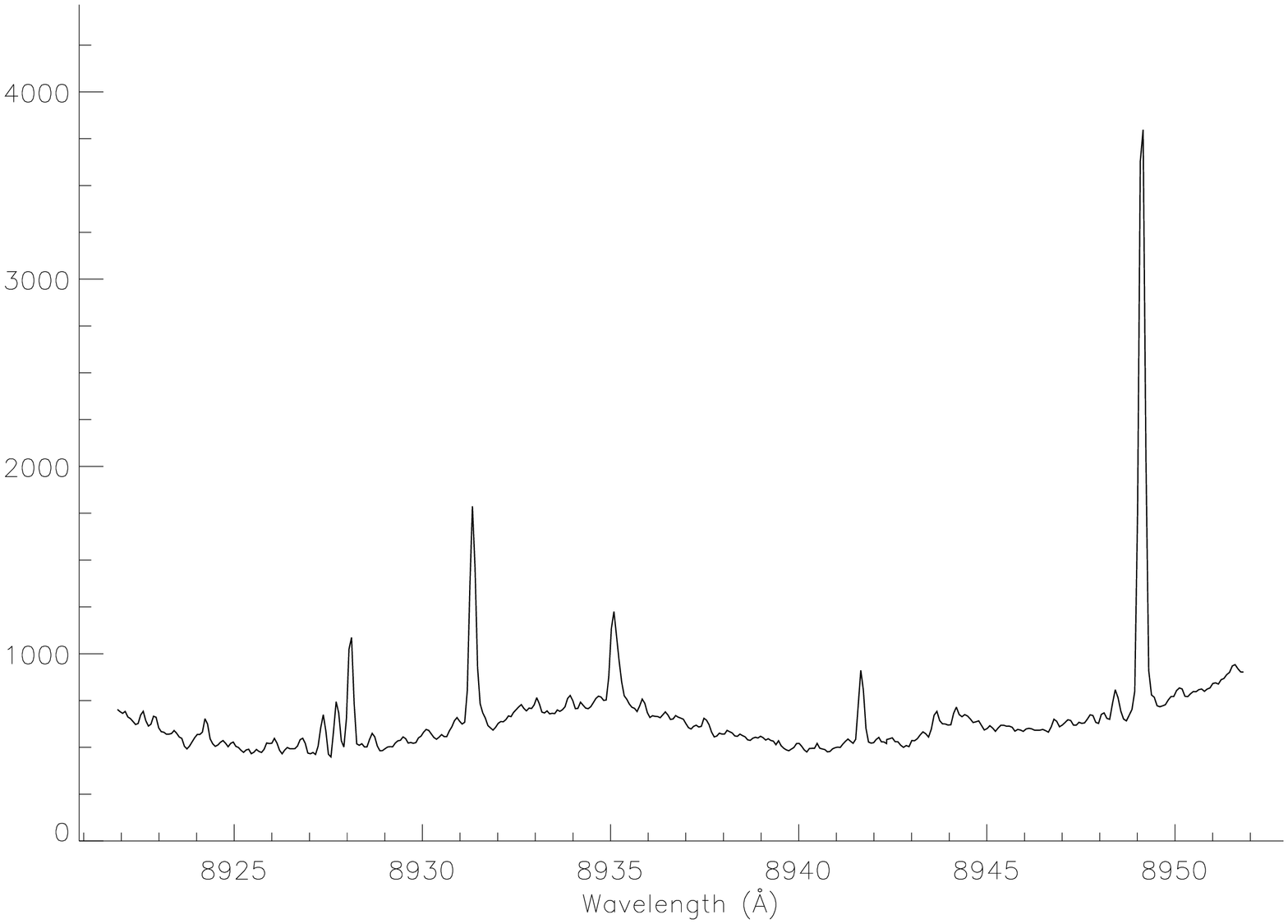}
\includegraphics[width=10cm,angle=90]{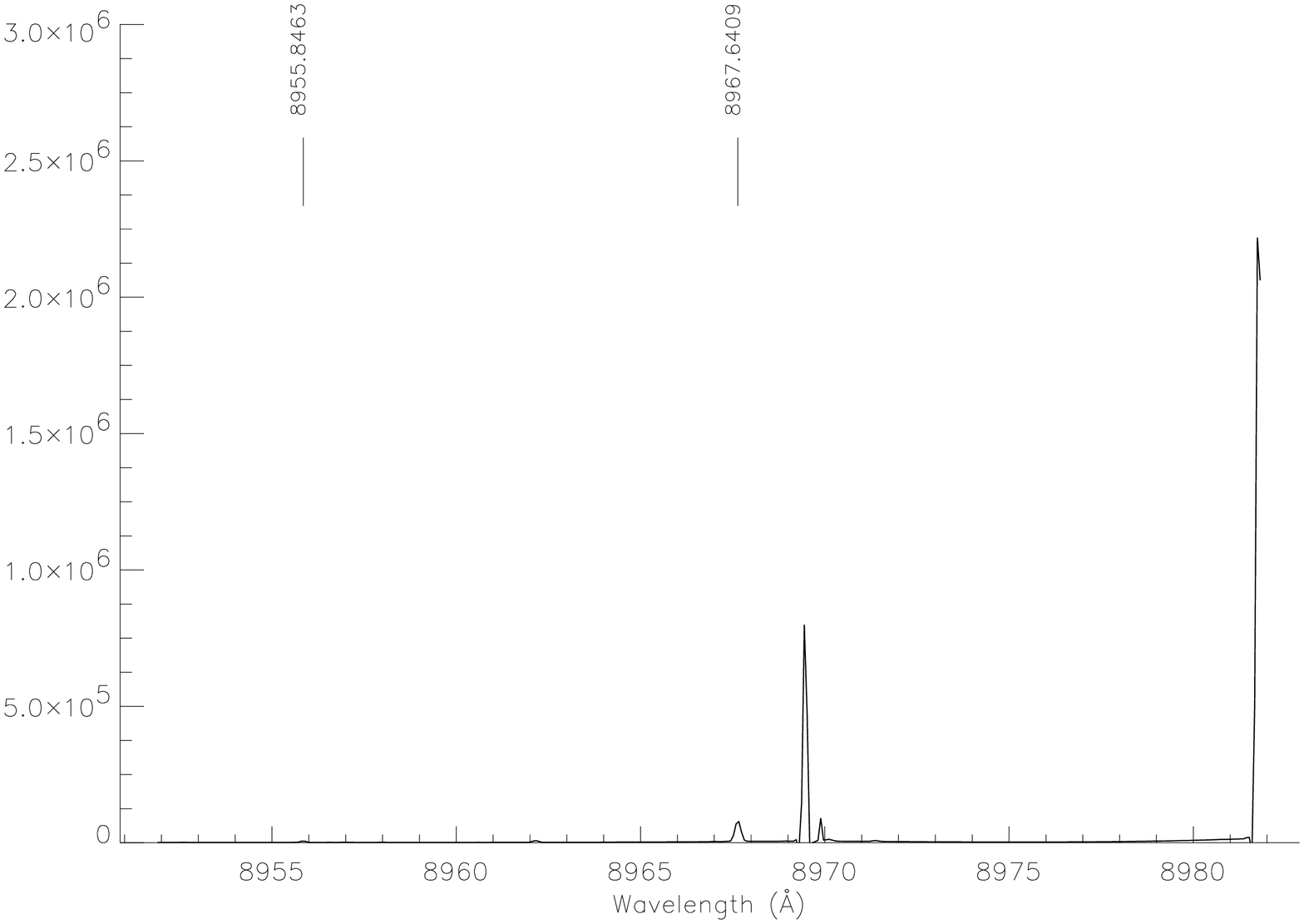}
\end{figure}
\clearpage
   
\begin{figure}
\centering
\includegraphics[width=10cm,angle=90]{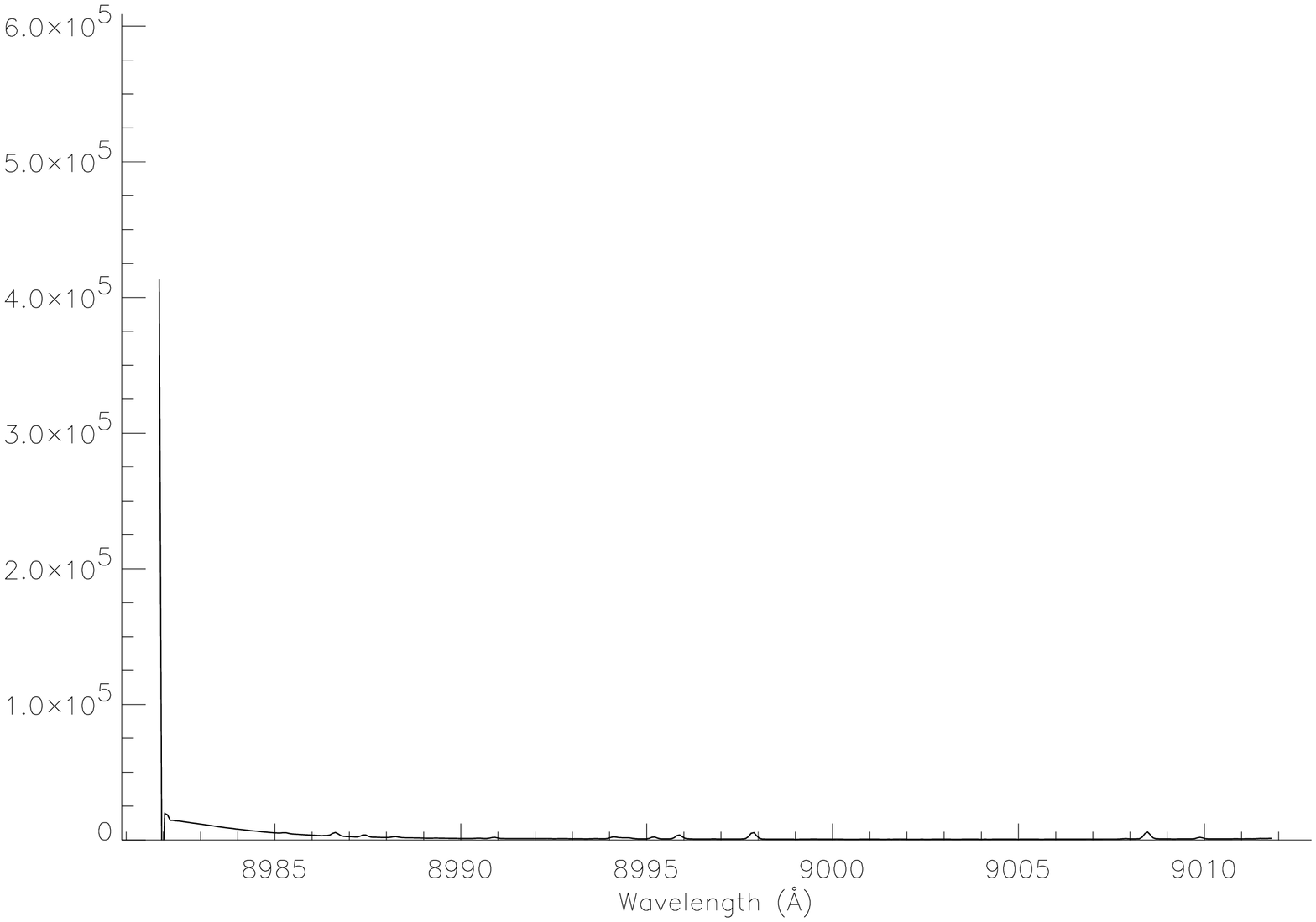}
\includegraphics[width=10cm,angle=90]{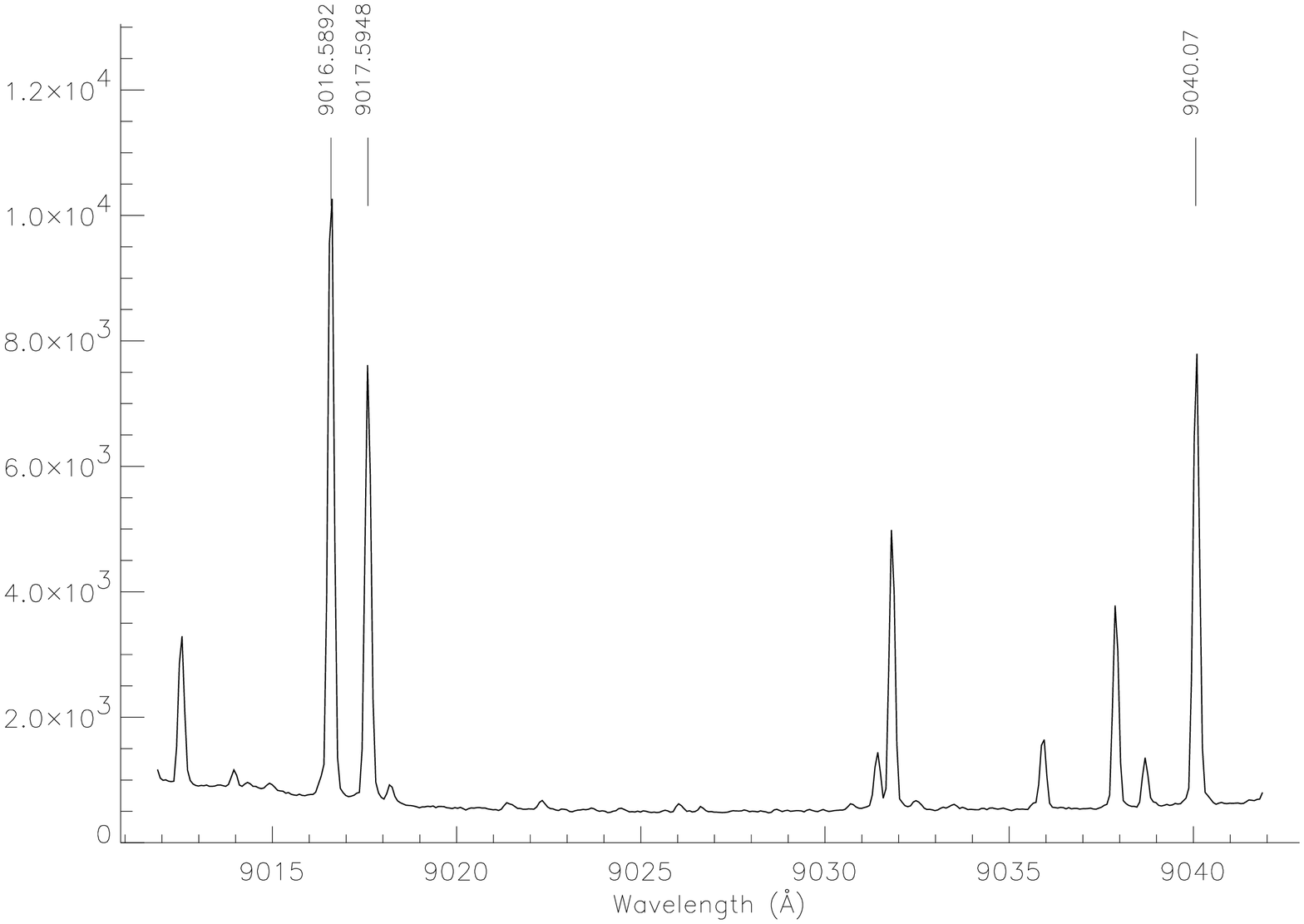}
\end{figure}
\clearpage
   
\begin{figure}
\centering
\includegraphics[width=10cm,angle=90]{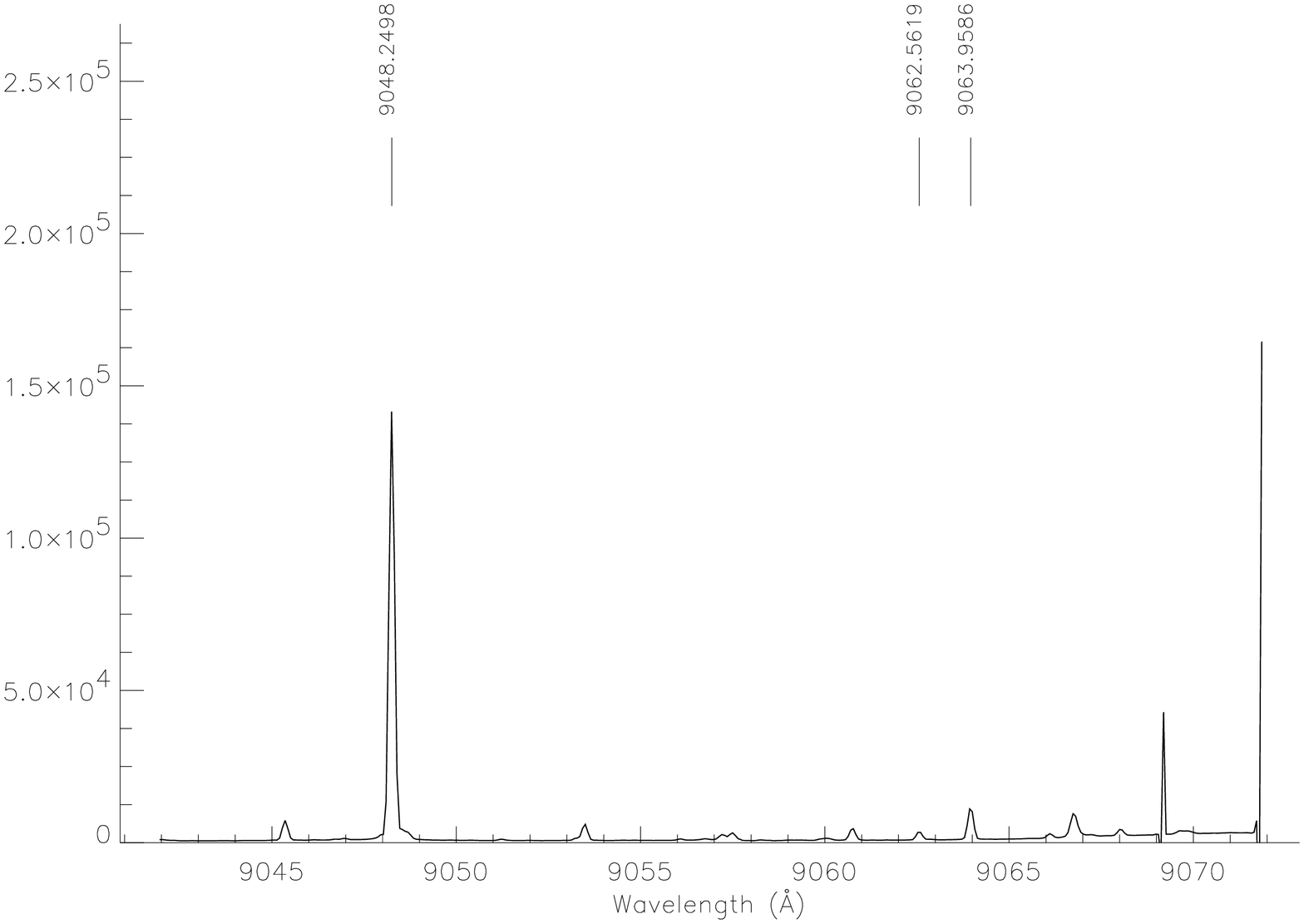}
\includegraphics[width=10cm,angle=90]{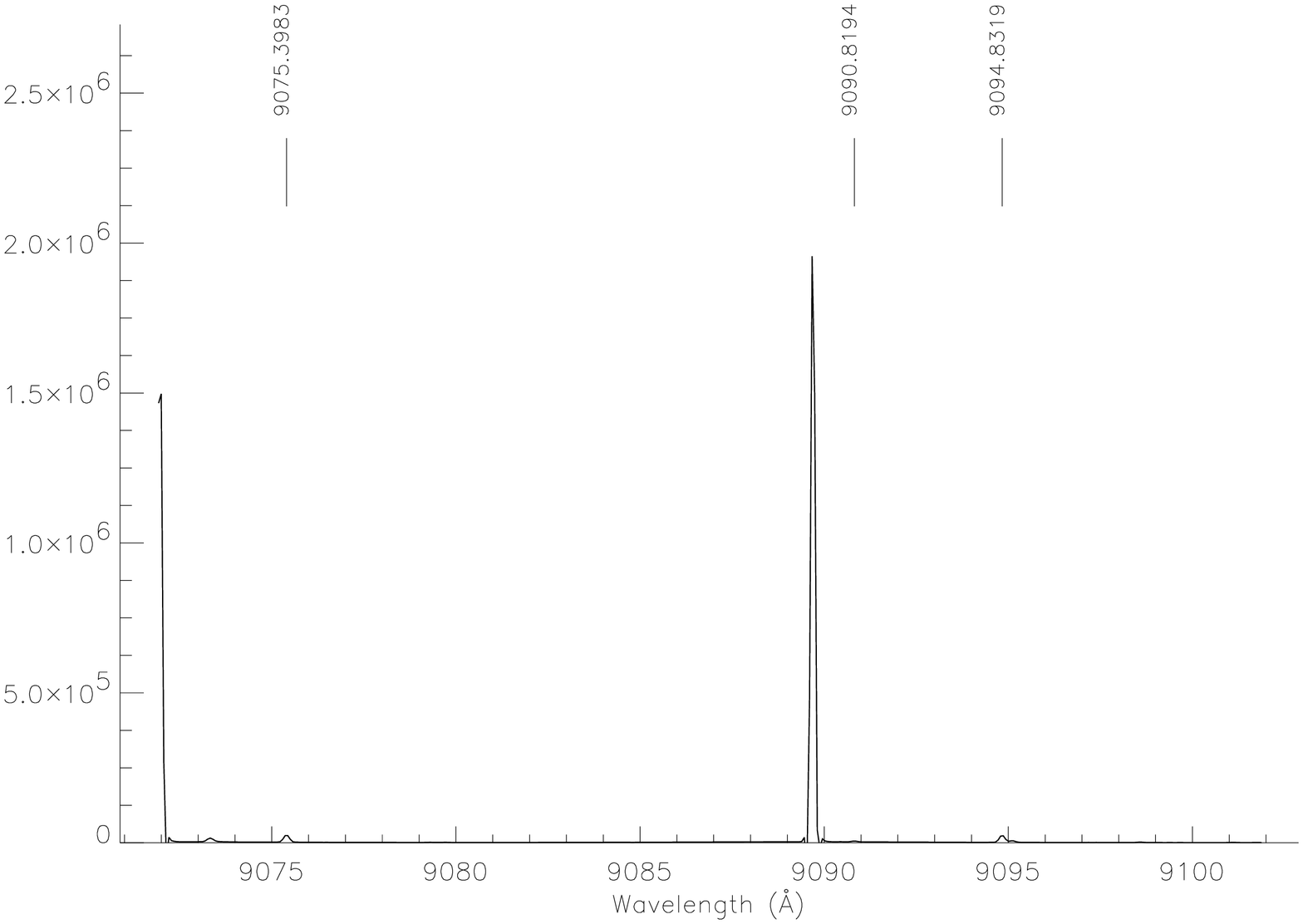}
\end{figure}
\clearpage
   
\begin{figure}
\centering
\includegraphics[width=10cm,angle=90]{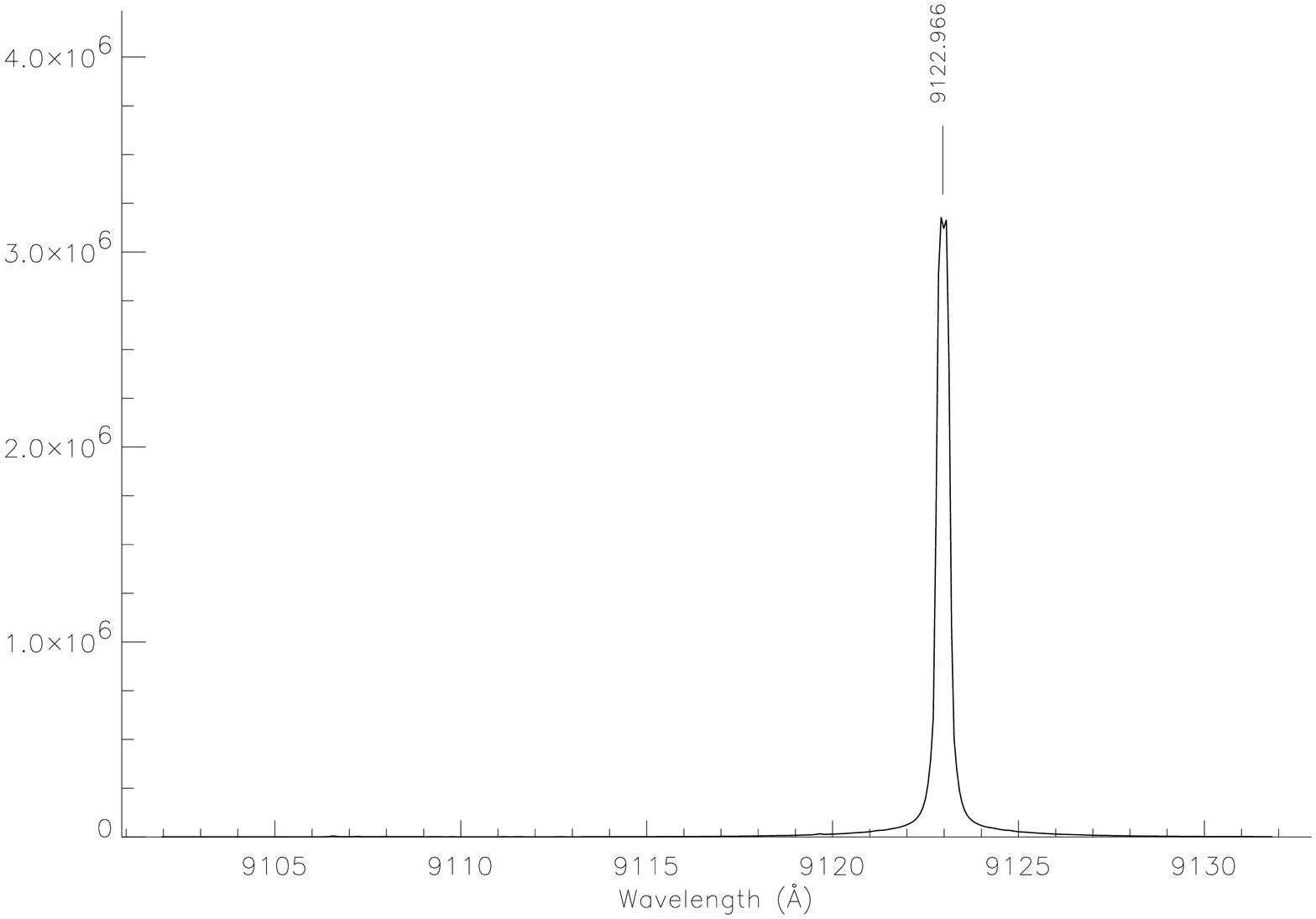}
\includegraphics[width=10cm,angle=90]{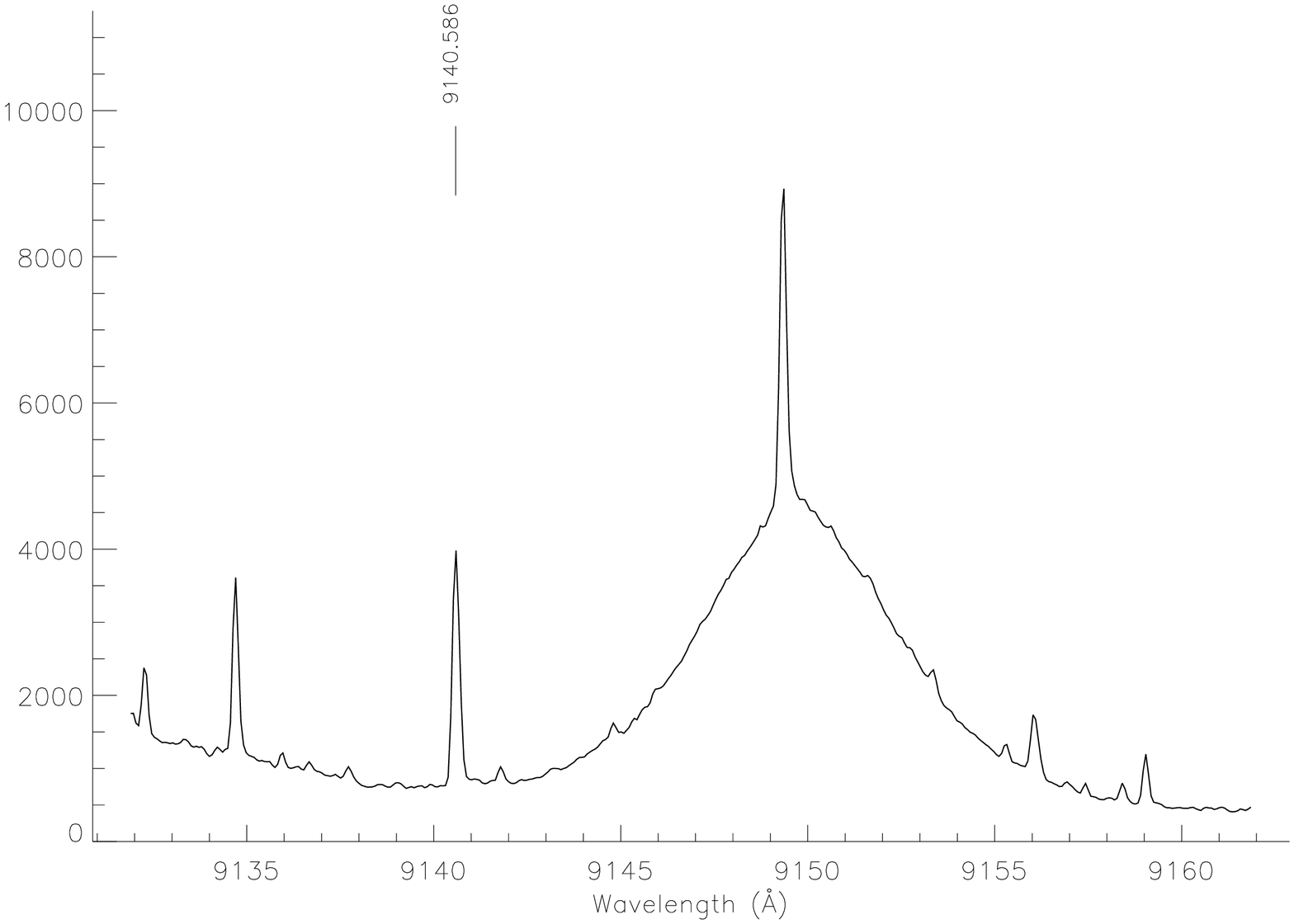}
\end{figure}
\clearpage
   
\begin{figure}
\centering
\includegraphics[width=10cm,angle=90]{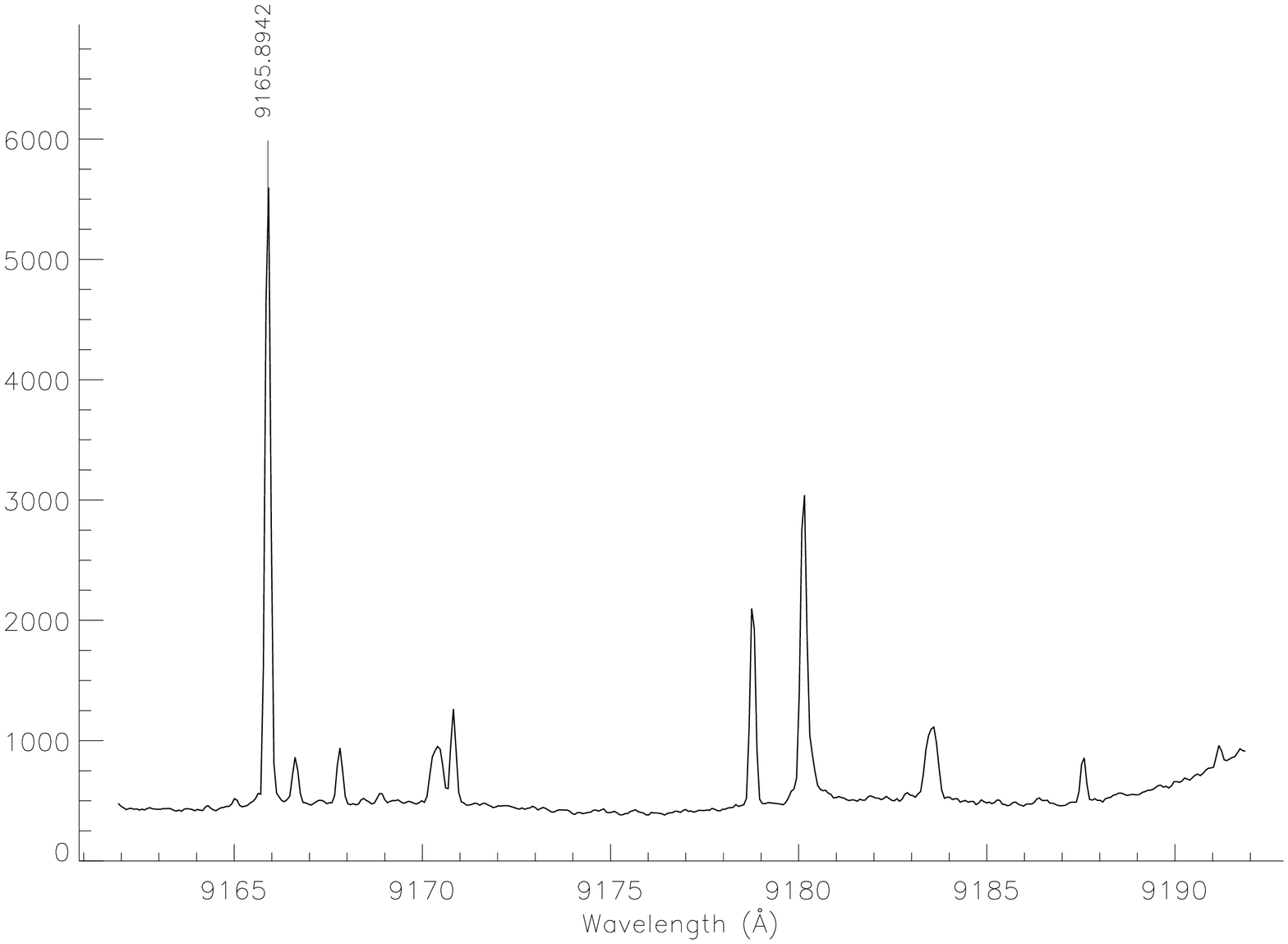}
\includegraphics[width=10cm,angle=90]{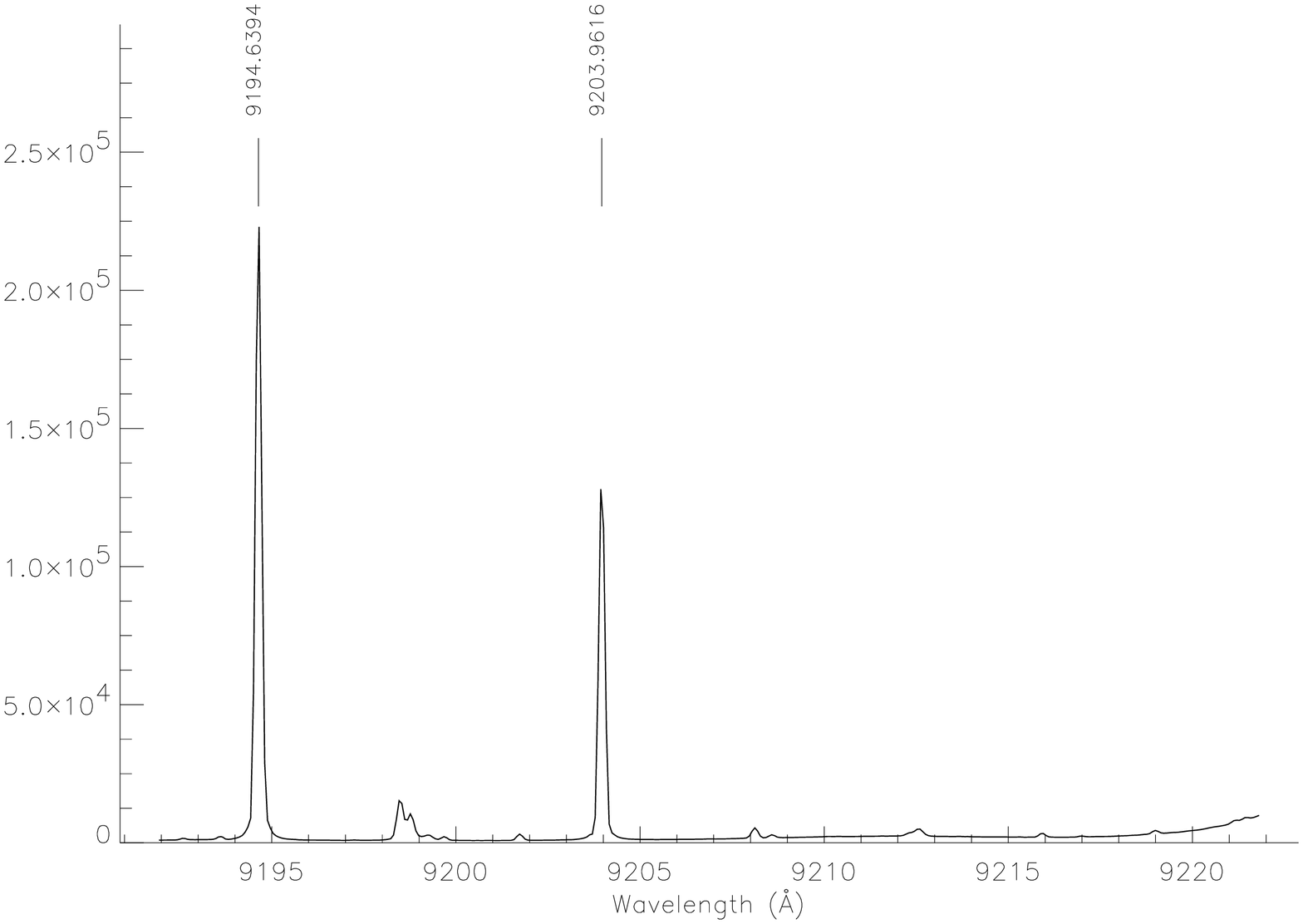}
\end{figure}
\clearpage
   
\begin{figure}
\centering
\includegraphics[width=10cm,angle=90]{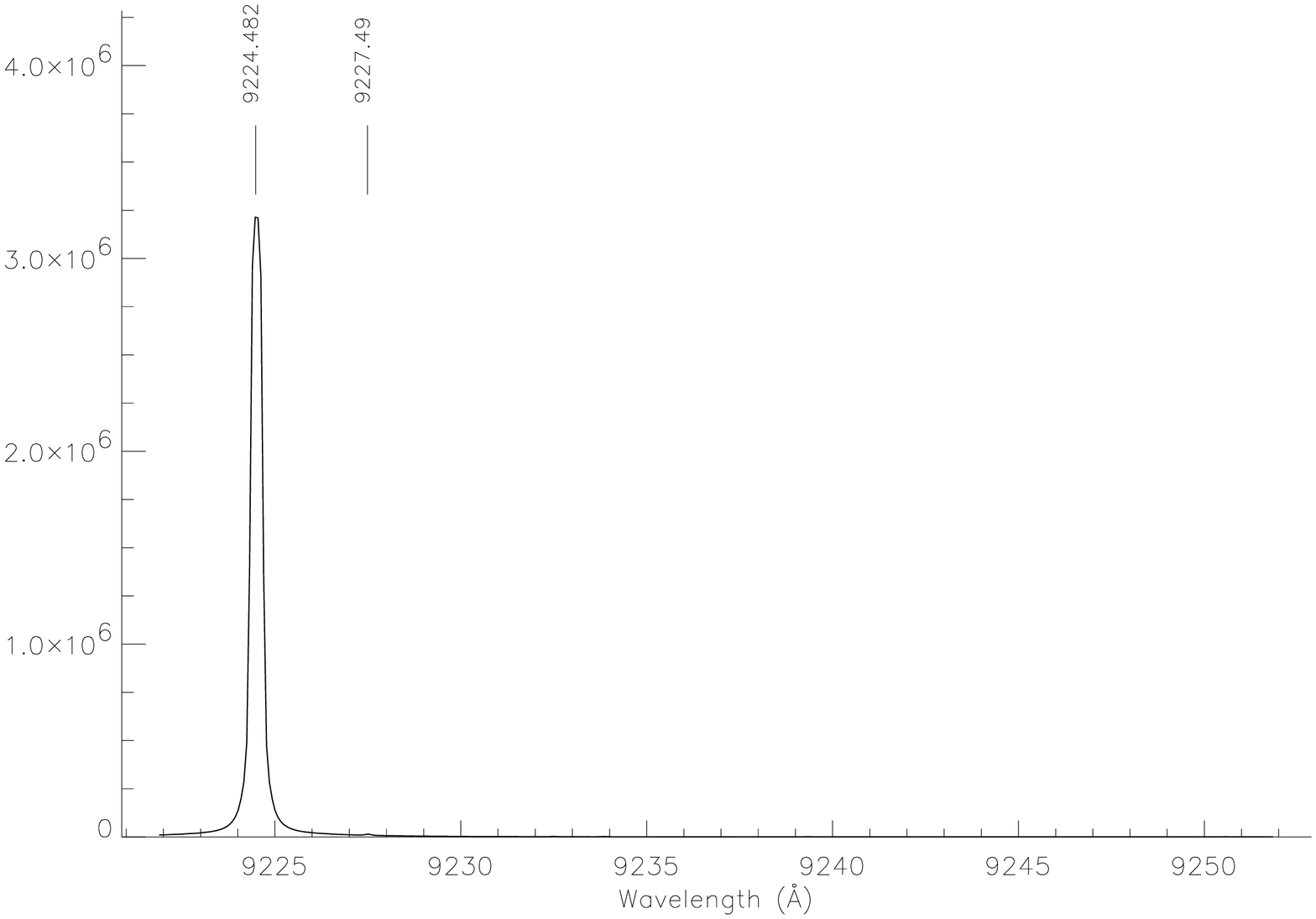}
\includegraphics[width=10cm,angle=90]{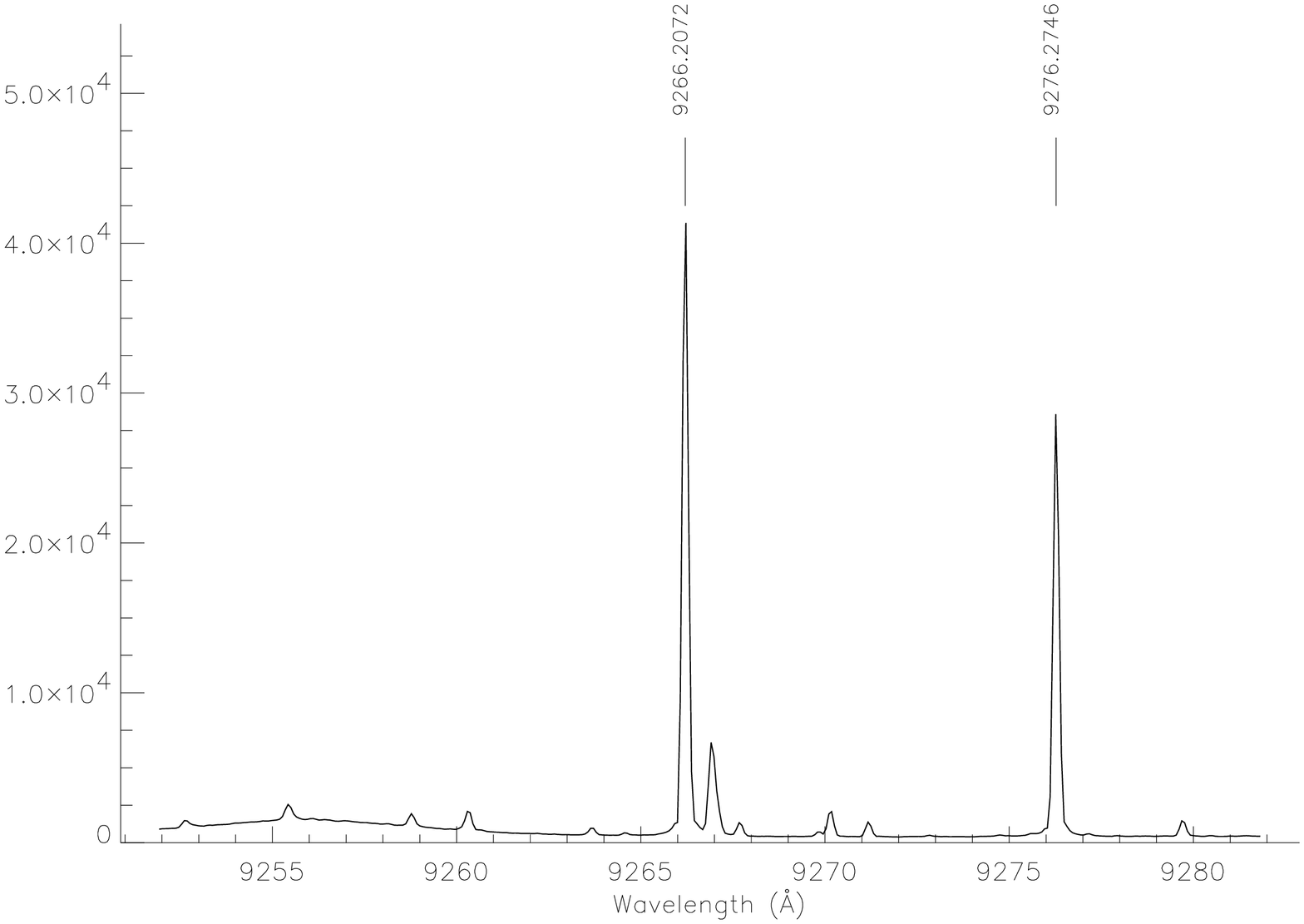}
\end{figure}
\clearpage
   
\begin{figure}
\centering
\includegraphics[width=10cm,angle=90]{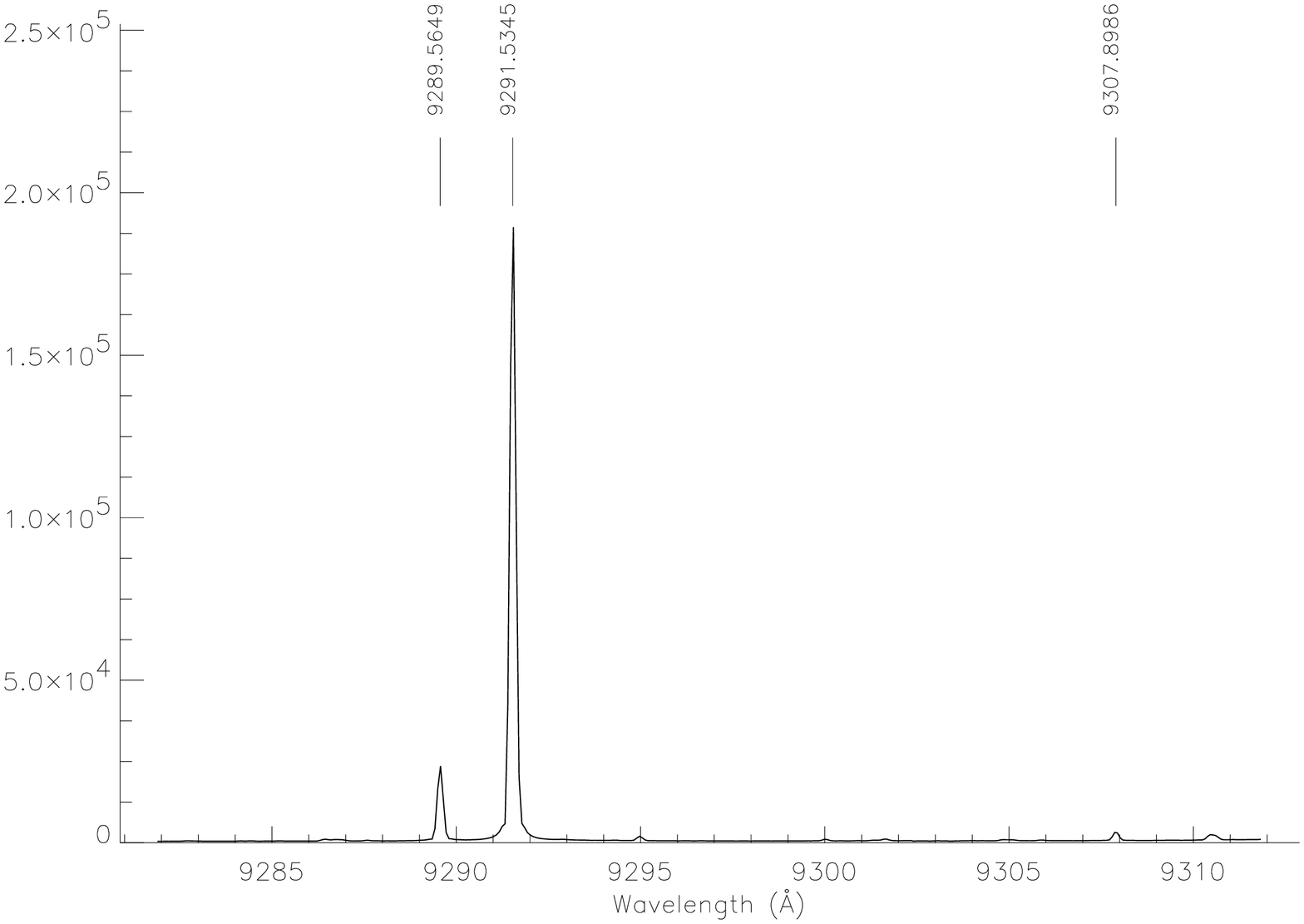}
\includegraphics[width=10cm,angle=90]{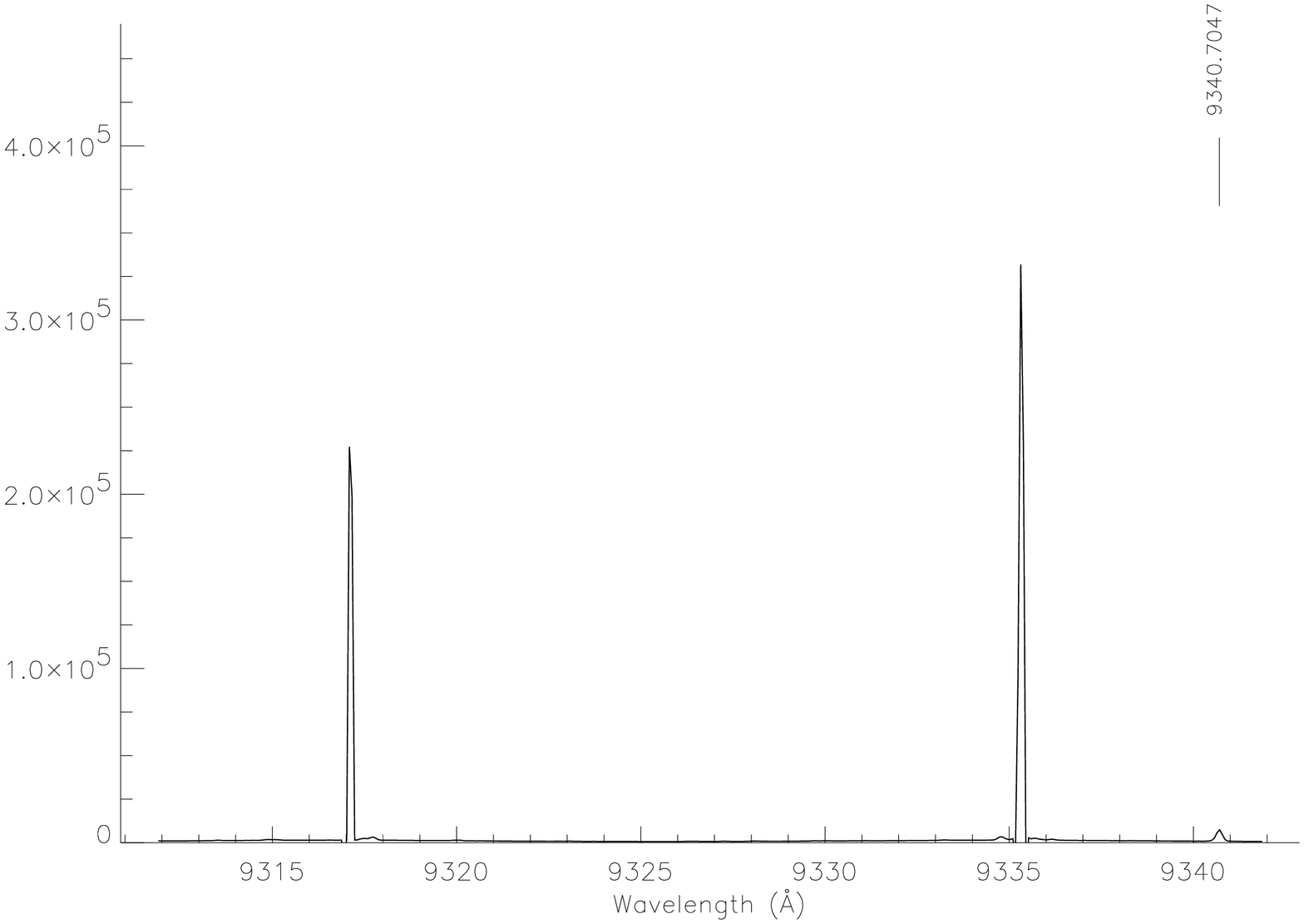}
\end{figure}
\clearpage
   
\begin{figure}
\centering
\includegraphics[width=10cm,angle=90]{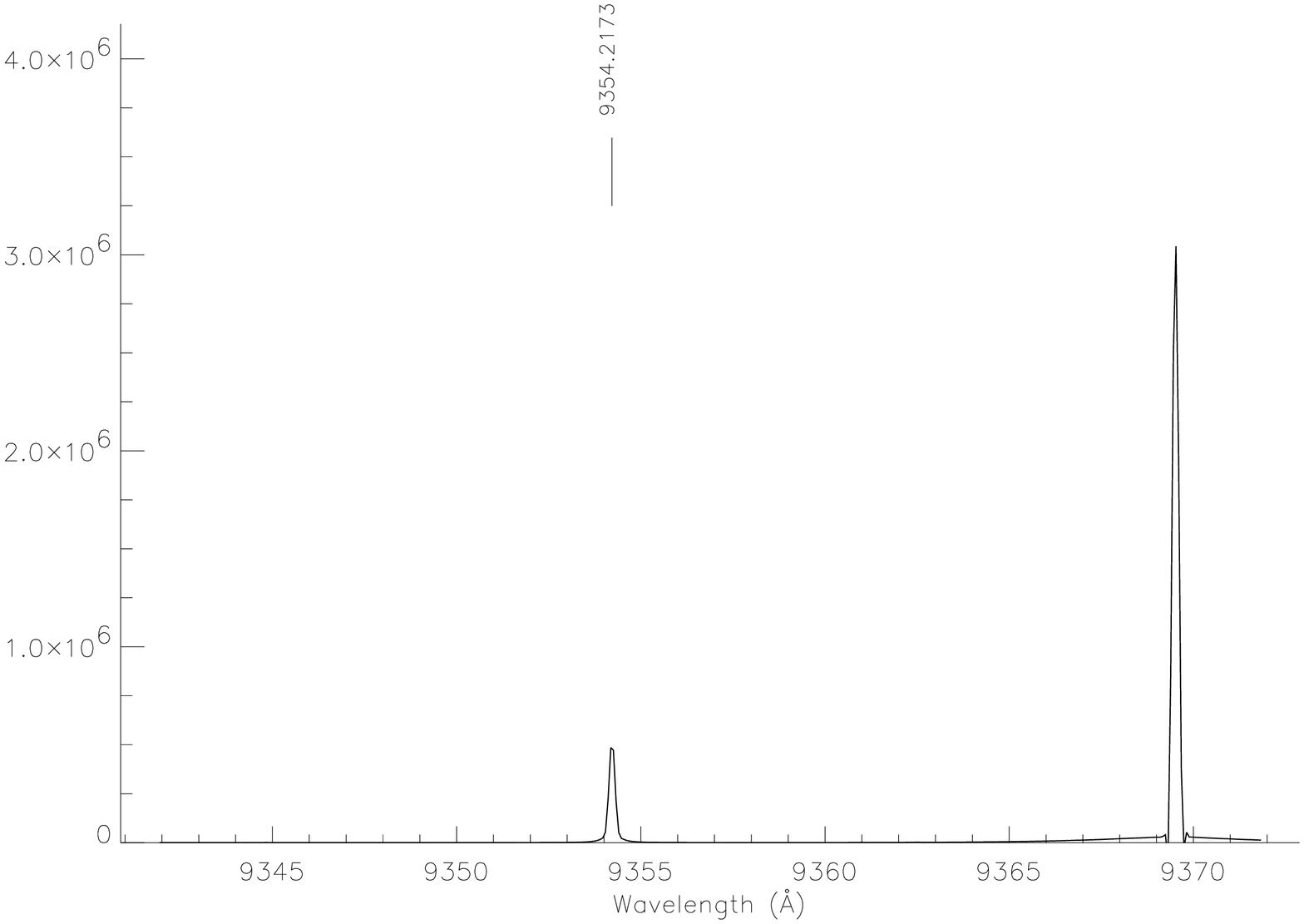}
\includegraphics[width=10cm,angle=90]{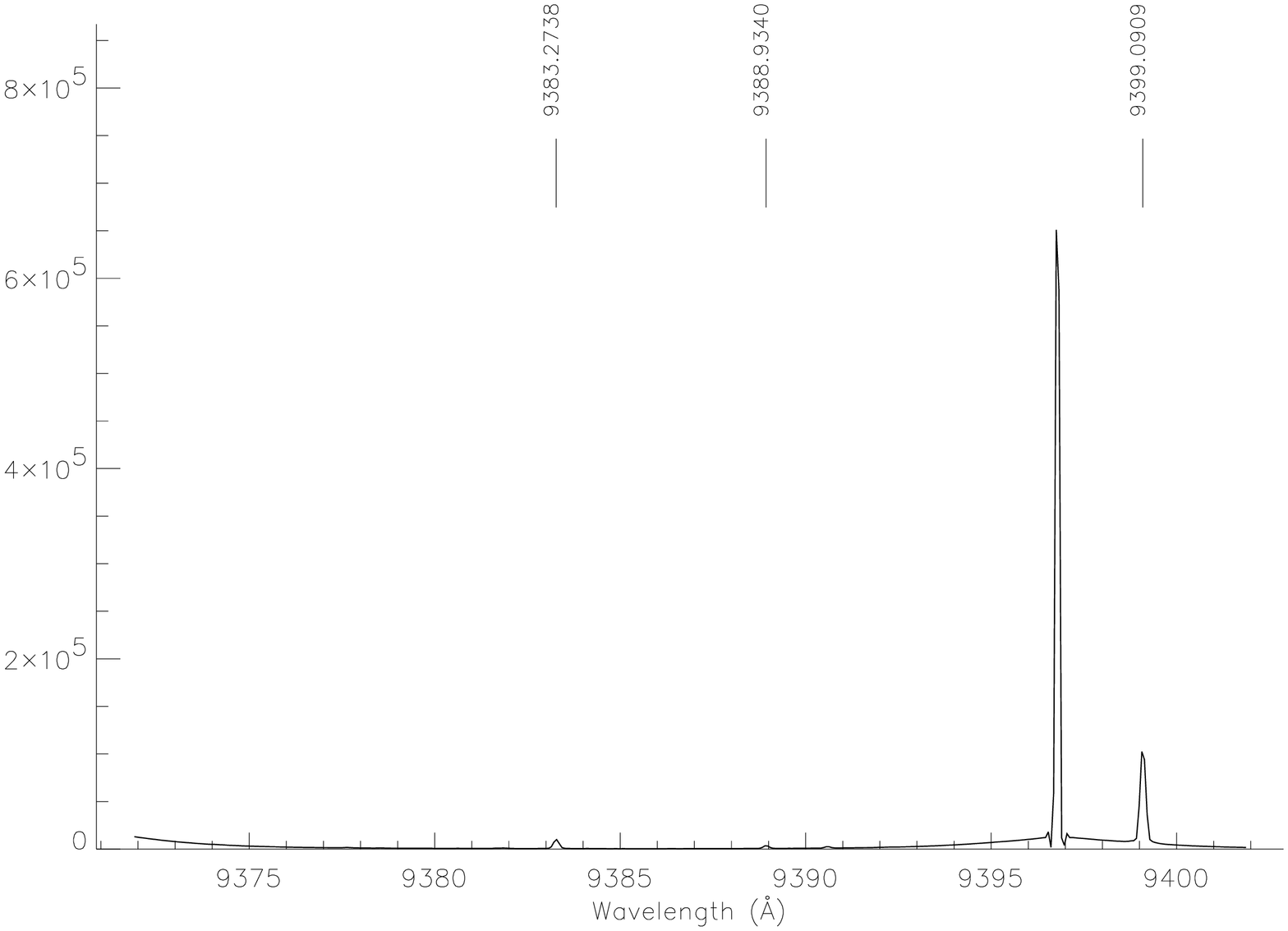}
\end{figure}
\clearpage
   
\begin{figure}
\centering
\includegraphics[width=10cm,angle=90]{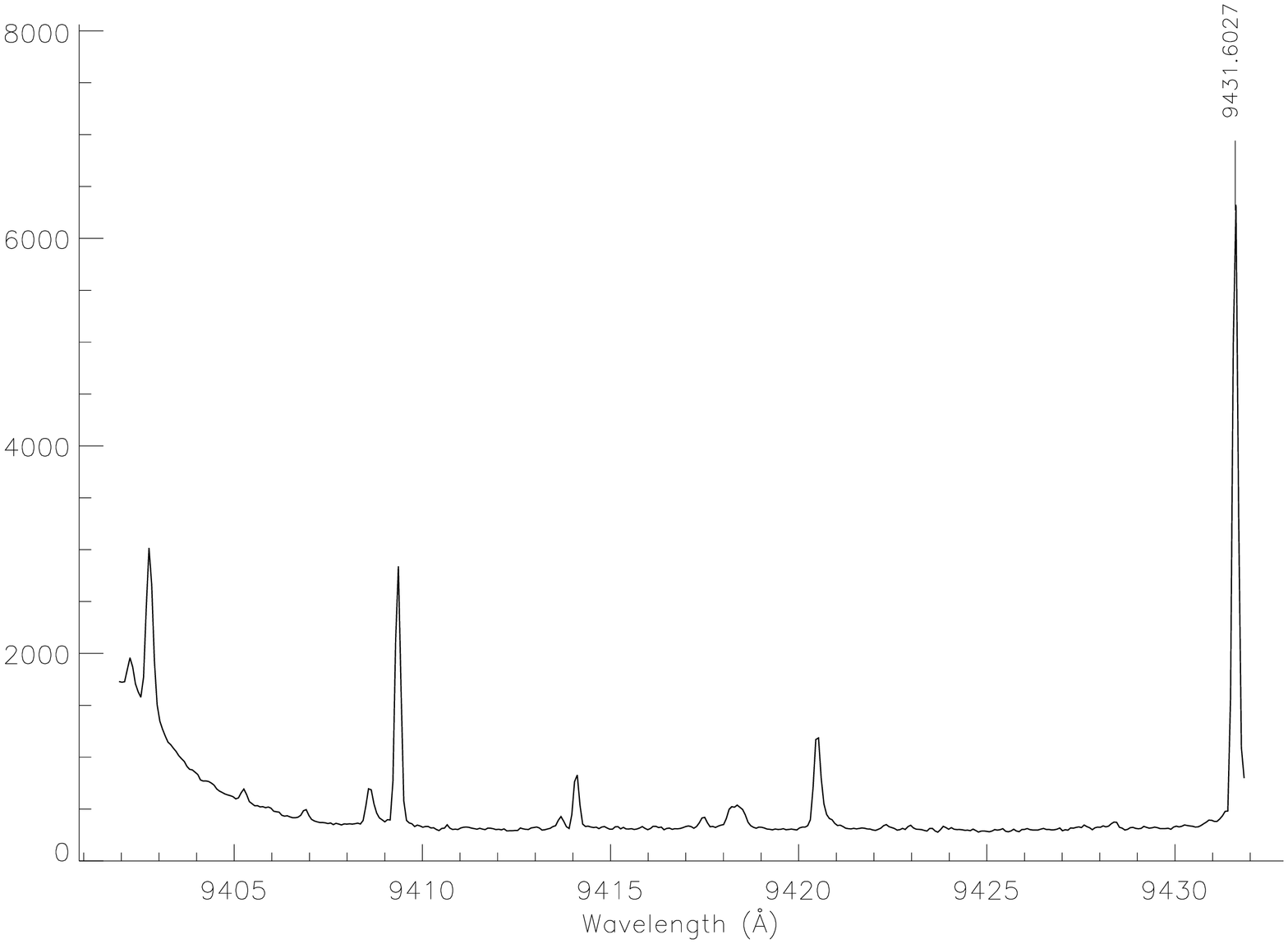}
\includegraphics[width=10cm,angle=90]{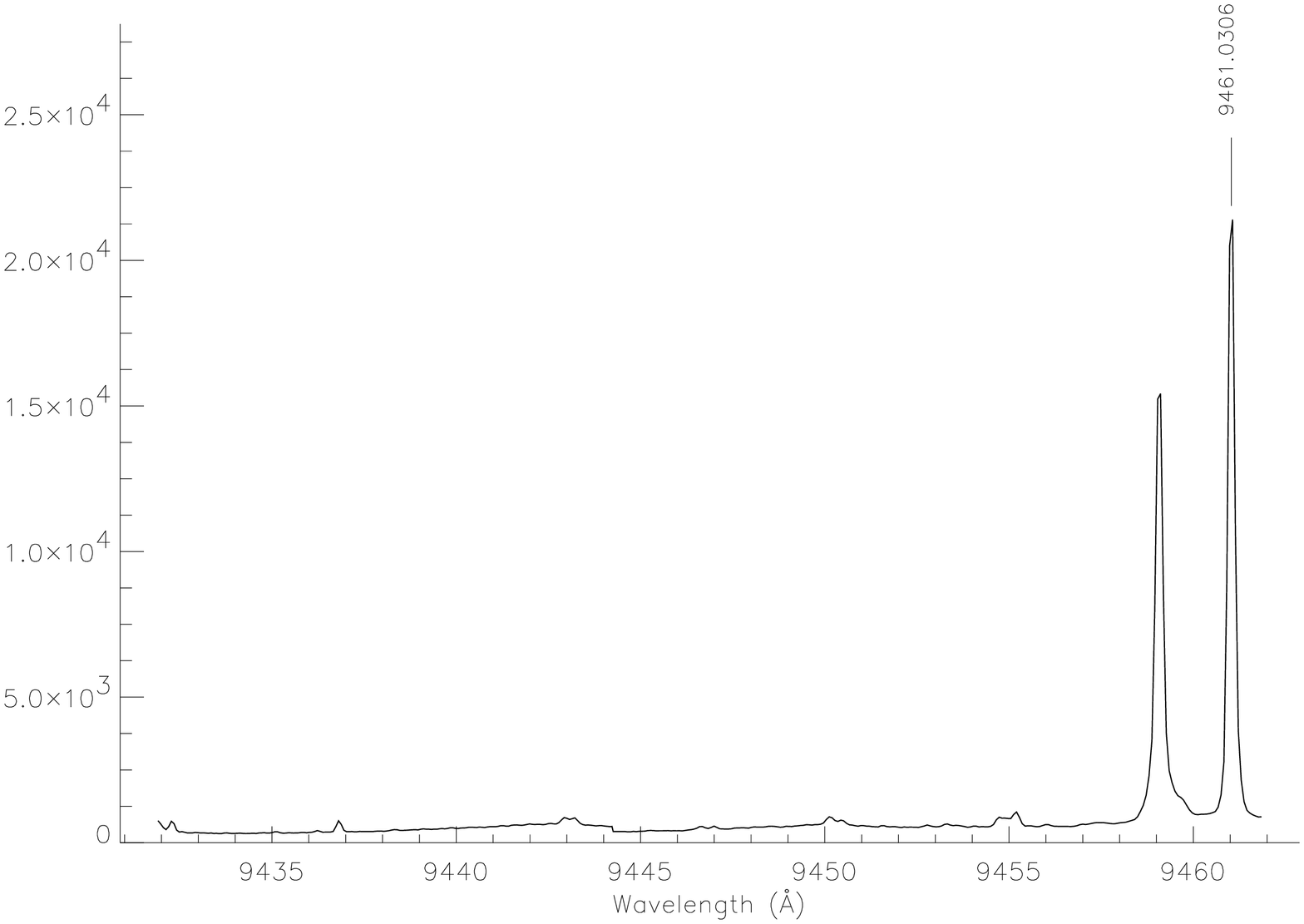}
\end{figure}
\clearpage
   
\begin{figure}
\centering
\includegraphics[width=10cm,angle=90]{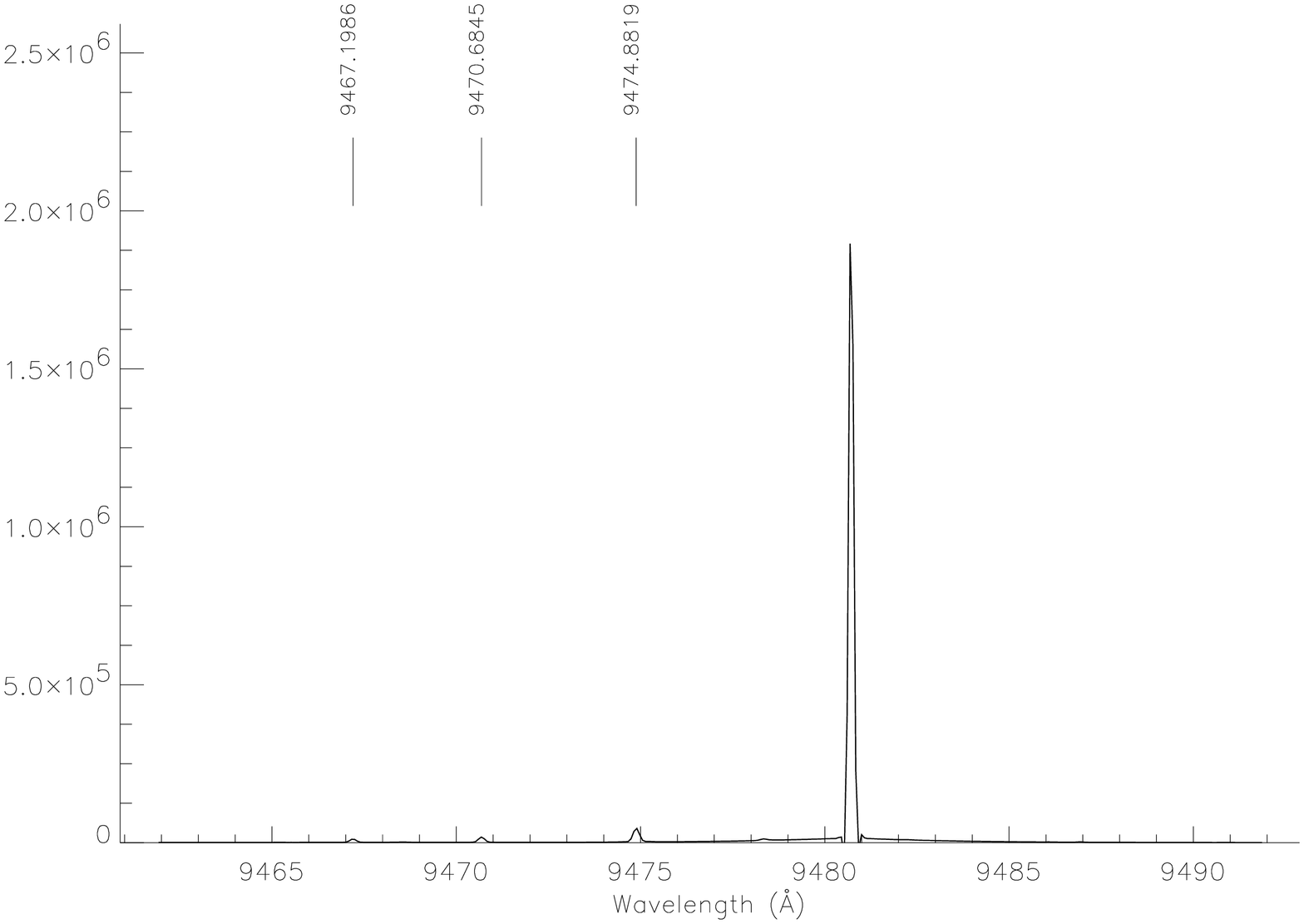}
\includegraphics[width=10cm,angle=90]{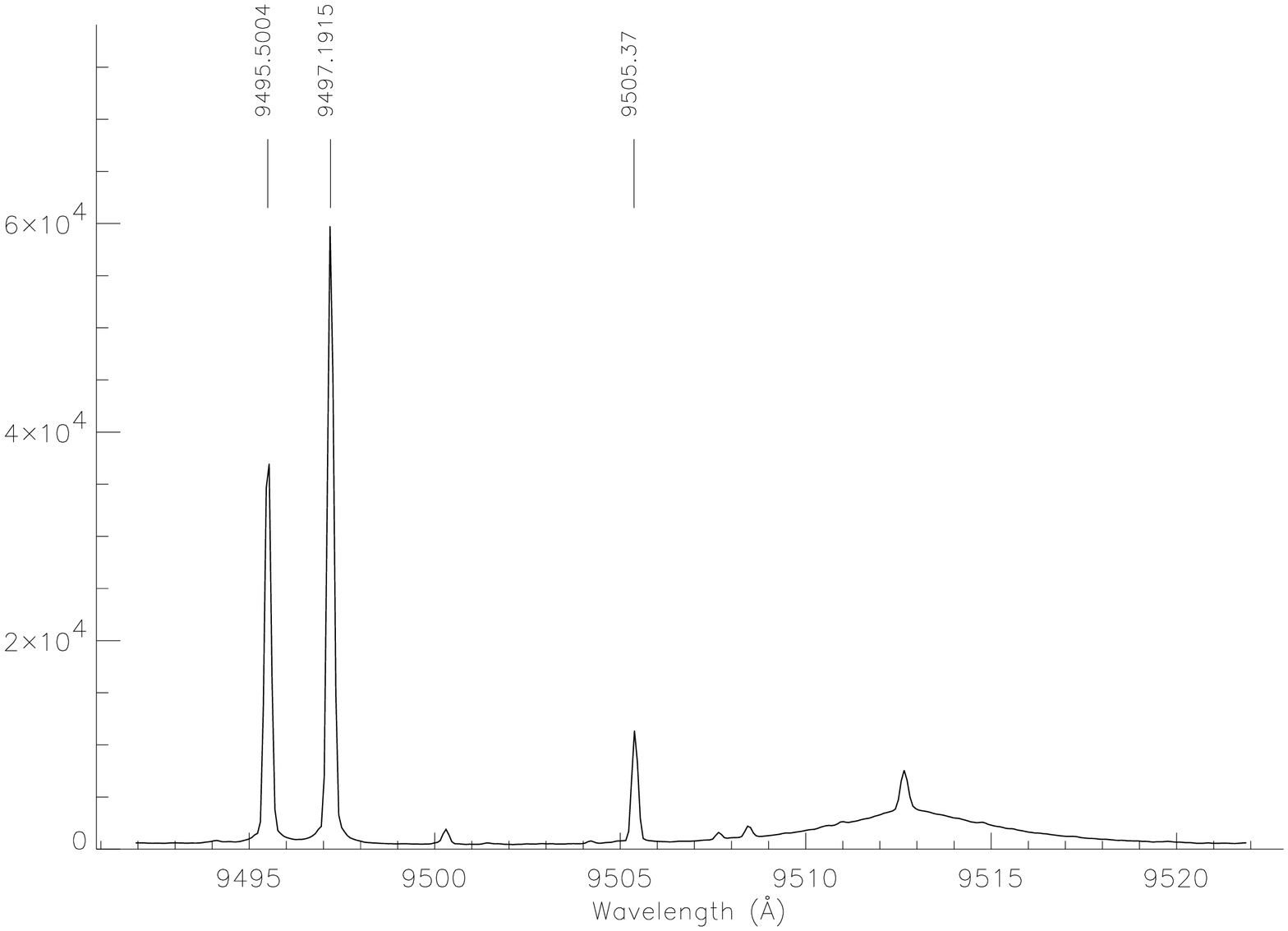}
\end{figure}
\clearpage
   
\begin{figure}
\centering
\includegraphics[width=10cm,angle=90]{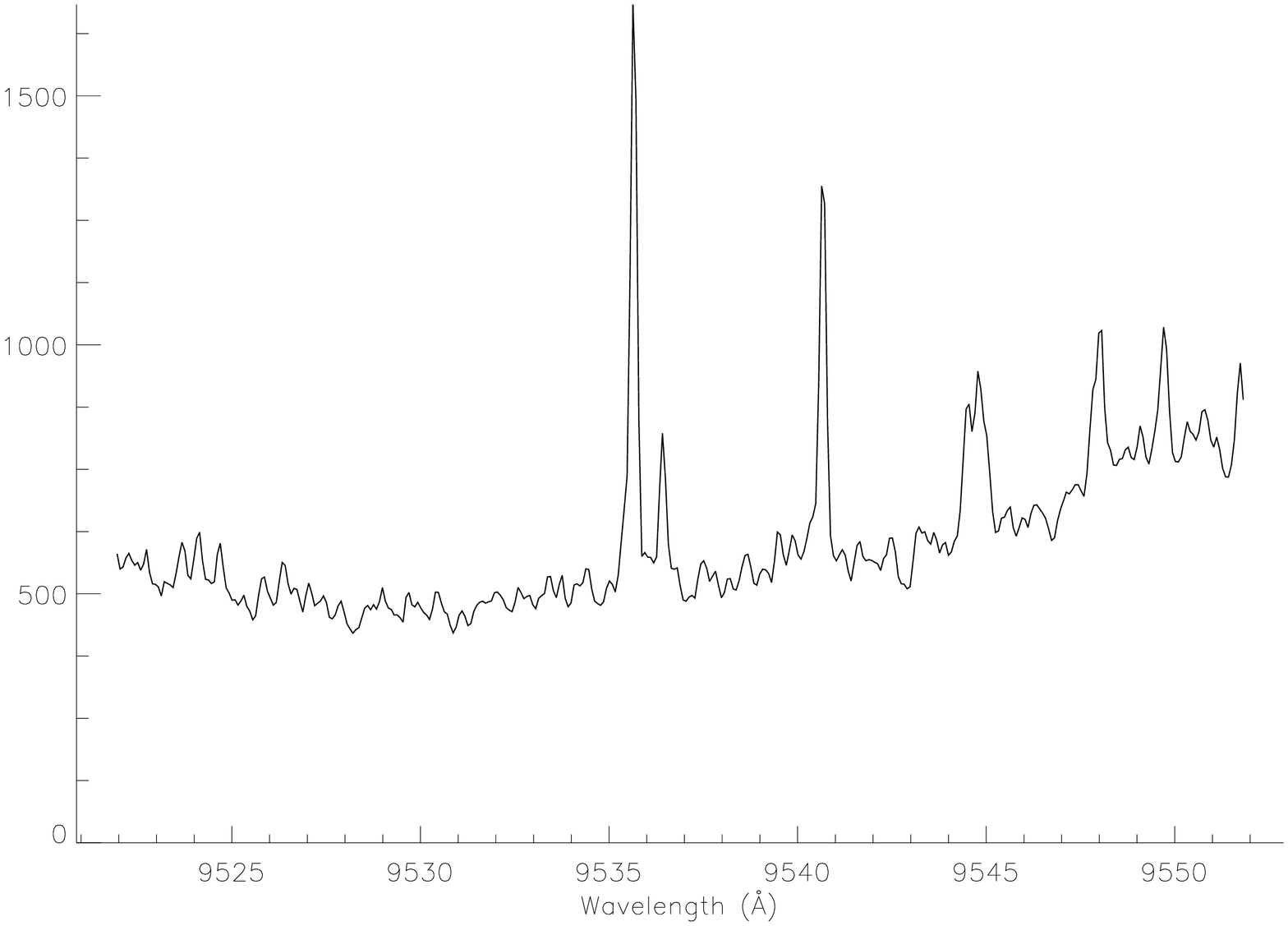}
\includegraphics[width=10cm,angle=90]{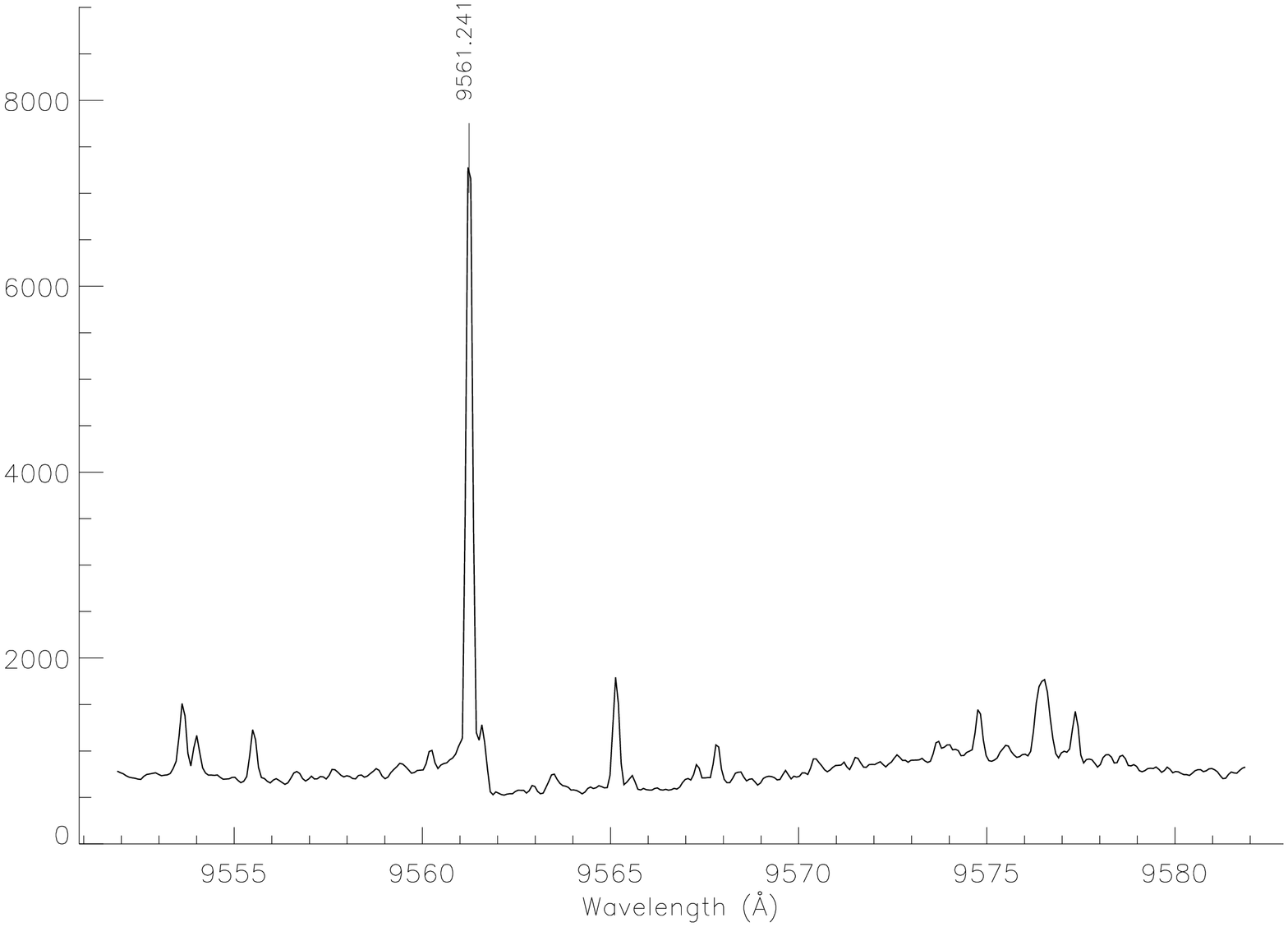}
\end{figure}
\clearpage
   
\begin{figure}
\centering
\includegraphics[width=10cm,angle=90]{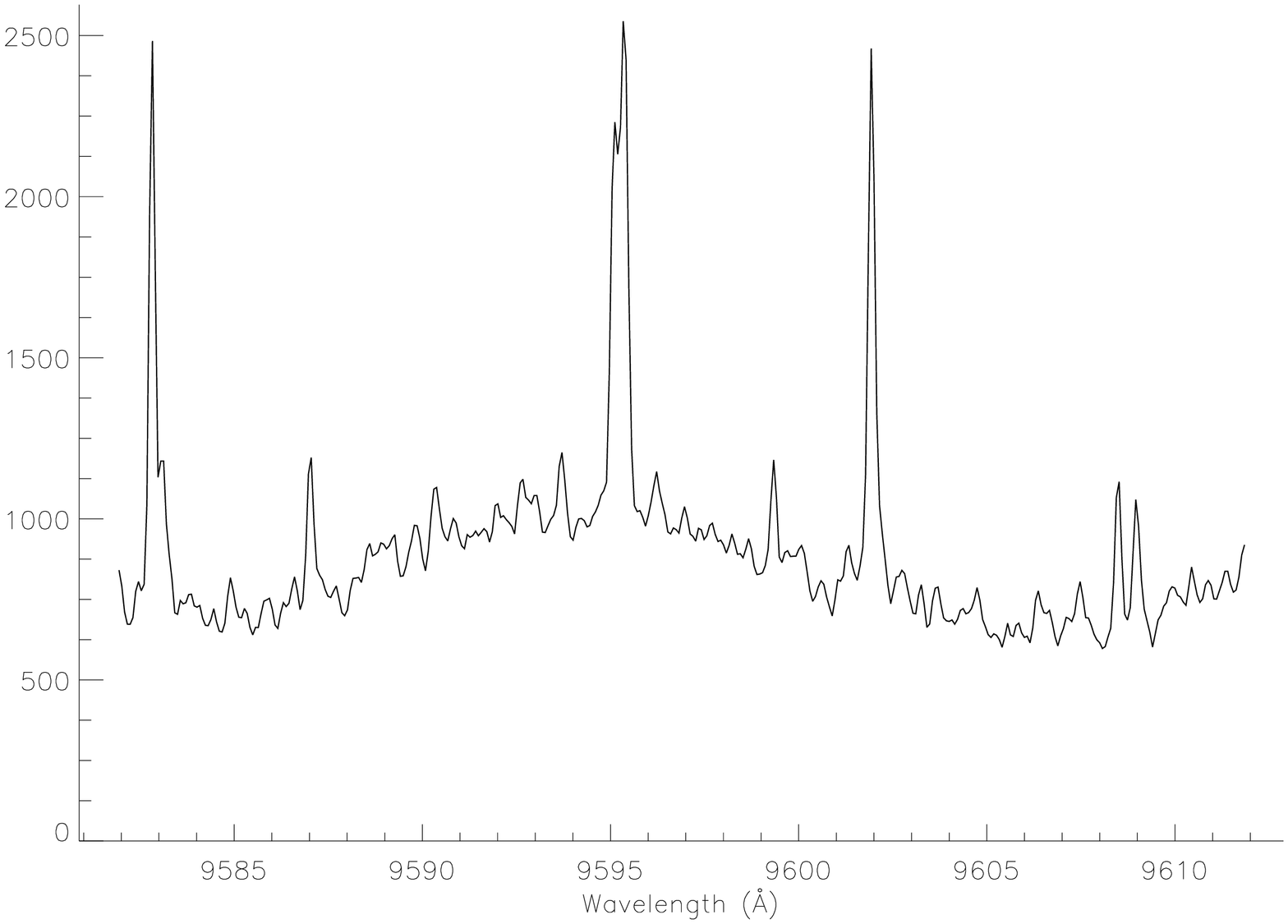}
\includegraphics[width=10cm,angle=90]{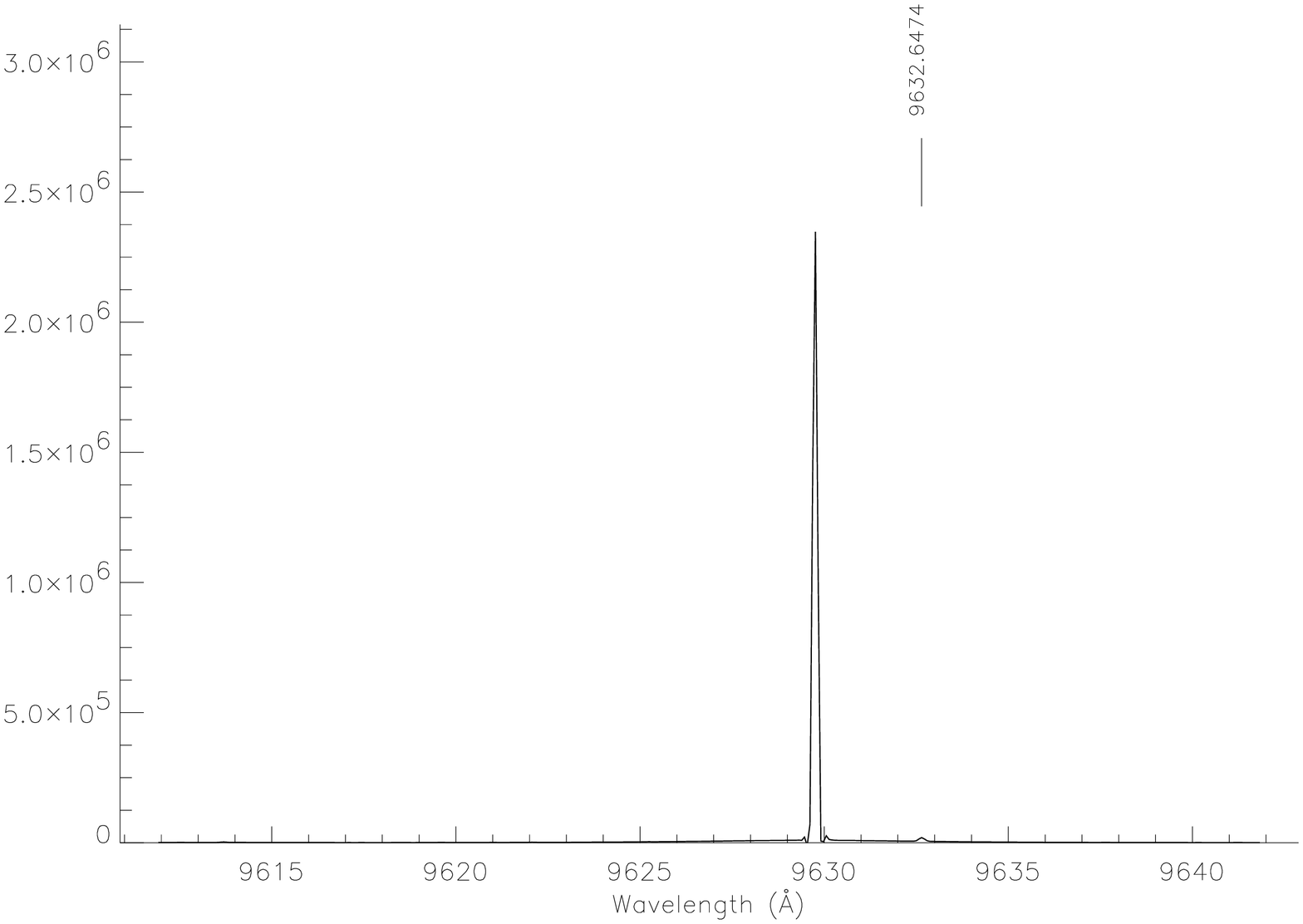}
\end{figure}
\clearpage
   
\begin{figure}
\centering
\includegraphics[width=10cm,angle=90]{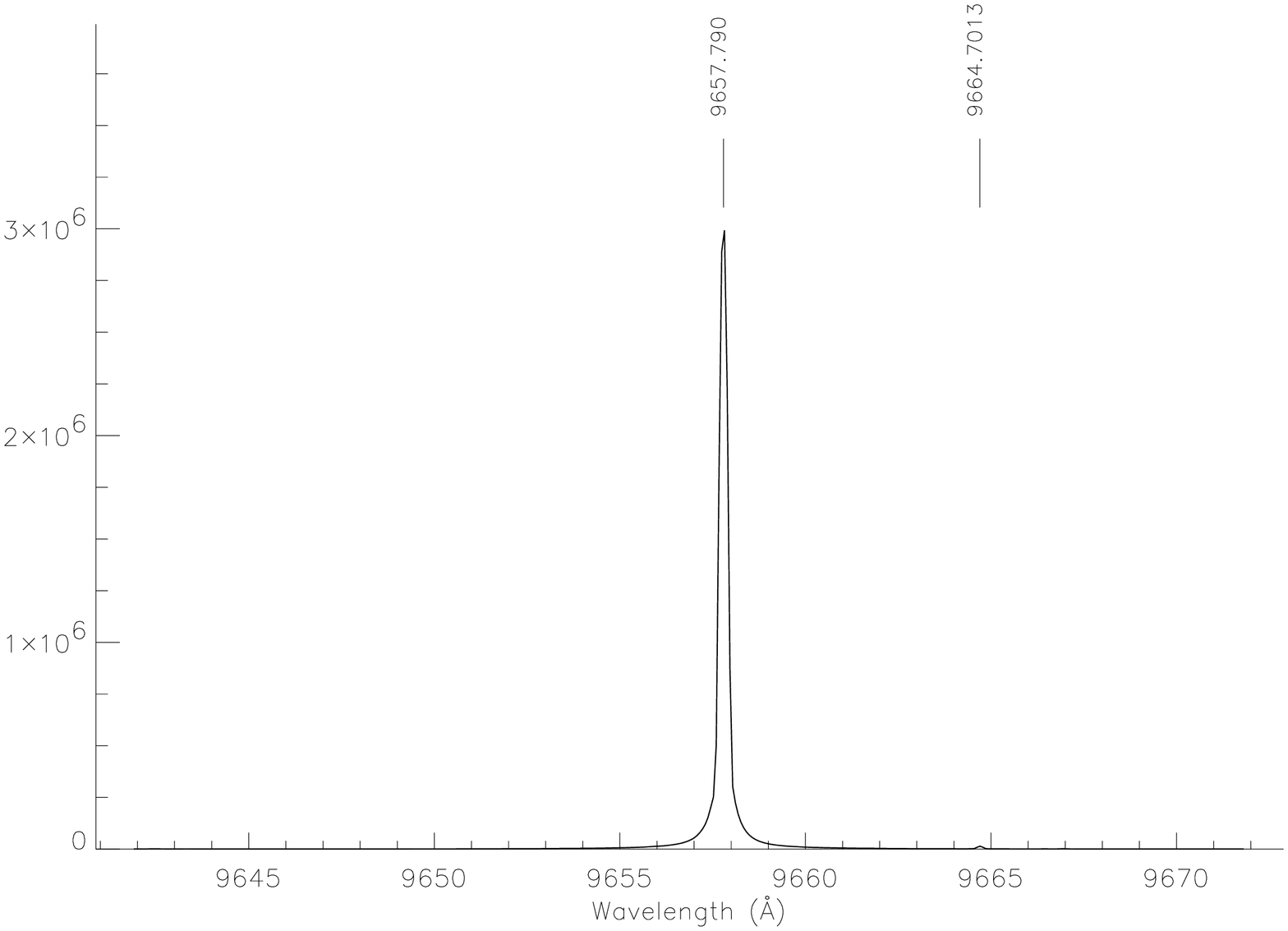}
\end{figure}

\clearpage

\part{Table of Wavelengths}

\tablecaption{Wavelengths of the emission lines identified in the {\it 2dcoud\'e} Th-Ar spectrum.}
\tablefirsthead{\hline Wavelength  & err & Wavelength  & err  & Wavelength  & err  \\
(\AA) &  (\AA) & (\AA) &  (\AA) & (\AA) &  (\AA) \\\hline}
\tablehead{\hline Wavelength  & err & Wavelength  & err  & Wavelength  & err  \\
(\AA) &  (\AA) & (\AA) &  (\AA) & (\AA) &  (\AA) \\\hline}
\tabletail{\hline}
\tablelasttail{\hline\hline}
\begin{supertabular}{cccccc}
3642.24853  &  0.00112	 & 3825.12491  &  0.00654	 & 3903.10183  &  0.00121  \\
3659.62650  &  0.00130	 & 3826.80854  &  0.00036	 & 3911.91417  &  0.00022  \\
3690.62366  &  0.00108	 & 3828.38477  &  0.00017	 & 3915.86455  &  0.01413  \\
3700.97999  &  0.00068	 & 3830.77388  &  0.00014	 & 3916.41750  &  0.00015  \\
3706.76779  &  0.00053	 & 3834.67914  &  0.00086	 & 3919.02226  &  0.00018  \\
3718.20606  &  0.00022	 & 3836.57938  &  0.00021	 & 3923.80175  &  0.00115  \\
3719.43450  &  0.00020	 & 3837.87551  &  0.00016	 & 3925.09364  &  0.00014  \\
3721.82535  &  0.00015	 & 3839.69598  &  0.00017	 & 3925.71947  &  0.00014  \\
3727.90288  &  0.00023	 & 3840.80021  &  0.00036	 & 3928.62424  &  0.00014  \\
3729.30875  &  0.00032	 & 3841.96102  &  0.00016	 & 3929.66923  &  0.00013  \\
3737.92561  &  0.01275	 & 3845.40428  &  0.00036	 & 3932.91121  &  0.00018  \\
3741.18349  &  0.00023	 & 3850.58136  &  0.00022	 & 3933.66144  &  0.00013  \\
3742.92365  &  0.00024	 & 3852.13598  &  0.00026	 & 3944.26888  &  0.00040  \\
3751.02291  &  0.00059	 & 3863.40900  &  0.00233	 & 3946.10027  &  0.00012  \\
3752.56886  &  0.00051	 & 3867.24425  &  0.04600	 & 3947.50427  &  0.00063  \\
3754.56993  &  0.02304	 & 3868.52835  &  0.00016	 & 3948.03039  &  0.00015  \\
3758.46779  &  0.00039	 & 3869.66350  &  0.00050	 & 3948.97726  &  0.00013  \\
3765.26845  &  0.00016	 & 3872.71423  &  0.00062	 & 3950.39461  &  0.00015  \\
3766.11705  &  0.00027	 & 3873.14831  &  0.00038	 & 3952.75460  &  0.00019  \\
3771.37037  &  0.00036	 & 3873.47211  &  0.00032	 & 3956.68939  &  0.00023  \\
3783.29770  &  0.00061	 & 3873.82169  &  0.00015	 & 3959.29976  &  0.00014  \\
3785.60968  &  0.00780	 & 3874.24485  &  0.00065	 & 3962.41985  &  0.00166  \\
3789.16546  &  0.00017	 & 3874.86203  &  0.00016	 & 3967.39246  &  0.00019  \\
3790.79348  &  0.00204	 & 3875.37316  &  0.00012	 & 3968.46750  &  0.00015  \\
3792.37284  &  0.00040	 & 3879.64350  &  0.00597	 & 3969.00320  &  0.00026  \\
3798.10282  &  0.00028	 & 3886.91636  &  0.00016	 & 3972.15111  &  0.00233  \\
3803.07444  &  0.00055	 & 3891.97950  &  0.00040	 & 3972.63889  &  0.00038  \\
3809.45611  &  0.00031	 & 3895.41889  &  0.00016	 & 3973.19662  &  0.00022  \\
3818.68780  &  0.00240	 & 3898.44045  &  0.00088	 & 3974.47744  &  0.00021  \\
3820.80147  &  0.00894	 & 3900.87941  &  0.00182	 & 3979.35600  &  0.00019  \\
\\
3980.08941  &  0.00061	 & 4049.94381  &  0.00022	 & 4100.34139  &  0.00010  \\
3990.49204  &  0.00017	 & 4050.88726  &  0.00017	 & 4102.61733  &  0.00015  \\
3991.73098  &  0.00025	 & 4052.92148  &  0.00011	 & 4103.90823  &  0.00013  \\
3994.54946  &  0.00025	 & 4053.52770  &  0.00019	 & 4104.38264  &  0.00042  \\
3994.79207  &  0.00017	 & 4059.25315  &  0.00011	 & 4105.91154  &  0.00081  \\
4001.05800  &  0.00017	 & 4063.40734  &  0.00051	 & 4107.05108  &  0.00041  \\
4001.89316  &  0.00021	 & 4064.33134  &  0.00017	 & 4107.85865  &  0.00048  \\
4003.30940  &  0.00022	 & 4067.45098  &  0.00020	 & 4108.41999  &  0.00010  \\
4005.09284  &  0.00013	 & 4069.20193  &  0.00014	 & 4109.32385  &  0.00013  \\
4007.02140  &  0.00029	 & 4069.39009  &  0.02187	 & 4110.83159  &  0.00030  \\
4008.21038  &  0.00015	 & 4070.23844  &  0.00008	 & 4112.75454  &  0.00013  \\
4009.04934  &  0.00010	 & 4071.74729  &  0.00024	 & 4115.75876  &  0.00014  \\
4011.74252  &  0.00015	 & 4072.00494  &  0.00010	 & 4116.71320  &  0.00012  \\
4012.49519  &  0.00013	 & 4072.38508  &  0.00010	 & 4127.41044  &  0.00020  \\
4013.85758  &  0.00022	 & 4075.50283  &  0.00012	 & 4131.01065  &  0.00705  \\
4014.71703  &  0.00034	 & 4076.62903  &  0.00012	 & 4131.72332  &  0.00010  \\
4018.10663  &  0.00037	 & 4076.94266  &  0.00019	 & 4132.75302  &  0.00023  \\
4019.12902  &  0.00012	 & 4079.57448  &  0.00012	 & 4134.07106  &  0.00013  \\
4022.07774  &  0.00013	 & 4080.68241  &  0.00051	 & 4140.23543  &  0.00025  \\
4024.80275  &  0.00016	 & 4081.36752  &  0.00025	 & 4142.70123  &  0.00014  \\
4027.00805  &  0.00012	 & 4082.38847  &  0.00026	 & 4143.64878  &  0.00029  \\
4029.82527  &  0.00029	 & 4083.46851  &  0.00016	 & 4148.18137  &  0.00027  \\
4030.29152  &  0.00082	 & 4085.04169  &  0.00060	 & 4154.72045  &  0.00112  \\
4030.84198  &  0.00013	 & 4085.43359  &  0.00051	 & 4156.51589  &  0.00015  \\
4035.46067  &  0.00022	 & 4086.52086  &  0.00013	 & 4158.59044  &  0.00012  \\
4036.04768  &  0.00013	 & 4087.28528  &  0.00036	 & 4159.51987  &  0.03208  \\
4036.56509  &  0.00015	 & 4088.72665  &  0.00023	 & 4161.73971  &  0.00033  \\
4038.23024  &  0.00105	 & 4089.13772  &  0.00016	 & 4162.50844  &  0.00025  \\
4042.89399  &  0.00013	 & 4094.74665  &  0.00038	 & 4163.64922  &  0.00045  \\
4043.39472  &  0.00014	 & 4096.07566  &  0.00025	 & 4164.18033  &  0.00011  \\
4044.41706  &  0.00025	 & 4097.74702  &  0.00012	 & 4165.76641  &  0.00013  \\
4045.95927  &  0.00584	 & 4098.73077  &  0.00035	 & 4168.63283  &  0.00005  \\
4049.52934  &  0.00614	 & 4098.93161  &  0.00044	 & 4170.53195  &  0.00012  \\
\\
4178.05924  &  0.00033	 & 4253.53882  &  0.00019	 & 4308.12225  &  0.00012  \\
4181.88383  &  0.00011	 & 4255.23699  &  0.00029	 & 4308.59488  &  0.00092  \\
4190.71300  &  0.00011	 & 4256.25389  &  0.00015	 & 4309.23879  &  0.00015  \\
4191.02972  &  0.00013	 & 4257.49609  &  0.00010	 & 4311.79937  &  0.00014  \\
4192.36171  &  0.00031	 & 4257.88808  &  0.00767	 & 4312.99285  &  0.00009  \\
4193.01636  &  0.00018	 & 4258.52038  &  0.00009	 & 4314.31853  &  0.00018  \\
4194.93574  &  0.00015	 & 4259.36182  &  0.00010	 & 4315.25435  &  0.00010  \\
4198.31668  &  0.00019	 & 4260.33282  &  0.00011	 & 4318.41523  &  0.00021  \\
4200.67454  &  0.00011	 & 4266.28641  &  0.00012	 & 4320.12636  &  0.00016  \\
4208.41051  &  0.00068	 & 4266.52763  &  0.00017	 & 4325.27282  &  0.00042  \\
4208.88743  &  0.00009	 & 4269.94248  &  0.00046	 & 4328.91500  &  0.00019  \\
4210.92313  &  0.00011	 & 4272.16888  &  0.00010	 & 4330.84369  &  0.00034  \\
4213.06738  &  0.00025	 & 4272.87452  &  0.00019	 & 4331.19990  &  0.00011  \\
4214.82838  &  0.00018	 & 4273.35715  &  0.00012	 & 4332.03000  &  0.00014  \\
4216.37326  &  0.00099	 & 4275.16072  &  0.00040	 & 4333.56094  &  0.00012  \\
4217.43185  &  0.00076	 & 4276.80736  &  0.00043	 & 4335.33774  &  0.00012  \\
4218.66520  &  0.00017	 & 4277.31373  &  0.00017	 & 4337.07093  &  0.00016  \\
4220.06882  &  0.00041	 & 4277.52848  &  0.00010	 & 4337.27698  &  0.00019  \\
4222.63812  &  0.00038	 & 4278.32336  &  0.00016	 & 4338.10770  &  0.00027  \\
4226.72669  &  0.00010	 & 4280.56758  &  0.00017	 & 4340.89577  &  0.00025  \\
4226.98784  &  0.00013	 & 4281.06707  &  0.00021	 & 4342.44395  &  0.00017  \\
4227.38689  &  0.00025	 & 4281.41395  &  0.00036	 & 4343.95127  &  0.00017  \\
4228.15820  &  0.00012	 & 4282.04361  &  0.00015	 & 4344.32631  &  0.00019  \\
4229.14752  &  0.00014	 & 4282.89872  &  0.00017	 & 4345.16818  &  0.00013  \\
4229.86954  &  0.00032	 & 4288.66904  &  0.00019	 & 4346.43418  &  0.00028  \\
4230.42614  &  0.00031	 & 4291.80996  &  0.00018	 & 4348.06425  &  0.00011  \\
4235.46357  &  0.00019	 & 4297.30778  &  0.00018	 & 4349.07241  &  0.00018  \\
4237.21991  &  0.00011	 & 4299.83958  &  0.00018	 & 4352.20521  &  0.00016  \\
4247.98844  &  0.00041	 & 4300.10080  &  0.00014	 & 4352.61636  &  0.00028  \\
4248.39049  &  0.00049	 & 4300.64974  &  0.00015	 & 4353.44712  &  0.00016  \\
4250.31664  &  0.00010	 & 4302.52594  &  0.00012	 & 4357.61226  &  0.00043  \\
4250.77613  &  0.00035	 & 4304.95619  &  0.00062	 & 4358.32005  &  0.00021  \\
4251.18518  &  0.00012	 & 4307.17631  &  0.00011	 & 4359.37202  &  0.00019  \\
\\
4362.06693  &  0.00011	 & 4430.99695  &  0.00012	 & 4487.45300  &  0.01585  \\
4363.79584  &  0.00099	 & 4432.25179  &  0.00020	 & 4488.31125  &  0.00076  \\
4365.93006  &  0.00012	 & 4432.96261  &  0.00016	 & 4488.67864  &  0.00014  \\
4367.83219  &  0.00015	 & 4433.83839  &  0.00015	 & 4489.66396  &  0.00024  \\
4369.87594  &  0.00010	 & 4434.95614  &  0.00016	 & 4490.98206  &  0.00016  \\
4370.75372  &  0.00009	 & 4435.68121  &  0.00190	 & 4493.33342  &  0.00012  \\
4371.32988  &  0.00018	 & 4438.74626  &  0.00091	 & 4498.53939  &  0.00033  \\
4374.12365  &  0.00012	 & 4439.12316  &  0.00046	 & 4498.94197  &  0.00014  \\
4374.79298  &  0.00045	 & 4439.46175  &  0.00014	 & 4499.98319  &  0.00012  \\
4375.95460  &  0.00011	 & 4439.87998  &  0.00079	 & 4502.92767  &  0.00018  \\
4378.17647  &  0.00019	 & 4440.57342  &  0.00037	 & 4505.21639  &  0.00015  \\
4379.66720  &  0.00011	 & 4440.86592  &  0.00031	 & 4506.47755  &  0.00044  \\
4381.40017  &  0.00017	 & 4443.66601  &  0.00053	 & 4510.52664  &  0.00065  \\
4381.85994  &  0.00013	 & 4445.89960  &  0.00019	 & 4510.73267  &  0.00018  \\
4385.05714  &  0.00012	 & 4447.83415  &  0.00032	 & 4513.22287  &  0.00016  \\
4391.11032  &  0.00012	 & 4448.87957  &  0.00013	 & 4513.67955  &  0.00018  \\
4392.97414  &  0.00019	 & 4450.79957  &  0.00047	 & 4515.11813  &  0.00014  \\
4397.91387  &  0.00043	 & 4452.56501  &  0.00016	 & 4519.25951  &  0.00017  \\
4400.09682  &  0.00017	 & 4454.77338  &  0.00015	 & 4521.19544  &  0.00015  \\
4400.98671  &  0.00011	 & 4458.00141  &  0.00016	 & 4522.32288  &  0.00015  \\
4401.58122  &  0.00018	 & 4460.55817  &  0.00027	 & 4530.31920  &  0.00012  \\
4402.92702  &  0.00016	 & 4461.52817  &  0.00017	 & 4530.55269  &  0.00012  \\
4408.88277  &  0.00012	 & 4463.66672  &  0.00042	 & 4532.25782  &  0.00043  \\
4414.48653  &  0.00010	 & 4465.34073  &  0.00020	 & 4533.07541  &  0.00015  \\
4414.77592  &  0.01360	 & 4469.52510  &  0.00019	 & 4533.30233  &  0.00014  \\
4416.23709  &  0.00031	 & 4470.98976  &  0.00063	 & 4534.12000  &  0.00044  \\
4416.84512  &  0.00018	 & 4474.76068  &  0.00011	 & 4535.25475  &  0.00011  \\
4421.54695  &  0.00071	 & 4475.22221  &  0.00011	 & 4535.50370  &  0.00053  \\
4422.04789  &  0.00016	 & 4478.59578  &  0.00027	 & 4537.64425  &  0.00025  \\
4423.72051  &  0.00029	 & 4479.63798  &  0.00054	 & 4540.99887  &  0.00028  \\
4424.83753  &  0.00046	 & 4481.81102  &  0.00010	 & 4544.51387  &  0.00024  \\
4426.00156  &  0.00010	 & 4483.34690  &  0.00017	 & 4545.05203  &  0.00011  \\
4430.18933  &  0.00011	 & 4486.89660  &  0.00016	 & 4545.91513  &  0.00016  \\
\\
4547.24951  &  0.00035	 & 4619.48172  &  0.00120	 & 4694.09109  &  0.00015  \\
4547.75946  &  0.00051	 & 4621.16062  &  0.00019	 & 4695.03717  &  0.00013  \\
4552.15552  &  0.00020	 & 4627.29726  &  0.00022	 & 4695.45335  &  0.00033  \\
4555.81194  &  0.00014	 & 4628.20180  &  0.00026	 & 4702.31601  &  0.00013  \\
4558.34549  &  0.00044	 & 4628.44084  &  0.00015	 & 4703.98938  &  0.00014  \\
4561.34857  &  0.00016	 & 4631.75982  &  0.00019	 & 4705.76026  &  0.00023  \\
4563.66494  &  0.00020	 & 4633.76969  &  0.00019	 & 4708.29380  &  0.00020  \\
4564.40584  &  0.00023	 & 4637.23295  &  0.00012	 & 4712.47941  &  0.00019  \\
4564.83434  &  0.00050	 & 4638.68421  &  0.00021	 & 4712.84121  &  0.00028  \\
4567.23994  &  0.00014	 & 4640.04650  &  0.00042	 & 4720.45845  &  0.00028  \\
4568.14268  &  0.00039	 & 4646.68605  &  0.00046	 & 4720.78221  &  0.00044  \\
4570.97184  &  0.00013	 & 4647.25051  &  0.00020	 & 4721.27635  &  0.00012  \\
4579.35021  &  0.00012	 & 4650.23576  &  0.00032	 & 4721.59088  &  0.00017  \\
4579.82764  &  0.00023	 & 4651.55700  &  0.00015	 & 4722.08844  &  0.00012  \\
4581.17528  &  0.00047	 & 4651.98883  &  0.00020	 & 4723.43801  &  0.00010  \\
4588.42535  &  0.00018	 & 4655.21239  &  0.00031	 & 4723.78377  &  0.00012  \\
4589.89824  &  0.00014	 & 4657.90169  &  0.00008	 & 4724.77646  &  0.00028  \\
4592.66617  &  0.00011	 & 4659.57100  &  0.00037	 & 4726.86863  &  0.00009  \\
4593.64239  &  0.00020	 & 4663.20227  &  0.00011	 & 4728.13445  &  0.00102  \\
4595.42072  &  0.00009	 & 4666.00548  &  0.00015	 & 4729.12739  &  0.00013  \\
4596.09693  &  0.00010	 & 4666.51579  &  0.00013	 & 4729.88004  &  0.00034  \\
4596.30760  &  0.00029	 & 4666.79878  &  0.00011	 & 4732.05400  &  0.00013  \\
4598.76336  &  0.00011	 & 4668.17152  &  0.00015	 & 4735.90645  &  0.00013  \\
4602.88563  &  0.00097	 & 4669.98366  &  0.00014	 & 4739.67669  &  0.00016  \\
4603.14419  &  0.00012	 & 4673.66062  &  0.00013	 & 4740.53176  &  0.00013  \\
4607.93463  &  0.00021	 & 4675.37569  &  0.00088	 & 4740.95412  &  0.00015  \\
4608.55814  &  0.01884	 & 4676.05657  &  0.00014	 & 4741.30633  &  0.00061  \\
4609.56750  &  0.00012	 & 4680.23671  &  0.00026	 & 4742.11827  &  0.00045  \\
4611.85864  &  0.00058	 & 4680.64144  &  0.00064	 & 4743.69258  &  0.00024  \\
4612.54617  &  0.00105	 & 4683.35160  &  0.00018	 & 4745.33146  &  0.00032  \\
4613.60495  &  0.00019	 & 4686.19425  &  0.00012	 & 4749.19958  &  0.00013  \\
4615.02466  &  0.00032	 & 4690.62187  &  0.00016	 & 4749.96976  &  0.00020  \\
4615.33360  &  0.00020	 & 4691.63231  &  0.00019	 & 4752.41319  &  0.00014  \\
\\
4761.10935  &  0.00063	 & 4823.99669  &  0.00016	 & 4881.20375  &  0.00019  \\
4764.34554  &  0.00016	 & 4826.69912  &  0.00013	 & 4882.24409  &  0.00021  \\
4764.86507  &  0.00013	 & 4829.79732  &  0.00032	 & 4889.04262  &  0.00013  \\
4765.59534  &  0.00022	 & 4831.12096  &  0.00012	 & 4892.76007  &  0.00034  \\
4766.60048  &  0.00012	 & 4831.59741  &  0.00013	 & 4893.44309  &  0.00072  \\
4773.24070  &  0.00027	 & 4832.80197  &  0.00018	 & 4894.95430  &  0.00013  \\
4775.31365  &  0.00038	 & 4840.47426  &  0.00028	 & 4899.24211  &  0.00045  \\
4775.79380  &  0.00020	 & 4840.84265  &  0.00015	 & 4902.04754  &  0.00021  \\
4777.19097  &  0.00024	 & 4843.92486  &  0.00024	 & 4902.79341  &  0.00029  \\
4778.29317  &  0.00012	 & 4844.16609  &  0.00029	 & 4904.75083  &  0.00021  \\
4779.72634  &  0.00024	 & 4845.16227  &  0.00035	 & 4910.15735  &  0.00018  \\
4782.76081  &  0.00029	 & 4847.81016  &  0.00013	 & 4910.79689  &  0.00023  \\
4783.86361  &  0.00015	 & 4848.36222  &  0.00013	 & 4911.37963  &  0.00028  \\
4786.53108  &  0.00013	 & 4849.13777  &  0.00017	 & 4912.52866  &  0.00021  \\
4787.14773  &  0.00017	 & 4849.86150  &  0.00018	 & 4919.81522  &  0.00012  \\
4789.38648  &  0.00012	 & 4850.43949  &  0.00012	 & 4921.61418  &  0.00013  \\
4792.08246  &  0.00046	 & 4852.86858  &  0.00016	 & 4922.94368  &  0.00021  \\
4793.24495  &  0.00020	 & 4858.33211  &  0.00017	 & 4924.42178  &  0.00026  \\
4795.91372  &  0.00013	 & 4861.21675  &  0.00014	 & 4925.42187  &  0.00091  \\
4800.17137  &  0.00065	 & 4861.71645  &  0.00016	 & 4925.94985  &  0.00035  \\
4806.02133  &  0.00012	 & 4863.16691  &  0.00010	 & 4927.29803  &  0.00071  \\
4808.13364  &  0.00013	 & 4865.47714  &  0.00013	 & 4927.77955  &  0.00015  \\
4809.61409  &  0.00014	 & 4865.91446  &  0.00018	 & 4929.08445  &  0.00036  \\
4812.37582  &  0.00075	 & 4867.55484  &  0.00017	 & 4929.98590  &  0.00015  \\
4813.00690  &  0.00100	 & 4868.88115  &  0.00021	 & 4933.20931  &  0.00011  \\
4813.72041  &  0.00058	 & 4871.28924  &  0.00018	 & 4933.85009  &  0.00028  \\
4813.89598  &  0.00016	 & 4872.91928  &  0.00012	 & 4936.77487  &  0.00020  \\
4819.19174  &  0.00027	 & 4874.36442  &  0.00014	 & 4937.82895  &  0.00020  \\
4820.46440  &  0.00038	 & 4876.49607  &  0.00042	 & 4938.50546  &  0.00080  \\
4820.88500  &  0.00014	 & 4877.81025  &  0.00032	 & 4939.64197  &  0.00016  \\
4821.58747  &  0.00016	 & 4878.00941  &  0.00015	 & 4943.06406  &  0.00015  \\
4822.85406  &  0.00015	 & 4878.73258  &  0.00014	 & 4945.45810  &  0.00016  \\
4823.60559  &  0.00014	 & 4879.86402  &  0.00014	 & 4946.66283  &  0.00019  \\
\\
4947.57191  &  0.00042	 & 5041.60720  &  0.00057	 & 5115.04406  &  0.00014  \\
4950.24872  &  0.00045	 & 5044.72013  &  0.00015	 & 5122.49794  &  0.00040  \\
4950.62541  &  0.00019	 & 5045.24743  &  0.00020	 & 5125.49003  &  0.00031  \\
4952.71868  &  0.00922	 & 5047.04335  &  0.00014	 & 5125.94829  &  0.00022  \\
4961.72560  &  0.00037	 & 5048.93238  &  0.00045	 & 5128.48953  &  0.00025  \\
4965.07985  &  0.00012	 & 5049.79714  &  0.00016	 & 5134.74547  &  0.00019  \\
4965.73033  &  0.00019	 & 5050.78383  &  0.00014	 & 5137.47354  &  0.00017  \\
4968.77257  &  0.00815	 & 5051.88833  &  0.00038	 & 5140.76410  &  0.00019  \\
4970.08385  &  0.00248	 & 5055.34649  &  0.00021	 & 5141.78314  &  0.00013  \\
4972.16127  &  0.00022	 & 5057.98555  &  0.00022	 & 5143.26758  &  0.00018  \\
4980.18557  &  0.00023	 & 5059.86060  &  0.00018	 & 5143.91591  &  0.00014  \\
4982.48636  &  0.00017	 & 5061.65593  &  0.00019	 & 5145.30889  &  0.00017  \\
4985.37142  &  0.00018	 & 5062.03748  &  0.00013	 & 5146.05606  &  0.00092  \\
4987.14683  &  0.00035	 & 5062.93166  &  0.00029	 & 5148.21191  &  0.00016  \\
4989.30823  &  0.00013	 & 5063.51462  &  0.00014	 & 5149.20665  &  0.00072  \\
4993.74878  &  0.00022	 & 5064.00660  &  0.00015	 & 5151.61137  &  0.00012  \\
4994.10668  &  0.00048	 & 5064.60152  &  0.00014	 & 5154.24280  &  0.00012  \\
4999.93980  &  0.00074	 & 5064.94510  &  0.00012	 & 5158.60442  &  0.00012  \\
5002.09648  &  0.00012	 & 5066.13565  &  0.00014	 & 5161.53993  &  0.00019  \\
5003.59786  &  0.00030	 & 5066.77781  &  0.00015	 & 5162.28755  &  0.00016  \\
5004.12596  &  0.00031	 & 5067.13887  &  0.00031	 & 5163.45810  &  0.00015  \\
5009.33558  &  0.00013	 & 5067.97358  &  0.00012	 & 5165.77350  &  0.00025  \\
5009.97422  &  0.00760	 & 5069.33388  &  0.00109	 & 5168.92125  &  0.00094  \\
5015.89003  &  0.00023	 & 5081.44671  &  0.00023	 & 5173.67123  &  0.00030  \\
5017.17430  &  0.00033	 & 5084.99374  &  0.00050	 & 5175.32407  &  0.00020  \\
5017.56235  &  0.01289	 & 5090.05130  &  0.00096	 & 5175.91068  &  0.00075  \\
5019.80600  &  0.00020	 & 5090.51420  &  0.00026	 & 5176.96170  &  0.00014  \\
5021.16500  &  0.01821	 & 5095.06292  &  0.00036	 & 5177.61618  &  0.00032  \\
5022.00649  &  0.00050	 & 5096.48440  &  0.00016	 & 5183.98797  &  0.00047  \\
5028.64948  &  0.00013	 & 5098.04337  &  0.00017	 & 5186.41290  &  0.00062  \\
5029.89112  &  0.00014	 & 5100.61837  &  0.00014	 & 5187.33734  &  0.00021  \\
5039.22982  &  0.00013	 & 5101.12941  &  0.00026	 & 5187.74726  &  0.00012  \\
5041.12231  &  0.00049	 & 5111.27767  &  0.00020	 & 5190.87169  &  0.00016  \\
\\
5193.82555  &  0.00022	 & 5277.49953  &  0.00022	 & 5369.28172  &  0.00026  \\
5194.45699  &  0.00016	 & 5281.06841  &  0.00022	 & 5370.70886  &  0.00020  \\
5195.81303  &  0.00016	 & 5286.88721  &  0.00039	 & 5372.70201  &  0.00024  \\
5198.80186  &  0.00015	 & 5291.81685  &  0.00022	 & 5374.82120  &  0.00021  \\
5199.16365  &  0.00016	 & 5294.39641  &  0.00020	 & 5375.35222  &  0.00021  \\
5203.84440  &  0.00039	 & 5296.27879  &  0.00014	 & 5375.76859  &  0.00051  \\
5205.15220  &  0.00021	 & 5297.74423  &  0.00015	 & 5376.12997  &  0.00020  \\
5209.72393  &  0.00021	 & 5298.28179  &  0.00016	 & 5376.77727  &  0.00021  \\
5211.22975  &  0.00014	 & 5300.52341  &  0.00015	 & 5378.83490  &  0.00014  \\
5213.34863  &  0.00020	 & 5301.40402  &  0.00036	 & 5379.11026  &  0.00013  \\
5216.59628  &  0.00014	 & 5306.98711  &  0.00028	 & 5382.92698  &  0.00014  \\
5218.52722  &  0.00016	 & 5307.46511  &  0.00018	 & 5384.03563  &  0.00020  \\
5219.10929  &  0.00014	 & 5310.26506  &  0.00014	 & 5386.61027  &  0.00017  \\
5220.70515  &  0.00058	 & 5312.00130  &  0.00012	 & 5388.05097  &  0.00015  \\
5220.92737  &  0.00017	 & 5312.52882  &  0.00013	 & 5392.57274  &  0.00015  \\
5221.28012  &  0.00015	 & 5312.90394  &  0.00014	 & 5393.97122  &  0.00017  \\
5228.22547  &  0.00017	 & 5317.49486  &  0.00020	 & 5394.76023  &  0.00013  \\
5231.15912  &  0.00012	 & 5320.76846  &  0.00085	 & 5397.51841  &  0.00028  \\
5233.22600  &  0.00023	 & 5325.14280  &  0.00026	 & 5398.70192  &  0.00022  \\
5234.10887  &  0.00103	 & 5325.43139  &  0.00037	 & 5398.91892  &  0.00025  \\
5238.81361  &  0.00015	 & 5326.27581  &  0.00022	 & 5399.17469  &  0.00022  \\
5239.55076  &  0.00017	 & 5326.97545  &  0.00016	 & 5399.62076  &  0.00077  \\
5240.19694  &  0.00023	 & 5329.37506  &  0.00089	 & 5400.14535  &  0.00038  \\
5247.65382  &  0.00017	 & 5330.07869  &  0.00024	 & 5402.60736  &  0.00030  \\
5258.35965  &  0.00014	 & 5337.01748  &  0.00150	 & 5403.20024  &  0.00025  \\
5260.10335  &  0.00017	 & 5343.58022  &  0.00016	 & 5407.34550  &  0.00032  \\
5261.47309  &  0.00053	 & 5347.97127  &  0.00022	 & 5407.65329  &  0.00015  \\
5265.55127  &  0.00020	 & 5349.46007  &  0.00017	 & 5410.76862  &  0.00016  \\
5266.70955  &  0.00018	 & 5351.12598  &  0.00019	 & 5415.51568  &  0.00019  \\
5270.26556  &  0.00018	 & 5355.63469  &  0.00047	 & 5417.48538  &  0.00014  \\
5272.92546  &  0.00029	 & 5360.14965  &  0.00018	 & 5421.35855  &  0.00020  \\
5274.11843  &  0.00017	 & 5361.15589  &  0.00049	 & 5424.00702  &  0.00026  \\
5277.14585  &  0.00039	 & 5362.57480  &  0.00021	 & 5425.67729  &  0.00013  \\
\\
5431.11142  &  0.00015	 & 5524.58810  &  0.00020	 & 5590.11345  &  0.00027  \\
5434.15071  &  0.00018	 & 5524.95986  &  0.00041	 & 5593.61279  &  0.00020  \\
5435.89123  &  0.00030	 & 5527.29559  &  0.00032	 & 5594.46056  &  0.00015  \\
5437.38634  &  0.00032	 & 5528.22926  &  0.00060	 & 5595.06352  &  0.00018  \\
5439.99248  &  0.00037	 & 5537.55654  &  0.00045	 & 5598.47925  &  0.00015  \\
5440.59978  &  0.00035	 & 5538.60527  &  0.00027	 & 5599.65380  &  0.00064  \\
5443.12166  &  0.00030	 & 5539.26123  &  0.00016	 & 5601.60285  &  0.00016  \\
5447.15333  &  0.00023	 & 5539.90999  &  0.00017	 & 5602.84575  &  0.00027  \\
5449.47893  &  0.00029	 & 5541.14488  &  0.00025	 & 5604.51443  &  0.00016  \\
5451.65459  &  0.00018	 & 5541.58287  &  0.00049	 & 5606.38608  &  0.00017  \\
5452.21827  &  0.00014	 & 5542.88973  &  0.00023	 & 5606.73537  &  0.00015  \\
5461.73682  &  0.00046	 & 5548.17498  &  0.00014	 & 5610.68027  &  0.00018  \\
5462.61255  &  0.00045	 & 5551.37121  &  0.00021	 & 5612.06761  &  0.00016  \\
5464.20517  &  0.00014	 & 5552.62182  &  0.00019	 & 5615.31909  &  0.00015  \\
5470.75872  &  0.00017	 & 5555.53118  &  0.00031	 & 5619.98246  &  0.00032  \\
5479.07596  &  0.00028	 & 5557.04518  &  0.00013	 & 5633.29476  &  0.00027  \\
5484.14420  &  0.00024	 & 5558.34277  &  0.00015	 & 5639.74573  &  0.00020  \\
5488.63106  &  0.00032	 & 5558.70396  &  0.00014	 & 5641.73436  &  0.00050  \\
5492.64292  &  0.00021	 & 5559.89126  &  0.00015	 & 5645.52770  &  0.00024  \\
5493.20295  &  0.00020	 & 5564.20156  &  0.00016	 & 5645.66478  &  0.00026  \\
5494.33090  &  0.00020	 & 5568.00573  &  0.00016	 & 5645.89043  &  0.00039  \\
5495.87880  &  0.00015	 & 5571.19371  &  0.00017	 & 5646.45182  &  0.00034  \\
5496.13658  &  0.00018	 & 5572.48204  &  0.00041	 & 5648.69478  &  0.00027  \\
5499.25518  &  0.00015	 & 5573.35344  &  0.00016	 & 5648.98922  &  0.00015  \\
5499.64682  &  0.00025	 & 5576.20481  &  0.00018	 & 5650.70626  &  0.00013  \\
5501.28035  &  0.00030	 & 5577.68526  &  0.00021	 & 5654.02370  &  0.00026  \\
5504.30142  &  0.00017	 & 5579.35786  &  0.00017	 & 5657.92557  &  0.00015  \\
5506.11880  &  0.00020	 & 5580.07662  &  0.00023	 & 5659.13584  &  0.00024  \\
5507.54095  &  0.00029	 & 5580.75540  &  0.00053	 & 5663.04243  &  0.00026  \\
5508.55830  &  0.00050	 & 5581.96215  &  0.00025	 & 5664.62029  &  0.00098  \\
5509.99315  &  0.00016	 & 5583.76116  &  0.00024	 & 5665.17978  &  0.00018  \\
5514.87287  &  0.00017	 & 5587.02649  &  0.00014	 & 5665.62860  &  0.00024  \\
5518.99023  &  0.00030	 & 5588.74874  &  0.00015	 & 5667.12822  &  0.00024  \\
\\
5673.83538  &  0.00030	 & 5800.82882  &  0.00018	 & 5916.72356  &  0.00034  \\
5674.99096  &  0.00023	 & 5802.08209  &  0.00022	 & 5918.94550  &  0.00023  \\
5677.05193  &  0.00018	 & 5804.14072  &  0.00017	 & 5925.40320  &  0.00025  \\
5681.90698  &  0.00035	 & 5812.96678  &  0.00492	 & 5928.81689  &  0.00023  \\
5685.19203  &  0.00021	 & 5815.42149  &  0.00022	 & 5929.93408  &  0.00023  \\
5700.91517  &  0.00014	 & 5830.82690  &  0.00027	 & 5936.38328  &  0.00033  \\
5707.10256  &  0.00016	 & 5832.37019  &  0.00019	 & 5937.16137  &  0.00031  \\
5717.17111  &  0.00046	 & 5834.26411  &  0.00017	 & 5937.66305  &  0.00016  \\
5719.62213  &  0.00017	 & 5838.94864  &  0.00028	 & 5938.45759  &  0.00017  \\
5720.18247  &  0.00016	 & 5840.64004  &  0.00017	 & 5938.82448  &  0.00017  \\
5724.46185  &  0.00028	 & 5843.80680  &  0.00021	 & 5944.64751  &  0.00018  \\
5725.38828  &  0.00017	 & 5845.91892  &  0.00021	 & 5948.79981  &  0.00020  \\
5736.02966  &  0.00022	 & 5852.68177  &  0.00019	 & 5955.56266  &  0.00030  \\
5739.52153  &  0.00016	 & 5853.47526  &  0.00019	 & 5969.73655  &  0.00023  \\
5741.16987  &  0.00017	 & 5854.12094  &  0.00024	 & 5973.66474  &  0.00014  \\
5741.82853  &  0.00016	 & 5857.44973  &  0.00024	 & 5975.06461  &  0.00017  \\
5742.08163  &  0.00070	 & 5860.31183  &  0.00015	 & 5986.26502  &  0.00051  \\
5748.74125  &  0.00014	 & 5863.71773  &  0.00018	 & 5987.30453  &  0.00018  \\
5749.38869  &  0.00016	 & 5868.37289  &  0.00023	 & 5989.04454  &  0.00015  \\
5749.78382  &  0.00017	 & 5869.85012  &  0.00023	 & 5991.00717  &  0.00016  \\
5753.02643  &  0.00022	 & 5870.54867  &  0.00027	 & 5994.12837  &  0.00016  \\
5760.55006  &  0.00018	 & 5871.18126  &  0.00025	 & 5999.00219  &  0.00024  \\
5763.52837  &  0.00016	 & 5882.62588  &  0.00015	 & 6001.20306  &  0.00018  \\
5767.77833  &  0.00024	 & 5885.70158  &  0.00020	 & 6005.16514  &  0.00019  \\
5768.18101  &  0.00018	 & 5886.53089  &  0.00029	 & 6007.07144  &  0.00017  \\
5771.75872  &  0.00045	 & 5888.58792  &  0.00014	 & 6010.16091  &  0.00017  \\
5772.11704  &  0.00031	 & 5891.45063  &  0.00017	 & 6015.42167  &  0.00022  \\
5773.94830  &  0.00017	 & 5895.28069  &  0.00024	 & 6021.03551  &  0.00017  \\
5777.39944  &  0.00030	 & 5899.84409  &  0.00017	 & 6025.15551  &  0.00024  \\
5789.64463  &  0.00017	 & 5905.57004  &  0.00019	 & 6030.44487  &  0.00025  \\
5792.42989  &  0.00019	 & 5912.08689  &  0.00013	 & 6032.12993  &  0.00023  \\
5796.06717  &  0.00018	 & 5914.38569  &  0.00017	 & 6032.86784  &  0.00443  \\
5798.47716  &  0.00028	 & 5914.67637  &  0.00015	 & 6035.19242  &  0.00024  \\
\\
6037.69770  &  0.00024	 & 6121.40961  &  0.00031	 & 6224.52658  &  0.00017  \\
6038.68023  &  0.00022	 & 6122.21475  &  0.00016	 & 6226.36959  &  0.00017  \\
6042.58974  &  0.00016	 & 6124.48113  &  0.00018	 & 6234.85468  &  0.00018  \\
6043.22648  &  0.00014	 & 6138.65313  &  0.00024	 & 6240.95247  &  0.00036  \\
6044.43617  &  0.00017	 & 6145.44387  &  0.00024	 & 6243.12129  &  0.00025  \\
6049.05118  &  0.00018	 & 6150.68244  &  0.00031	 & 6257.42342  &  0.00024  \\
6050.98221  &  0.00015	 & 6151.99255  &  0.00017	 & 6258.60455  &  0.00035  \\
6052.72420  &  0.00019	 & 6154.06864  &  0.00018	 & 6261.41715  &  0.00015  \\
6053.38100  &  0.00014	 & 6154.51618  &  0.00017	 & 6266.17281  &  0.00031  \\
6055.59209  &  0.00048	 & 6155.24347  &  0.00018	 & 6271.54485  &  0.00018  \\
6059.37466  &  0.00021	 & 6155.58023  &  0.00016	 & 6274.11670  &  0.00015  \\
6061.53539  &  0.00044	 & 6157.08734  &  0.00021	 & 6276.16339  &  0.00021  \\
6069.02017  &  0.00024	 & 6161.35072  &  0.00018	 & 6277.23959  &  0.00027  \\
6073.10168  &  0.00022	 & 6162.17075  &  0.00014	 & 6279.16383  &  0.00020  \\
6077.10420  &  0.00025	 & 6164.47948  &  0.00018	 & 6279.95815  &  0.02006  \\
6077.87264  &  0.00016	 & 6169.82270  &  0.00024	 & 6285.27663  &  0.00127  \\
6078.42064  &  0.00023	 & 6170.17949  &  0.00021	 & 6287.25504  &  0.00031  \\
6079.22189  &  0.00021	 & 6172.27964  &  0.00022	 & 6291.19032  &  0.00057  \\
6085.37309  &  0.00020	 & 6173.10040  &  0.00021	 & 6292.89014  &  0.00021  \\
6087.26165  &  0.00031	 & 6178.43005  &  0.00019	 & 6293.24105  &  0.00024  \\
6088.03072  &  0.00021	 & 6180.70447  &  0.00024	 & 6296.87351  &  0.00016  \\
6098.12136  &  0.00035	 & 6182.62060  &  0.00018	 & 6300.91548  &  0.00036  \\
6098.80757  &  0.00017	 & 6188.12400  &  0.00016	 & 6303.24983  &  0.00023  \\
6099.08327  &  0.00027	 & 6191.90566  &  0.00018	 & 6307.66070  &  0.00017  \\
6101.72476  &  0.00018	 & 6193.85553  &  0.00028	 & 6310.81016  &  0.00035  \\
6105.63759  &  0.00019	 & 6198.22297  &  0.00017	 & 6315.77577  &  0.00030  \\
6107.53309  &  0.00019	 & 6200.43240  &  0.00028	 & 6317.18415  &  0.00043  \\
6112.83657  &  0.00018	 & 6203.49284  &  0.00016	 & 6326.36631  &  0.00019  \\
6114.53839  &  0.00028	 & 6205.86084  &  0.00027	 & 6327.27676  &  0.00018  \\
6114.92457  &  0.00016	 & 6207.22028  &  0.00017	 & 6337.62094  &  0.00020  \\
6116.16586  &  0.00024	 & 6208.68747  &  0.00031	 & 6339.66814  &  0.00021  \\
6119.68735  &  0.00519	 & 6215.94022  &  0.00017	 & 6342.85884  &  0.00018  \\
6120.55610  &  0.00029	 & 6220.01159  &  0.00032	 & 6348.73711  &  0.00016  \\
\\
6355.63063  &  0.00032	 & 6512.36386  &  0.00017	 & 6662.26723  &  0.00019  \\
6355.91060  &  0.00020	 & 6522.04265  &  0.00034	 & 6664.05185  &  0.00022  \\
6364.88931  &  0.00550	 & 6531.34069  &  0.00020	 & 6666.35982  &  0.00019  \\
6369.13373  &  0.00030	 & 6538.10300  &  0.00018	 & 6668.81741  &  0.00032  \\
6369.57728  &  0.00026	 & 6551.70566  &  0.00032	 & 6673.58170  &  0.00142  \\
6371.94376  &  0.00026	 & 6554.16121  &  0.00020	 & 6674.69727  &  0.00020  \\
6376.92960  &  0.00027	 & 6558.87597  &  0.00031	 & 6677.28267  &  0.00016  \\
6379.67332  &  0.00029	 & 6564.44542  &  0.00035	 & 6678.70706  &  0.00022  \\
6384.71862  &  0.00018	 & 6577.21371  &  0.00021	 & 6683.36804  &  0.00020  \\
6387.39657  &  0.00018	 & 6577.65272  &  0.00033	 & 6684.29492  &  0.00028  \\
6388.81349  &  0.00019	 & 6583.90583  &  0.00017	 & 6697.71165  &  0.00024  \\
6389.39022  &  0.00030	 & 6588.53872  &  0.00018	 & 6711.25077  &  0.00032  \\
6394.04855  &  0.00032	 & 6591.48330  &  0.00025	 & 6713.96925  &  0.00022  \\
6399.20918  &  0.00040	 & 6593.46310  &  0.00032	 & 6719.20842  &  0.00022  \\
6400.70010  &  0.00021	 & 6593.93918  &  0.00019	 & 6727.45692  &  0.00029  \\
6406.44530  &  0.00027	 & 6599.48251  &  0.00033	 & 6728.11878  &  0.00034  \\
6411.89776  &  0.00020	 & 6604.85513  &  0.00030	 & 6733.74893  &  0.00034  \\
6413.61336  &  0.00018	 & 6605.41664  &  0.00028	 & 6752.83453  &  0.00019  \\
6416.30827  &  0.00016	 & 6613.38169  &  0.00027	 & 6753.65887  &  0.00036  \\
6437.76094  &  0.00022	 & 6618.16486  &  0.00031	 & 6756.45413  &  0.00030  \\
6439.07058  &  0.00019	 & 6619.94388  &  0.00023	 & 6757.10840  &  0.00028  \\
6446.77133  &  0.00042	 & 6632.08648  &  0.00039	 & 6758.20311  &  0.00026  \\
6457.28144  &  0.00017	 & 6638.22298  &  0.00028	 & 6766.61445  &  0.00019  \\
6483.08320  &  0.00025	 & 6638.91179  &  0.00019	 & 6772.18758  &  0.00025  \\
6490.73704  &  0.00028	 & 6639.74143  &  0.00016	 & 6778.31194  &  0.00027  \\
6493.19749  &  0.00026	 & 6643.70019  &  0.00026	 & 6780.12490  &  0.00027  \\
6493.77771  &  0.00027	 & 6644.66317  &  0.00027	 & 6780.41222  &  0.00024  \\
6499.64525  &  0.00018	 & 6646.54026  &  0.00026	 & 6787.73546  &  0.00026  \\
6501.99214  &  0.00020	 & 6648.49457  &  0.00033	 & 6788.83897  &  0.00029  \\
6503.51070  &  0.00032	 & 6648.95744  &  0.00029	 & 6791.23392  &  0.00034  \\
6506.98714  &  0.00019	 & 6654.36651  &  0.00044	 & 6824.67783  &  0.00034  \\
6508.36040  &  0.00036	 & 6658.67582  &  0.00023	 & 6827.25189  &  0.00025  \\
6509.05150  &  0.00018	 & 6660.67681  &  0.00023	 & 6829.03478  &  0.00027  \\
\\
6834.92380  &  0.00019	 & 7036.28056  &  0.00019	 & 7206.48347  &  0.00124  \\
6854.10972  &  0.00028	 & 7045.79659  &  0.00021	 & 7206.98230  &  0.00025  \\
6861.27018  &  0.00031	 & 7053.61867  &  0.00025	 & 7208.00456  &  0.00029  \\
6863.53598  &  0.00032	 & 7054.42986  &  0.00043	 & 7212.68817  &  0.00037  \\
6866.36215  &  0.00471	 & 7058.48837  &  0.00031	 & 7218.05109  &  0.00026  \\
6866.76329  &  0.00033	 & 7060.04077  &  0.00030	 & 7219.14938  &  0.00030  \\
6868.45137  &  0.00030	 & 7060.65320  &  0.00025	 & 7230.15410  &  0.00158  \\
6871.29027  &  0.00029	 & 7061.39273  &  0.00041	 & 7230.85929  &  0.00050  \\
6874.75253  &  0.00029	 & 7064.45018  &  0.00026	 & 7233.53434  &  0.00050  \\
6879.58367  &  0.00024	 & 7067.21482  &  0.00404	 & 7242.09254  &  0.00036  \\
6886.40775  &  0.00022	 & 7068.73683  &  0.00027	 & 7244.69650  &  0.00026  \\
6887.08903  &  0.00021	 & 7072.39289  &  0.00026	 & 7255.35518  &  0.00042  \\
6888.17461  &  0.00019	 & 7075.33176  &  0.00033	 & 7258.17713  &  0.00040  \\
6889.30231  &  0.00024	 & 7084.16842  &  0.00019	 & 7265.17418  &  0.00024  \\
6909.84840  &  0.00026	 & 7086.70711  &  0.00026	 & 7270.66199  &  0.00021  \\
6911.22682  &  0.00024	 & 7089.33875  &  0.00033	 & 7272.93621  &  0.00023  \\
6916.12760  &  0.00042	 & 7107.48057  &  0.00025	 & 7284.90131  &  0.00034  \\
6925.01202  &  0.00035	 & 7124.55894  &  0.00028	 & 7285.44659  &  0.00041  \\
6937.66520  &  0.00020	 & 7125.82190  &  0.00025	 & 7311.71720  &  0.00024  \\
6942.53635  &  0.00024	 & 7142.33139  &  0.00045	 & 7316.00613  &  0.00030  \\
6943.61029  &  0.00020	 & 7147.04243  &  0.00028	 & 7324.80657  &  0.00088  \\
6945.49050  &  0.00028	 & 7148.55867  &  0.00030	 & 7326.14847  &  0.00020  \\
6951.48094  &  0.00020	 & 7150.28452  &  0.00028	 & 7328.28361  &  0.00019  \\
6954.65671  &  0.00040	 & 7156.93525  &  0.00030	 & 7341.15102  &  0.00017  \\
6960.25246  &  0.00019	 & 7158.83995  &  0.00025	 & 7342.57825  &  0.00042  \\
6965.43053  &  0.00132	 & 7159.93729  &  0.00034	 & 7350.81673  &  0.00024  \\
6989.65517  &  0.00019	 & 7162.55991  &  0.00029	 & 7353.29306  &  0.00024  \\
6992.20887  &  0.00661	 & 7168.89453  &  0.00029	 & 7372.11899  &  0.00027  \\
6993.02748  &  0.00780	 & 7173.37091  &  0.00019	 & 7376.87597  &  0.00037  \\
7000.80384  &  0.00027	 & 7176.72263  &  0.00030	 & 7380.42486  &  0.00029  \\
7002.88334  &  0.00036	 & 7191.21965  &  0.00304	 & 7383.98007  &  0.00210  \\
7018.56696  &  0.00027	 & 7200.04543  &  0.00026	 & 7385.49880  &  0.00025  \\
7030.25248  &  0.00014	 & 7202.19014  &  0.00032	 & 7392.98242  &  0.00020  \\
\\
7393.43702  &  0.00034	 & 7618.33739  &  0.00740	 & 7847.53805  &  0.00021  \\
7402.25238  &  0.00035	 & 7625.70435  &  0.00030	 & 7848.45556  &  0.00040  \\
7412.33824  &  0.00021	 & 7628.88264  &  0.00035	 & 7861.91474  &  0.00188  \\
7418.54959  &  0.00019	 & 7630.30957  &  0.00037	 & 7864.02084  &  0.00039  \\
7425.29315  &  0.00016	 & 7635.09038  &  0.00894	 & 7865.96753  &  0.00025  \\
7428.93895  &  0.00021	 & 7647.37820  &  0.00028	 & 7868.19662  &  0.00023  \\
7430.25331  &  0.00024	 & 7652.31691  &  0.00034	 & 7886.28177  &  0.00029  \\
7435.36906  &  0.00022	 & 7653.82650  &  0.00020	 & 7891.07597  &  0.00029  \\
7436.29816  &  0.00020	 & 7654.69804  &  0.00035	 & 7899.74621  &  0.00141  \\
7444.73883  &  0.00830	 & 7660.88562  &  0.00163	 & 7900.31764  &  0.00028  \\
7447.83970  &  0.00712	 & 7670.05996  &  0.00019	 & 7916.44318  &  0.00040  \\
7469.13112  &  0.00432	 & 7676.21805  &  0.00052	 & 7937.73350  &  0.00046  \\
7471.16233  &  0.00035	 & 7678.12619  &  0.00027	 & 7941.72630  &  0.00034  \\
7481.35258  &  0.00038	 & 7685.30561  &  0.00045	 & 7948.16952  &  0.00490  \\
7483.61823  &  0.00653	 & 7693.70909  &  0.00087	 & 7972.59469  &  0.00044  \\
7484.32757  &  0.00036	 & 7699.78372  &  0.00153	 & 7978.97178  &  0.00031  \\
7487.97285  &  0.00043	 & 7704.81806  &  0.00032	 & 7987.97377  &  0.00030  \\
7500.65736  &  0.00022	 & 7710.26810  &  0.00032	 & 7993.67897  &  0.00049  \\
7503.86889  &  0.00093	 & 7712.25355  &  0.01838	 & 8006.15066  &  0.00132  \\
7508.48582  &  0.00070	 & 7713.93678  &  0.00067	 & 8014.78272  &  0.00253  \\
7510.41083  &  0.00029	 & 7723.76161  &  0.00205	 & 8022.20821  &  0.00033  \\
7511.34918  &  0.00062	 & 7724.20358  &  0.00225	 & 8024.25334  &  0.00043  \\
7511.79109  &  0.00045	 & 7728.95089  &  0.00025	 & 8030.19900  &  0.00041  \\
7514.64648  &  0.00744	 & 7742.56189  &  0.00072	 & 8032.43517  &  0.00589  \\
7523.13350  &  0.00032	 & 7782.30221  &  0.01090	 & 8037.22179  &  0.00032  \\
7525.50449  &  0.00028	 & 7788.93215  &  0.00030	 & 8046.11798  &  0.00033  \\
7531.14070  &  0.00150	 & 7798.33466  &  0.01343	 & 8053.30881  &  0.00023  \\
7549.31259  &  0.00024	 & 7814.31631  &  0.00814	 & 8062.62858  &  0.00046  \\
7567.74122  &  0.00021	 & 7817.76829  &  0.00040	 & 8075.64915  &  0.00034  \\
7569.50997  &  0.00046	 & 7834.45599  &  0.00197	 & 8085.21832  &  0.00041  \\
7589.31400  &  0.00022	 & 7840.30431  &  0.00042	 & 8093.62265  &  0.00035  \\
7598.19133  &  0.01118	 & 7841.79020  &  0.00021	 & 8094.05854  &  0.00038  \\
7607.81417  &  0.00712	 & 7842.26577  &  0.00055	 & 8103.68081  &  0.00295  \\
\\
8115.29404  &  0.00421	 & 8510.62203  &  0.00061	 & 8955.84626  &  0.00039  \\
8119.18040  &  0.00024	 & 8516.55260  &  0.00044	 & 8967.64092  &  0.00042  \\
8129.40427  &  0.00028	 & 8521.43880  &  0.00232	 & 9016.58923  &  0.00034  \\
8138.45512  &  0.01231	 & 8539.79200  &  0.00037	 & 9017.59481  &  0.00049  \\
8143.12447  &  0.01095	 & 8554.94195  &  0.00038	 & 9040.06594  &  0.01217  \\
8159.72579  &  0.00036	 & 8573.11869  &  0.00039	 & 9048.24982  &  0.00037  \\
8169.78545  &  0.00038	 & 8605.77640  &  0.00028	 & 9062.56189  &  0.00044  \\
8186.90997  &  0.00029	 & 8620.46163  &  0.00022	 & 9063.95860  &  0.00027  \\
8231.40608  &  0.00048	 & 8639.43962  &  0.00035	 & 9075.39825  &  0.00024  \\
8252.39320  &  0.00030	 & 8645.30800  &  0.00036	 & 9090.81937  &  0.00062  \\
8254.74043  &  0.00059	 & 8662.14057  &  0.00059	 & 9094.83187  &  0.00027  \\
8259.51060  &  0.00070	 & 8665.48488  &  0.00031	 & 9122.96613  &  0.00981  \\
8264.51084  &  0.00544	 & 8667.94386  &  0.00121	 & 9140.58609  &  0.00220  \\
8275.60738  &  0.01353	 & 8678.40931  &  0.00040	 & 9165.89420  &  0.00035  \\
8320.85458  &  0.00026	 & 8709.23307  &  0.00049	 & 9194.63939  &  0.00041  \\
8330.44762  &  0.00031	 & 8723.71188  &  0.01298	 & 9203.96161  &  0.00052  \\
8358.72415  &  0.00051	 & 8732.42418  &  0.00052	 & 9224.48188  &  0.00509  \\
8384.72476  &  0.00035	 & 8748.03016  &  0.00031	 & 9227.48883  &  0.01661  \\
8401.98907  &  0.00047	 & 8758.24142  &  0.00027	 & 9266.20720  &  0.00044  \\
8403.80056  &  0.00078	 & 8761.68807  &  0.00025	 & 9276.27463  &  0.00037  \\
8408.19936  &  0.00604	 & 8766.74271  &  0.00029	 & 9289.56488  &  0.00045  \\
8411.89995  &  0.00082	 & 8771.86105  &  0.00029	 & 9291.53446  &  0.00044  \\
8416.72734  &  0.00025	 & 8772.79392  &  0.01001	 & 9307.89861  &  0.00045  \\
8417.99749  &  0.00033	 & 8775.57120  &  0.00029	 & 9340.70467  &  0.00045  \\
8421.22566  &  0.00023	 & 8784.54554  &  0.00949	 & 9354.21727  &  0.00063  \\
8424.62916  &  0.00795	 & 8799.08923  &  0.00040	 & 9383.27381  &  0.00037  \\
8445.48570  &  0.00032	 & 8841.18314  &  0.00044	 & 9388.93399  &  0.00036  \\
8446.50974  &  0.00032	 & 8842.07056  &  0.00038	 & 9399.09087  &  0.00037  \\
8456.48348  &  0.00337	 & 8849.91144  &  0.00026	 & 9431.60272  &  0.00086  \\
8464.23685  &  0.00029	 & 8868.83358  &  0.00034	 & 9461.03063  &  0.00063  \\
8471.82321  &  0.00035	 & 8889.06206  &  0.00466	 & 9467.19863  &  0.00064  \\
8478.35795  &  0.00038	 & 8892.98159  &  0.00051	 & 9470.68455  &  0.00044  \\
8490.30694  &  0.00045	 & 8905.65688  &  0.00039	 & 9474.88188  &  0.00040  \\
\\
9495.50044  &  0.00038		 & & & & \\
9497.19152  &  0.00043		 & & & & \\		
9505.37191  &  0.01645		 & & & & \\
9561.24113  &  0.00889		 & & & & \\		
9632.64737  &  0.00058		 & & & & \\	
9657.79030  &  0.00379		 & & & & \\		
9664.70129  &  0.00074		 & & & & \\
\end{supertabular}

\end{document}